\documentclass[longauth]{aa}  

\usepackage{graphicx}
\usepackage{amsmath,amssymb} 
\usepackage{natbib}
\usepackage{soul}
\usepackage{xcolor}
\usepackage{txfonts}
\usepackage{hyperref}
\usepackage[caption=false]{subfig}
\usepackage{pdflscape}
\usepackage{dblfloatfix}
\usepackage{amsmath}
\def\oiii{[\ion{O}{iii}]~}
\def\sii{[\ion{S}{ii}]~}
\def\oii{[\ion{O}{ii}]~}
\def\nii{[\ion{N}{ii}]~}

\def\hb{\mathrm{H\beta}~}
\def\ha{\mathrm{H\alpha}~}
\def\arcsec{^{\prime\prime}}
\begin{document} 

\title{SUPER-II: Spatially resolved ionized gas kinematics and scaling relations in $z \sim$ 2 AGN host galaxies}

   \author{D. Kakkad\inst{1,2},
          V. Mainieri\inst{2}, G. Vietri\inst{2,3,4}, S. Carniani\inst{5}, C. M. Harrison\inst{6}, M. Perna\inst{7,8}, J. Scholtz\inst{2,9,10}, C. Circosta\inst{2,11}, G. Cresci\inst{8}, B. Husemann\inst{12},M. Bischetti\inst{13,14}, C. Feruglio\inst{15}, F. Fiore\inst{15}, A. Marconi\inst{8,16}, P. Padovani\inst{2}, M. Brusa\inst{17,18}, C. Cicone\inst{19}, A. Comastri\inst{18}, G. Lanzuisi\inst{18}, F. Mannucci\inst{8}, N. Menci\inst{13}, H. Netzer\inst{20}, E. Piconcelli\inst{13}, A. Puglisi\inst{9}, M. Salvato\inst{21}, M. Schramm\inst{22}, J. Silverman\inst{23,24}, C. Vignali\inst{17,18}, G. Zamorani\inst{18}, L. Zappacosta\inst{13}
          }
   \institute{European Southern Observatory, Alonso de Cordova 3107, Vitacura, Casilla 19001, Santiago de Chile, Chile\\        		\email{dkakkad@eso.org}
   \and
   European Southern Observatory, Karl-Schwarzschild-Strasse 2, Garching bei M\"{u}nchen, Germany
   \and 
   Cluster of Excellence, Boltzmann-Str. 2, 85748 Garching bei M\"{u}nchen, Germany
   \and
   INAF IASF-Milano, Via Alfonso Corti 12, 20133 Milano
   \and
   Scuola Normale Superiore, Piazza dei Cavalieri 7, I-56126 Pisa, Italy
   \and 
   School of Mathematics, Statistics and Physics, Newcastle University, Newcastle upon Tyne, NE1 7RU, UK
   \and
   Centro de Astrobiolog\'ia (CAB, CSIC--INTA), Departamento de Astrof\'\i sica, Cra. de Ajalvir Km.~4, 28850 -- Torrej\'on de Ardoz, Madrid, Spain
   \and
   INAF - Osservatorio Astrofisico di Arcetri, Largo E. Fermi 5, I-50125, Firenze, Italy
   \and
   Centre for Extragalactic Astronomy, Department of Physics, Durham University, South Road, Durham DH1 3LE, UK
   \and
   Chalmers University of Technology, Department of Earth and Space Sciences, Onsala Space Observatory, 43992, Onsala, Sweden
   \and
   Department of Physics \& Astronomy, University College London, Gower Street, London WC1E 6BT, United Kingdom
   \and
   Max-Planck-Institut f\"{u}r Astronomie, K\"{o}nigstuhl 17, D-69117 Heidelberg, Germany
   \and 
   INAF – Osservatorio Astronomico di Roma, Via Frascati 33, 00078 Monte Porzio Catone (Roma), Italy
   \and 
   Universit\'{a} degli Studi di Roma “Tor Vergata”, Via Orazio Raimondo 18, 00173 Roma, Italy
   \and
   INAF – Osservatorio Astronomico di Trieste, via G.B. Tiepolo 11, 34143 Trieste, Italy
   \and
   Dipartimento di Fisica e Astronomia, Universit\'{a} di Firenze, Via G. Sansone 1, I-50019, Sesto Fiorentino (Firenze), Italy
   \and 
   Dipartimento di Fisica e Astronomia dell’Universit\'{a} degli Studi di Bologna, via P. Gobetti 93/2, 40129 Bologna, Italy	
   \and 
   INAF/OAS, Osservatorio di Astrofisica e Scienza dello Spazio di Bologna, via P. Gobetti 93/3, 40129 Bologna, Italy
   \and
   Institute of Theoretical Astrophysics, University of Oslo, P.O. Box 1029, Blindern, 0315 Oslo, Norway
   \and
   School of Physics and Astronomy, Tel-Aviv University, Tel-Aviv 69978, Israel 
   \and
   Max-Planck-Institut f\"{u}r extraterrestrische Physik (MPE), Giessenbachstrasse 1, D-85748 Garching bei M\"{u}nchen, Germany
   \and
   National Astronomical Observatory of Japan, Mitaka, 181-8588 Tokyo, Japan
   \and 
   Kavli Institute for the Physics and Mathematics of the Universe, The University of Tokyo, Kashiwa, Japan 277-8583 (Kavli IPMU, WPI)
   \and 
   Department of Astronomy, School of Science, The University of Tokyo, 7-3-1 Hongo, Bunkyo, Tokyo 113-0033, Japan
             }

   \date{Received ?; accepted ?}

 
  \abstract
    {}
  {The SINFONI survey for Unveiling the Physics and Effect of Radiative feedback (SUPER) aims at tracing and characterizing ionized gas outflows and their impact on star formation in a statistical sample of X-ray selected Active Galactic Nuclei (AGN) at z$\sim$2. We present the first SINFONI results for a sample of 21 Type-1 AGN spanning a wide range in bolometric luminosity (log $\mathrm{L_{bol}}$ = 45.4--47.9 erg/s). The main aims of this paper are determining the extension of the ionized gas, characterizing the occurrence of AGN-driven outflows, and linking the properties of such outflows with those of the AGN.}
  {We use Adaptive Optics-assisted SINFONI observations to trace ionized gas in the extended narrow line region using the \oiii$\lambda$5007 line. We classify a target as hosting an outflow if its non-parametric velocity of the \oiii line, $\mathrm{w_{80}}$, is larger than 600 km/s. We study the presence of extended emission using dedicated point-spread function (PSF) observations, after modelling the PSF from the Balmer lines originating from the Broad Line Region.}
  {We detect outflows in all the Type-1 AGN sample based on the $\mathrm{w_{80}}$ value from the integrated spectrum, which is in the range $\sim$650--2700 km/s. There is a clear positive correlation between $\mathrm{w_{80}}$ and the AGN bolometric luminosity (99\% correlation probability), but a weaker correlation with the black hole mass (80\% correlation probability). A comparison of the PSF and the \oiii radial profile shows that the \oiii emission is spatially resolved for $\sim$35\% of the Type-1 sample and the outflows show an extension up to $\sim$6 kpc. The relation between maximum velocity and the bolometric luminosity is consistent with model predictions for shocks from an AGN driven outflow. The escape fraction of the outflowing gas increase with the AGN luminosity, although for most galaxies, this fraction is less than 10\%.}
   {}
   \keywords{galaxies: active -- galaxies: evolution -- galaxies: high-redshift -- quasars: emission lines -- techniques: imaging spectroscopy
               }
    \titlerunning{SUPER-III: Spatially resolved ionized gas kinematics in $z \sim$ 2 AGN}
   \authorrunning{D. Kakkad et al.}	
   \maketitle
%
\section{Introduction}\label{sect1}    

Quasars represent some of the most energetic sources in the Universe which may regulate 
the gas flows in and out their host galaxies. A manifestation of the impact that 
super-massive black holes (SMBHs) may have on the galaxy are the well established black 
hole-host galaxy scaling relations such as the BH mass, $\mathrm{M_{BH}}$, vs. 
galaxy mass, $\mathrm{M_{\ast}}$, \citep[e.g. ][]{magorrian98,laesker16,schutte19} and the 
$\mathrm{M_{BH}}$ vs. stellar velocity dispersion, $\sigma_{\ast}$, relations 
\citep[e.g.][]{gebhardt00, batiste17,caglar20}. 

One promising physical phenomenon to link the growth of the SMBH and the evolution of its host is that of fast winds launched from the  AGN accretion disk \citep[>1000 km/s, e.g.][]{king03, begelman03, menci08, zubovas12,faucher12,choi14,nims15,hopkins16}. These winds are hypothesized to shock against the surrounding gas and drive outflows that propagate out to large distances from the AGN, heat the interstellar medium (ISM) and potentially eject large amount of gas out of the system \citep[e.g.][]{ishibashi16,zubovas18}. Such fast winds are now observed in a vast number of AGN host galaxies at both low as well as high redshift, in different phases of gas -  neutral phase using sodium absorption lines \citep[e.g.][]{krug10, rupke11, cazzoli16,concas19,roberts-borsani20}, cold molecular gas phase using different transitions of CO, HCN and [\ion{C}{ii}] for instance \citep[e.g.][]{garcia-burillo14,tadhunter14,feruglio17, aladro18, michiyama18,zschaechner18, aalto19, husemann19, cicone20, veilleux20}, warm and hot molecular gas phase using transitions in the mid- to near-infrared \citep{veilleux09, davies14, hill14, riffel15,emonts17,may18,petric18, riffel20}, and ionized gas phase observed using the rest-frame optical emission lines such as the forbidden transitions of \oiii$\lambda$5007 \citep[e.g.][]{harrison14, kakkad16, zakamska16, fiore17,venturi18, baron19, coatman19, forster-schreiber19}. Due to the higher surface brightness of the ionized gas traced by the forbidden transition \oiii$\lambda$5007 relative to other optical transitions (e.g. \oii$\lambda$3727, \sii$\lambda$6716), outflows in this phase can be studied in detail for a large number of galaxies. In AGN, these forbidden ionized transitions such as \oiii, \nii, and \sii are emitted from the  Extended Narrow Line Region (ENLR), making these transitions ideal to trace kiloparsec scale ionized gas outflows from the AGN \citep[e.g.][]{bennert02,hainline14, dempsey18}. 

It is particularly important to study the impact that such galactic scale AGN-driven outflows may have on their host galaxies at z$\sim$2 where both the volume-averaged star-formation rate and the black hole growth rate peak \citep[e.g.][]{shankar09, madau14, curran19, wilkins19,tacconi20}. Tremendous progress has been made through integral field unit (IFU) spectroscopy that provides spatially resolved information on the structure and extension of the outflows \citep[e.g.][]{riffel13,mcelroy15,thomas17,revalski18,davies19,radovich19}. Compared to the classical narrow band imaging and/or slit spectroscopy, IFU spectroscopy allows to identify the emission from the host galaxy by subtracting the contribution from the AGN. A few IFU studies have also claimed the presence of ``negative" as well as ``positive'' feedback in the presence of outflows i.e. ionized outflows suppressing as well as enhancing star formation within the AGN host galaxies \citep[e.g.][]{cano-diaz12, cresci15, carniani16, maiolino17,gallagher19}; however, the full intepretation of these results can be complicated by effects such as dust obscuration \citep[e.g.][]{whitaker14,brusa18,scholtz20}. Most of the literature is however focused on a limited number of targets which have been pre-selected to have a higher probability to show the presence of outflows such as targets selected based on their colors, high Eddington ratio or high luminosity \citep[e.g.][]{brusa15a,perna15,kakkad16,bischetti17,perrotta19}. The lack of studies in a wider parameter space of the properties of the AGN and their host galaxies has so far prevented the assessment in a systematic way of a possible trend between the activity of the AGN itself (e.g. quantified by its luminosity) and the presence of outflows and consequently whether all outflows have an impact on their host galaxies. We are therefore in need of an unbiased sample where a wider range in Eddington ratio and/or bolometric luminosity is used for follow-up outflow studies. 

We are currently in an era of large IFU surveys of galaxies \citep{sanchez12,bundy14, bryant15, stott16,forster-schreiber18,wisnioski19,den-brok20}. Such statistical samples are now able to eliminate the selection biases from previous studies and give an overall picture of the ISM dynamics in the presence of both star formation and AGN processes. Recent work targeting star forming galaxies at high redshift, such as SINS/zC-SINF survey \citep[e.g.][]{forster-schreiber18, davies19} with SINFONI indicate that the majority of  galaxies ($\sim$70\%) show ordered disk rotation while the rest of the targets either show turbulent disk structure or the presence of outflows in addition to the disk rotation. The mass loading factor, which is the ratio between the outflow mass and the star formation rate, is correlated with the level of star formation within the host galaxies. The outflow fraction in SINS/zC-SINF survey is similar to KMOS-3D survey \citep{forster-schreiber19} where 30\% of the 599 observed targets show the presence of outflowing gas, inferred as being driven by both star formation and AGN processes. Star formation-driven winds on average show low mass loading factors when compared to the AGN-driven outflows. Coupled with the reported results in the KROSS survey \citep[e.g.][]{swinbank19}, a fraction of the total gas mass is believed to escape in the low mass galaxies while all of the outflowing gas is retained in galaxies at high masses. IFU surveys of star forming galaxies come to a common conclusion that the outflow velocities are typically enhanced in the presence of an AGN.

Among AGN surveys, \citet{leung19} target optically-selected AGN using single-slit spectroscopy from the MOSDEF survey with bolometric luminosities in the range $\mathrm{10^{44} - 10^{47}}$ erg/s at z$\sim$1.4--3.8 and find outflows in 17\% of their sample of $\sim$160 AGN. Moreover, it is claimed that the ionized gas mass outflow rates correlate positively with the luminosity of the AGN, but do not depend on the galaxy stellar mass, in contrast with the findings of \citet{forster-schreiber18} on star forming galaxies. AGN from MOSDEF survey have higher incidence of outflows compared to redshift and stellar mass matched star forming galaxies. Although MOSDEF uses single-slit spectroscopy which, as mentioned earlier, has its own limitations, a similar result was found by \citet{harrison16} with the KASHz survey targeting X-ray selected AGN (2--10 keV) at z$\sim$1.1--2.5 with KMOS. The more luminous KASHz AGN are found to be more likely to host outflows with velocities >600 km/s and these ionized gas velocities are 10 times more prevalent in the AGN host galaxies than star forming galaxies at similar redshift and similar $\ha$ luminosity distribution. 

Most of the the current AGN surveys at z>1 described above are based on seeing-limited observations. Consequently, both host galaxies as well as the possible outflows are usually spatially unresolved, which prevents one to have a complete understanding of how extended these outflows are. In addition, the presence of unresolved outflows implies that further assumptions have to be made (e.g. on the scale and morphology) which propagate in larger uncertainties in derived quantities such as mass loading factor and outflow kinetic power.  Therefore, the way forward is to use Adaptive Optics (AO) assisted observations to resolve smaller physical scales also at z>1 \citep[e.g.][]{daviesR20}. A higher spatial resolution enables one to study the impact of AGN from kiloparsec scales to Mpc scales. In this paper, we present the first results from the Type-1 AGN sample from the SINFONI survey for Unveiling the Physics and Effect of Radiative feedback\footnote{Observations taken as a part of the following ESO program 196.A-0377} \citep[SUPER\footnote{http://www.super-survey.org}][]{circosta18}. SUPER is designed to overcome the limitation of low spatial resolution in order to understand the true spatial extent of outflows in AGN host galaxies at high redshift. Apart from the extension of the outflows and the ionized gas, SUPER aims at answering some of the fundamental questions such as: how prevalent are ionized outflows in X-ray selected AGN host galaxies? Do the outflow properties e.g. velocity, mass outflow rates show a correlation with those of the AGN and of the host galaxy (e.g. AGN bolometric luminosity, host star formation rate)? And if they do, what is the scaling relation between the corresponding parameters? Lastly, do these outflows have any effect on the host (e.g. its star formation rate or gas mass)?
 
This paper is arranged as follows: in sect. \ref{sect2}, we describe the sample used and its characteristics and in sect. \ref{sect3} we discuss the observing strategy followed by the data reduction procedure and stacking of data cubes in sect. \ref{sect4}. We describe in detail the analysis procedure and the  corresponding results obtained in sect. \ref{sect5}, discuss the implications of these results and derive scaling relations in sect. \ref{sect6}. Concluding remarks are reported in sect. \ref{sect7}.

Throughout this paper, we adopt the following $\Lambda$CDM cosmology parameters: $\mathrm{H_{o}}$ = 70 km/s, $\mathrm{\Omega_{M}}$ = 0.3, $\mathrm{\Omega_{\Lambda}}$ = 0.7.

\section{Sample description}   \label{sect2}

\begin{figure}
\centering
\includegraphics[scale=0.44]{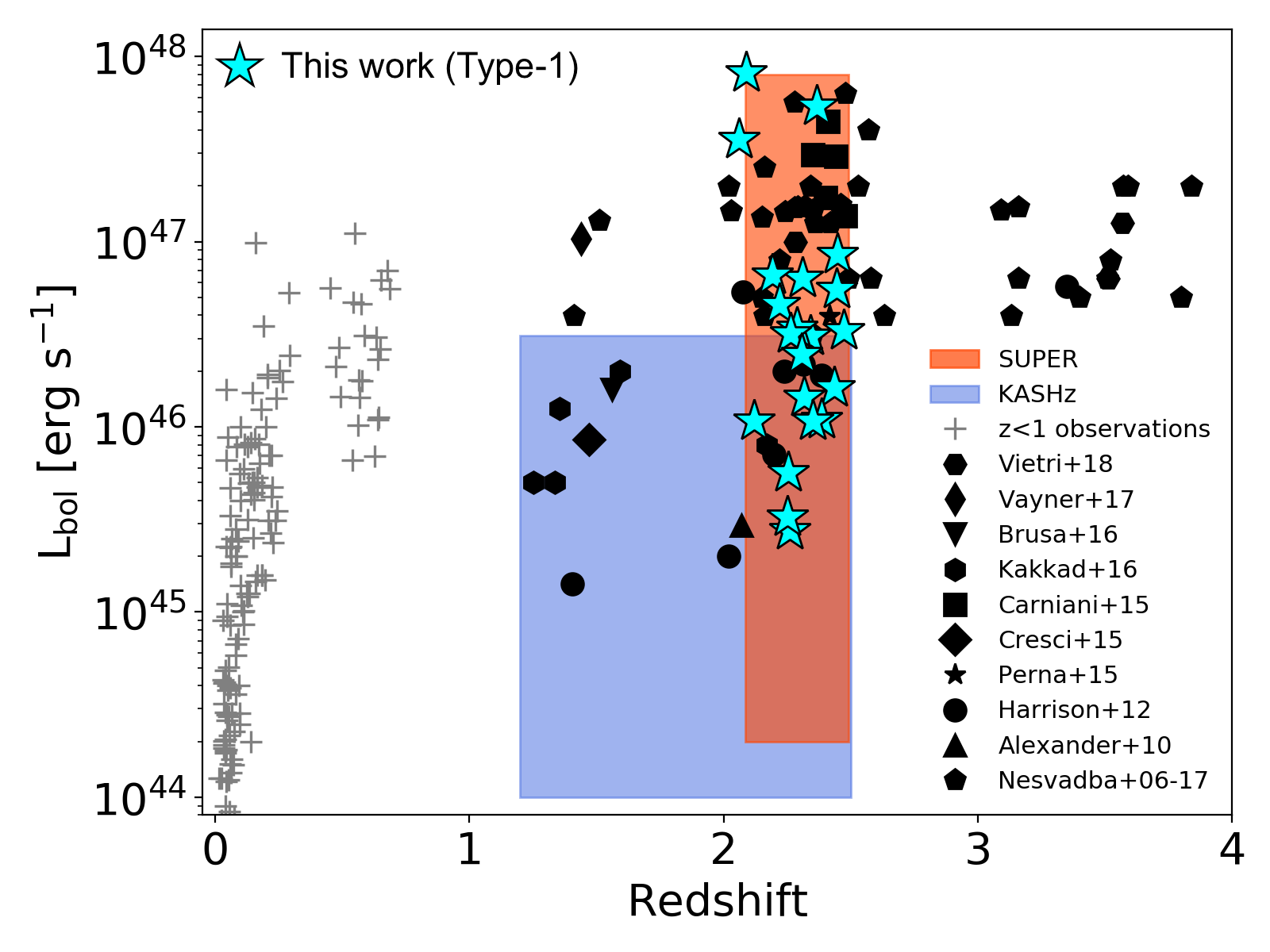}
\caption{Past and on-going IFU surveys targeting the \oiii$\lambda$5007 emission line to study the properties of outflows in AGN host galaxies in the $\mathrm{L_{bol}}$ vs. redshift plane. The grey data points represent surveys at $z<1$ \citep{husemann13,husemann14,husemann17,liu13,harrison14,karouzos16,bae17,rupke17}, while the solid black symbols are collected from $z>1$ studies \citep{nesvadba06, alexander10, harrison12, cresci15, carniani15, brusa16, kakkad16, nesvadba17,vayner17,vietri18}. The blue shaded area covers the parameter space of the KASHz survey \citep{harrison16} and the red shaded area that of the overall SUPER survey \citep{circosta18}. This paper presents an analysis on the Type 1 sub-sample of the SUPER survey, the data points of which are shown as {\it cyan} star symbols. Adapted from \citet{circosta18}, to whom we refer the reader for further details.
\label{fig:sample_selection}}
\end{figure}

The sample presented in this paper is derived from the SUPER survey, a large program with the Spectrograph for INtegral Field Observations in the Near Infrared \citep[SINFONI,][]{eisenhauer03} mounted on the Cassegrain focus of Unit Telescope 4 (UT4) at the Very Large Telescope (VLT). We briefly discuss the overall properties of the SUPER sample in this section, in particular the Type-1 sub-sample used in this paper. For more details on sample selection and the survey characteristics we refer the reader to \citet{circosta18}.

The SUPER survey consists of a sample of 39 AGN selected in the X-rays ($\mathrm{L_{2-10~kev} > 10^{42}~erg/s}$) in the Chandra Deep Field South \citep[e.g.][]{luo17}, {\it COSMOS-Legacy} \citep[e.g.][]{civano16}, the wide area XMM-XXL  \citep[e.g.][]{georgakakis11, liu16, menzel16}, and Stripe 82 X-ray  \citep[e.g.][]{lamassa16} surveys and from the WISE/SDSS selected Hyper-luminous quasar sample \citep[e.g.][]{bischetti17}. The X-ray selection with the luminosity cut ensures a pure AGN selection as there is no contamination from the host galaxy and/or X-ray binaries at these energies \citep[see e.g.][]{brandt15, padovani17}. The sample covers a redshift range of 2.1 -- 2.5, which is the epoch of maximal activity of the volume averaged star formation in galaxies and the growth of black holes in the universe, making it ideal to study effects of radiative feedback from the black hole on the host galaxy \citep[e.g.][]{madau14}. Owing to the presence of ancillary multi-wavelength data sets, we are able to derive accurate measurements of the black hole and the host galaxy properties via spectral energy distribution fitting of UV-to-FIR photometry and X-ray spectral fitting (details in \citealt{circosta18} for the analysis of the multi-wavelength data sets and \citealt{vietri20} for the black hole mass estimations). Of the 39 targets, 22 were classified as Type-1 (56\%) and the remaining 17 as Type-2 (44\%), based on the presence or absence of broad emission lines such as MgII or CIV in the optical spectra. One target, cid\_1205, which was previously reported as a Type-2 AGN in \citet{circosta18} is now classified as a Type-1 AGN based on the presence of BLR emission in $\ha$ line in the K-band SINFONI spectrum. So, 23 galaxies (58\%) from the SUPER sample are now classified as Type-1 AGN. The overall SUPER sample span a wide range in AGN and host galaxy properties which allows us to identify any existing correlation between the outflow properties derived from the SINFONI data and those derived from the multi-wavelength ancillary data set.

In this paper, we focus on the analysis of the H-band SINFONI data of the Type-1 targets from the SUPER survey. The selected sample in the context of other SUPER targets and on-going IFU surveys is shown as $cyan$ star symbols in Fig. \ref{fig:sample_selection}.  These targets populate the moderate-to-high luminosity range and span $\sim$2.5 order of magnitude in bolometric luminosity among the SUPER sample. We explore the following range of black hole and host galaxy properties for the spatially resolved data set presented in this paper: log M$_{\ast}$/M$_{\odot}$~$\sim$10.38--11.20, SFR <94--686 M$_{\odot}$/yr, log L$_{\rm bol}$/[erg/s] ~$\sim$ 45.4--47.9, log M$_{\rm BH}$/M$_{\odot}$ ~ $\sim$8.3--10.7 and log N$_{\rm H}$/cm$^{-2}$ ~ <21.25-- >24.1. The properties of the individual targets obtained from the SED fits and the available optical data are reported in Table \ref{table:SUPER_properties}.

\section{Observations} \label{sect3}

The SUPER SINFONI observations have been carried out between November 2015 and December 2018 as  part of the ESO large program 196.A-0377 in both service mode as well as visitor mode. We use SINFONI for AO assisted observations in the H-band (1.45-1.85 $\mu$m) to trace the rest-frame optical lines H$\beta$ and \oiii$\lambda$5007, and the K-band (1.95-2.45 $\mu$m) to trace \nii, H$\alpha$ and \sii$\lambda$6716, 6731, all of which can be ionized by both AGN and star formation processes. Due to the lack of an appropriate Natural Guide Star (NGS), most of the observations were carried out with the Laser Guide Star (LGS) in clear sky conditions. Hence, all but three targets were observed using the seeing enhancer (SE) mode, which offers an improvement in image quality compared to the natural seeing, since no suitable bright tip-tilt star (TTS) was available close to the chosen targets, given their location in deep fields. We used a plate scale of 3$\arcsec\times$ 3$\arcsec$ with a spatial sampling of 0.05$\arcsec\times$0.1$\arcsec$, which gets re-sampled to 0.05$\arcsec\times$0.05$\arcsec$ in the final data cube. Three targets were observed in seeing-limited mode during the visitor-mode runs when the conditions were not ideal to close AO loop. The plate scale for observations without AO is 8$\arcsec\times$ 8$\arcsec$ with a spatial sampling of 0.25$\arcsec\times$0.25$\arcsec$ in the final reduced data cube. The average spectral resolution in the H-band and K-band is $\sim$ 3000 and $\sim$ 4000 respectively which translates to a channel width of $\sim$ 2 \text{\AA} and $\sim$ 2.5 \text{\AA} respectively. 

\begin{landscape}
\begin{table}
\centering
\caption{Properties of the 20 Type-1 SUPER targets presented in this paper.}
\label{table:SUPER_properties}
\begin{tabular}{ ccccccccccc }
    \hline
Target & RA & DEC & $\mathrm{z_{opt}}$& log M$_{\ast}$ & SFR & log L$_{bol}$ & log $\mathrm{L_{[2-10~keV]}}$ & log N$_{H}$ & log M$_{BH}$\\
& (1) & (2) & (3) & (4) & (5) & (6) & (7) & (8) & (9) \\
 & (h:m:s) & (d:m:s) & & M$_{\odot}$ & M$_{\odot}$/yr & erg/s & erg/s & cm$^{-2}$ & M$_{\odot}$ \\ \hline\hline
X\_N\_160\_22   & 02:04:53.81 & -06:04:07.82 &   2.445 & - & -  & 46.74$\pm$0.02 & 44.77$^{+0.14}_{-0.19}$ & <22.32 & 9.05$\pm$0.30\\
X\_N\_81\_44 & 02:17:30.95 & -04:18:23.66 & 2.311 & 11.04$\pm$0.37 & 229$\pm$103 & 46.80$\pm$0.03 & 44.77$^{+0.07}_{-0.09}$ & <21.86 & 9.02$\pm$0.30\\
X\_N\_53\_3 & 02:20:29.84 & -02:56:23.41 & 2.434 & - & 686$\pm$178 & 46.21$\pm$0.03 & 44.80$^{+0.10}_{-0.13}$ & 22.77$^{+0.37}_{-0.67}$ & 8.51$\pm$0.30\\
X\_N\_66\_23 & 02:22:33.64 & -05:49:02.73 & 2.386 & 10.96$\pm$0.29 & <268 & 46.04$\pm$0.02 & 44.71$^{+0.06}_{-0.08}$ & <21.51 & 8.92$\pm$0.30\\
X\_N\_35\_20 & 02:24:02.71 & -05:11:30.82 & 2.261 & - & - & 45.44$\pm$0.02 & 44.00$^{+0.07}_{-0.40}$ & <22.27 & 8.38$\pm$0.37\\
X\_N\_12\_26 & 02:25:50.09 & -03:06:41.16 & 2.471 & - & - & 46.52$\pm$0.02 & 44.56$^{+0.13}_{-0.12}$ & <20.90 & 8.84$\pm$0.30\\
X\_N\_44\_64 & 02:27:01.46 & -04:05:06.73 & 2.252 & 11.09$\pm$0.25 & 229$\pm$80 & 45.51$\pm$0.07 & 44.21$^{+0.11}_{-0.17}$ & <21.97 & 8.74$\pm$0.31\\
X\_N\_4\_48 & 02:27:44.63 & -03:42:05.46 & 2.317 & - & - & 46.16$\pm$0.02 & 44.52$^{+0.09}_{-0.16}$ & <21.85 & 8.88$\pm$0.31\\
X\_N\_102\_35 & 02:29:05.94 & -04:02:42.99 & 2.190 & - & - & 46.82$\pm$0.02 & 45.37$^{+0.05}_{-0.11}$ & <22.17 & 8.82$\pm$0.30\\
X\_N\_115\_23   & 02:30:05.66 & -05:08:14.10 &  2.342 & -  & -  & 46.49$\pm$0.02 & 44.93$^{+0.08}_{-0.10}$ & <22.26 & 9.08$\pm$0.30\\
cid\_166  & 09:58:58.68 & +02:01:39.22 & 2.448 & 10.38$\pm$0.22  & <224 & 46.93$\pm$0.02 & 45.15$^{+0.03}_{-0.02}$ & <21.25 & 9.30$\pm$0.30\\
cid\_1605 & 09:59:19.82 & +02:42:38.73 & 2.121 & - & <94 & 46.03$\pm$0.02 & 44.69$^{+0.06}_{-0.04}$ & 21.77$^{+0.51}_{-0.75}$ & 8.52$\pm$0.31\\
cid\_346  & 09:59:43.41 & +02:07:07.44 & 2.219 & 11.01$\pm$0.22 & 362$\pm$49 & 46.66$\pm$0.02 & 44.47$^{+0.08}_{-0.09}$ & 23.05$^{+0.17}_{-0.19}$ & 9.15$\pm$0.30\\
cid\_1205 & 10:00:02.57 & +02:19:58.68 & 2.255 & 11.20$\pm$0.10 & 384$\pm$33 & 45.75$\pm$0.17 & 44.25$^{+0.21}_{-0.23}$ & 23.50$\pm$0.27 & 8.94$\pm$0.31\\
cid\_467 & 10:00:24.48 & +02:06:19.76 & 2.288 & 10.10$\pm$0.29 & <147 & 46.53$\pm$0.04 & 44.87$^{+0.04}_{-0.05}$ & 22.31$^{+0.23}_{-0.32}$ & 9.26$\pm$0.31\\
J1333+1649  & 13:33:35.79 & +16:49:03.96 & 2.089 & - & - & 47.91$\pm$0.02 & 45.81$^{+0.07}_{-0.06}$ & 21.81$^{+0.22}_{-0.34}$ & 9.96$\pm$0.30\\
J1441+0454  & 14:41:05.54 & +04:54:54.96 & 2.059 & - & - & 47.55$\pm$0.02 & 44.77$^{+0.10}_{-0.11}$ & 22.77$^{+0.18}_{-0.21}$ & 9.24$\pm$0.30\\
J1549+1245   & 15:49:38.73 & +12:45:09.20 & 2.365 & - & - & 47.73$\pm$0.04 & 45.38$^{+0.02}_{-0.02}$ & 22.69$^{+0.09}_{-0.11}$ & 10.66$\pm$0.30\\
S82X1905 & 23:28:56.35 & -00:30:11.74 & 2.263 & - & - & 46.50$\pm$0.02 & 44.91$^{+0.50}_{-0.50}$ & 22.95$^{+0.35}_{-0.17}$ & 8.87$\pm$0.30\\
S82X1940 & 23:29:40.28 & -00:17:51.68 & 2.351 & - & - & 46.03$\pm$0.02 & 44.72$^{+0.30}_{-0.30}$ & <20.50 & 8.33$\pm$0.30\\
S82X2058 & 23:31:58.62 & -00:54:10.44 & 2.308 & - & - & 46.39$\pm$0.02 & 44.67$^{+0.30}_{-0.30}$ & <20.50 & 9.09$\pm$0.30\\
    \hline
\end{tabular}
\vspace{1ex}\\
{\raggedright {\bf Notes:}\newline 
(1) \& (2): Right ascension and declination of the optical counterpart of the target (J2000).\newline 
(3): Spectroscopic redshift obtained from archival optical spectra.\newline 
(4): Galaxy stellar mass obtained from SED fitting, wherever applicable.\newline 
(5): Star formation rate derived from the far-infrared (8--1000 $\mu$m) luminosity.\newline 
(6): AGN bolometric luminosity derived from SED fitting.\newline 
(7): Hard band (2-10 keV) X-ray luminosity corrected for absorption with 90\% confidence level error.\newline
(8): Absorbing hydrogen column density with 90\% confidence limits.\newline 
(9): Latest black hole mass estimates from the SINFONI data. \newline
All the error values (except N$_{H}$) are 1$\sigma$ uncertainties. Further details about the derivation of these properties is given in \citet{circosta18} and Vietri et al. in prep.\par}
\end{table}
\end{landscape}

Most of the targets in the SUPER survey are too faint for direct acquisition during the observation ($23.49 < {\rm H}_{\rm mag} <15.72$, $22.27 < {\rm K}_{\rm mag} <15.34$) and therefore a blind offset from a nearby bright star was used. Before each science observation, a dedicated PSF star was observed to get an estimate of the image quality after the AO correction and compare it to the natural seeing at the observatory. The PSF observation lasted 30--60 s on source along with a sky exposure of a similar duration. This PSF observation was however not strictly used to select or discard the observation blocks to be used for the final cubes because occasionally the conditions varied during the one hour observation. In case the conditions degraded or improved significantly, the exposures were discarded or included accordingly. The AO-assisted observations gave an average image quality of up to $\sim$0.22$\arcsec$ (median = 0.3$\arcsec$) in H-band and $\sim$0.2$\arcsec$ (median = 0.3$\arcsec$) in K-band, inferred from the FWHM of the dedicated PSF star observation. The PSF of the observations for each band is reported in Table \ref{table:SUPER_observations}. 

During science exposures, we used a dithering pattern where the target was moved within the SINFONI field of view (FoV) so the sky for a particular frame was obtained from the subsequent frame and vice-versa. In the case of extended targets, the sky subtraction following such a dithering pattern would lead to the subtraction of the object signal itself and therefore a dedicated sky exposure was taken in a pattern ``O-S-O-O-S-O'' (S = sky; O = Object).  Each object and sky exposure was limited to 10 minutes long to minimize the variation of the infrared sky. The total on-source exposure time for the targets in either bands (H or K) ranges between 1 hour to $\sim$6 hours. Observing patterns and exposure times in either bands for each target are summarized in Table \ref{table:SUPER_observations}.

To correct for atmospheric absorption and to flux calibrate the final co-added science cube telluric stars were observed with the same setup as the science observations within 0.2 airmass and 2 hours of the science observations. Each telluric star exposure lasted 2-3 s with number of integration (NDIT) of 5 along with a sky observation of similar exposure time as that of the star. The stars were selected to have a K-band magnitude between 7 and 8.5 and a stellar type of B2V, B3V, B4V or B5V.

Out of the 23 Type 1 AGN in the SUPER sample, 21 were observed in both H-band and K-band (Table \ref{table:SUPER_observations}), 18 were detected in H-band, and all 21 were detected in the K-band. X\_N\_53\_3 was detected neither in continuum nor in emission lines in the H-band, but detected in K-band. X\_N\_44\_64, although detected in continuum in both bands, lacks emission lines in the H-band. Two targets (lid\_206 and S82X2106) from the list of \citet{circosta18}  could not be observed since part of the survey was executed in visitor mode rather than service mode as originally planned, and therefore we had a higher fraction of time lost for bad weather conditions.

\begin{table*}
\centering
\caption{Observation parameters of the targets presented in this paper for each band.}
\label{table:SUPER_observations} 
\begin{tabular}{c|c|c|c|cc}
    \hline
Target  & Observing mode$^{a}$ & \multicolumn{2}{c}{PSF$^{b}$ ($^{\prime\prime}$, kpc)} & \multicolumn{2}{c}{$\tau_{exp}^{c}$ (h)}  \\
    & &   H  &   K  &   H  &   K \\\hline\hline
X\_N\_160\_22   & AO & 0.31, 2.5 & 0.34, 2.8 & 3.0 & 1.0\\
X\_N\_81\_44     & AO & 0.27, 2.2 & 0.24, 2.0 & 7.0 & 1.0\\
X\_N\_53\_3     & AO   & 0.47, 3.8 & 0.44, 3.6 & 1.0 & 1.0\\
X\_N\_66\_23   & AO       & 0.24, 1.9  & 0.45, 3.7 & 1.0 & 0.7\\
X\_N\_35\_20   & AO       & 0.27, 2.2 & 0.26, 2.1 & 1.0 & 1.0\\
X\_N\_12\_26   &  AO      & 0.30, 2.4  & 0.30, 2.4 & 6.0 & 2.0\\
X\_N\_44\_64 * & AO        & 0.66, 5.4  & >0.20, >1.6  & 1.0 & 1.0\\
X\_N\_4\_48*    & AO        & 0.36, 2.9    & >0.20, >1.6  & 3.0   & 1.0\\
X\_N\_102\_35 & noAO   & 0.92, 7.6  & 0.85, 7.0 & 1.0 & 1.0\\
X\_N\_115\_23   & AO  & 0.30, 2.4 & 0.27, 2.2 & 2.0 & 2.0\\
cid\_166            & AO   & 0.29, 2.4 & 0.28, 2.3 & 3.5 & 1.5\\
cid\_1605        & noAO    & 0.7, 5.8  & 0.65, 5.4    & 2.0    & 2.0\\
cid\_346           & AO   & 0.30, 2.5 & 0.30, 2.5 & 3.7 & 2.0\\
cid\_1205 & AO & 0.3, 2.5 & 0.3, 2.5 & 1.0 & 1.0\\
cid\_467          & noAO    & 1.1, 9.0 & 0.98, 8.0          & 2.0    & 2.0\\ 
J1333+1649     & AO  & 0.50, 4.2 & 0.40, 3.3 & 1.0 & 1.0\\
J1441+0454     & AO  & 0.34, 2.8  & 0.21, 1.8 & 1.0 & 1.0\\
J1549+1245     & AO  & 0.22, 1.8 & 0.30, 2.4 & 1.0 & 1.0\\
S82X1905         & AO & 0.34, 2.5 & 0.35, 2.9 & 5.0 & 2.0\\
S82X1940       & AO        & 0.30, 2.4 & 0.30, 2.4  & 4.3  & 3.0\\
S82X2058       & AO       & 0.27, 2.2  & 0.32, 2.6  & 6.0  & 2.0\\\hline
\end{tabular}\\
\vspace{1ex}
{\raggedright {\bf Notes:} $^{a}$Mode of observation AO = H and K band observations were taken separately with Adaptive Optics corrections and noAO = Observations were taken with HK grating under bad weather conditions during the visitor mode runs with no corrections; $^{b}$ FWHM of the dedicated PSF star before the science observation in arcsec; $^{c}$Exposure time (hours) in each band. *An accurate estimation of the PSF could not be obtained for X\_N\_44\_64 and X\_N\_4\_48 as the PSF star was at the edge of the SINFONI field-of-view.\par}
\end{table*}

\section{Data reduction} \label{sect4}

We used the latest version of the ESO pipeline (3.1.1) to reduce the SINFONI data. The pipeline corrects for the presence of non-linear and hot pixels, flags the pixels which have flat lamp intensities higher than a given threshold and performs a flat field correction, computes optical distortions and slitlet distances and performs wavelength calibration using exposures from xenon+argon arc lamp in the H-band and neon+argon arc lamp in the K-band. Science exposures, PSF and telluric star observations are reduced using the recipe {\it sinfo\_rec\_jitter} which outputs re-sampled data cubes of the individual exposures of the science frames as well as the PSF and telluric cubes corrected for the distortions, bad pixels and calibrated for wavelength using the above mentioned steps. 

The sky subtraction was performed externally using the improved sky subtraction procedure described in \citet{davies07}. During this procedure, 10-30\% of the pixels from an object-free region were used to create a model sky spectrum which was shifted in wavelength space to match the wavelength axis of the object frames. The processed sky spectrum was then subtracted from the object exposure across the FoV. 

To remove the telluric absorption features and to flux calibrate the data cubes, first the hydrogen features were removed from the observed telluric star spectrum, divided by a black body spectrum and normalized it to get the response function of the instrument. Both the science and the telluric cubes were divided by this response curve to correct for the telluric features. The spectrum extracted from the corrected telluric cube was then convoluted with the appropriate filter (H-band or K-band) from the 2MASS  catalogue (Two Micron All-Sky Survey: \citealt{skrutskie06}) to get the required flux per unit count to be applied to the entire data cube. 

Lastly, the flux calibrated individual frames from multiple exposures were combined using the pipeline recipe  $sinfo\_utl\_cube\_combine$ with a sigma clipping parameter (\texttt{--ks\_clip=TRUE}) and scaling the sky (using \texttt{--scale\_sky=TRUE}) within individual exposures. It has been verified that the sigma clipping does not remove signal from the original target itself. By setting the sky scaling parameter, the spatial median of each exposure is subtracted from the contributing exposure to remove sky background which might not have been removed in the previous steps of the reduction. For observations within the same night, the offsets given by the header keywords \texttt{CUMOFFSETX/Y} are verified to be reliable by comparing the stacked cube with manually calculated offsets. Observations taken on a different night might have shifts in the centroid of the image. We performed a two-dimensional Gaussian fit to the combined cubes obtained from individual observing blocks during each night and the difference in the centroid of the Gaussian fits gave the relative offsets between the observations from different nights. After the determination of these offsets, each contributing cube is aligned and co-added such that the intensity of a pixel is given by the weighted mean of the intensity of the corresponding overlapping pixels from the individual cubes, where the weight depends on the exposure time of the individual frames. Any residual cosmic ray signal within the final co-added cube is then removed using a sigma clipping procedure. The total on-source exposure time of each target in the H- and  K-bands is reported in Table \ref{table:SUPER_observations}.

The WCS coordinates of the co-added cube resulting from the SINFONI pipeline are inaccurate. We therefore applied an astrometric correction registering the peak of the continuum emission from the AGN with the optical/near-infrared coordinates reported in \citet{circosta18}. Further, we noted that the SINFONI observations of target J1549+1245 show a systematic spatial shift in the centroid of the continuum location as a function of wavelength. This could be due to an inaccurate correction of the atmospheric dispersion and/or to a rotation in the grating between the time of the science observation and when the wavelength calibration was obtained. To correct for such spatial shifts, we measure the offset along the X and Y directions as a function of wavelength and the derived offset functions were used to re-align the cubes using the drizzle algorithm \citep{fruchter02}.

\begin{figure}
\centering
\subfloat{\includegraphics[scale=0.45]{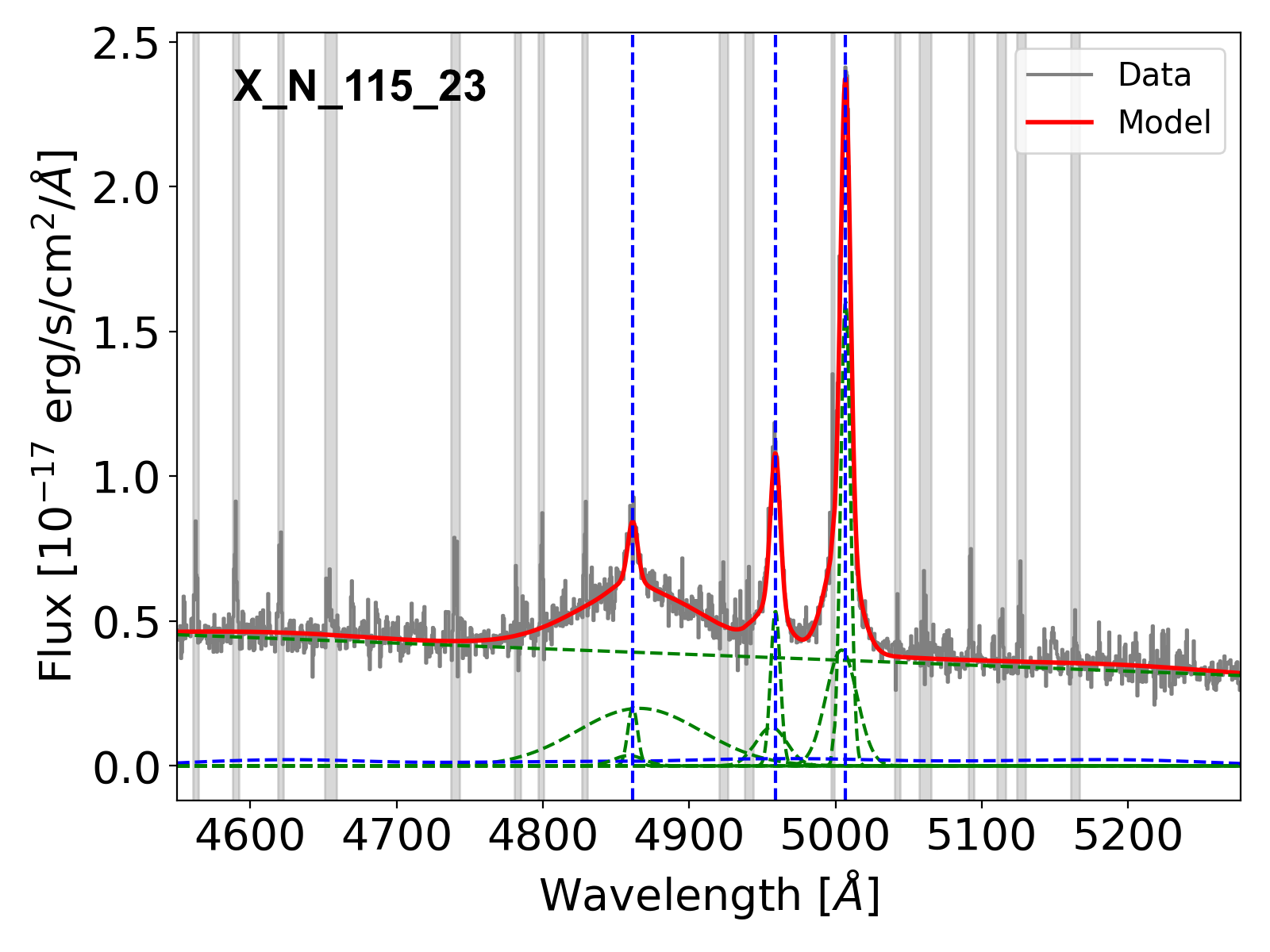}}
\caption{Integrated H-band spectrum of X\_N\_115\_23 shown here as an example. The grey curve shows the observed spectrum, the red curve shows the reproduced overall emission line model,  the blue dashed curve shows the iron emission and the dashed green curves show the continuum emission and the individual Gaussian components (Narrow, Broad and BLR) used to reproduce the profiles of various emission lines. The blue vertical lines indicate the location of $\hb$, \oiii$\lambda$4959 and \oiii$\lambda$5007. The vertical grey regions mark the channels with strong sky lines which were masked during the fitting procedure. The X-axis shows the rest frame wavelength after correcting for the redshift of the target and the Y-axis shows the observed flux density. The integrated spectra of the rest of the targets are shown in Appendix \ref{sect:app}. \label{fig:intspec_example_xn_115_23}}
\end{figure}

\section{Analysis and Results} \label{sect5}
The overall analysis of the reduced SINFONI cubes consists of three steps: (1) derive the properties of the ionized gas from modelling the integrated spectrum; (2) determine if the \oiii emission is resolved after subtracting the AGN PSF and/or comparing the spatial profiles with a suitable PSF model; (3) measure the extension of the ionized gas at various velocity slices to determine the extent of the outflows. For the purpose of this paper, we focus the analysis only on the H-band SINFONI data to trace the kinematics of the ionized gas using the forbidden \oiii transition. The results from the analysis of the K-band spectrum will be presented in a later publication. 

\subsection{Modelling the integrated spectrum} \label{sect5.1}
In this section, we describe the line fitting procedure on the integrated spectrum. While this paper will focus on the NLR properties, a more in-depth discussion on the BLR properties will be presented in a forthcoming publication (Vietri et al. in prep).

The integrated spectrum for each object was extracted from a circular aperture centered on the target which includes at least $\sim$95\% of the total emission. The aperture used to extract the spectrum in each object is reported in Table \ref{table:SINFONI_properties1}. The target center was calculated using a two-dimensional Gaussian fit on the H-band continuum image obtained by collapsing the cube over all the spectral channels. The error on the spectrum was estimated creating an $rms$ spectrum obtained from an object-free region. The analysis of the H-band spectrum was restricted to the region spanned by $\hb$, \oiii$\lambda$4959 and  \oiii$\lambda$5007. The residual sky lines were masked from the spectrum during the fitting procedure.

We modeled the extracted spectra using the \texttt{scipy.curve-fit} package in python, which uses the principle of least squares to find the optimal set of parameters for a given fitting model. We used a simple linear model to fit the AGN continuum. The iron emission was modeled using observed FeII templates from the literature \citep{boroson92, veron-cetty04, tsuzuki06}. The \oiii and H$\beta$ emission lines were modeled with Gaussian functions and the kinematic components of the two lines were coupled with each other. The number of Gaussian components for the \oiii emission line was restricted to two, and the addition of the second Gaussian depended on whether it minimizes the reduced chi-square value of the overall model. We will refer to these individual Gaussian components as ``narrow'' or ``broad'' according to the values of the line widths (FWHM), which are left as free parameters in the fitting procedure. We will not associate a physical meaning to each single Gaussian component, and consequently to the terms narrow and broad, but we will rather use a non-parametric approach in the paper to define velocity as described further down in this section. For $\hb$, a third Gaussian component or a broken power law was required to reproduce the Broad Line Region emission\footnote{Hereafter, unless differently specified, a broad Gaussian refers to the non-BLR component.}, which has been used to infer the black hole masses of the SUPER targets (more details in Vietri et al. (in prep)). The line centroid and width of the narrow and broad components of \oiii$\lambda\lambda$4959,5007 and $\hb$ are tied to each other, based on the assumption of a common origin for these emission lines. Furthermore, the emission line ratio \oiii$\lambda$5007:\oiii$\lambda$4959 is set equal to $\sim$3:1 based on theoretical values \citep[e.g.][]{storey00, dimitrijevic07}. In order to estimate the uncertainty on the derived parameters, we created 100 mock spectra by adding $rms$ noise to the modeled spectrum and repeated the line fitting procedure on these mock spectra. The errors reported in Table \ref{table:SINFONI_properties1} are the standard deviation for each parameter obtained with this procedure. Since we lack sufficient constraints to correct for dust reddening, we do not attempt to correct the emission line luminosities for extinction effects.

As an example of the fit performed on the integrated H-band SINFONI spectra, we show the object X\_N\_115\_23 along with the spectral model in Fig. \ref{fig:intspec_example_xn_115_23}. The integrated spectra for the rest of the Type-1 SUPER targets can be found in Appendix \ref{sect:app}. The line fitting parameters of the H-band spectrum are reported in Table \ref{table:SINFONI_properties1}. Out of the 21 Type-1 AGN presented in Table 2, we were able to derive line properties for 19 objects.  X\_N\_53\_3 is not detected in continuum and it does not show any emission lines in the H-band either. X\_N\_44\_64 is detected in the continuum but does not have any emission lines. Finally, J1441+0454 is well detected in $\hb$, but the spectrum does not show the presence of \oiii emission at the expected observed wavelength. This object is one of the brightest AGN in our sample and the possible lack of \oiii emission at high bolometric luminosities has been previously reported in the literature  \citep[e.g. WISSH quasars:][]{bischetti17}. On the other hand, while the line fitting suggests that the spectrum around the expected location of \oiii is dominated by strong iron emission, it also found a significant highly blue-shifted ($\sim$ -3000 km/s) \oiii line with FWHM$\sim$1956 km/s. Such extreme blue-shift has also been observed before in the literature in high-redshift Extremely Red Quasars \citep[ERQs, see][]{perrotta19}. Due to the degeneracy between the iron component and the \oiii emission and the highly blended nature of the observed H$\beta$-\oiii emission (see Fig. \ref{fig:intspec_alltargets3}) we will limit the analysis of the \oiii properties to the integrated spectum for this object and will not attempt to characterize the extended nature of the \oiii emission in the next sections.

\begin{figure}
\centering
\includegraphics[scale=0.5]{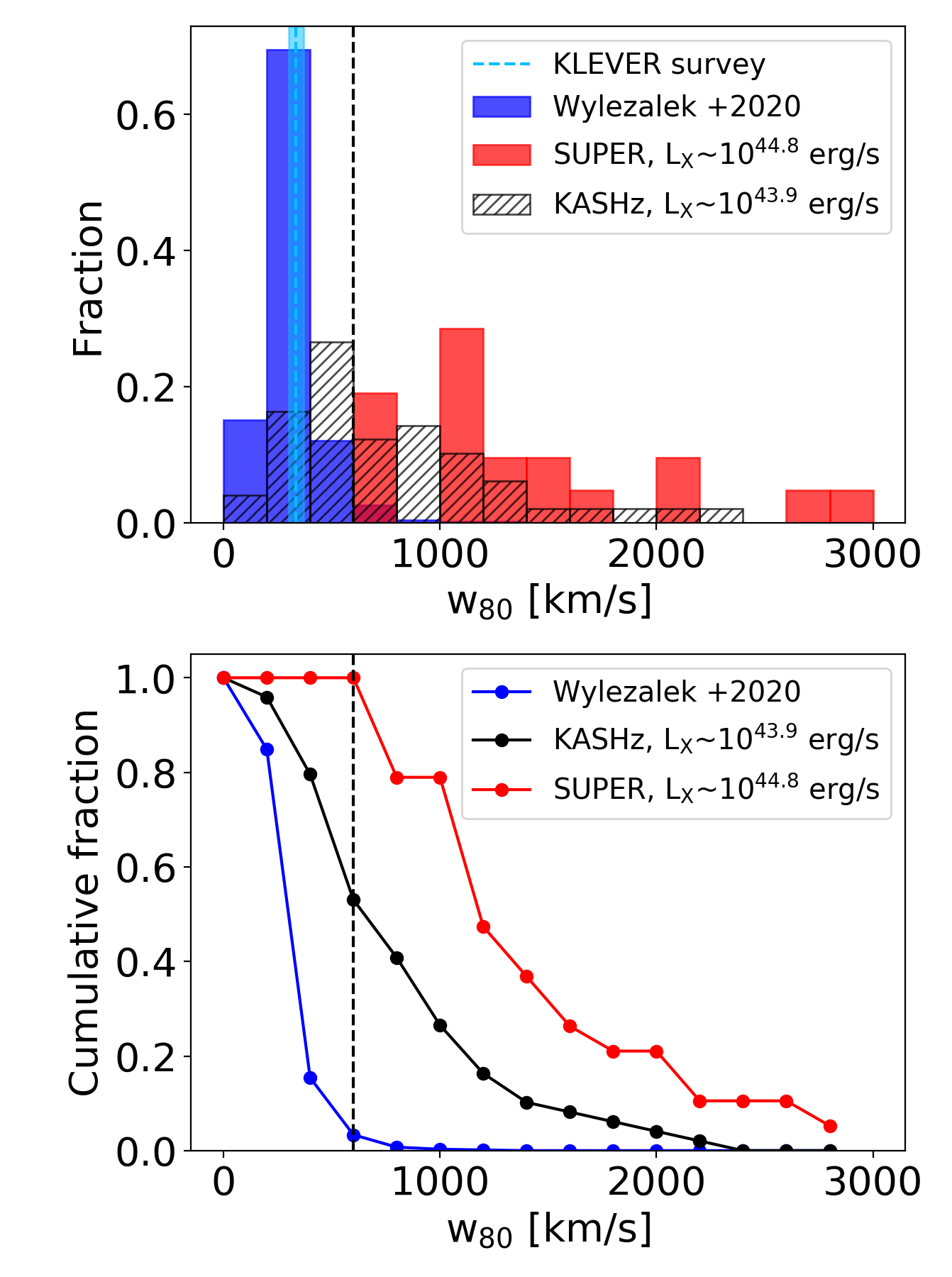}
\caption{The top panel shows the distribution of the non-parametric velocity, $\mathrm{w_{80}}$ measured for \oiii$\lambda$5007 line, for AGN and star forming samples from different surveys. The AGN sample consists of targets from the SUPER survey in red (this paper) and targets from KASHz survey matched in redshift in hatched black \citep{harrison16}. The Type-1 sample used in this paper has a median hard X-ray luminosity of 10$^{44.8}$ erg/s and the redshift-matched KASHz targets have a median X-ray luminosity of 10$^{43.9}$ erg/s. The blue histogram shows the  $\mathrm{w_{80}}$ distribution of mass-matched low redshift star forming sample from \citet{wylezalek20}, while the ocean-blue vertical bar shows the $\mathrm{w_{80}}$ value from the redshift and mass-matched star forming sample from the KLEVER survey (334$\pm$35 km/s, \citealt{curti20}). The dashed black-line at 600 km/s corresponds to the $\mathrm{w_{80}}$ value used in this work to define if a target hosts an AGN-driven outflow. The lower panel shows the inverse cumulative $\mathrm{w_{80}}$ distribution with the same color coding as in the top panel. It is clear from both the panels that targets from the SUPER survey occupy the higher end of the $\mathrm{w_{80}}$ distribution. Based on the $\mathrm{w_{80}}$ criteria, all the targets in the SUPER survey show the presence of outflows, and $\sim$52\% of the redshift matched targets in the KASHz survey show outflows. \label{fig:w80_distribution}}
\end{figure}

\begin{landscape}
\begin{table}
\centering
\caption{Narrow Line Region properties from the line fitting of the integrated H-band SINFONI spectra of the SUPER sample.}\label{table:SINFONI_properties1}
\begin{tabular}{ccc|c|cc|ccccc|cc}
\hline
Target & $\mathrm{\lambda_{range}^{a}}$ & {\rm $D_{ext}^{b}$} & $\mathrm{z_{\oiii}^{c}}$ & \multicolumn{2}{c}{log L$_{\oiii}^{d}$}  & \multicolumn{5}{c}{$\mathrm{v_{\oiii}^{e}}$} & \multicolumn{2}{c}{ log L$_{H\beta}^{d}$}\\
&  & &  & narrow & broad & FWHM$_{\rm narrow}$ & FWHM$_{\rm broad}$ & $\mathrm{v_{10}}$ & $\mathrm{w_{80}}$ & $\mathrm{v_{max}}$ & narrow & broad\\
& \text{\AA} & arcsec & & erg/s & erg/s & km/s & km/s & km/s & km/s & km/s & erg/s & erg/s\\
\hline\hline
X\_N\_160\_22 & 4550--5200 & 1.0 & 2.442 & 43.11$\pm$0.04 & 43.30$\pm$0.07 & 869$\pm$50 & 3035$\pm$185 & -2333$\pm$146 & 2816$\pm$160 & 3637$\pm$222 & 42.60$\pm$0.08 & -- \\
X\_N\_81\_44  & 4500--5420 & 0.9 & 2.317 & 42.68$\pm$0.04 & 42.19$\pm$0.13 & 494$\pm$30 & 1851$\pm$130 & -336$\pm$40 & 775$\pm$108 & 1682$\pm$117 & 41.43$\pm$0.05 & 43.24$\pm$0.05\\
X\_N\_66\_23 &  4600--5250 & 0.7 & 2.384 & --& 43.16$\pm$0.03 & --& 900$\pm$44 & -495$\pm$36 & 1001$\pm$51 & 764$\pm$38 & 42.84$\pm$0.05 & --\\
X\_N\_35\_20 & 4600--5200 & 0.3 & 2.260 & -- & 41.87$\pm$0.45 & -- & 643$\pm$106 & -330$\pm$78 & 681$\pm$209 & 546$\pm$90 & -- & --\\
X\_N\_12\_26 & 4450--5180 & 0.8 & 2.471 & -- & 42.41$\pm$0.07 & -- & 933$\pm$184 & -680$\pm$57 & 1044$\pm$89 & 792$\pm$156 & -- & 42.06$\pm$0.07\\
X\_N\_4\_48  & 4500--5250 & 0.5 & 2.314 & -- & 42.24$\pm$0.07 & -- & 1126$\pm$151 & -648$\pm$84 & 1198$\pm$100 & 956$\pm$129 & -- & 41.50$\pm$0.06\\
X\_N\_102\_35 & 4620--5400 & 0.3 & 2.190 & 42.51$\pm$0.13 & 42.76$\pm$0.13 & 550$\pm$112 & 1473$\pm$261 & -1076$\pm$140 & 1501$\pm$170 & 1761$\pm$295 & 41.18$\pm$0.10 & 42.14$\pm$0.09\\
X\_N\_115\_23 & 4550--5280 & 0.7 & 2.340 & 43.27$\pm$0.03 & 43.18$\pm$0.02 & 471$\pm$24 & 1495$\pm$62 & -623$\pm$57 & 1015$\pm$67 & 1437$\pm$56 & 42.38$\pm$0.06 & 42.13$\pm$0.28\\
cid\_166  & 4430--5185 & 0.6 & 2.460 & 42.60$\pm$0.11 & 43.17$\pm$0.07 & 503$\pm$91 & 1703$\pm$150 & -1502$\pm$41 & 1755$\pm$109 & 2136$\pm$155 & 41.77$\pm$0.04 & 42.89$\pm$0.10\\
cid\_1605  & 4650--5200 & 0.4 & 2.117 & -- & 42.43$\pm$0.43 & -- & 1095$\pm$244 & -573$\pm$98 & 1153$\pm$142 & 929$\pm$208 & -- & --\\
cid\_346  & 4550--5450 & 0.7 & 2.216 & -- & 42.97$\pm$0.04 & -- & 1989$\pm$212 & -1343$\pm$166 & 2142$\pm$231 & 1689$\pm$181 & -- & 42.74$\pm$0.14\\
cid\_1205 & 4700--5200 & 0.3 & 2.256 & -- & 42.66$\pm$0.04 & -- & 636$\pm$72 & -223$\pm$36 & 717$\pm$48 & 540$\pm$61 & -- & --\\
cid\_467   & 4600--5200 & 0.4 & 2.284 & -- & 42.87$\pm$0.43 & -- & 1256$\pm$132 & -702$\pm$112 & 1368$\pm$154 & 1067$\pm$112 & -- & --\\
J1333+1649 &4750--5350 & 1.1 & 2.098 & 44.00$\pm$0.02 & 44.58$\pm$0.04 & 602$\pm$22 & 2700$\pm$80 & -2271$\pm$78 & 2714$\pm$96 & 3248$\pm$87 & -- & 44.08$\pm$0.12\\
J1441+0454* & 4730--5350 & 1.0 & 2.053 & -- & 43.41$\pm$0.03 & -- & 1956$\pm$96 & -3698$\pm$70 & 2161$\pm$102 & 1661$\pm$82 & 42.29$\pm$0.13 & 42.79$\pm$0.19\\
J1549+1245 & 4450--5250 & 1.0 & 2.367 & 43.15$\pm$0.10 & 44.49$\pm$0.05 & 327$\pm$49 & 1362$\pm$25 & -613$\pm$27 & 1457$\pm$38 & 1413$\pm$32 & 41.40$\pm$0.10 & 43.47$\pm$0.06\\
S82X1905 & 4550--5450 & 0.8 & 2.272 & -- & 42.82$\pm$0.04 & -- & 607$\pm$59 & -304$\pm$34 & 678$\pm$55 & 515$\pm$54 & 42.18$\pm$0.15 & --\\
S82X1940  & 4550--5400 & 0.6 & 2.349 & 42.28$\pm$0.03 & 42.54$\pm$0.03 & 354$\pm$20 & 1336$\pm$71 & -820$\pm$51 & 1186$\pm$39 & 1373$\pm$71 & 41.30$\pm$0.17 & 42.09$\pm$0.09\\
S82X2058  & 4600--5400 & 0.6 & 2.314 & 42.44$\pm$0.08 & 42.61$\pm$0.08 & 539$\pm$61 & 1442$\pm$109 & -972$\pm$88 & 1340$\pm$95 & 1701$\pm$234 & 42.01$\pm$0.09 & 41.67$\pm$0.40\\
    \hline
\end{tabular}
\vspace{1ex}\newline
{\raggedright{\bf Notes:}\newline
$^{a}$The wavelength range used for the emission line fitting.\newline
$^{b}$The diameter (in arcsec) of the circular aperture centered on the target used to extract the spectrum.\newline
$^{c}$Redshift of the target determined from the peak location of the \oiii$\lambda$5007 in the integrated spectrum. \newline
$^{d}$The luminosity of the individual Gaussian components in erg/s, not corrected for reddening. The emission lines were modeled using multiple Gaussian components and the terms ``narrow'' and ``broad'' refer to these individual Gaussian components parameters. In case of single Gaussian fits, the Gaussian component is classified as broad if the width (FWHM) > 600 km/s.``--'' means that there was no detection of the corresponding component. The errors indicate 1$\sigma$ uncertainty. \newline
$^{e}$Line widths (FWHM) of the individual Gaussian components. $\mathrm{v_{10}}$ and $\mathrm{w_{80}}$ are the non-parametric velocities and $\mathrm{v_{max}}$ the maximum velocity as defined in Sect. \ref{sect5.1}. \newline
*The fit does not constrain the NLR properties due to contamination from telluric absorption and strong contribution from FeII emission. The reported \oiii emission is highly blue-shifted ($\sim$-3000 km/s) and the \oiii redshift corresponds to this component.\newline
There is no detection of lines in X\_N\_53\_3 and X\_N\_44\_64.\par}
\end{table}
\end{landscape}

We adopt non-parametric measures for the line properties, which has the advantage that the parameter values do not depend on the fitting function adopted (e.g. the number of Gaussian components) that may strongly depend on the signal-to-noise of the spectrum under investigation (see e.g. \citealt{zakamska14}, \citealt{harrison14} for more details). In particular, we measure the velocity at the 10th percentile of the overall \oiii line profile ($\mathrm{v_{10}}$) and the velocity width of the line that contains 80$\%$ of the line flux ($\mathrm{w_{80}=v_{90}-v_{10}}$). For a Gaussian profile, the value of $\mathrm{w_{80}}$ approximately corresponds to the FWHM of the emission line. We also compute the maximum velocity, $\mathrm{v_{max}}$, following the definition by \citet{rupke13} as the shift between the narrow and the broad Gaussian components of \oiii plus twice the sigma of the broad Gaussian. For fits with single Gaussian, we estimate $\mathrm{v_{max}}$ as twice the sigma of the broad Gaussian. The non-parametric measures of velocities and the line widths are reported in Table \ref{table:SINFONI_properties1}.

In the following, we will use $\mathrm{w_{80}}$ to identify AGN with clear signatures of outflows. Fig. \ref{fig:w80_distribution} shows the distribution of $\mathrm{w_{80}}$ for the AGN sample from SUPER and KASHz surveys in red and black histograms respectively. The KASHz sample is matched in redshift (z$\sim$2.0--2.5) to the SUPER sample. For comparison, we also plot the $\mathrm{w_{80}}$ distribution of low-redshift mass-matched sample of MANGA galaxies from \citet{wylezalek20}, which is shown in blue in Fig. \ref{fig:w80_distribution}. The light blue vertical area denotes the $\mathrm{w_{80}}$ value from a stacked spectrum of mass-matched and redshift-matched galaxies from the KLEVER survey \citep{curti20}, a high redshift survey targeting star forming galaxies. The stellar mass of the overall SUPER sample has been used to mass-match the comparison samples, as most of the Type-1 AGN do not have reliable stellar mass measurements \citep{circosta18}.

\begin{figure}
\centering
\includegraphics[scale=0.44]{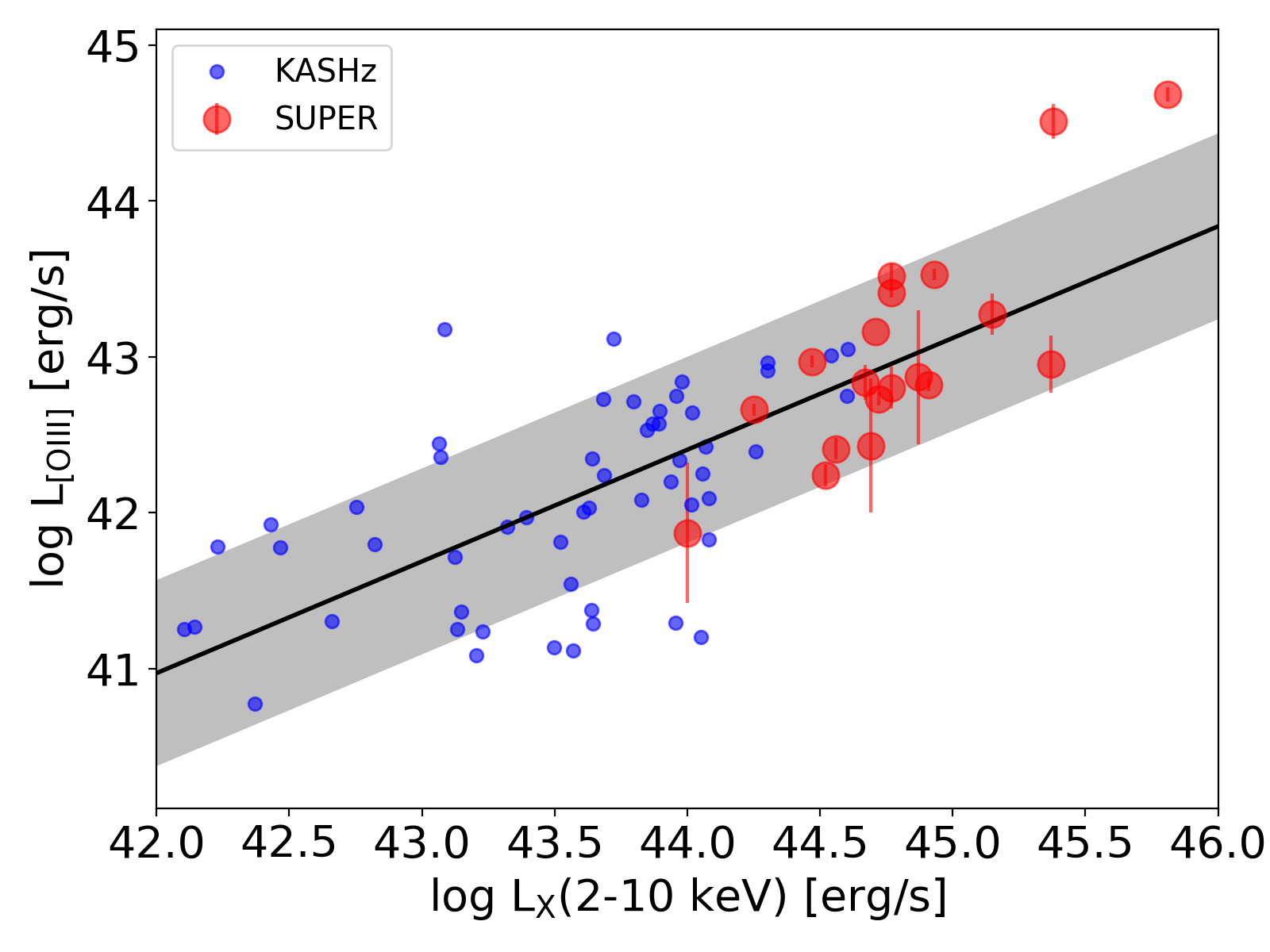}
\caption{Total \oiii luminosity vs. hard X-ray luminosity ($\mathrm{L_{X}}$(2-10 keV) for the SUPER (red circles) and KASHz targets (blue data points), the latter from a similar survey targeting X-ray selected AGN at high redshift. The combined dataset of KASHz and SUPER show a positive correlation between the two quantities, shown by the black line. The shaded region represents the 1$\sigma$ scatter on the relation. KASHz targets occupy the lower X-ray luminosity range compared to SUPER targets, which may explain the difference in the distribution of $\mathrm{w_{80}}$ in the two surveys in Fig. \ref{fig:w80_distribution}. \label{fig:Loiii-Lx-relation}}
\end{figure}

From the histogram and the cumulative distributions in Fig. \ref{fig:w80_distribution}, it is clear that the AGN sample from the KASHz and SUPER surveys occupy the higher end of $\mathrm{w_{80}}$ values compared to mass-matched samples at low as well as high redshift star forming galaxies. A $\mathrm{w_{80}}$ value of 600 km/s lies at the high end of the tail of the $\mathrm{w_{80}}$ distribution for low-redshift star forming galaxies and it is well above the average values obtained from KLEVER. Consequentely, we will consider in this paper a $\mathrm{w_{80}}$ value greater than 600 km/s as a signature of an AGN-driven outflow. We note that similar cut in $\mathrm{w_{80}}$ was also used in previous works \citep[e.g.][]{harrison16} and is a conservative estimate when compared to the cuts used in other works \citep[e.g. 500 km/s in ][]{wylezalek20}.

Based on the above definition of outflows, all the observed Type-1 AGN from  SUPER detected in \oiii show the presence of AGN-driven outflows (90\% outflow detection rate if we include targets which were observed and not detected in \oiii). The KASHz survey, on the other hand, reported $\sim$50\% of their sample having $\mathrm{w_{80}}$ larger than 600 km/s \citep{harrison16}, therefore a lower fraction of AGN with clear outflow signatures according to the adopted definition. If we restrict the KASHz sample to the same redshift range of SUPER, $2<z<2.5$, the fraction of KASHz targets hosting outflows results to be at $\sim 52\%$ as shown in Fig. \ref{fig:w80_distribution}.The difference between the $\mathrm{w_{80}}$ distributions for SUPER and KASHz surveys is probably due to the different luminosity range of the AGN sampled by these surveys. Fig. \ref{fig:Loiii-Lx-relation} shows the relation between total \oiii luminosity and hard X-ray luminosity for the targets presented in this paper as well as for the redshift-matched AGN sample from the KASHz survey. We have eight objects in common between the two surveys (for both Type-1 and Type-2 targets) and the $\mathrm{w_{80}}$ values derived from SINFONI and KMOS for these objects are perfectly consistent with each other. As it can be clearly seen from this figure, there is little overlap between the luminosity ranges covered by the two surveys: $\approx 90\%$ of the Type-1 sample presented in this paper has L$_{\rm [2-10 kev]}> 3 \times 10^{44}$ erg s$^{-1}$, while $\approx 90\%$ of the KASHz AGN are at L$_{\rm [2-10 kev]}< 3 \times 10^{44}$ erg s$^{-1}$ (Fig. \ref{fig:w80_distribution}, blue data points). As already reported in previous works \citep[e.g.][]{fiore17} and presented for the SUPER Type-1 sample in sect. \ref{sect6}, there is a positive correlation between the velocity associated with the outflow and the bolometric luminosity of the AGN. Therefore, the detection of a higher fraction of outflows in the SUPER sample compared to the KASHz survey could naturally be explained with the prevalence of the faster outflows at higher bolometric luminosity (see Fig. \ref{fig:w80_distribution}).

Although we will use non-parametric measures along the paper to characterize the line properties, we also report here the incidence of outflows in our sample based on the presence of a broad Gaussian component in the line fit. For the 21 Type-1 AGN presented in this paper, 14 targets require a broad component with width $\sim$1000-3000 km/s in their \oiii profile. We can therefore claim based on this measure that $67\%$ ($\sim 75\%$ if we limit to the objects with detected \oiii) of the SUPER Type-1 sample have fast ionized outflows. This outflow incidence fraction is much higher than some previous works for star forming as well as AGN host galaxies \citep[e.g.][]{harrison16,forster-schreiber18, swinbank19} and comparable to other lower redshift studies targeting AGN \citep[e.g.][]{rakshit18, davies20}. As we mentioned earlier, this method of classification for outflows is highly dependent on the signal-to-noise of the spectra and models used for line fitting, and therefore these comparisons are subject to biases.

Apart from the detection of the bright emission lines already discussed in this section, we also detect $\hb$ in all Type-1 sources, except cid\_1205 which shows a strong skyline at the location of $\hb$ emission. All targets but X\_N\_66\_23 show the detection of a BLR $\hb$ component. The non-detection of the BLR component of H$\beta$ in X\_N\_66\_23 is possibly a consequence of the short exposure time leading to a low S/N in the spectrum. Compared to the \oiii profile where the broad component was detected in $\sim$75\% of the sample, 62\% of the sample required an additional broad Gaussian component for $\hb$ line. The $\hb$ component flux is usually 10 times fainter than that of \oiii line in Type-1 AGN \citep[e.g.][]{leighly99,rodriguez-ardila00}, hence for the current sample the non-detection of broad $\hb$ could be simply due to the lower S/N. Although the kinematic components of the \oiii and the $\hb$ line are coupled to each other, the relatively low S/N of the $\hb$ line means that the addition of second Gaussian in $\hb$ profile does not change the $\chi^{2}$ of the fit in some targets. Based on this consideration, we will use the results on the \oiii line profile to assess the incidence of AGN-driven outflows in our sample. We refer the reader to Vietri et al. (2020) for further discussion on $\hb$ line profile. 

\subsection{Extension of the \texorpdfstring{\oiii}~emission line region} \label{sect5.2}

\begin{figure*}[h!]
\centering
\subfloat{\includegraphics[scale=0.31]{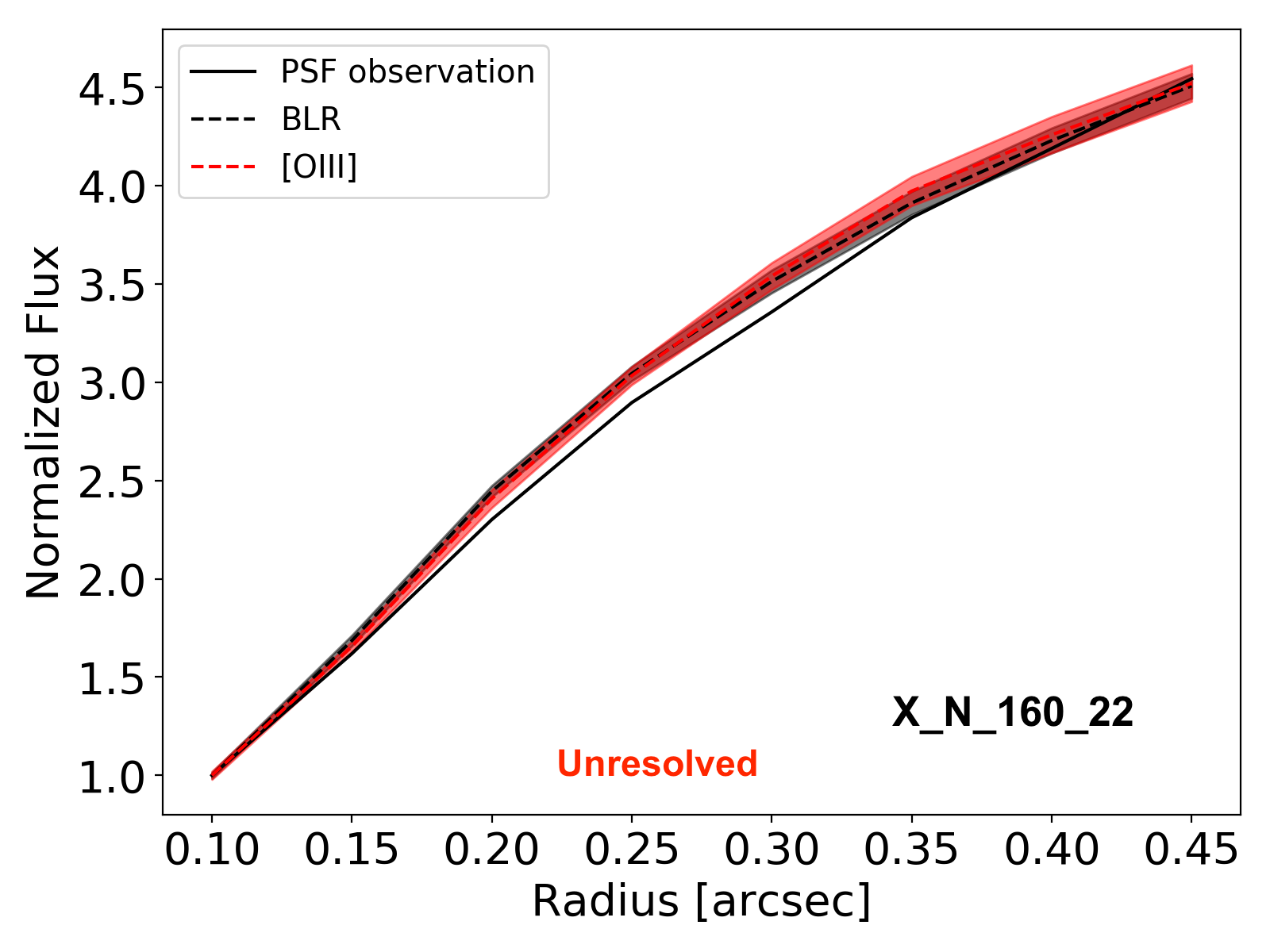}}
\subfloat{\includegraphics[scale=0.31]{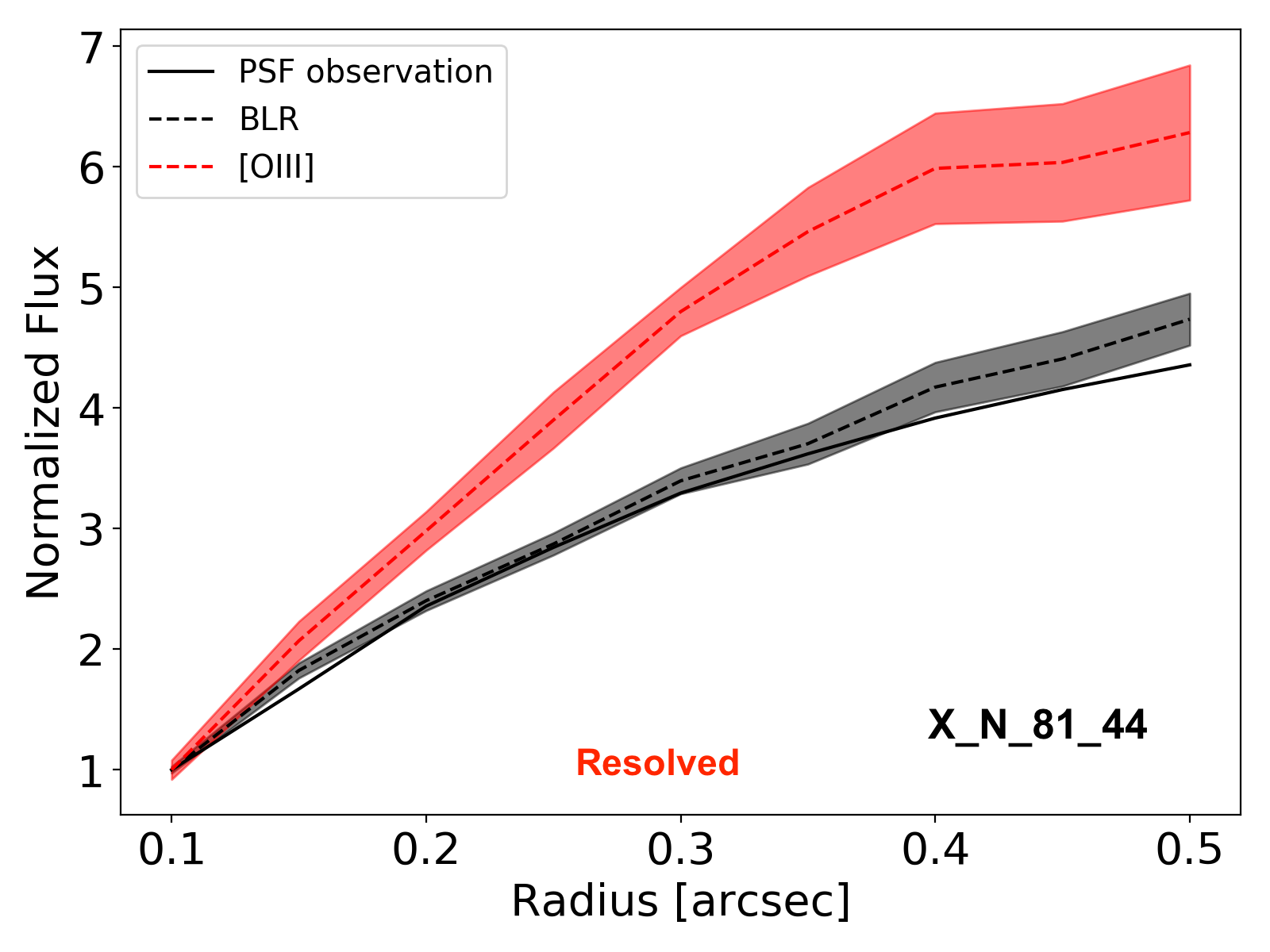}}
\subfloat{\includegraphics[scale=0.31]{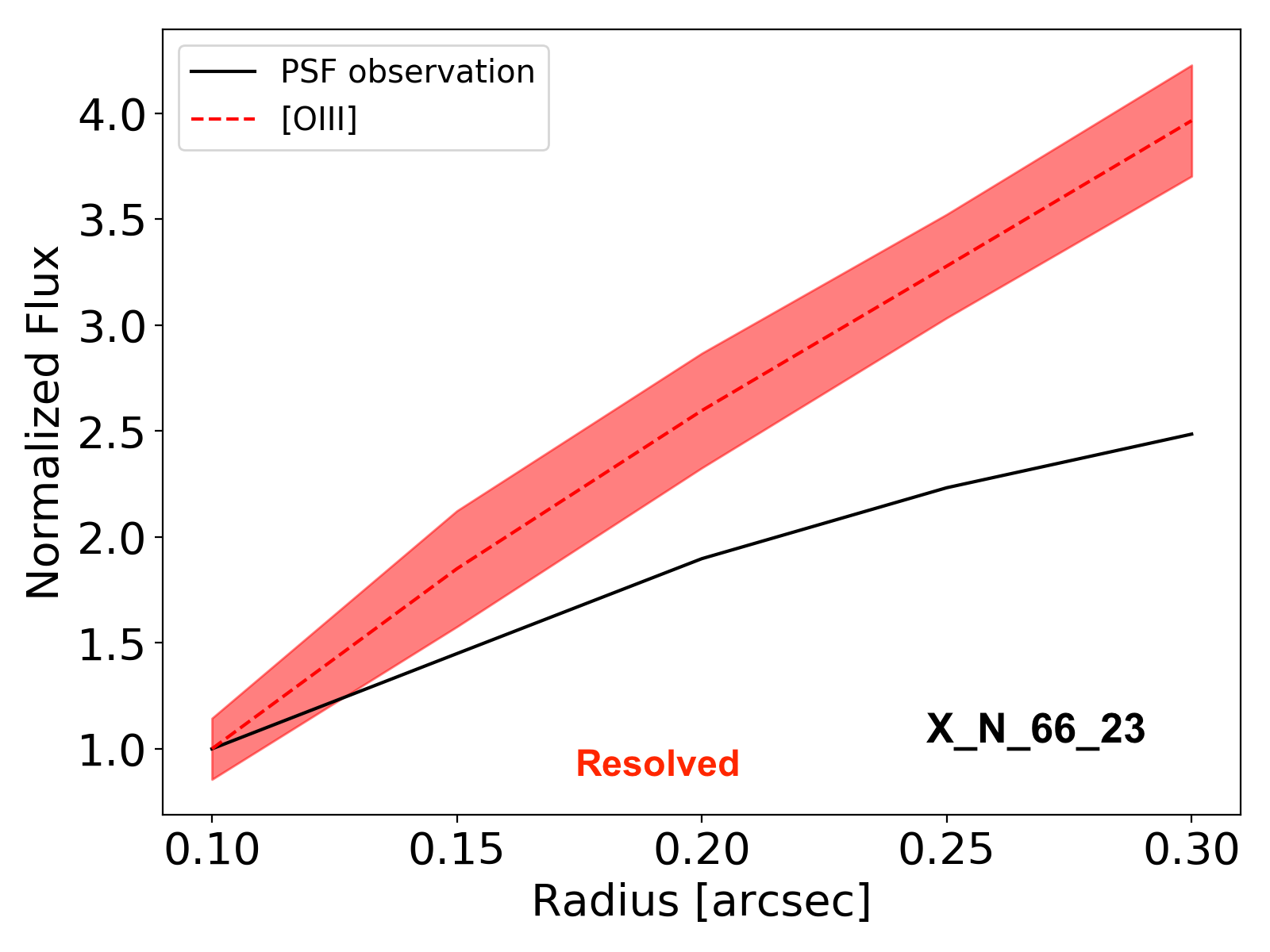}}\\
\subfloat{\includegraphics[scale=0.31]{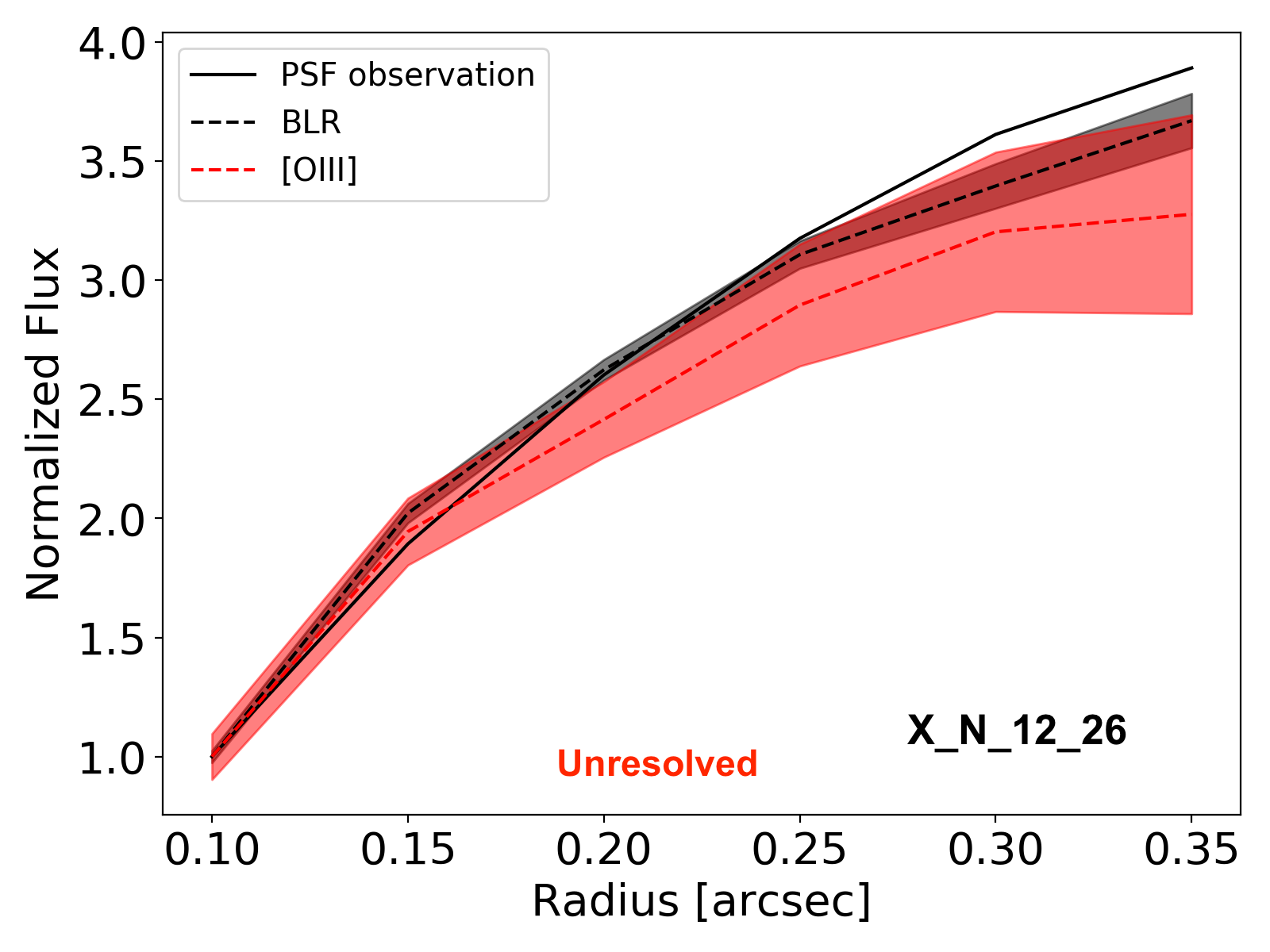}}
\subfloat{\includegraphics[scale=0.31]{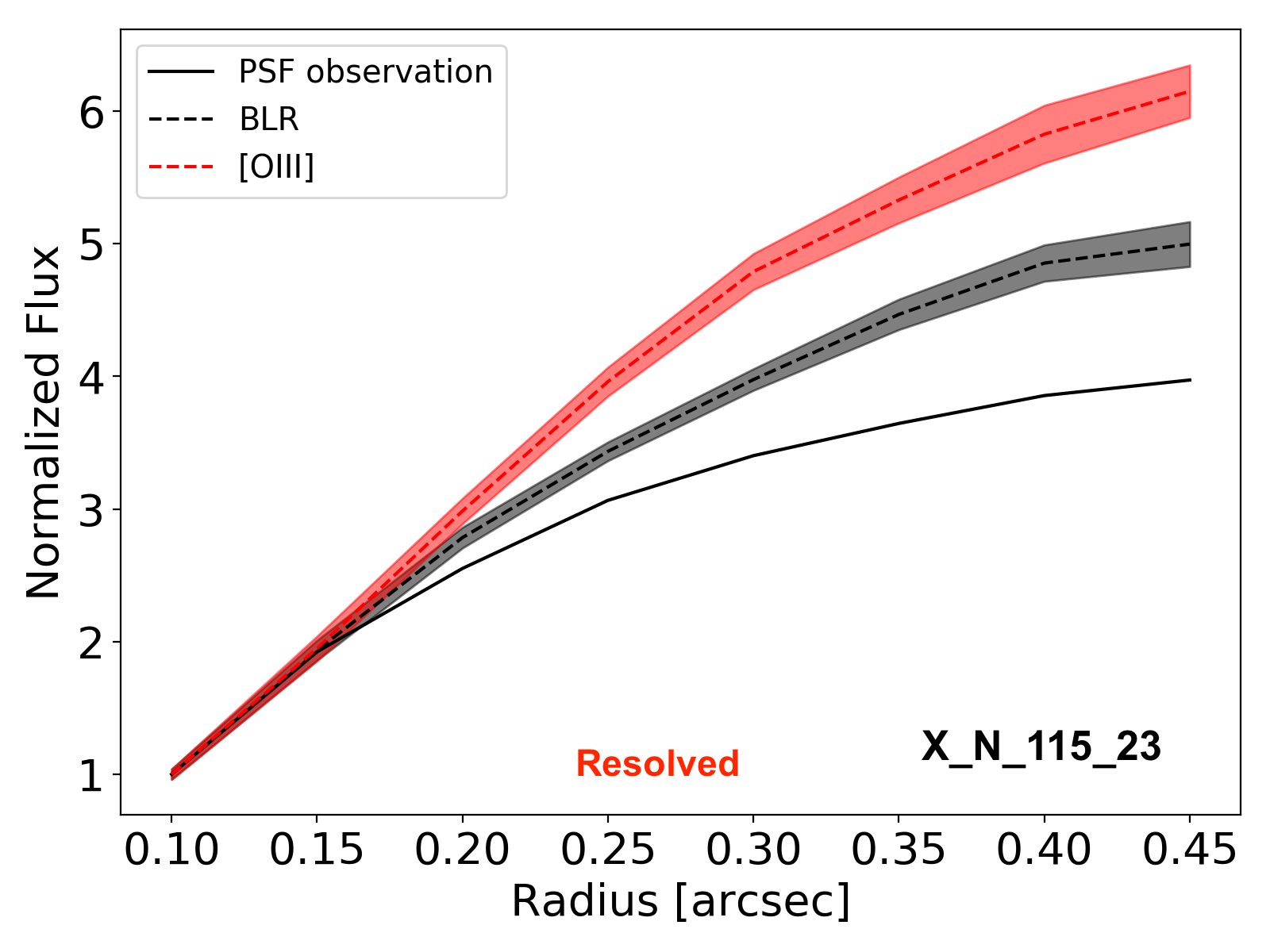}}
\subfloat{\includegraphics[scale=0.31]{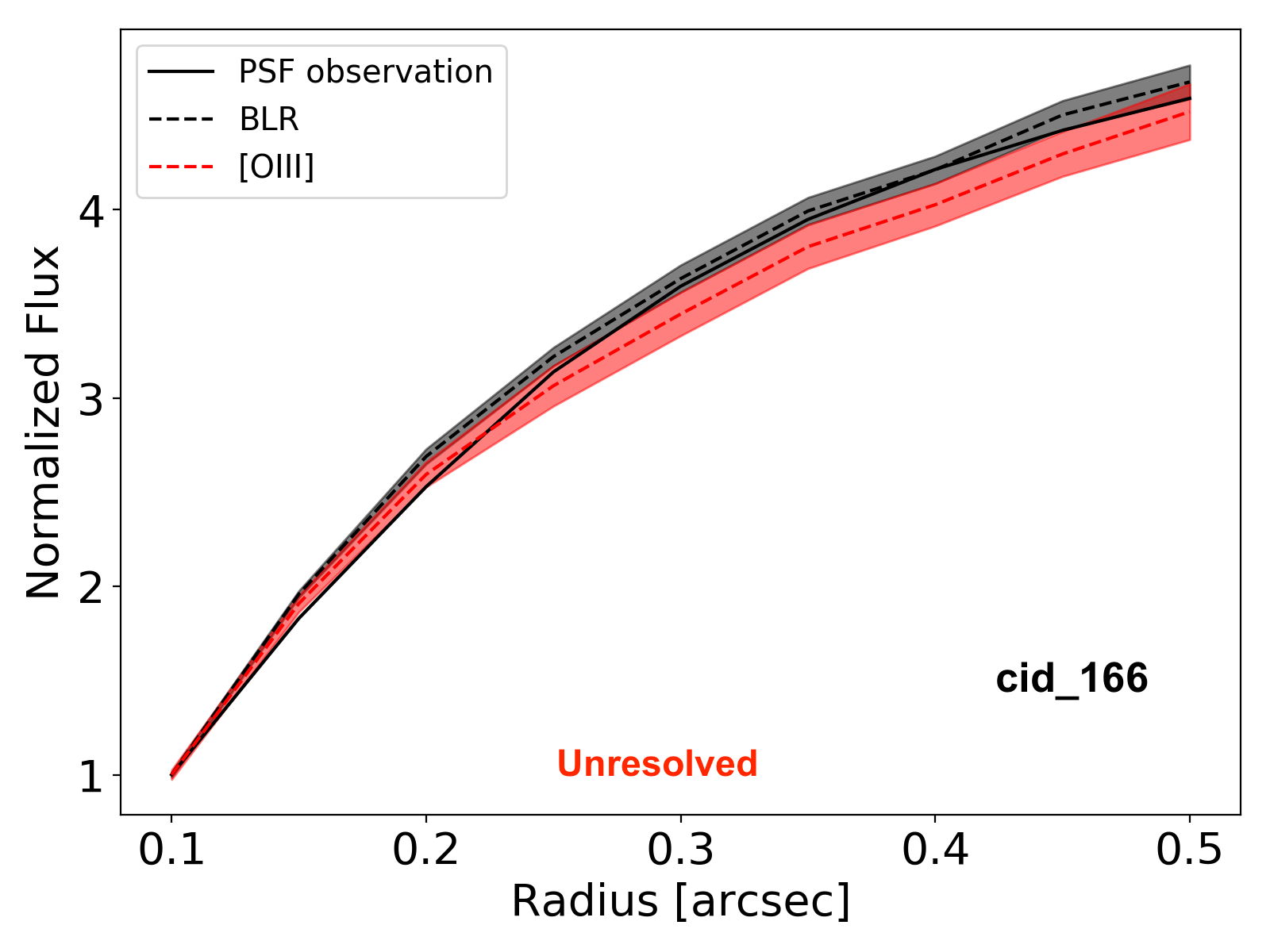}}\\
\subfloat{\includegraphics[scale=0.31]{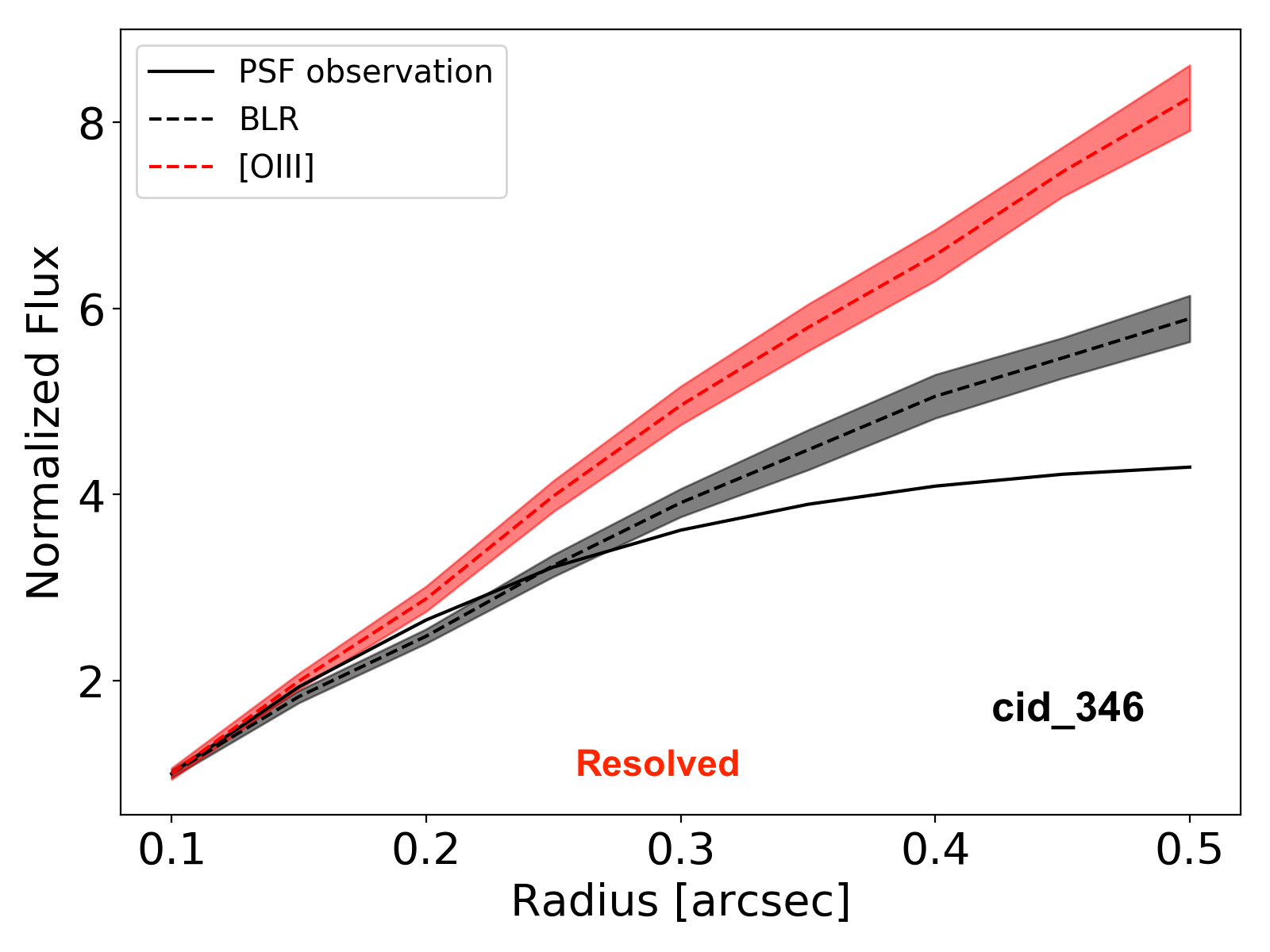}}
\subfloat{\includegraphics[scale=0.31]{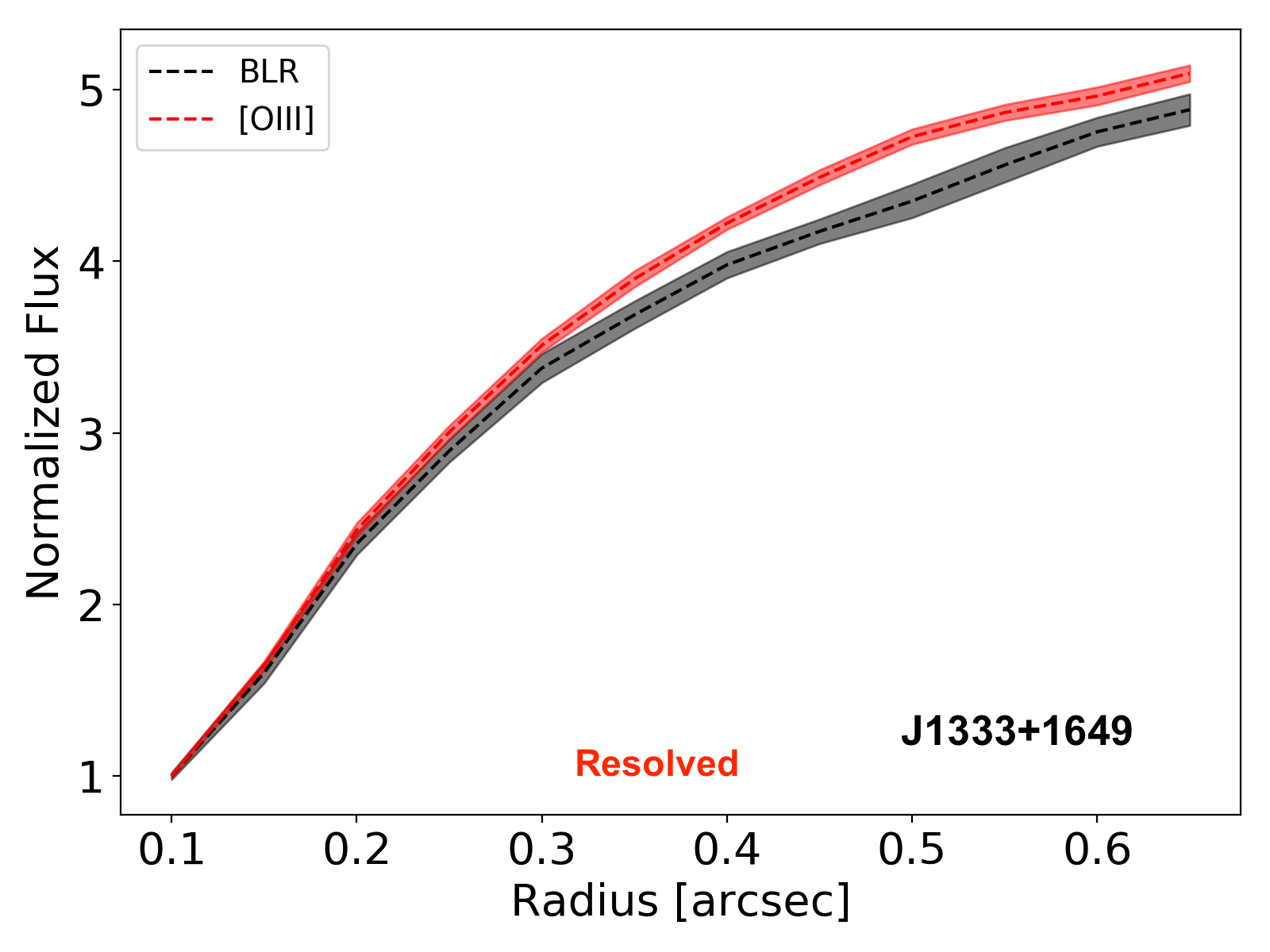}}
\subfloat{\includegraphics[scale=0.31]{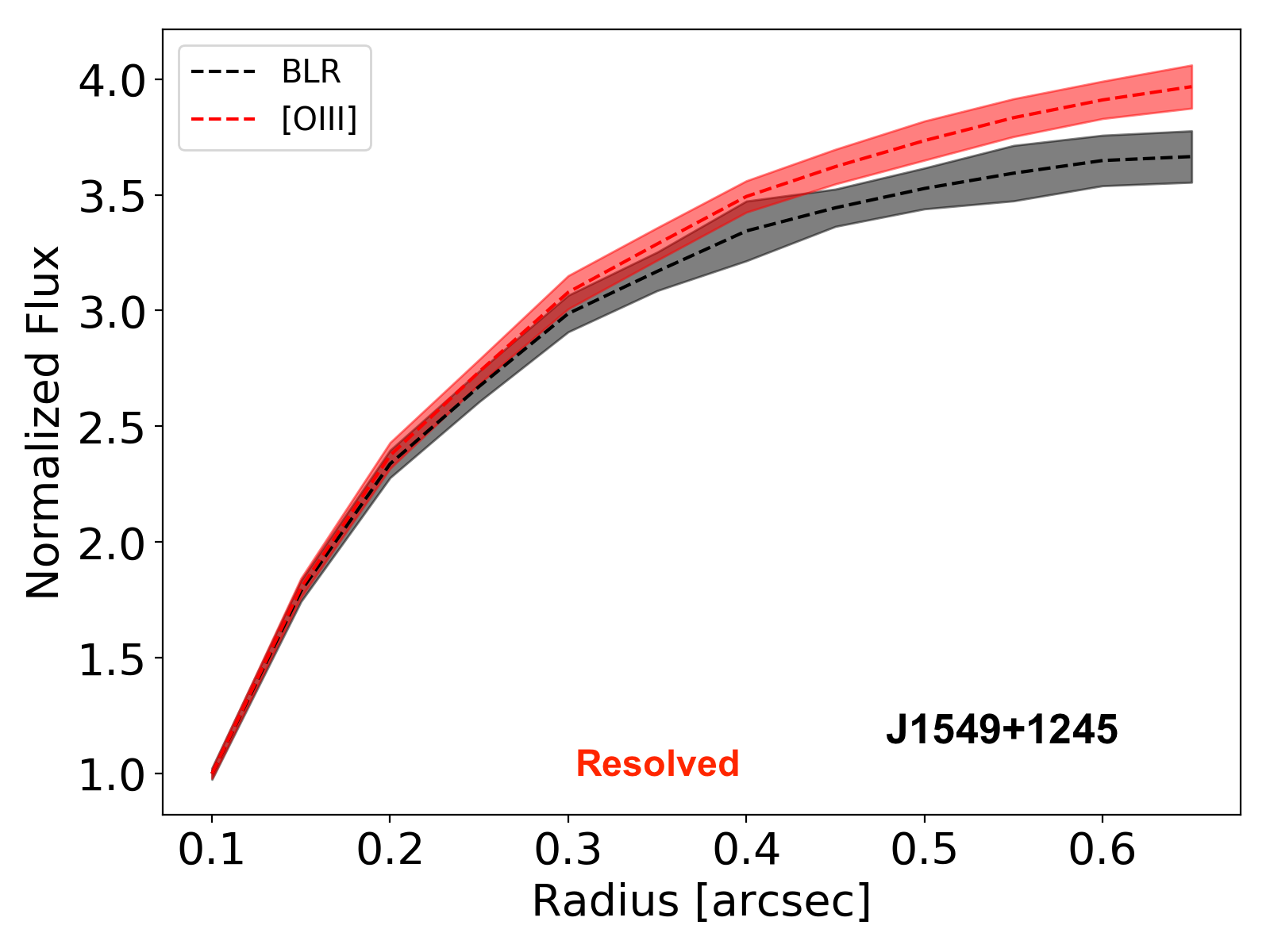}}\\
\subfloat{\includegraphics[scale=0.31]{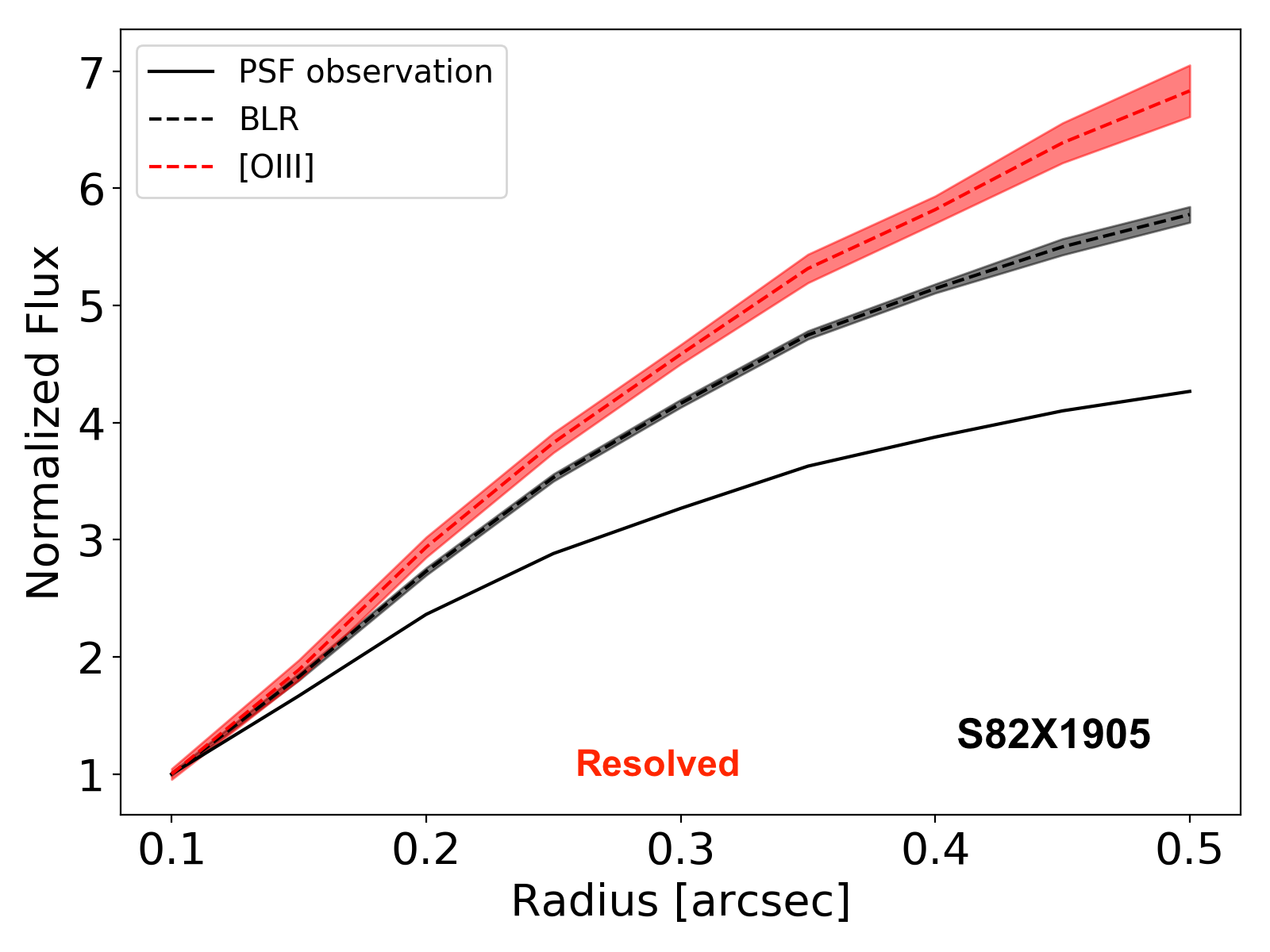}}
\subfloat{\includegraphics[scale=0.31]{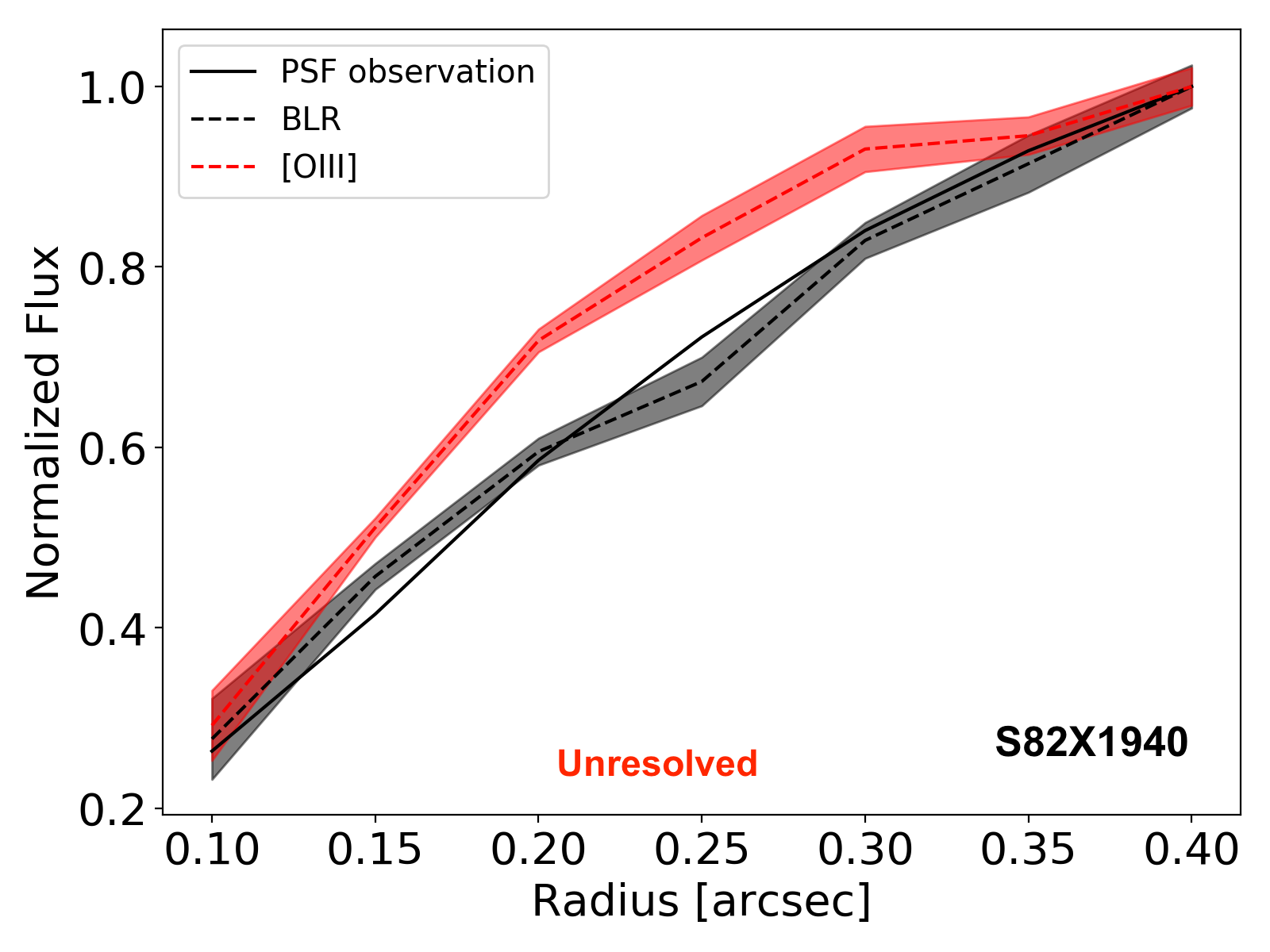}}
\caption{Curves-of-growth for the Type-1 sample as described in section \ref{sect5}. The solid black line shows the curve from the image obtained by collapsing the PSF-observation cube at the location of \oiii$\lambda$5007 channels. The $\hb$ BLR component is shown as a dashed black curve with the grey area showing the 1$\sigma$ uncertainly. The BLR component is unresolved and serves as a proxy to determine the PSF during the observations which is also compared with the dedicated PSF star observation. The red curve shows the curve-of-growth for the total \oiii emission i.e. narrow and broad components. All the uncertainty are at 1$\sigma$ levels. The BLR component is not plotted for X\_N\_66\_23 as it remains undetected. For the rest of the Type 1 sample in the SUPER survey, the S/N is not enough to perform such a curve-of-growth analysis. \label{fig:growth_curves}} 
\end{figure*}

\begin{figure*}
\centering
\includegraphics[scale=0.33]{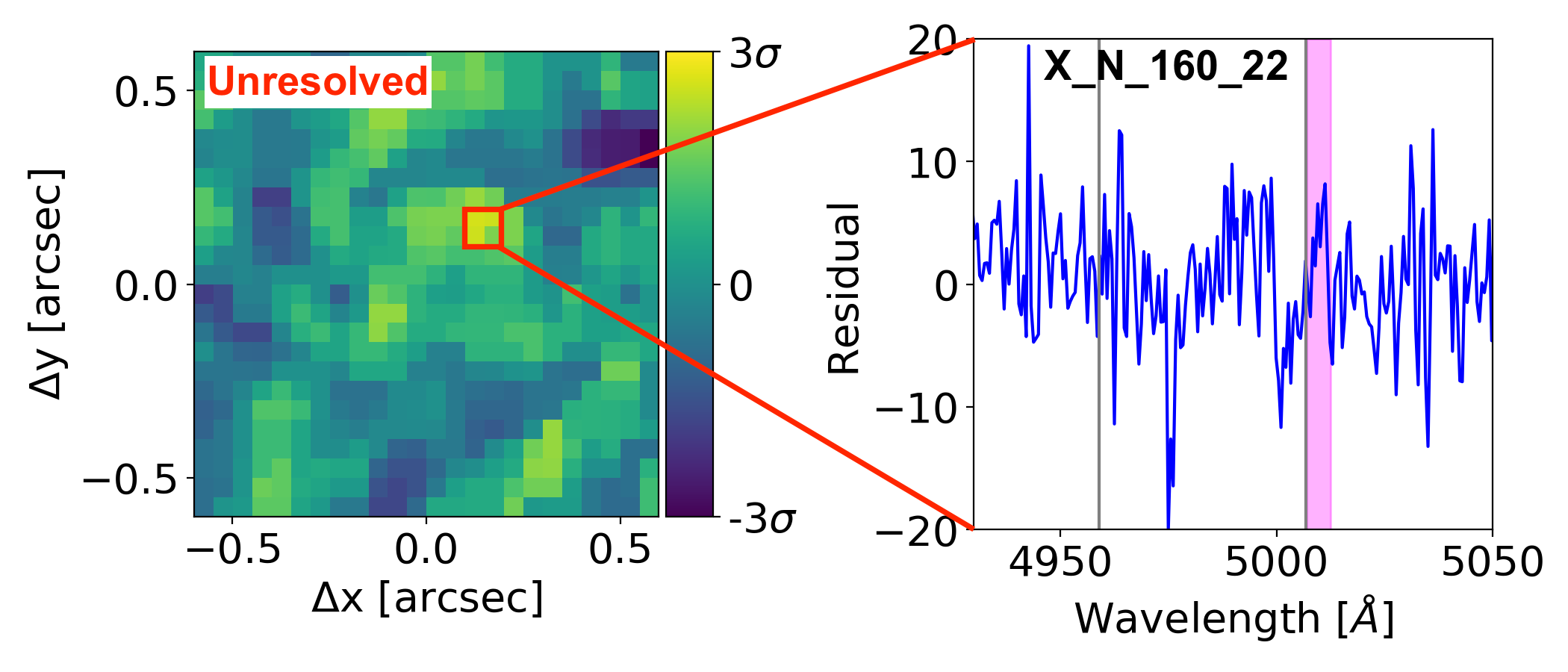}
\includegraphics[scale=0.33]{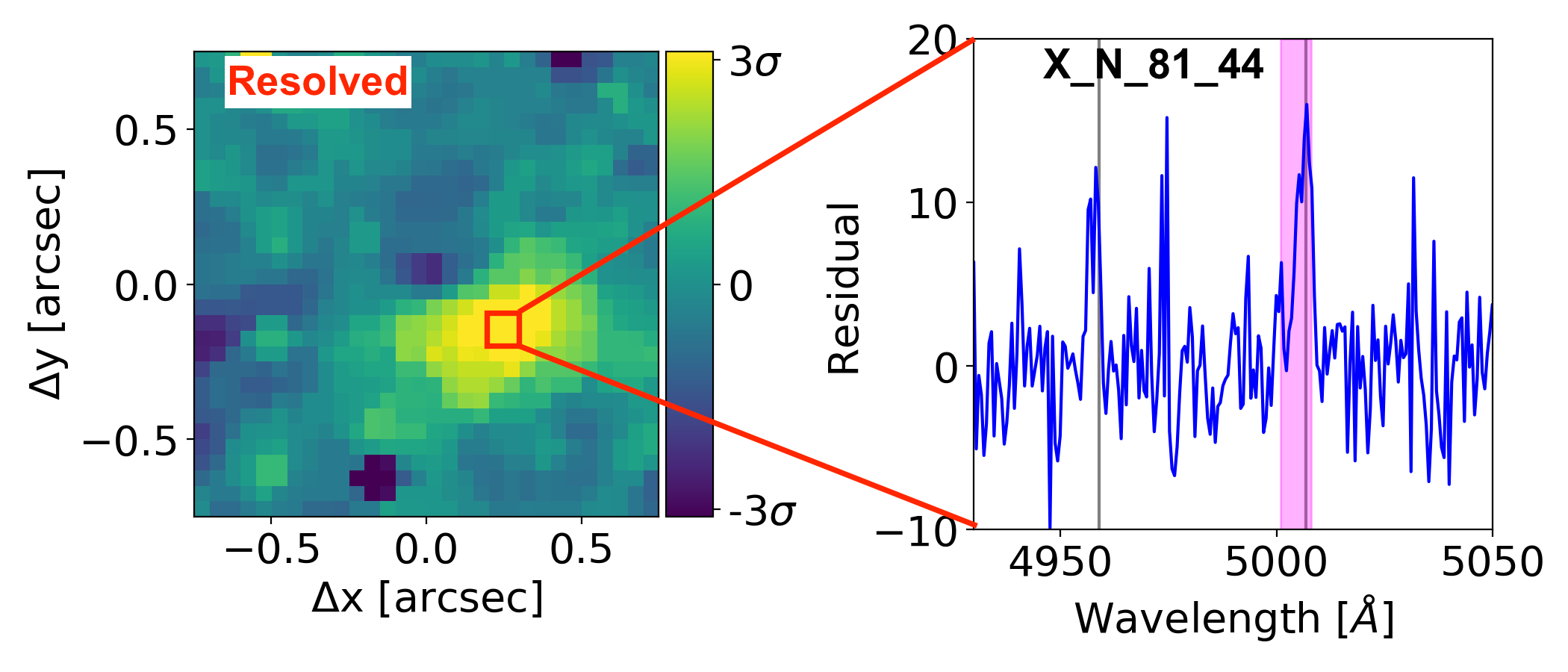}\\
\includegraphics[scale=0.33]{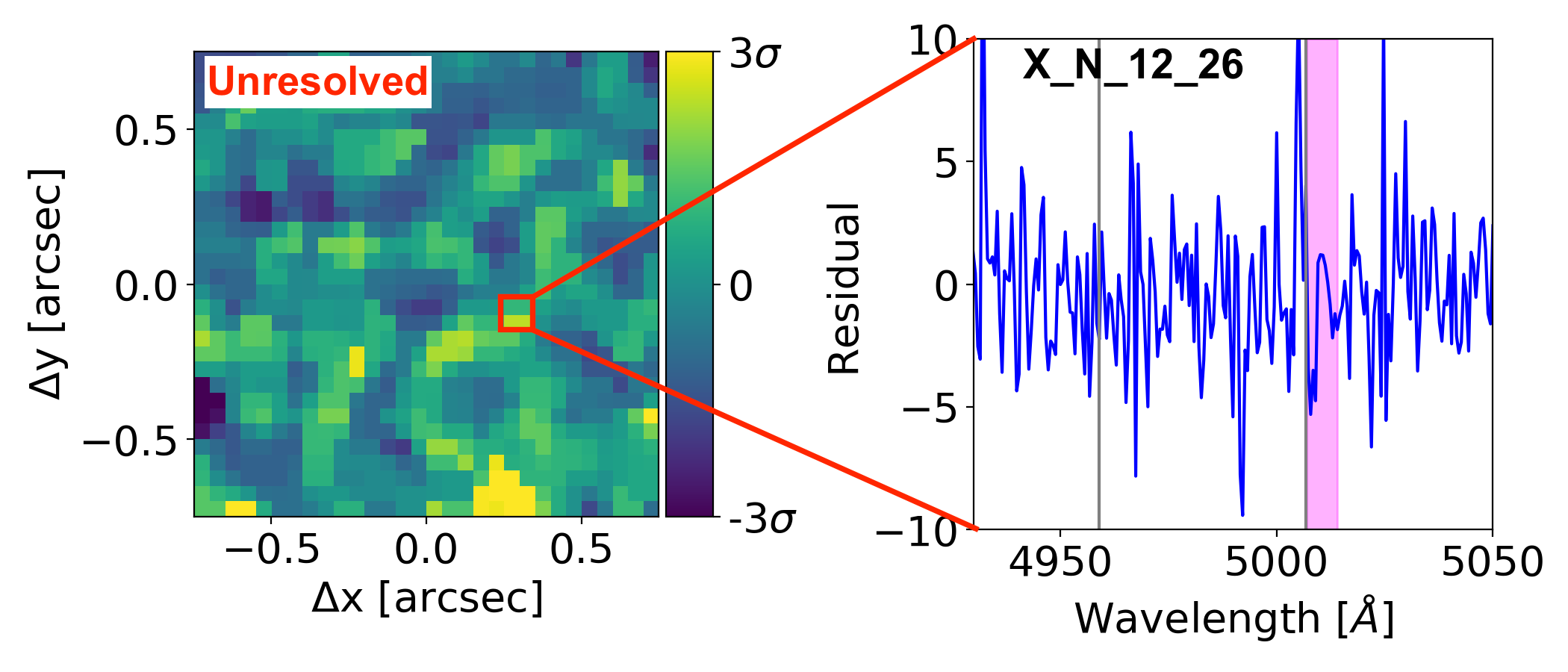}
\includegraphics[scale=0.33]{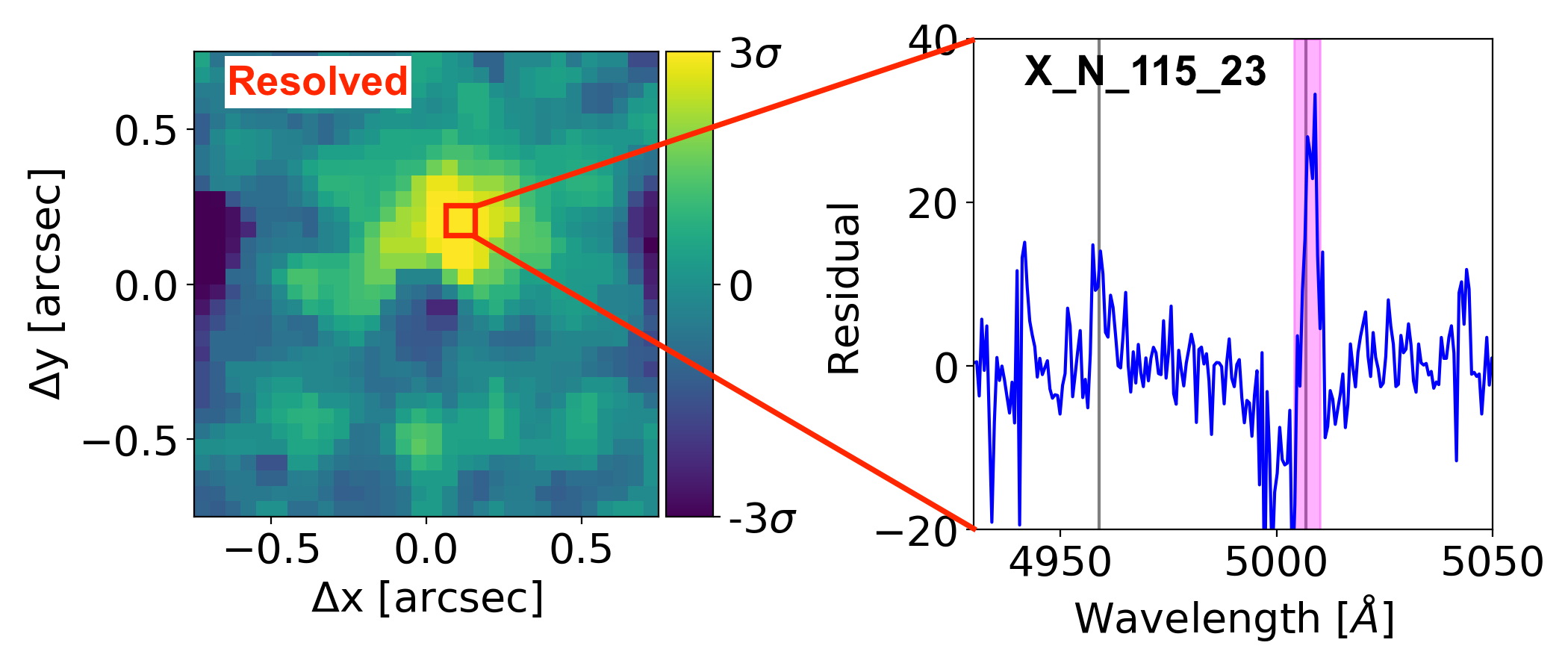}\\
\includegraphics[scale=0.33]{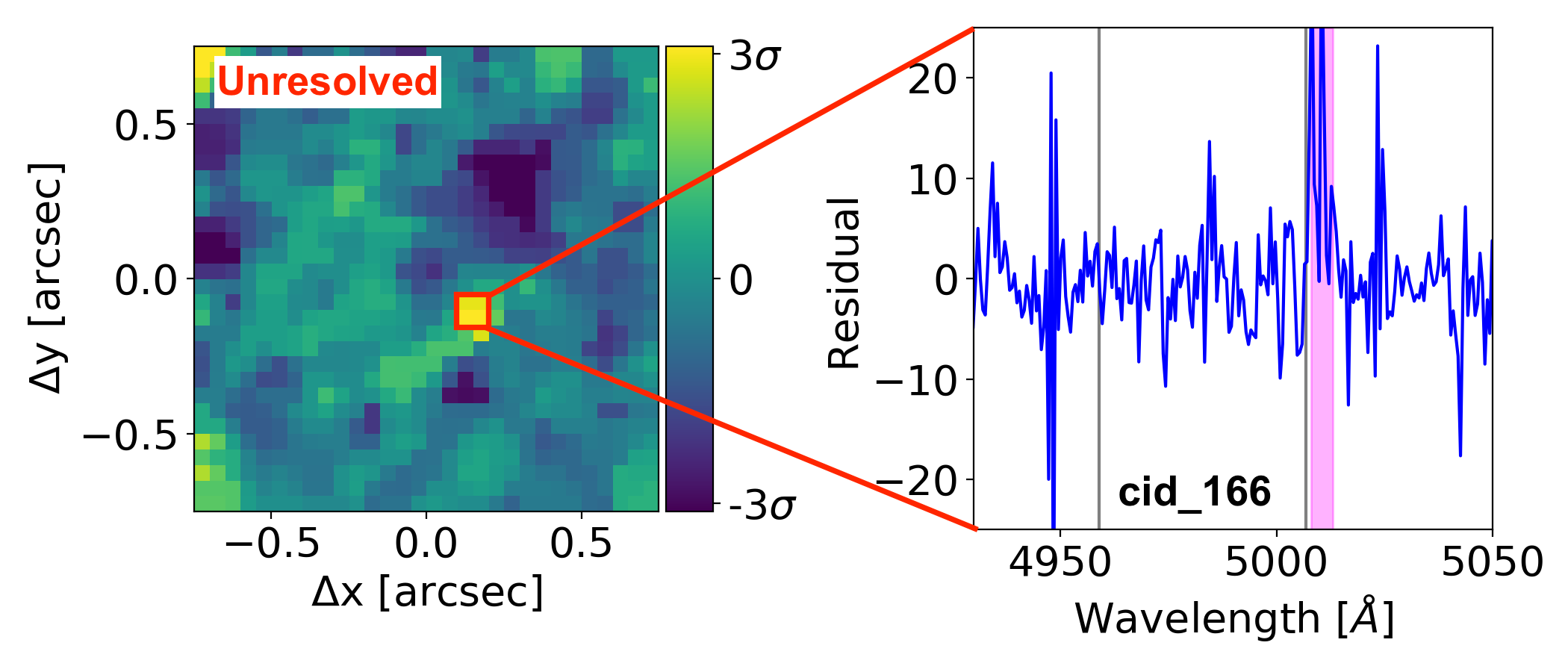}
\includegraphics[scale=0.33]{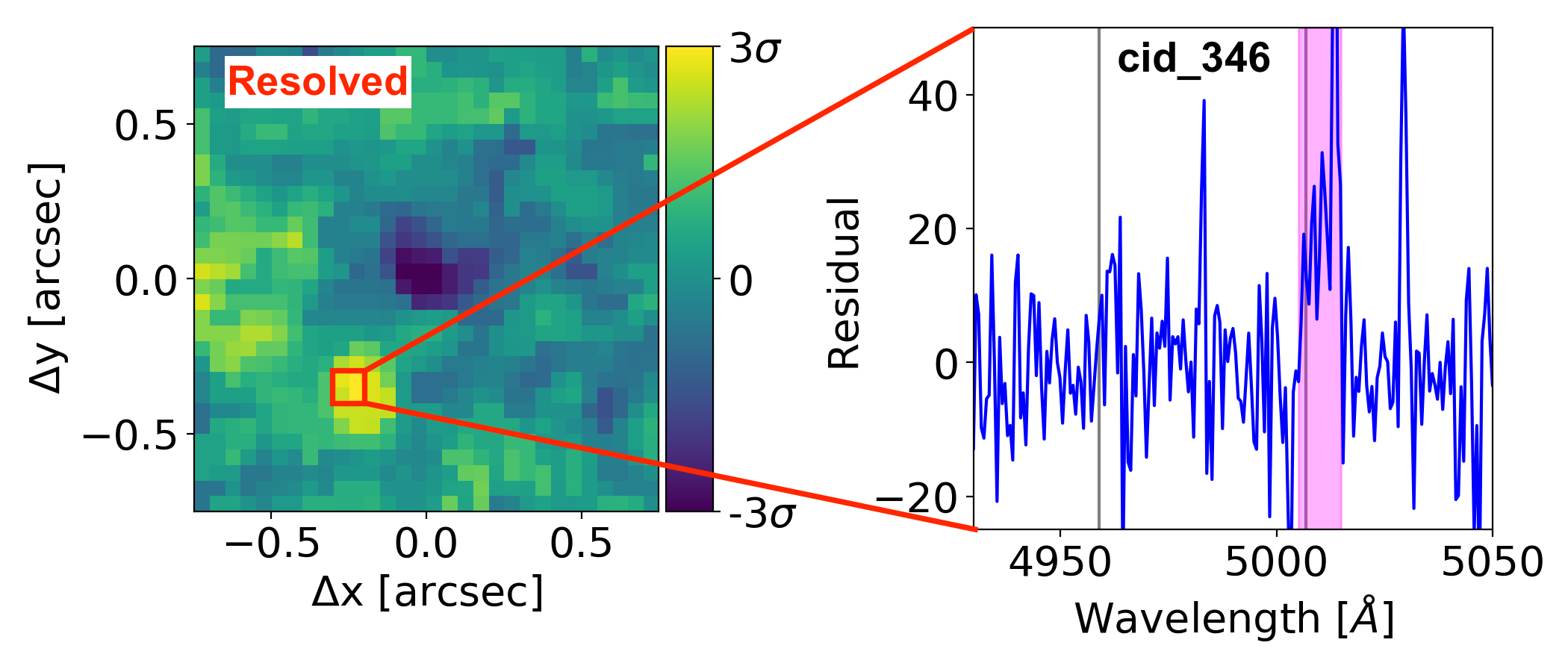}\\
\includegraphics[scale=0.33]{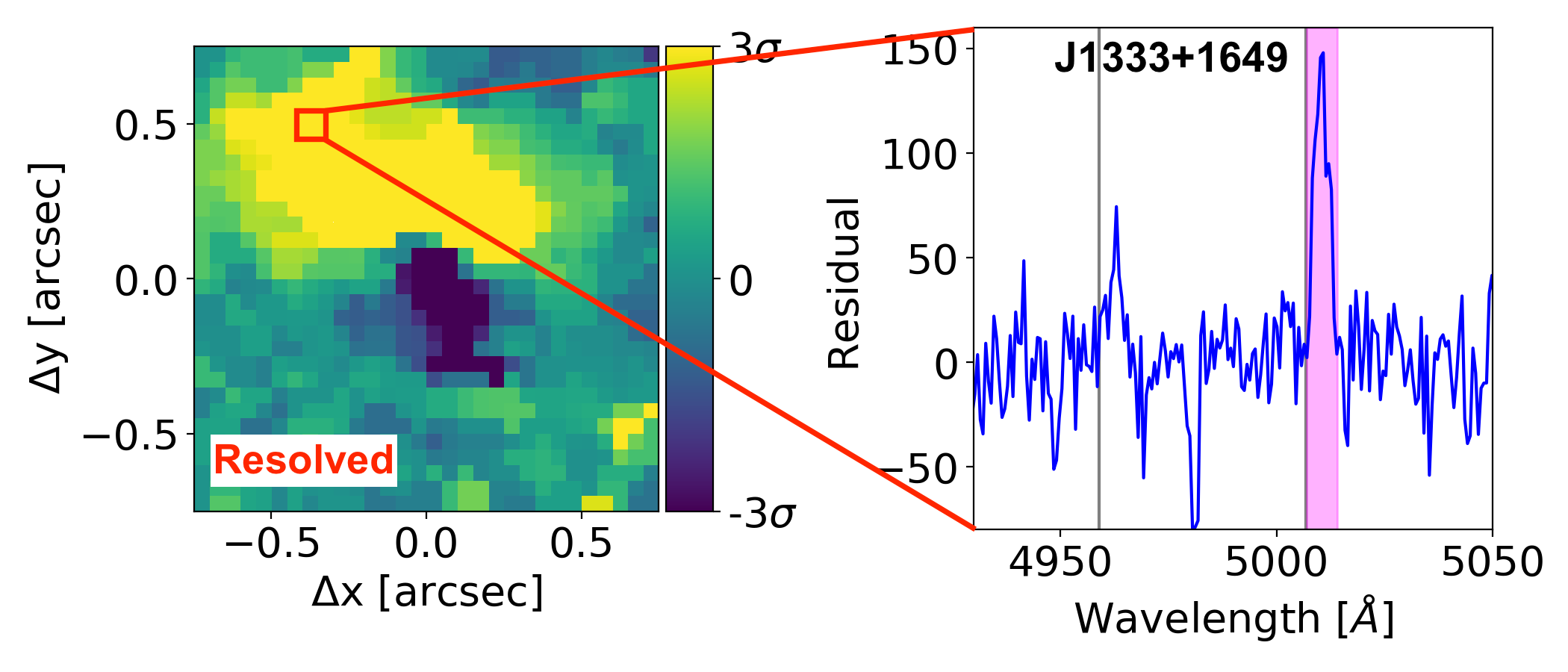}
\includegraphics[scale=0.33]{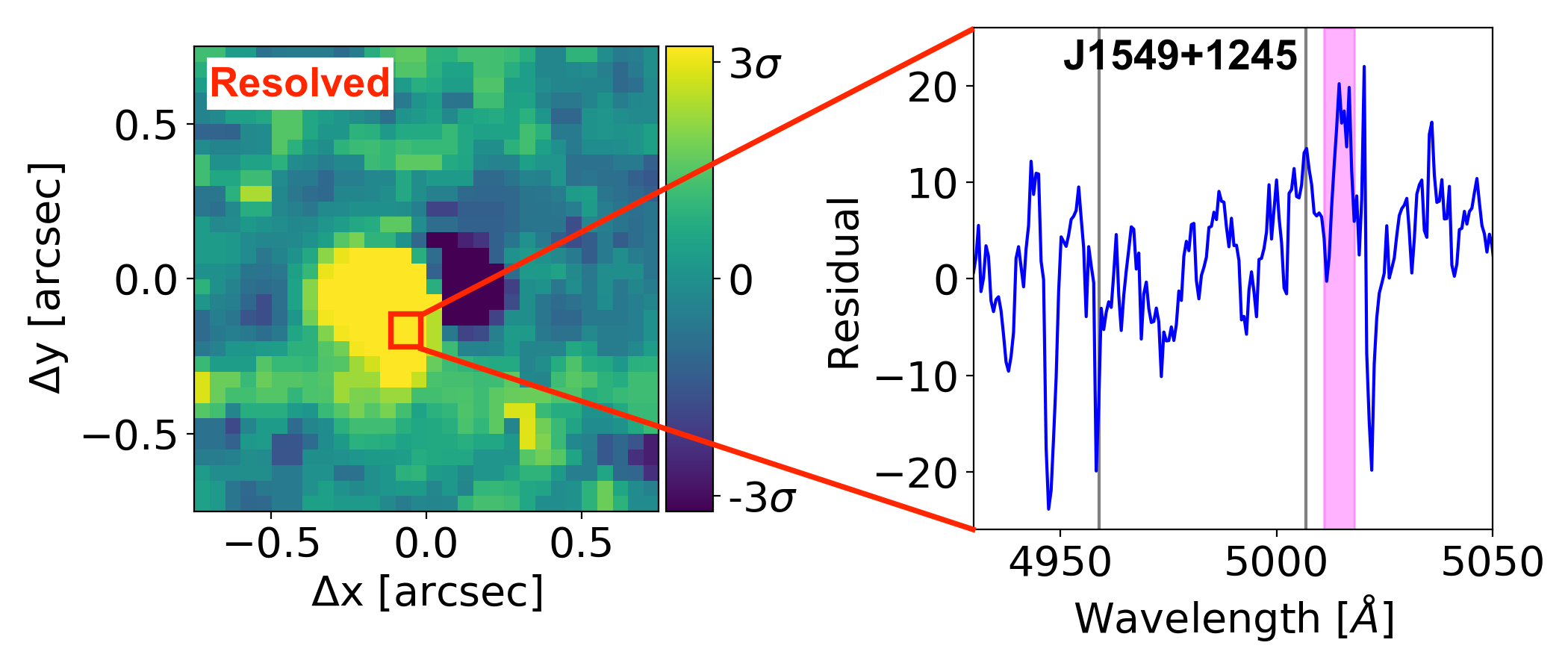}\\
\includegraphics[scale=0.33]{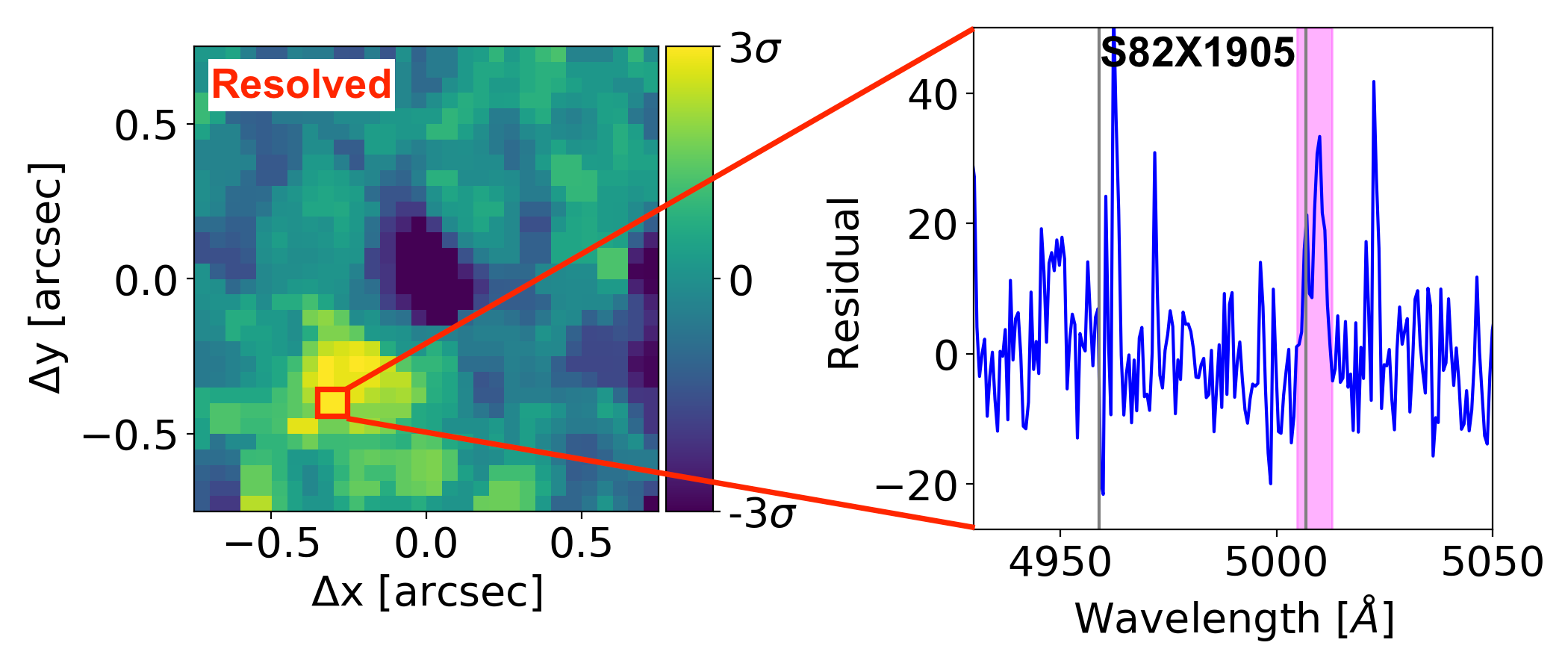}
\includegraphics[scale=0.33]{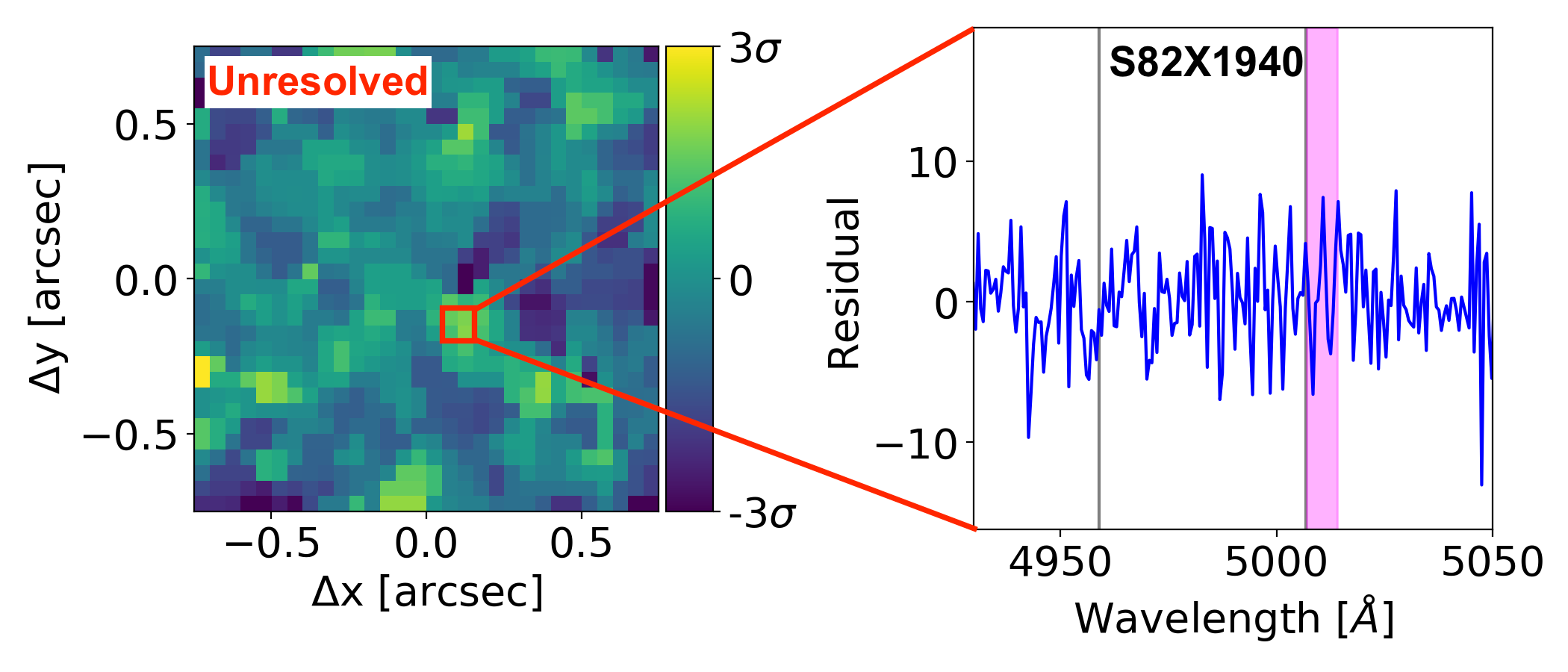}\\
\caption{Residuals obtained using the "PSF-subtraction" method across the SINFONI FoV as described in section \ref{sect5} for the Type-1 sample. The left panel shows the residual image obtained by collapsing the channels at the location of \oiii$\lambda$5007 after subtracting the emission from the AGN. The color map is on a linear scale with the yellow regions showing excess emission (i.e. $>3 \sigma$). North is up and East is to left. The right panel shows the spectrum extracted at the location of excess emission, with an aperture of 0.1$^{\arcsec}\times$0.1$^{\arcsec}$ shown by the red square in the left panel. The vertical grey lines in the residual spectrum shows the location of the \oiii$\lambda$4959 and \oiii$\lambda$5007 respectively, while the channels used to create the residual map on the left are indicated with a magenta box. Non-zero residuals in the images show that \oiii emission is resolved, while a noisy pattern is indicative of unresolved emission, which are respectively marked in the left panels. X\_N\_66\_23 is not included in this analysis due to non-detection of the BLR component.
\label{fig:astrometry1}}
\end{figure*}

Before discussing the kinematics of the ionized gas, we now focus on the actual extension and morphology of the emission line region. First, we need to verify that the \oiii emission is actually extended or rather unresolved and what it may erroneously be interpreted as extended emission being nothing else than beam smearing effects \citep[e.g.][]{husemann16, villar-martin16}. We use two techniques to asses if the emission is truly extended. The first consists in comparing the curve-of-growth (COG) of the total \oiii emission of the AGN with that of an observed and modeled PSF. The second technique, which we will call "PSF-subtraction" method, consists in producing a residual \oiii map after subtracting the nuclear \oiii spectrum spaxel-by-spaxel following a 2D PSF profile.

To construct the curves of growth, we start by extracting the spectrum from an aperture of radius 0.1$\arcsec$ centered at the peak location of the AGN continuum emission and perform the line fitting as described in Sect. \ref{sect5.1}. From the best fit we derive the flux value for the total \oiii emission, and repeat this procedure increasing the extraction radius in steps of 0.05$\arcsec$. We finally reconstruct the curve of growth plotting the line fluxes derived as a function of the aperture radius. We only plot the total \oiii flux and not the individual components in these plots as we do not give any physical significance to individual Gaussian components. The curves of growth are derived for 11 targets (out of the 21 in the Type-1 sample observed) for which the signal-to-noise in the integrated spectrum extracted from their respective apertures (Table \ref{table:SINFONI_properties1}) is greater than 5. The H-band data cubes of cid\_1205 and S82X2058 are contaminated by a bright stripe, due to which spatially resolved analysis of these targets is not performed.

We then need to have an accurate description of the PSF to compare with the \oiii emission. We had designed our SINFONI observations to have dedicated PSF star observations close in time and space for each single science observations, and we can therefore repeat the COG procedure described above for the AGN also for the dedicated PSF star, in order to derive the PSF profile as a function of the distance from the center. We were not able to construct the growth curve of PSF observation for J1333+1649 and J1549+1245 as the PSF star was at the edge of the data cube. Nevertheless, since the AGN in this paper are Type-1 we have access to the spatially unresolved BLR emission as traced by $\hb$ which we can use as an alternative method to trace the PSF profile (the only exception being X$\_$N$\_$66$\_$23 for which we do not have a clear detection of the $\hb$ line). Out of the eight objects for which we can use both methods to trace the PSF, both the dedicated PSF star and  the $\hb$ BLR component give consistent PSF profiles for five objects so we can use both to trace the spatial resolution of the data cubes. For the remaining three objects (X\_N\_115\_23, cid\_346, S82X1905) the PSF profile as traced by the dedicated star observation is narrower than those derived from the $\hb$ profile. These differences between the PSF traced by the BLR emission and the PSF star could be due to the difference in the AO correction, which might change in long exposure observations. Also, as the PSF star observations were performed at the beginning of the $\sim 1$ hour long science OBs, for these three objects the conditions got relatively worse during the science observations and therefore we consider the PSF profile derived from the BLR the correct one to compare since they trace the conditions simultaneously with the science observations. We finally estimate errors on the COGs by repeating the fitting procedure 100 times on mock spectra obtained by adding {\it rms} errors on an object-free location of the spectrum on the original models. The curves of growth for the 11 targets with a S/N$>5$ in each apertures are presented in Fig. \ref{fig:growth_curves}. 

We can now compare the COG of each single target with the PSF. As discussed above whenever possible we consider the PSF curve obtained from the unresolved BLR the best representations of the observing conditions during the science observations. From Fig. \ref{fig:growth_curves} we find that for 7 targets ($\approx$ 63\% of the targets for which this analysis has been performed) there is a 2$\sigma$ excess in the total \oiii emission compared to the PSF at the maximum radius sampled by the COG, and therefore we will claim that in these objects the \oiii is spatially resolved. As done also in previous works, we could compare the half light radii (the radius containing $50\%$ of the flux) of the \oiii and the PSF emission reported in Table \ref{table:outflow_properties} to asses the spatially resolved nature of the \oiii emission. The half light radii were estimated using linear spline to interpolate between the data points of COG. For five objects (X\_N\_66\_23, X\_N\_115\_23, cid\_346, J1549+1245 and S82X1905) the \oiii half light radius is larger than the PSF half light radius within $\approx$2$\sigma$. For the remaining targets the COG of the \oiii emission and the half light radius is consistent with the PSF within 2$\sigma$. In particular, for the two targets X\_N\_81\_44 and J1333+1649 the half-light radius method may not be sensitive to the fainter very extended emission beyond the PSF, contrary to the overall COG and PSF-method described below.\newline

As mentioned at the beginning of this section, we use a second method to asses the spatial extension of the ionized gas emission, namely the ``PSF-subtraction" method as described in \citet{carniani15}. The basic principle behind the PSF-subtraction method is that if the observed emission is unresolved and/or is a result of beam smearing from the PSF of the AGN, the spectrum at any given distance from the location of the AGN is the same as the spectrum at the location of the AGN, except for a scaling factor across the spectrum \citep[e.g.][]{jahnke04}. In this case, the BLR component of $\hb$ smears the emission across the FoV and is therefore AGN dominated. If the emission does not mimic the spectrum at the location of AGN, then the observed spectral line is resolved. Following this principle for our sample, we first model the spectrum extracted at the AGN location (within a radius of 0.1$\arcsec$), determined using the center of the continuum in the H-band cube. We will refer to this spectrum as the ``nuclear model'' which we then subtract from every pixel across the SINFONI FoV, after allowing a variation in the overall normalization factor of the spectrum while keeping the rest of the kinematic components fixed. This is followed by creation of channel maps at the location of \oiii$\lambda$5007 emission in the nuclear-spectrum-subtracted cube. The channels used to create the residual maps were 6$\AA$ wide, except cid\_346 for which a channel width of 10$\AA$ was used. Following the principle described above, a noisy residual map will indicate an unresolved emission, while a map with structures having non-zero residual will indicate a resolved emission. Similar to the case of the COG, we restrict this analysis only to targets which have a S/N larger than 5 in the integrated spectrum (extracted from the respective  apertures reported in Table \ref{table:SINFONI_properties1}). We also restrict the analysis to targets which show BLR $\hb$ emission which is used to scale the overall normalization of the nuclear spectrum. Since we do not detect a BLR component in X\_N\_66\_23, we perform the PSF-subtraction analysis for 10 SUPER targets. The results of this method are shown in Fig. \ref{fig:astrometry1}. The left panels in these figures show the residual maps obtained after collapsing the nuclear-spectrum subtracted cubes along the \oiii channels, while the right panels show the residual spectrum extracted at the region of excess emission. From Fig. \ref{fig:astrometry1}, we see excess residuals at a level of >2$\sigma$ for 6 SUPER targets - X\_N\_81\_44, X\_N\_115\_23, cid\_346, J1333+1649, J1549+1245 and S82X1905. 

In summary, from the joint analysis based on the COG and "PSF-subtraction" method we have identified seven sources from our sample which show extended \oiii emission. Further analysis on the spectra across the SINFONI FoV will therefore be restricted to the 7 targets, which show extended gas emission with at least one of the methods described above.

\subsection{Modeling ionized gas kinematics across the FoV} \label{sect5.3}

In the following section, we describe the pixel-by-pixel analysis for the H-band raw data cubes of the 7 targets which are resolved from the methods described in Sect. \ref{sect5.2}. For the modeling of the spectrum across every pixel, we use the parameters obtained in the integrated spectrum as a prior. Apart from the constraints set while modeling the integrated spectrum, we also fix the centroid and the width of the unresolved BLR component of $\hb$ and only allowed a variation in its peak. In the H-band, the Gaussian parameters of other emission lines are allowed to vary. The emission line widths (FWHM) were also kept greater than $\sim$100 km/s to avoid any spurious fit to a sky line. The validity of the line fitting across each spaxel was checked by subtracting the emission line model from the raw data and making a collapsed map along the channels spanned by the \oiii$\lambda$5007 emission. Due to the low S/N in each spaxel in case of X\_N\_66\_23 and cid\_346, nearby spaxels were averaged within a radius of 0.1$\arcsec$ to produce binned spectra. This results in a higher S/N of the spaxels without compromising the respective resolution of the observations.

\begin{figure*}
\centering
\subfloat{\includegraphics[scale=0.53]{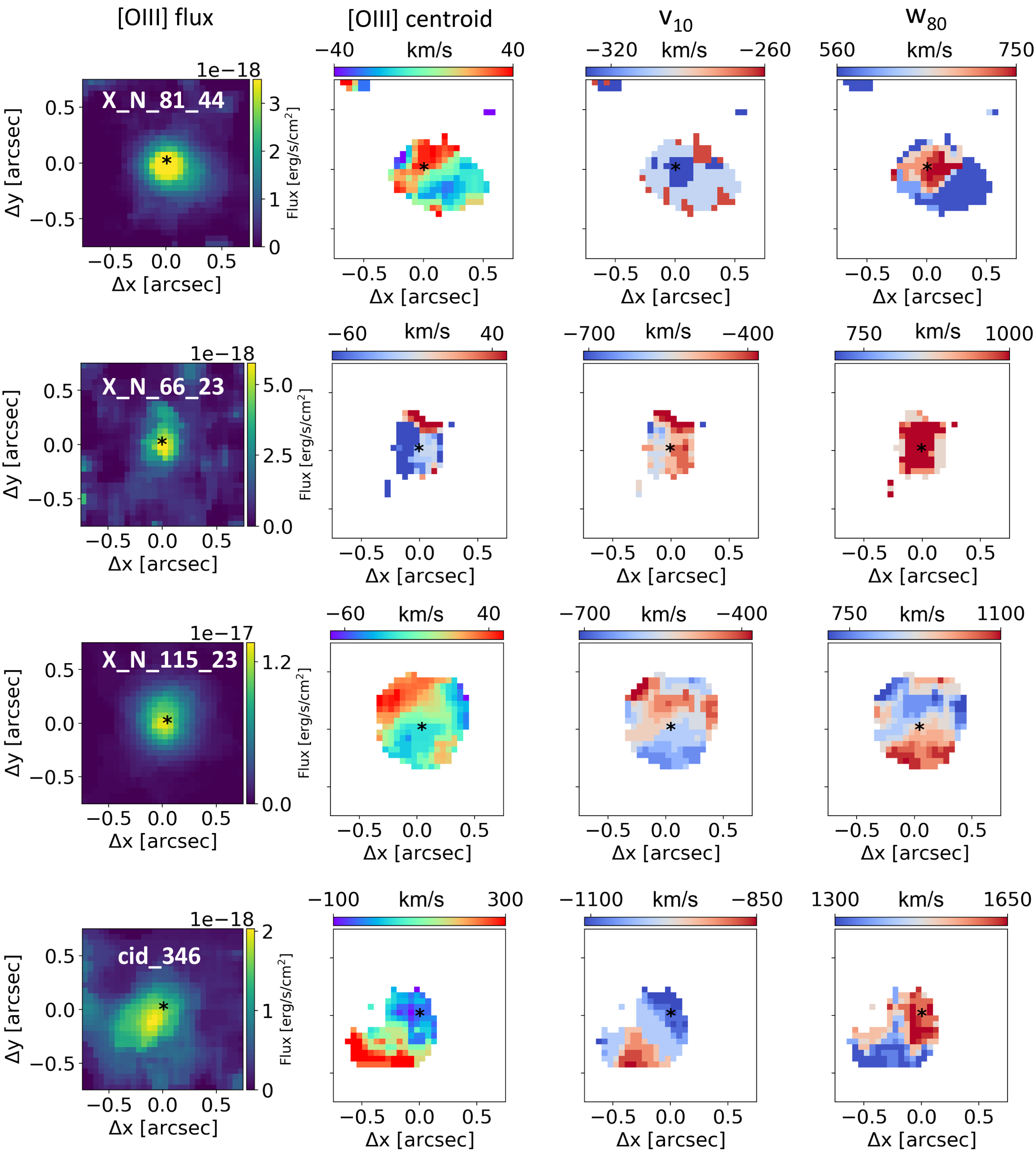}}\\
\caption{Flux and velocity maps of the targets showing resolved/extended emission inferred from the curve-of-growth and PSF-subtraction method (From top to bottom: X\_N\_81\_44, X\_N\_66\_23, X\_N\_115\_23 and cid\_346). {\it Column 1}: Background image shows the flux map of total \oiii emission (narrow+broad).; {\it Column 2}: \oiii line centroid map; {\it Column 3}: $\mathrm{v_{10}}$ map and {\it Column 4}: $\mathrm{w_{80}}$ map. In all the maps, the black star denotes the center position of the H-band continuum, as a proxy for AGN location. North is up and East is to left. \label{fig:flux_velocity1}}
\end{figure*}

\begin{figure*}
\centering
\subfloat{\includegraphics[scale=0.53]{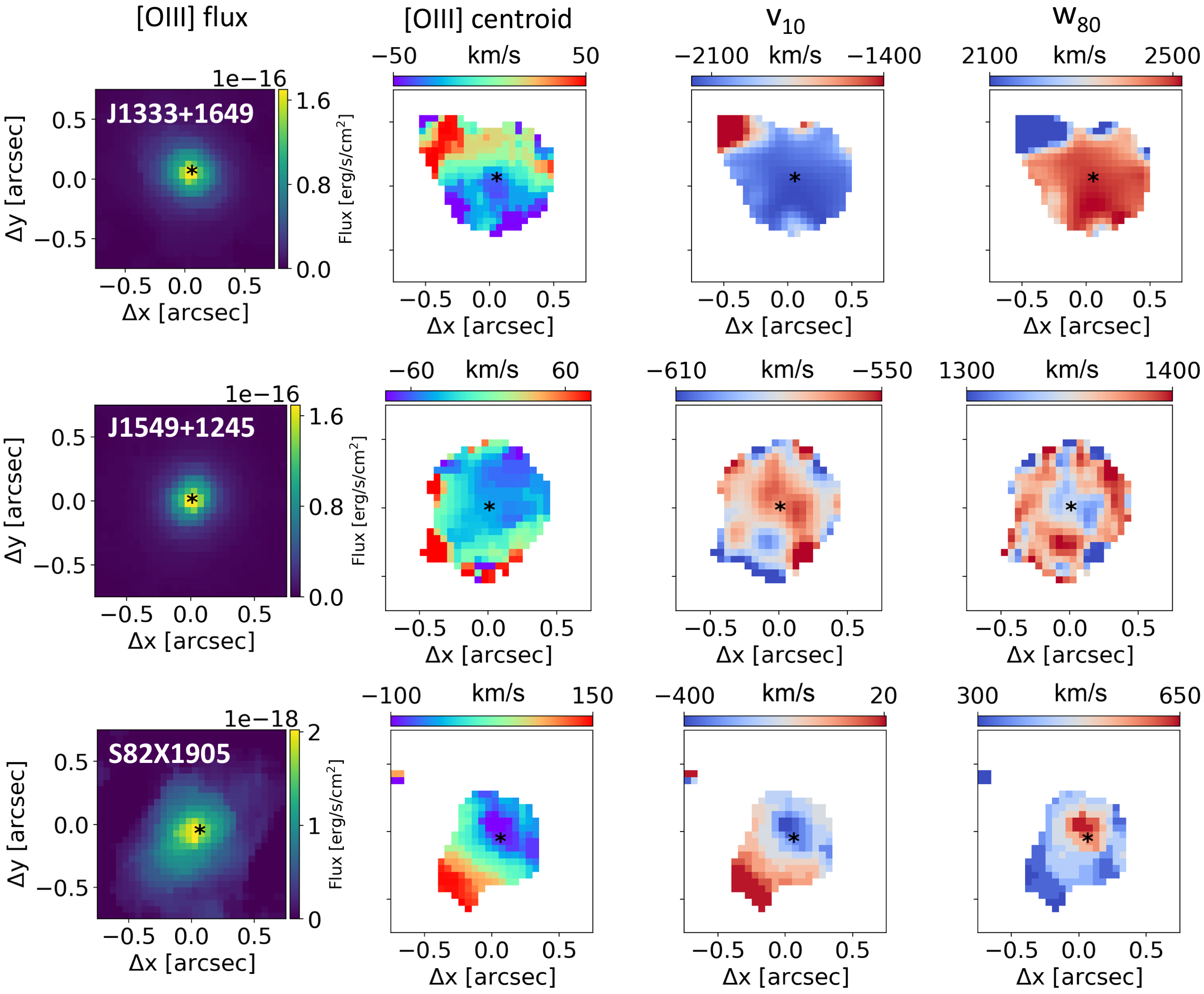}}\\
\caption{Same as Fig. \ref{fig:flux_velocity1} for J1333+1649, J1549+1245 and S82X1905. \label{fig:flux_velocity2}}
\end{figure*}

As a result of the pixel-by-pixel analysis, various line and continuum maps are created which we describe below. These maps are shown in Figs. \ref{fig:flux_velocity1} and \ref{fig:flux_velocity2}. All the maps are smoothed to the spatial resolution of the observations derived from the PSF and a S/N cut of 2 is employed in the \oiii profile. First, we created broad band continuum maps, masking the emission lines, and used a 2-D Gaussian fit to determine the photo-centroid of the resulting image which we will use to mark the AGN position.

The total \oiii line flux maps are shown in the first column of Figs. \ref{fig:flux_velocity1} and \ref{fig:flux_velocity2}, while the line centroid maps obtained from the peak of the total \oiii profile are presented in the second column. A smooth gradient is observed in some of the line centroid maps, e.g. in X\_N\_81\_44, X\_N\_66\_23, cid\_346 and S82X1905, which could be  indicative of a rotating disk in the host galaxy or a bi-polar ionized outflow. The observed extreme high velocities in the $\mathrm{w_{80}}$ maps ($\sim$1000 km/s) for most of these galaxies suggest that the kinematics of the ionized gas are also likely affected by the central AGN. For several of our targets, we do detect extended features in emission, which are not produced by wings of the PSF as demonstrated by the analysis performed in the previous sections (e.g. X\_N\_115\_23, CID\_346, J1333+1649, J1549+1245, S82X1905). A possible explanation for these irregular features in the gas kinematics could be the presence of a merger event between two galaxies. As far as a major merger is concerned, we do not see clear evidence of multiple galaxies in the available ancillary data. We would, therefore, favor alternative scenarios in which the extended features could be interpreted as an inflow or outflow. Inflows are usually detected in absorption and are predicted to show lower velocities than what we measure for the SUPER sample \citep[e.g.][]{bouche13}. Also, since models predict a low covering fraction for the inflows \citep[e.g.][]{steidel10}, we are inclined to interpret these features as outflows. In order to further investigate the kinematic features of the ionized gas in these AGN, we also generated the maps with the spatial variation of $\mathrm{v_{10}}$ and $\mathrm{w_{80}}$ which are shown in the third and the fourth column in Figs. \ref{fig:flux_velocity1} and \ref{fig:flux_velocity2}. The inspection of both maps reinforces our conclusion that most of our objects do not present the kinematic signatures of an undisturbed rotating disk: the $\mathrm{v_{10}}$ map shows regions with extremely blue shifted velocity of more than -1000 km/s with $\mathrm{w_{80}}$ that peaks at the location of maximum velocity shift. Although a proper modeling of the velocity field of these galaxies is beyond the scope of this paper, we will discuss here the morphology and size of the outflows signatures.

Following the discussion of the integrated spectra in Sec 5.1, we will use the cut $\mathrm{w_{80}}>$600 km/s to identify regions dominated by outflowing ionized gas. Projected velocities (e.g. $\mathrm{v_{10}}$ maps in the third column in Figs. \ref{fig:flux_velocity1} \& \ref{fig:flux_velocity2}) will be more sensitive to the galaxy inclination while the $\mathrm{w_{80}}$ maps are more likely to be less sensitive to inclination (e.g. see discussion in \citealt{harrison12}). For each source, we measure the maximum projected spatial extent of the $\mathrm{w_{80}}>$600 km/s component of the \oiii line profile (we will name this quantity D$_{600}$, following \citealt{harrison14}). From the $\mathrm{D_{600}}$ values in the $\mathrm{w_{80}}$ maps we find that the outflows are extended from $\sim$2 kpc up to 6 kpc (see Table \ref{table:outflow_properties}).

\begin{table*}
\centering
\caption{Properties of the ionized gas obtained from the analysis of the curves-of-growth, spectroastrometry and the flux and velocity maps. J1441+0454 has not been mentioned in this table due to the lack of \oiii emission.\label{table:outflow_properties}}
\begin{tabular}{cccccccccc}
\hline 
Target & $\mathrm{r_{1/2}^{\oiii}}$ & $\mathrm{r_{1/2}^{PSF}}$ & $\mathrm{v_{o}}$  & $\mathrm{R_{o}}$ & $\mathrm{D_{600}}$ & $\mathrm{L_{\oiii}^{o}}$ & $\mathrm{\dot{M}_{cone}}$ & $\mathrm{\dot{M}_{thin-shell}}$ & $\mathrm{f_{esc}}$\\
& (1) & (2) & (3) & (4) & (5) & (6) & (7) & (8) & (9)\\
& kpc & kpc & km/s & kpc & kpc & erg/s & $\mathrm{M_{\odot}/yr}$ & $\mathrm{M_{\odot}/yr}$ & \%\\
\hline\hline
X\_N\_160\_22 & 1.54$\pm$0.02 & 1.52$\pm$0.02 & -1700 & 0.4 & -- & 43.36 & 4--79 & 5--105& 38\\
X\_N\_81\_44  & 1.84$\pm$0.16 & 1.61$\pm$0.05 & +500 & 0.9 & 2.4 & 42.26 & 0.1--1 & 0.1--2 & 6\\
X\_N\_66\_23  & 1.29$\pm$0.12 & 1.03$\pm$0.02 & 600  & 0.4 & 2.4 & 42.88 & 0.4--7 & 0.5--10 & 4\\
X\_N\_35\_20**  & --            & --            & 681   & <2.2 & -- & 41.31 & 0.01--0.2 & 0.01--0.2 & 0\\
X\_N\_12\_26  & 1.08$\pm$0.07 & 1.14$\pm$0.01 & -400 & 0.3 & -- & 42.06 & 0.1--1 & 0.1--2 & 2\\
X\_N\_4\_48*  & -- & -- & 1198 & <2.9 & -- & 41.97 & 0.1--1 & 0.1--2 & 5\\
X\_N\_102\_35*& -- & -- & 1501 &  <7.6 & -- & 42.68 & 0.4--9 & 0.6--12 & 16\\
X\_N\_115\_23 & 1.67$\pm$0.04 & 1.49$\pm$0.04 & +900 & 0.9 & 4.0 & 43.09 & 0.3--6 & 0.4--8 & 7\\
cid\_166      & 1.42$\pm$0.03 & 1.42$\pm$0.02 & -2100 & 0.7 & -- & 43.10 & 1--27 & 2--36 & 30\\
cid\_1605*    & -- & -- & 1153 &  <5.8 & -- & 42.15 & 0.1--2 & 0.1--3 & 4\\
cid\_346      & 2.12$\pm$0.07 & 1.90$\pm$0.05 & +600 & 2.8 & 5.6 & 42.83 & 0.6--11 & 0.8--15 & 27\\
cid\_1205$^{\dagger}$ & -- & -- & 717 & <2.5 & -- & 42.08 & 0.05--1 & 0.1--1 & 0\\
cid\_467*     & -- & -- & 1368 & <9.0 & -- & 42.63 & 0.4--7 & 0.5--10 & 8\\
J1333+1649    & 1.76$\pm$0.03 & 1.74$\pm$0.05 & -2900 & 0.3 & 6.5 & 44.54 & 51--1021 & 68--1361 & 43\\
J1441+0454** & -- & -- & 2161 & <2.8 & -- & 43.41 & 3--68 & 4--91 & 97\\
J1549+1245    & 1.35$\pm$0.03 & 1.25$\pm$0.03 & +1300 & 0.2 & 4.0 & 44.29 & 16--326 & 22--435 & 12\\
S82X1905      & 1.86$\pm$0.04 & 1.73$\pm$0.01 & +300 & 2.7 & 1.6 & 42.21 & 0.04--0.8 & 0.1--1 & 0\\
S82X1940      & 1.23$\pm$0.03 & 1.29$\pm$0.04 & -600 & 0.6 & -- & 42.35 & 0.2--3 & 0.2--4 & 8\\
S82X2058$^{\dagger}$     & -- & -- & 1340 & <2.2 & --  & 42.54 & 0.3--6 & 0.4--8 & 13\\
\hline
\end{tabular}
\vspace{1ex}\newline
{\raggedright{\bf Notes}:
(1) \& (2): $\mathrm{r_{1/2}^{\oiii}}$ and $\mathrm{r_{1/2}^{PSF}}$ are the half-light radius of the BLR PSF and \oiii emission respectively, derived from the curve-of-growth analysis for 11 targets with S/N> 5 in the integrated spectrum of \oiii, as described in Sect. \ref{sect5.2}. $^{**}\mathrm{r_{1/2}^{\oiii}}$ is not computed for X\_N\_35\_20 and J1441+0454 as the \oiii S/N is low. Targets marked by * were observed without AO and are not resolved. H-band data cubes of cid\_1205 and S82X1058 (marked by $^{\dagger}$) have a bright stripe which interferes in the spatially resolved analysis and therefore the half-light radius have not been reported (see Sect. \ref{sect5.1}). \newline 
(3) \& (4): $\mathrm{v_{o}}$ is the outflow velocity of the bulk of the ionized gas, which is at the maximum distance, $\mathrm{R_{o}}$ , from the AGN. $\mathrm{v_{o}}$ and $\mathrm{R_{o}}$ are derived from the spectroastrometry analysis of the 11 targets with S/N>5 in the integrated spectrum of \oiii, as described in Sect. \ref{sect5.4}. For targets marked by *, ** or $^{\dagger}$, we report the $\mathrm{w_{80}}$ value of the integrated spectrum and the PSF value as a proxy for $\mathrm{v_{o}}$ and $\mathrm{R_{o}}$.\newline
(5): D$_{600}$ is the maximum projected spatial extent of $\mathrm{w_{80}}$ > 600 km/s from the maps in Figs. \ref{fig:flux_velocity1} and \ref{fig:flux_velocity2}. D$_{600}$ is calculated for 7 targets which show extended ionized gas emission from COG and PSF-subtraction methods.\newline 
(6): $\mathrm{L_{\oiii}^{o}}$ is the outflow luminosity computed from \oiii$\lambda$5007 emission line for channels with |v|>300 km/s (Sect. \ref{sect5.5}).\newline
(7) \& (8): $\mathrm{\dot{M}_{cone}}$ and $\mathrm{\dot{M}_{thin-shell}}$ are the masss outflow rates assuming a bi-conical and a thin-shell geometry described in Sect. \ref{sect5.5}. The two values in each column correspond to the outflow rates assuming an electron density of 10$^{4}$ cm$^{-3}$ and 500 cm$^{-3}$.\newline
(9): $\mathrm{f_{esc}}$ is the fraction of the outflowing gas which has the capability to escape the gravitational potential of the host galaxy (Sect. \ref{sect6}).\par}
\end{table*}

From the $\mathrm{w_{80}}$ maps in Figs. \ref{fig:flux_velocity1} and \ref{fig:flux_velocity2}, it is clear that the ionized outflows detected in these AGN present a diversity of projected structures. In five sources (X\_N\_81\_44, X\_N\_115\_23, cid\_346, S82X1905) there is a velocity gradient along a particular axis while for two objects (J1333+1649 and J1549+1245) the maps are more spherically symmetric. Some of the kinematic structures present in the $\mathrm{w_{80}}$ maps could be consistent with a bi-conical outflow, where the inclination of the outflow with respect to the disk of the galaxy and the line-of-sight of the observer could explain the observed velocity maps: i.e. a high inclination of the outflow with respect to the line-of-sight will result in a clear velocity gradient along the outflow axis, with possibly both the blue-shifted and red-shifted part of the outflowing gas detected; a low inclination of the outflow with respect to the line of sight would produce a more uniform and spherically symmetric velocity field. A proper kinematic modeling would be required to constraint the ionized gas kinematics, and we will present this analysis in an upcoming paper on the full survey sample.

\subsection{Radial distance of high-velocity ionized gas using Spectroastrometry} \label{sect5.4}

The analysis presented in section \ref{sect5.2} is useful in determining the presence or absence of extended ionized gas. However, the methods described there do not determine the bulk velocity of the observed extended gas i.e. if the extended component of the gas is in outflow. Conventionally, the velocity maps derived in section \ref{sect5.3} have been used to quantify the extension of the outflowing gas. These maps can only be used for targets which are well extended beyond the width (FWHM) of the observed PSF. For marginally resolved or unresolved targets, the velocity maps can be affected by PSF smearing. For the latter case, the {\it spectroastrometry} technique \citep{carniani15} is useful to determine the radial distance of the bulk\footnote{Bulk here should be interpreted in a luminosity weighted sense, namely we refer to the location where most of the emitted luminosity is produced.} of the gas moving at certain velocity from the AGN location.

\begin{figure*}
\centering
\subfloat{\includegraphics[scale=0.35]{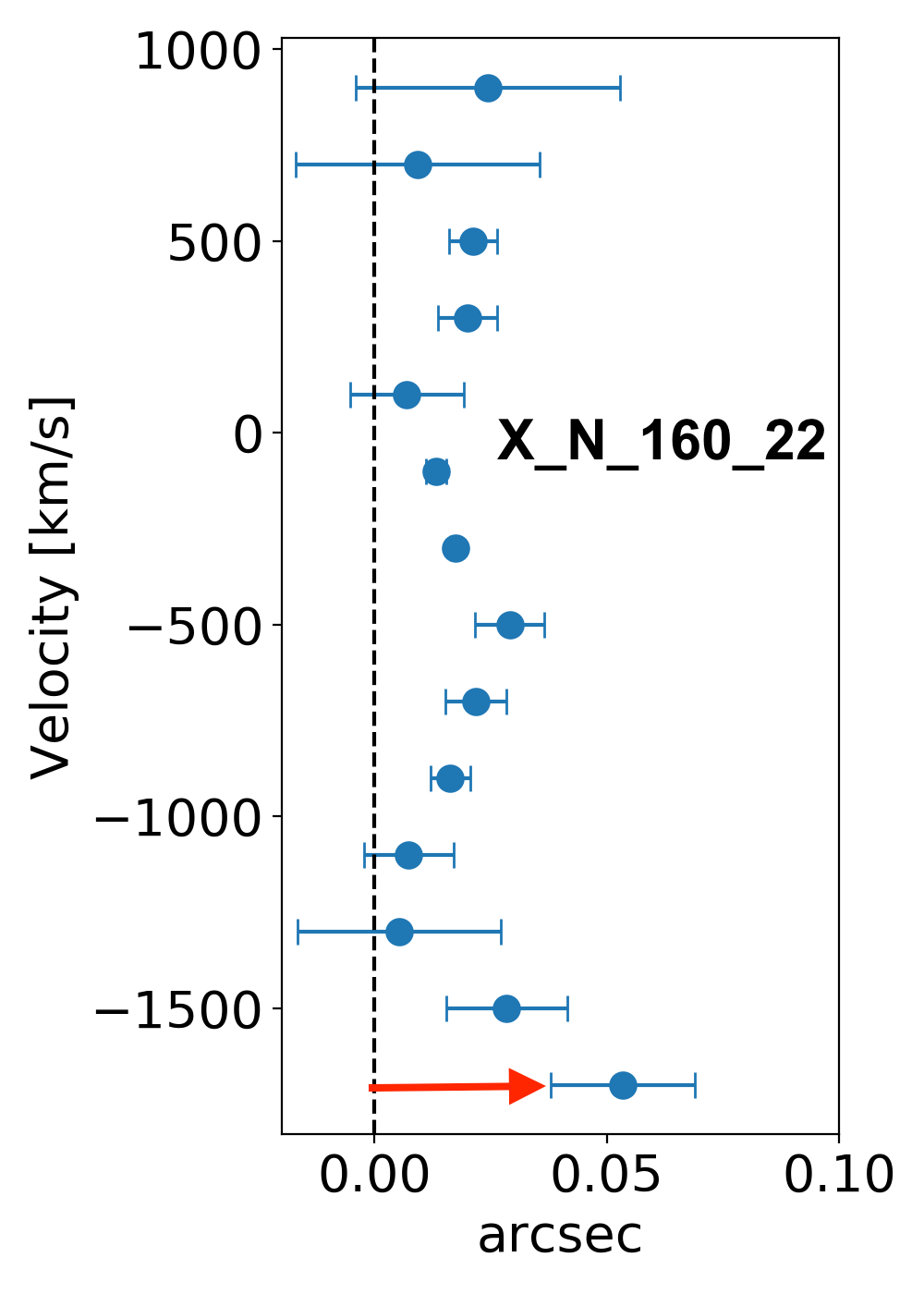}}
\subfloat{\includegraphics[scale=0.35]{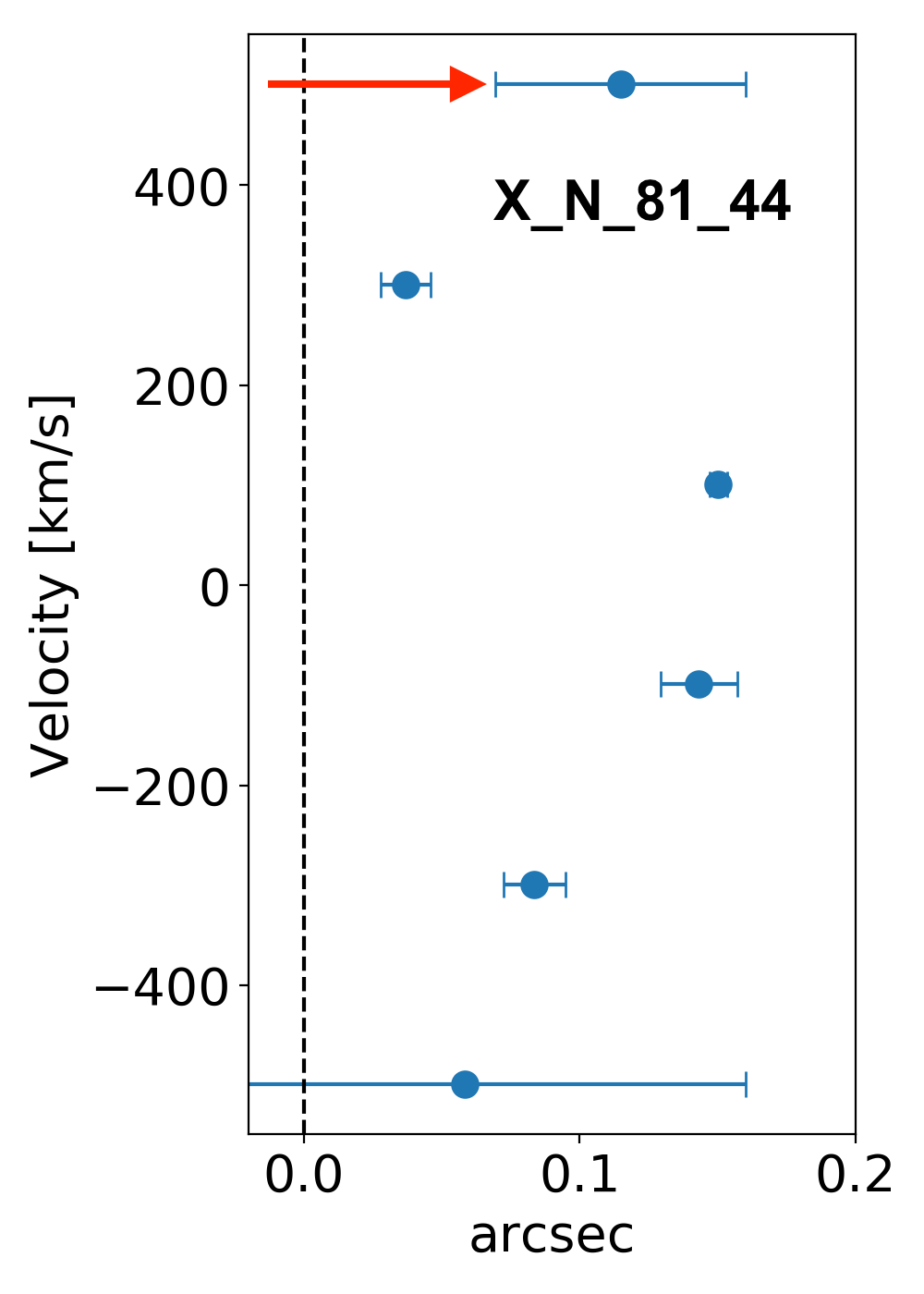}}
\subfloat{\includegraphics[scale=0.35]{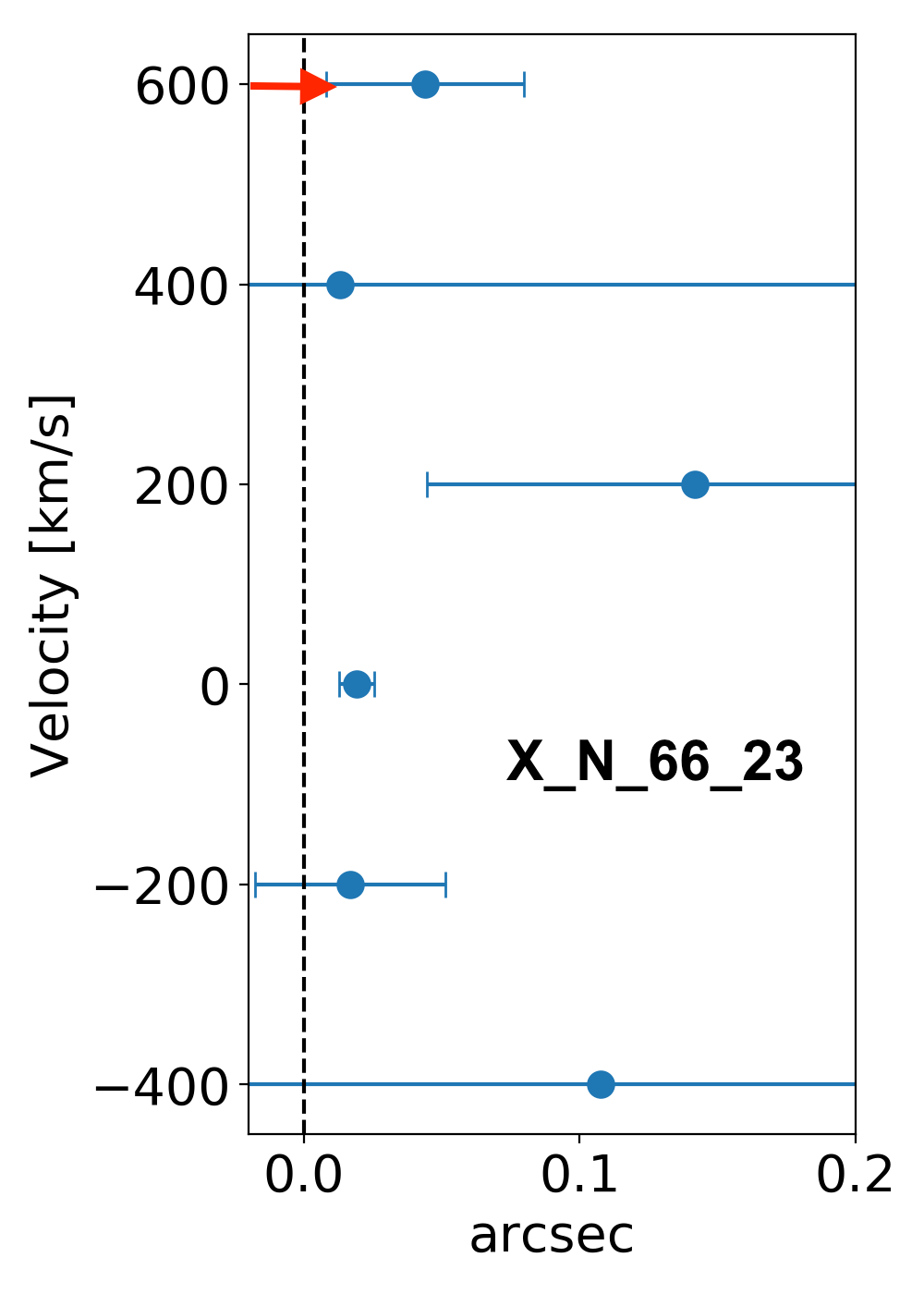}}
\subfloat{\includegraphics[scale=0.35]{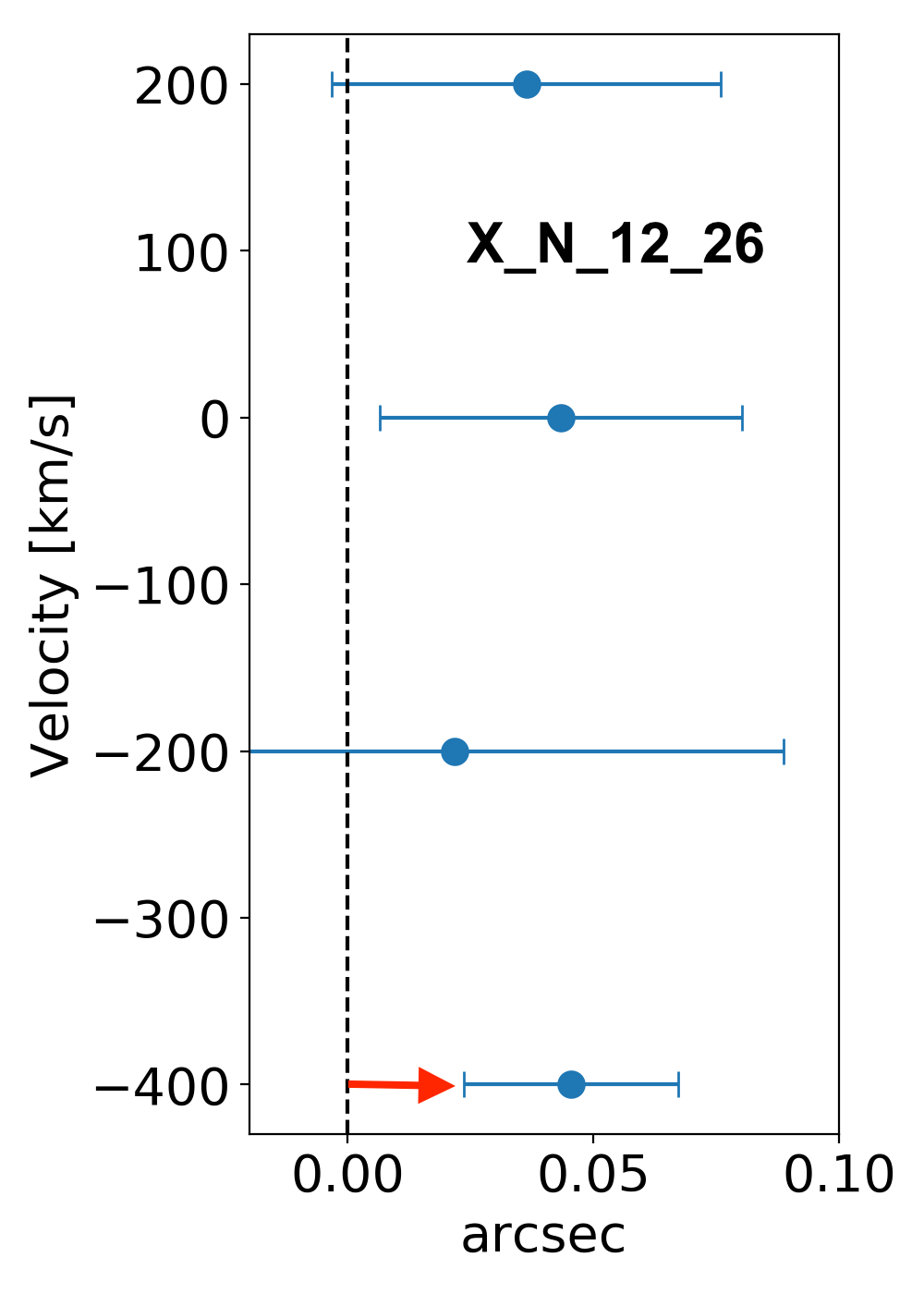}}\\
\subfloat{\includegraphics[scale=0.35]{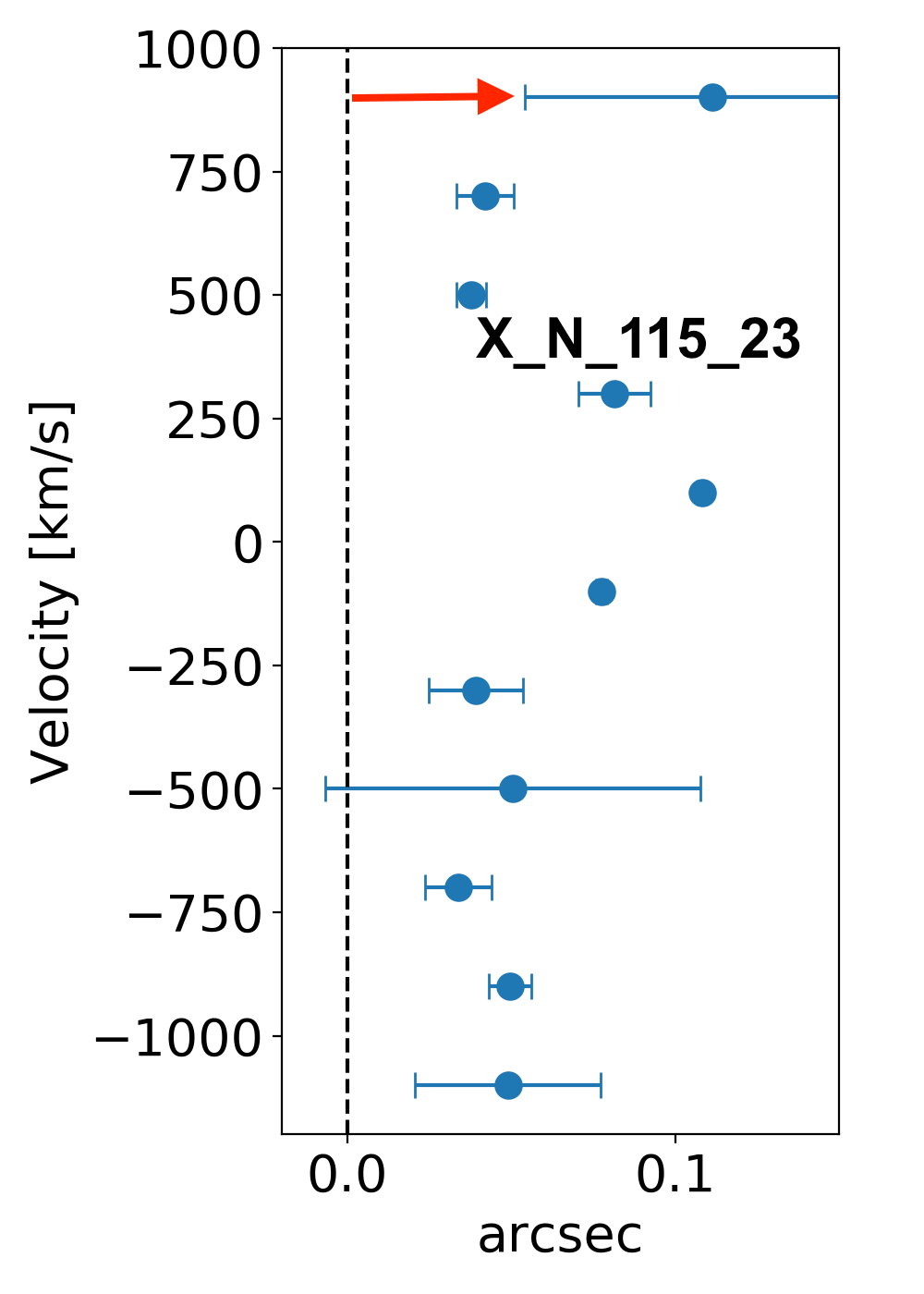}}
\subfloat{\includegraphics[scale=0.35]{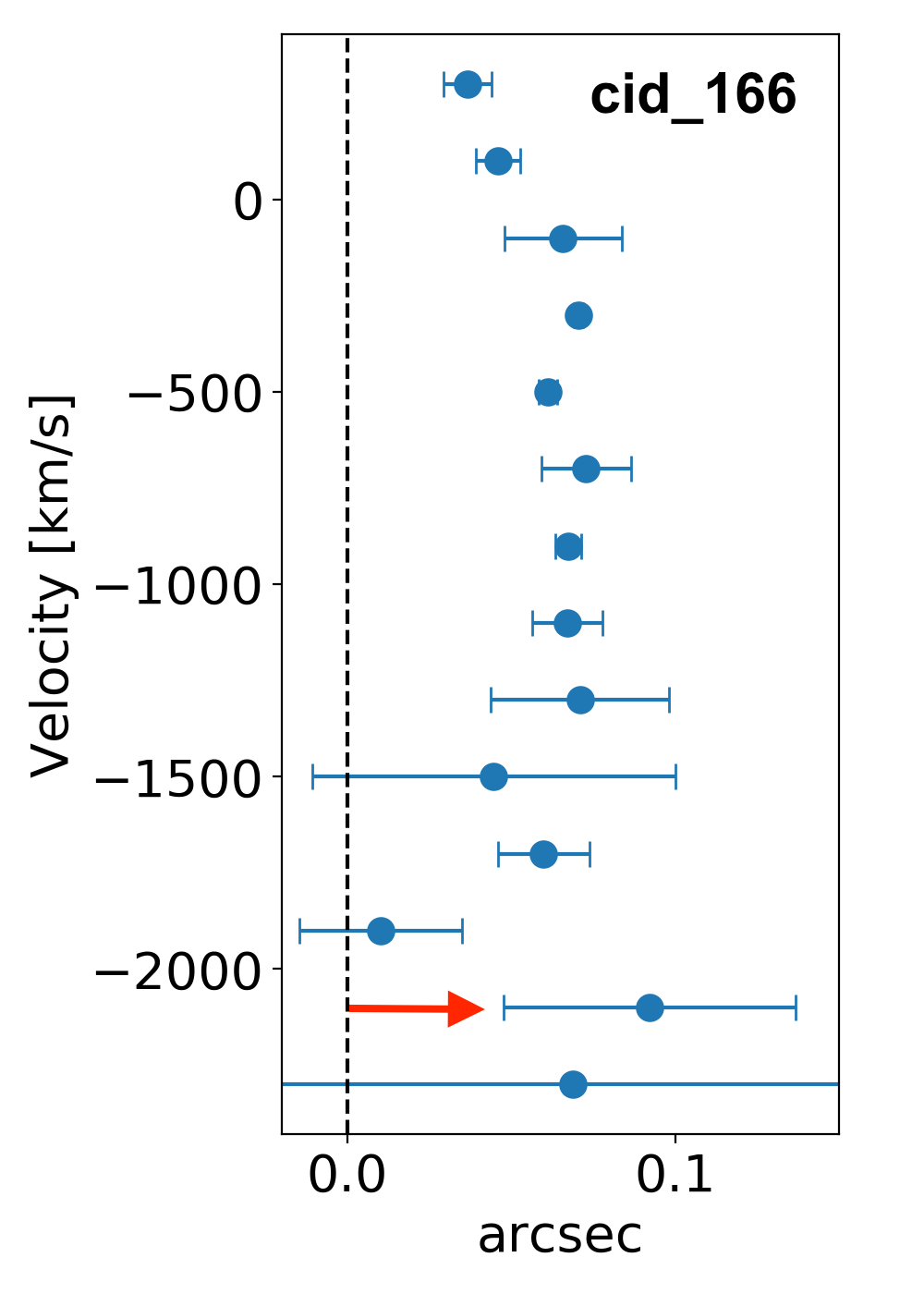}}
\subfloat{\includegraphics[scale=0.35]{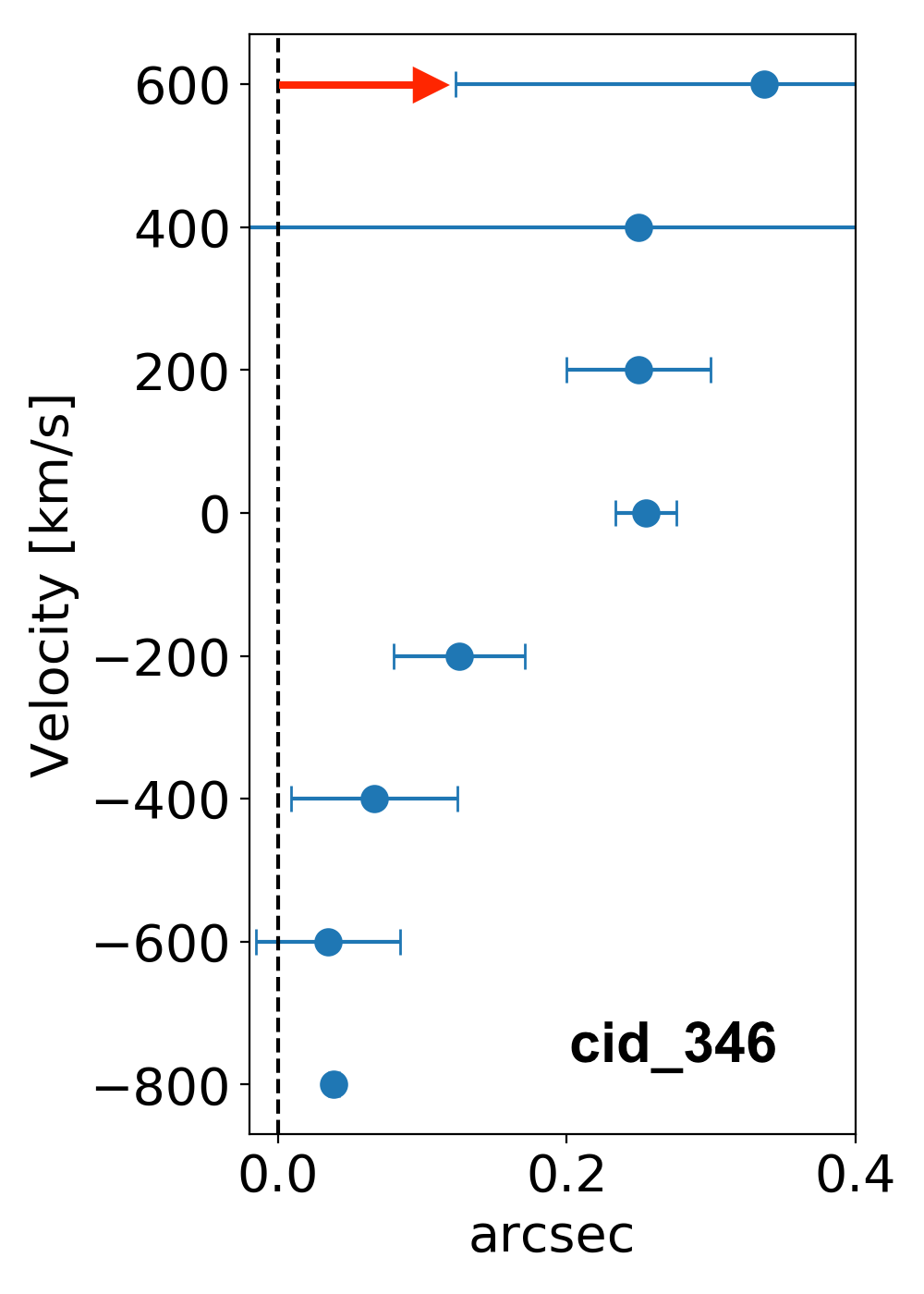}}
\subfloat{\includegraphics[scale=0.35]{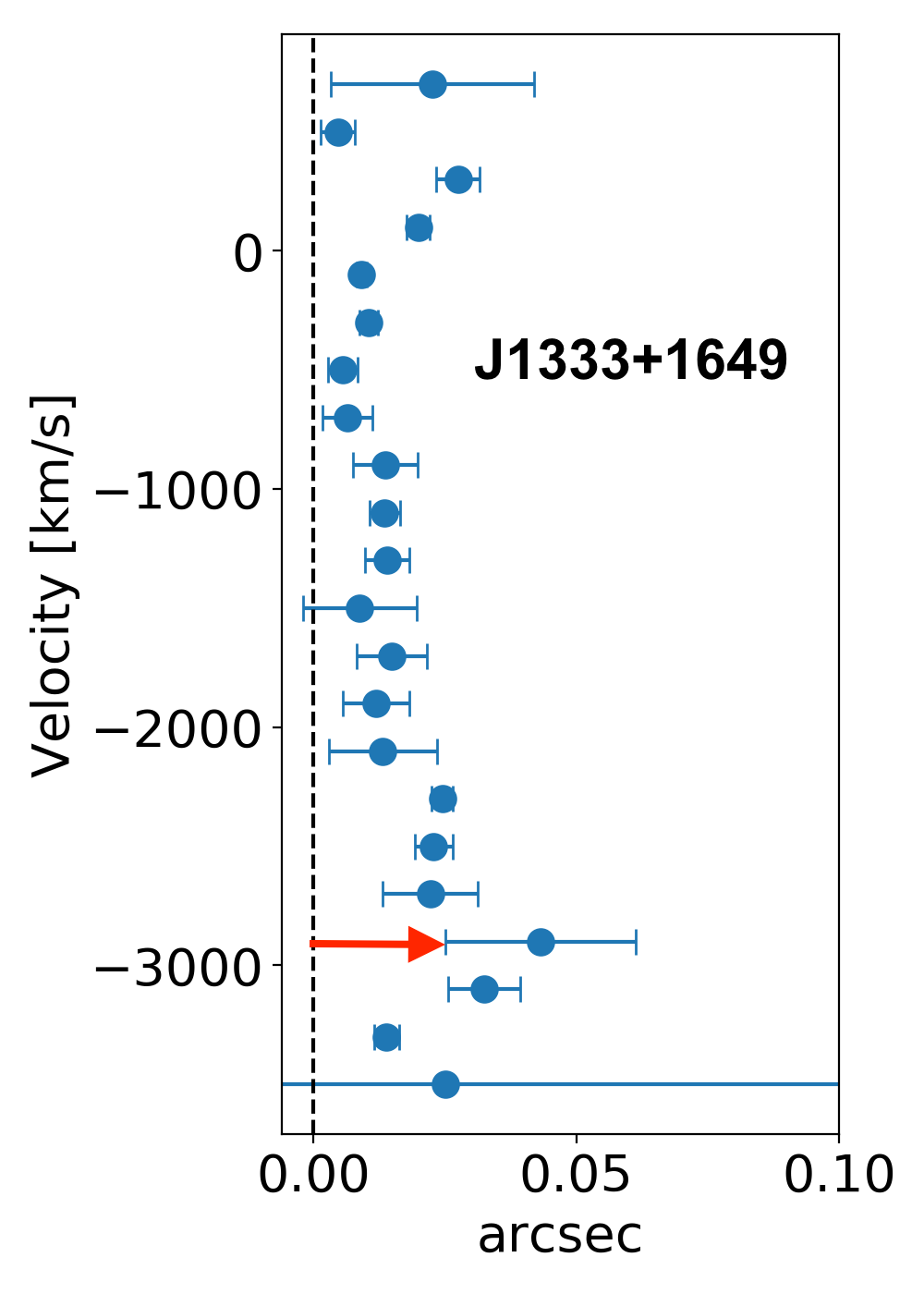}}\\
\subfloat{\includegraphics[scale=0.35]{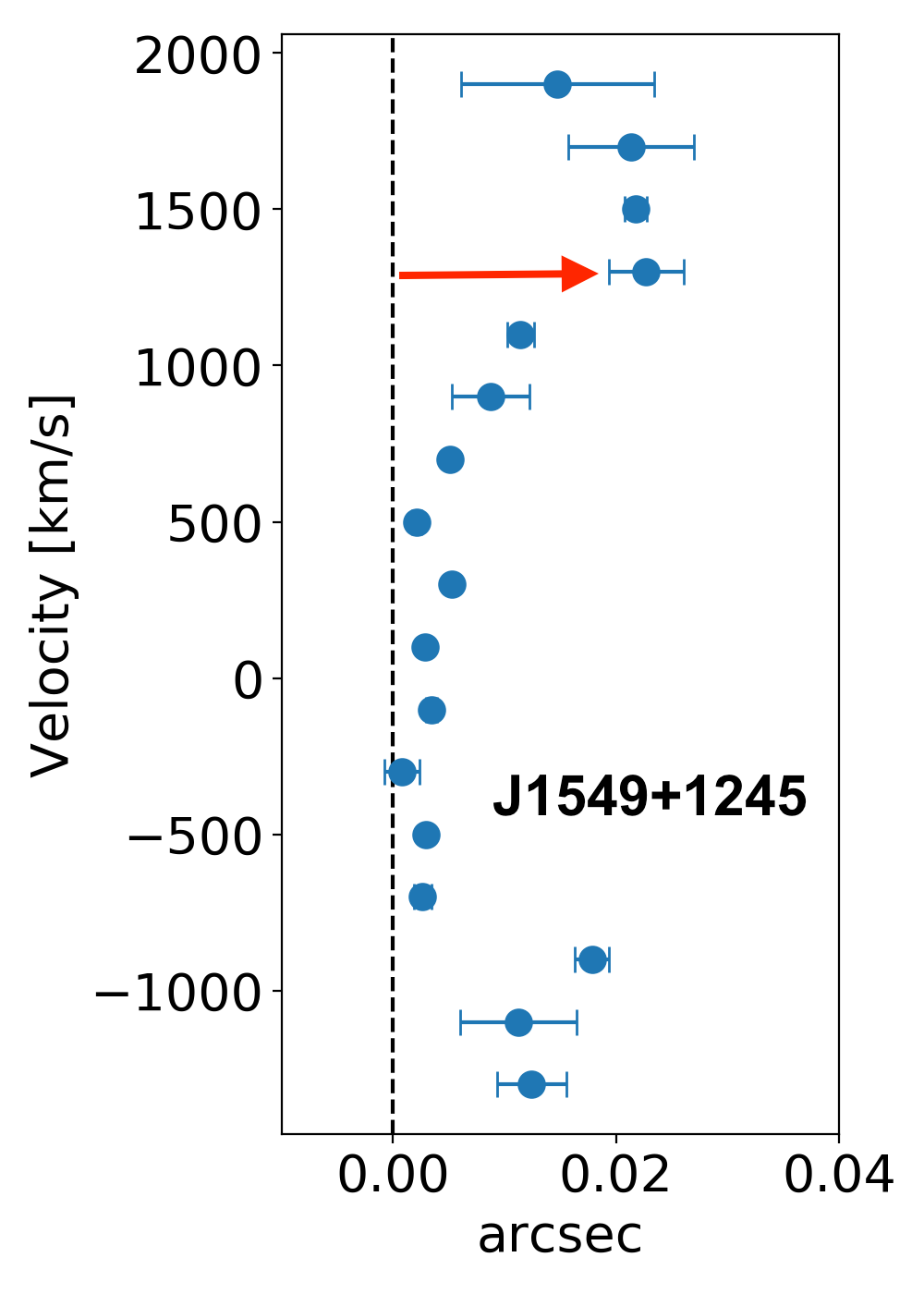}}
\subfloat{\includegraphics[scale=0.35]{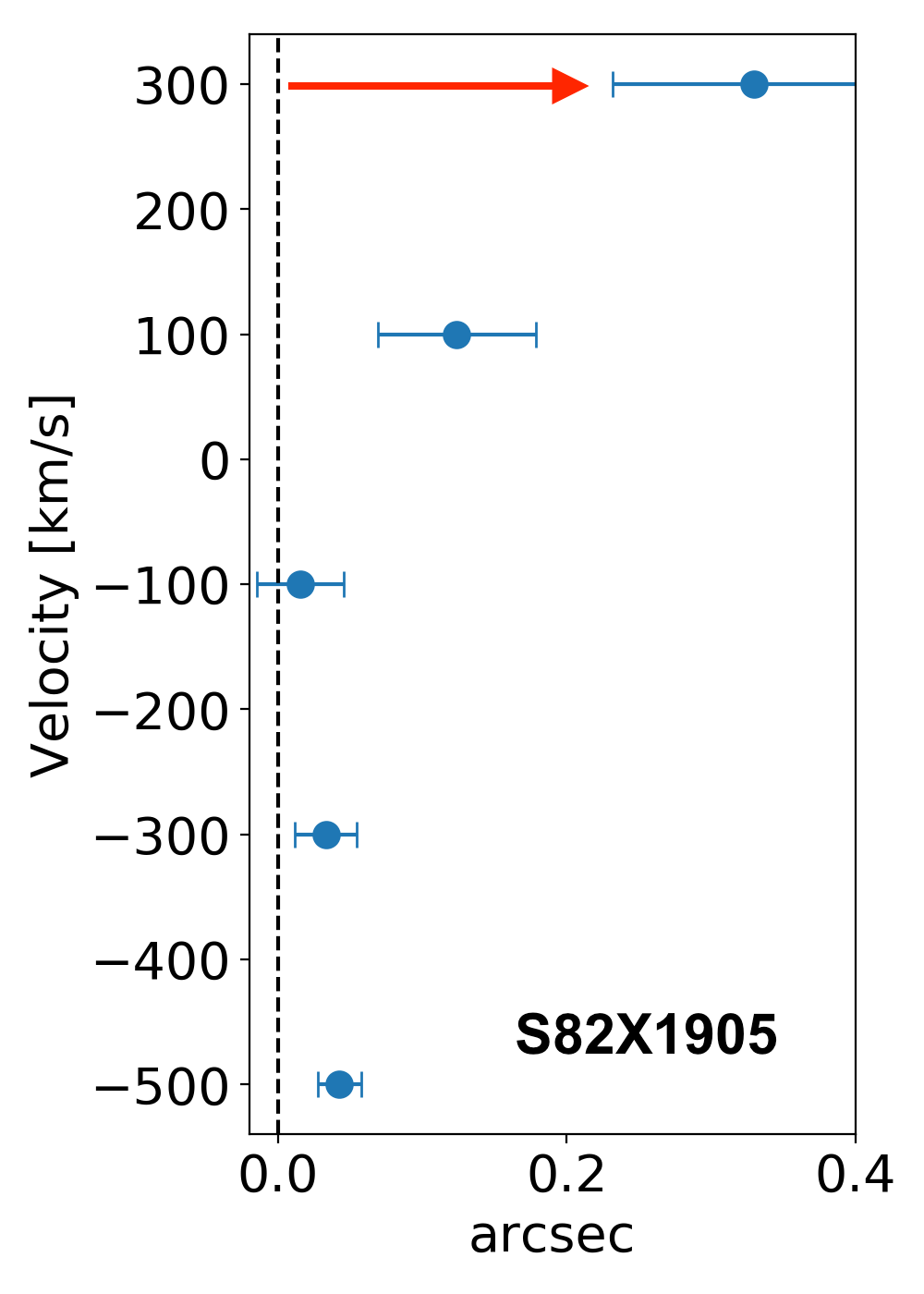}}
\subfloat{\includegraphics[scale=0.35]{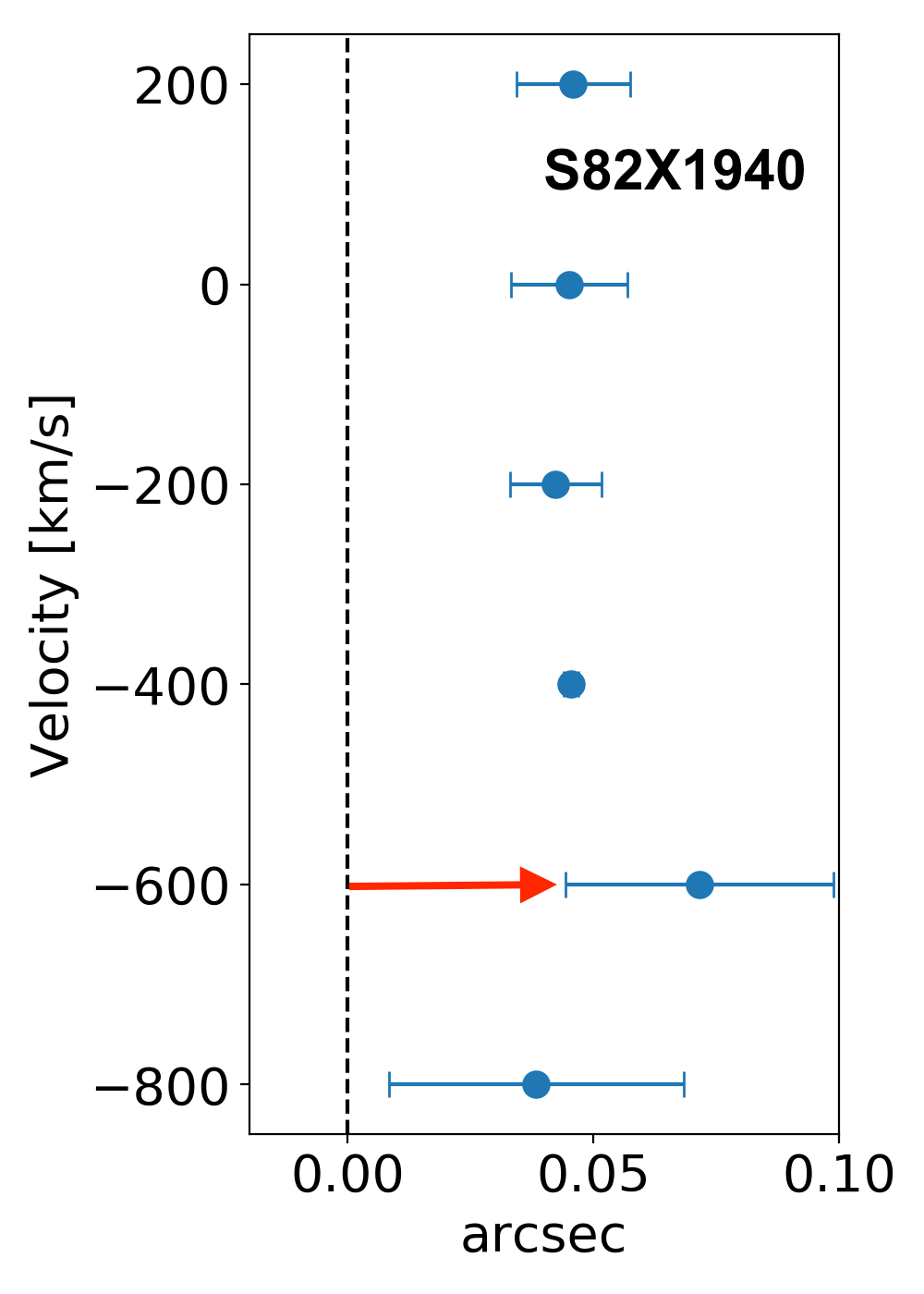}}
\caption{Results of spectroastrometry applied on the type-1 SUPER targets. The X-axis represents the distance between the continuum center (vertical dashed line) and the position of the photo-center of the \oiii line at the corresponding velocity channel given in the Y-axis. The error bars represent 1$\sigma$ uncertainty on the values. The red arrow marks the cloud with velocity greater than 600 km/s (300 km/s in the \oiii line profile) whose centroid is at the maximum distance from the AGN.\label{fig:spectroastro1}}
\end{figure*}

\begin{figure*}
\centering
\subfloat{\includegraphics[scale=0.7]{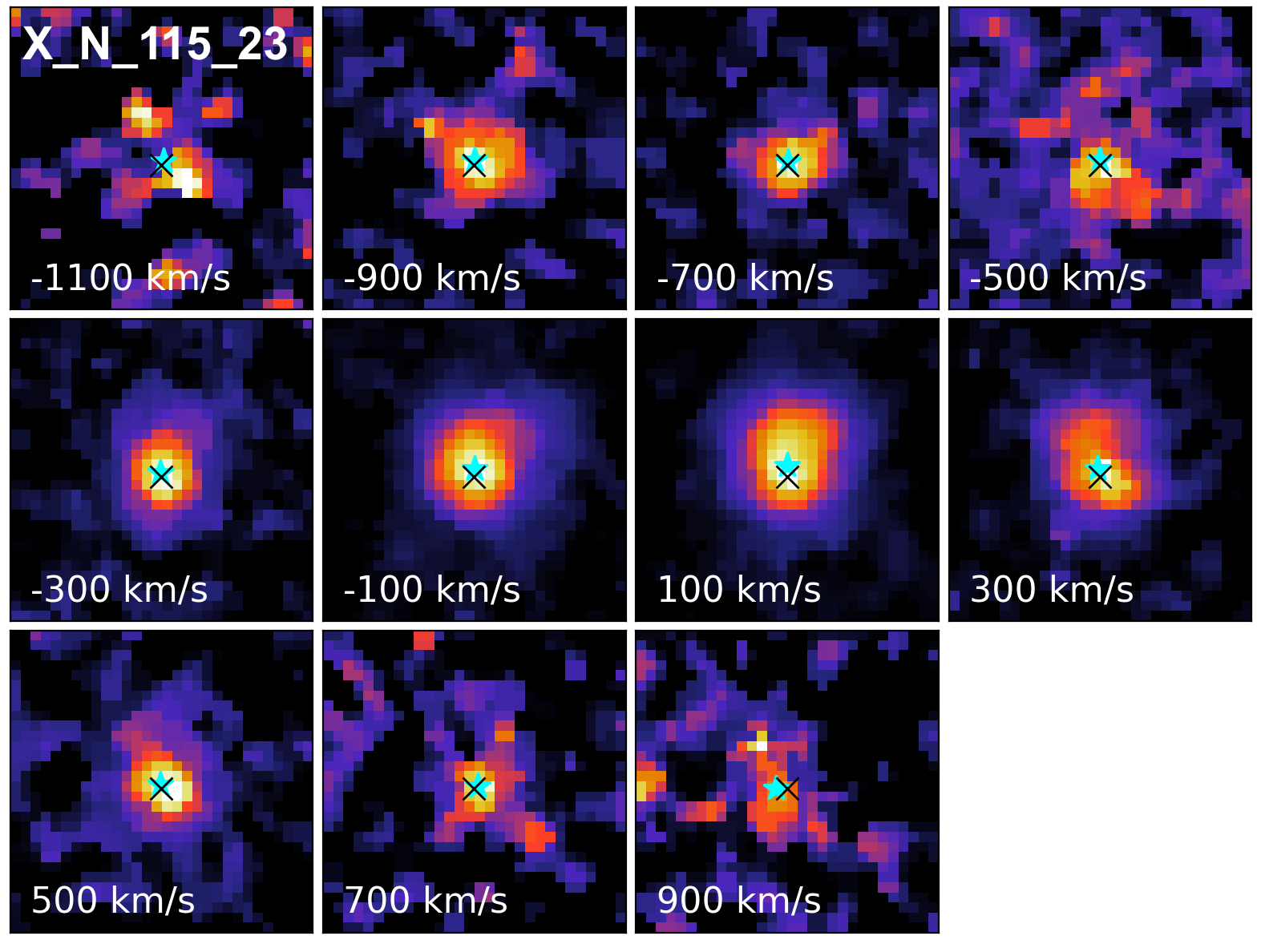}}
\caption{1.5$\arcsec$ x 1.5$\arcsec$ \oiii$\lambda$5007 channel maps of X\_N\_115\_23 at different velocity slices, after subtracting the $\hb$, \oiii$\lambda$4959 and iron models from the raw cube. Each velocity slice is 200 km/s wide and the displayed value is the center velocity of the respective channel. Black cross marks the location of the H-band continuum peak, used as a proxy for the AGN position and the cyan star marks the location of the \oiii centroid determined from a 2-D Gaussian fit. North is up and East is to the left. \label{fig:spectroastrometry_images}}
\end{figure*}

The principle of the spectroastrometry technique is that if the gas is moving at a distance R from the AGN position at a velocity different than the systemic velocity of the host galaxy (systemic here is defined by the position of the peak of the \oiii line in the spectrum), then the photo-centroid of the line at that velocity is also shifted by a distance R from the AGN position. In this method, we first collapse the original data cube along channels containing AGN continuum and the photo-centroid of the resulting image is determined with a 2-D Gaussian fit, similar to the method described in Sect. \ref{sect5.3}. The photo-centroid of the continuum is used as an indicator of the AGN location. We then use the spectrum models derived in section \ref{sect5.3} to subtract all the components from the raw spectrum across every pixel except those of the \oiii$\lambda$5007. The resulting cube containing only the \oiii$\lambda$5007 emission is then collapsed along wavelength channels with \oiii emission in bins of $\sim$200 km/s such that we achieve a S/N of $\sim$2 in the \oiii flux for each spectral channel. Similar to the continuum image, we then calculate the photo-centroid of each of the images obtained from the collapsed channels using a 2-D Gaussian fit. The velocity bins of $\sim$200 km/s ensures there are minimal errors in determining the centroid due to the spectral resolution of the instrument. The distance between the photo-centroid in each velocity channel and the AGN location is then plotted as a function of the center velocity of the channels. Note that spectroastrometry does not calculate the size of the outflowing gas but the spatial offsets between ionized gas clouds at different velocities with respect to the systemic velocity of the host galaxy. The offsets can be measured at scales smaller than the spatial resolution of the observations, and therefore the method can provide the distance of the bulk of the gas moving at a given velocity in unresolved or marginally resolved targets.

We apply the spectroastrometry method to all the 11 SUPER Type-1 AGN which have S/N > 5 in the \oiii line of the integrated spectrum. The results from the spectroastrometric analysis are shown in Fig. \ref{fig:spectroastro1}. The blue data points show the measured offsets from the AGN location, marked by the vertical dashed line at R = 0. The errors shown in the plot are the 1$\sigma$ uncertainty obtained from the output of the 2-D Gaussian fitting procedure. For each target we also produced the corresponding maps for each velocity channel included in the spectroastrometry analysis. Fig. \ref{fig:spectroastrometry_images} shows the channel maps for X\_N\_115\_23 as an example, while the channel maps for the rest of the targets have been moved to Appendix \ref{sect:app}. 

As presented in Sec. 5.1, we have used a cut of $\mathrm{w_{80}}$>600 km/s to identify AGN with ionized outflows and in the previous section we have used the $\mathrm{w_{80}}$ maps, for the object with spatially resolved emission, to characterize the maximum extension of the outflowing gas ({\rm D$_{600}$}), similarly to previous studies \citep[e.g.][]{harrison14, cresci15}. We can now use the spectroastrometry analysis to characterize, also for the objects which are not spatially resolved, at which distance is located the bulk of the outflowing gas. We have marked with a red arrow in the plots in Fig. \ref{fig:spectroastro1} the velocity channel with $|{\rm v}| > 300$ km/s with the maximum radial distance from the AGN location. The corresponding velocity ($\mathrm{v_{o}}$) and distance ($\mathrm{R_{o}}$) are reported in Table \ref{table:outflow_properties}. We infer a maximum spatial offset of 0.30$\arcsec$ and a median offset of $\sim$0.1$\arcsec$ between the AGN location and bulk of the outflowing gas. This translates into a maximum physical physical distance of $\sim$2.2 kpc (0.8 kpc median), which is smaller than the spatial resolution of of the observations in the H-band. These numbers are consistent with \citet{carniani15} where the maximum offset of outflowing gas is $\lesssim$2 kpc. Note that there could be multiple gas clouds at velocities above 600 km/s. We only highlight the ones which are at the maximum distance from the AGN. Therefore, generally the bulk of the high velocity outflowing gas ($|{\rm v}| > 300$ km/s) is concentrated within $\sim$2 kpc from the AGN location in the majority of the high-redshift AGN host galaxies. Note that the actual extent of the outflowing gas might be larger than the values obtained from the spectroastrometry, which aims to identify the distance from the bulk of the gas moving at a given velocity.

\begin{figure}
\centering
\includegraphics[scale=0.5]{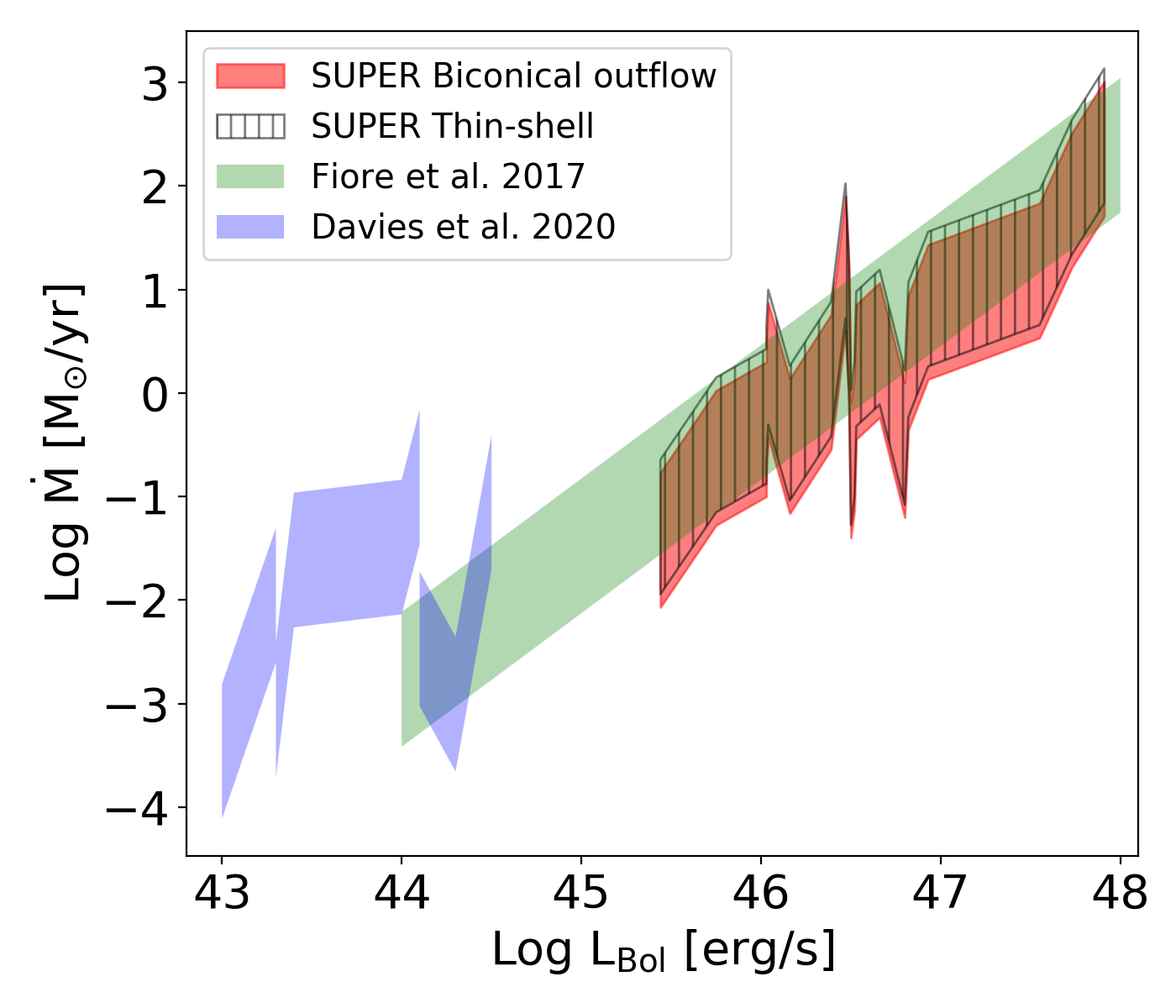}
\caption{Ionized gas \oiii mass outflow rate vs. the bolometric luminosity of the AGN for SUPER Type-1 sample presented in this paper and literature data of low as well as high redshift AGN. The red shaded area and the black hatched area shows the mass outflow rates for the SUPER targets assuming a bi-conical outflow model and a thin-shell model, as described in Sect. \ref{sect5.5}. The green shaded area shows the outflow rates for ionized gas from literature data compiled in \citet{fiore17} (after rescaling the relation with the same assumption as the Type-1 SUPER targets, see Sect. \ref{sect5.5}) and the blue shaded region shows the outflow rates for low redshift X-ray AGN sample from \citet{davies20}. The shaded region in all the studies correspond to mass outflow rates assuming an electron density from 500 cm$^{-3}$ -- 10,000 cm$^{-3}$.}
\label{fig:outflow_rates}
\end{figure}

There are a couple of targets with notably interesting features from the spectroastrometry analysis. In the case of cid\_346, the spectroastrometry method reveals the presence of gas moving at $\sim$600 km/s at a distance of $\sim$0.3$\arcsec$ equivalent to a physical distance of $\sim$4 kpc from the AGN location. Similarly, also in the case of S82X1905 we find that the bulk of the gas moving at $\sim 300$ km/s is located at $\sim$4 kpc from the center. We argue that a possible explanation for such extended redshifted emission is that the receding part of the outflow is not obscured by the dust of the host galaxy in these two cases, possibly due to the larger extent of the outflow itself.

We also note from the spectroastrometry maps that the bulk of the low velocity \oiii component i.e. the non-outflowing component of the \oiii emission is not necessarily emitted at the same location as that of the AGN in a few targets. This is expected as even in low redshift galaxies the \oiii emission is dominated in the extended NLR in the form of ionization cones, which are not co-spatial with the AGN location.

We now summarize the different methods used to determine the extension of the ionized gas and the outflows associated with the gas. Using COG and PSF-subtraction methods, we compared the spatial distribution of the ionized gas with the observed PSF and inferred whether the ionized gas is extended or not. The COG and PSF-subtraction methods rely on robust PSF measurements either from the BLR emission or if the BLR emission is not available, a dedicated PSF observation. Using either of these methods, we find that the ionized gas is extended in X\_N\_81\_44, X\_N\_66\_23, X\_N\_115\_23, cid\_346, J1333+1649, J1549+1245 and S82X1905.   In these targets with extended ionized gas, we constructed the flux and velocity maps and calculated $\mathrm{D_{600}}$ value which is the maximum projected spatial extent of $\mathrm{w_{80}}$ > 600 km/s. The $\mathrm{D_{600}}$ parameter quantifies the spatial extent of the outflow associated with the ionized gas which is consistent with the outflow definition we have adopted throughout this paper i.e. $\mathrm{w_{80}}$ > 600 km/s. The $\mathrm{D_{600}}$ parameter is in the range $\approx$1.5--6.5 kpc from the AGN location. The $\mathrm{D_{600}}$ parameter should only be determined for targets which show the presence of extended ionized gas emission beyond the observed PSF. Lastly, we determined the radial distance at which the bulk of the ionized gas (weighted by luminosity) is moving at velocity greater than 600 km/s using spectroastrometry technique. We find that for most galaxies the bulk of the high velocity gas (|v|>600 km/s) is contained within $\sim$1 kpc from the AGN location.  The spectroastrometry technique has the advantage that it can be used for targets which are unresolved or marginally resolved. However, the technique does not provide the true spatial extent of the ionized gas, but the centroid of the bulk gas moving at a specific velocity.

\begin{figure*}
\centering
\subfloat{\includegraphics[scale=0.4]{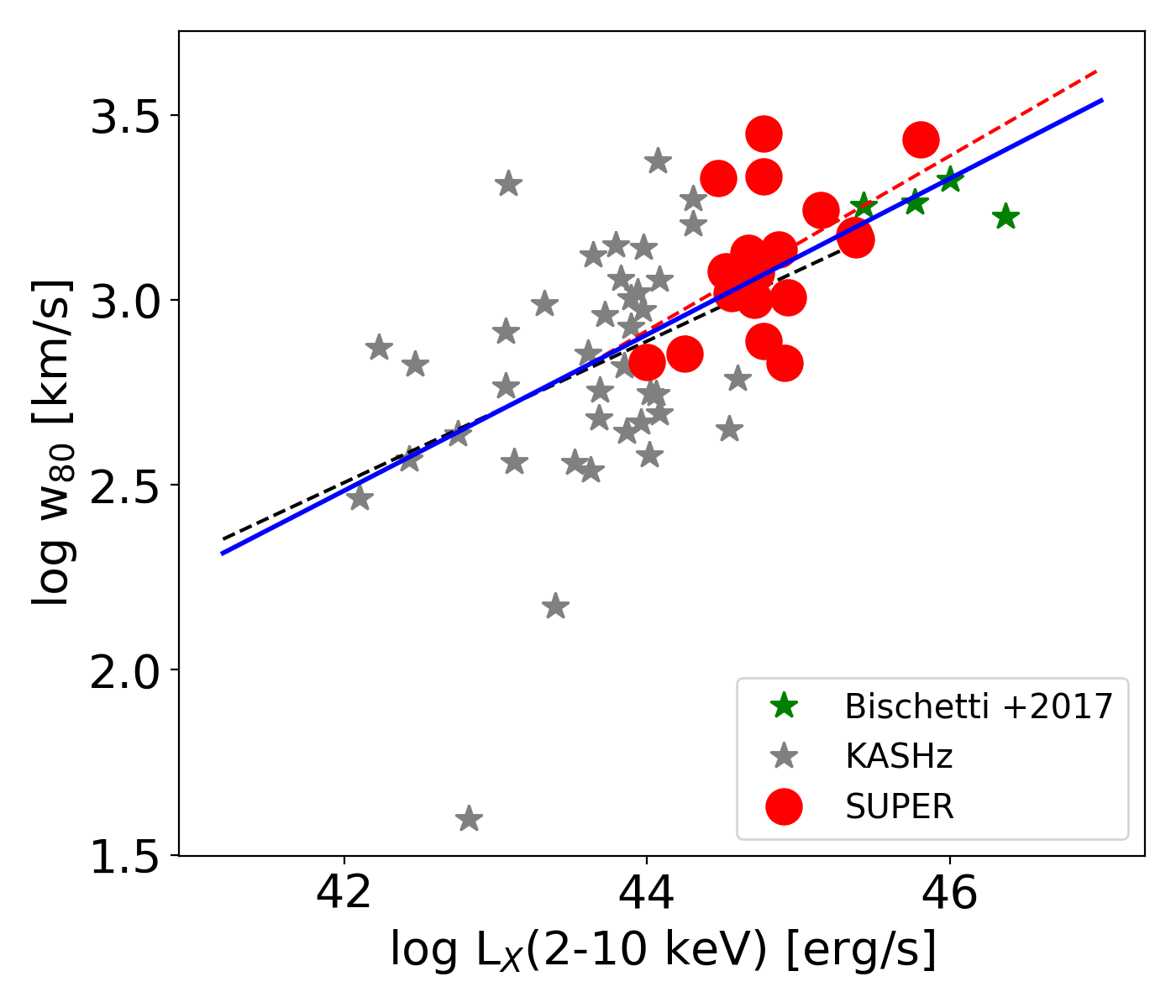}}
\subfloat{\includegraphics[scale=0.4]{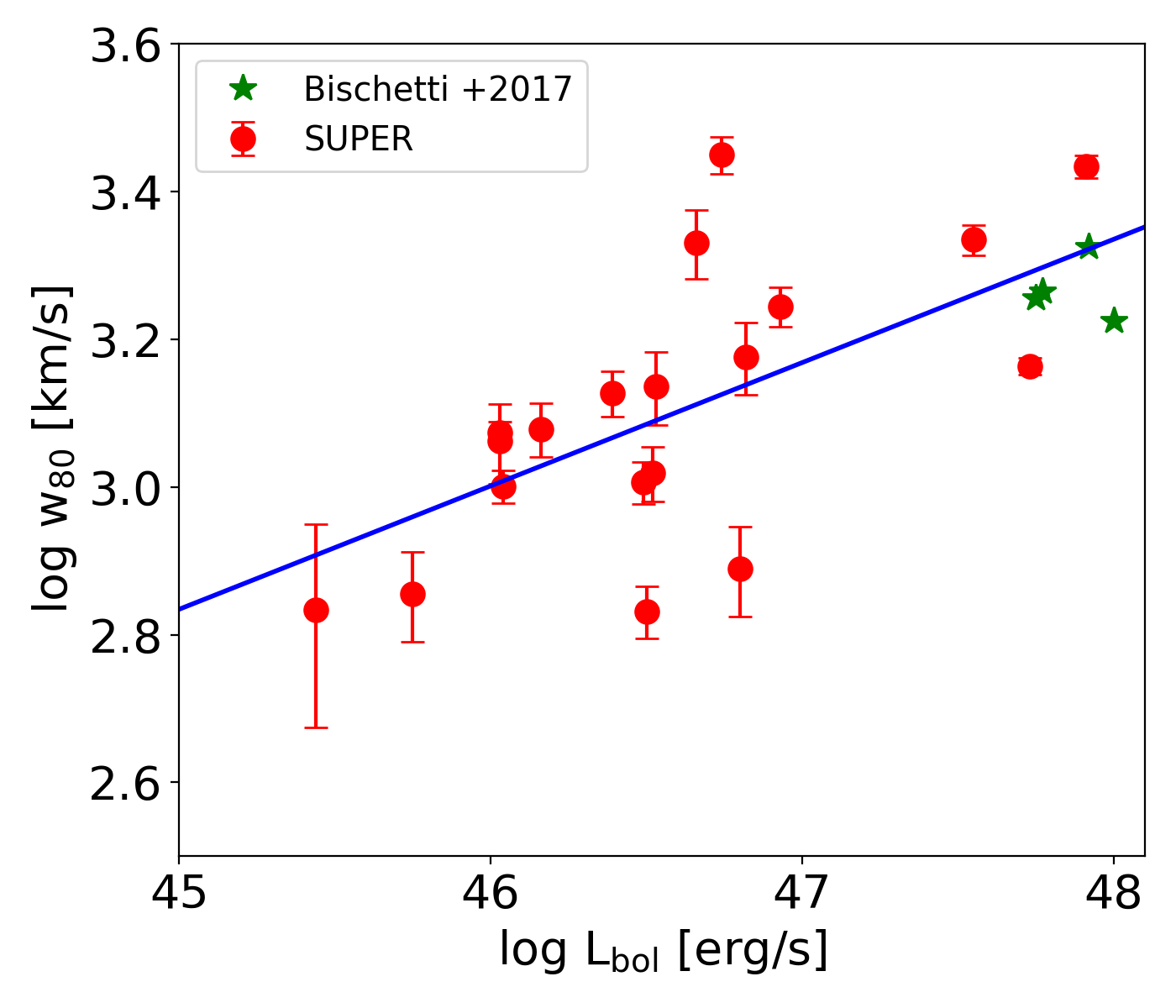}}\\
\subfloat{\includegraphics[scale=0.4]{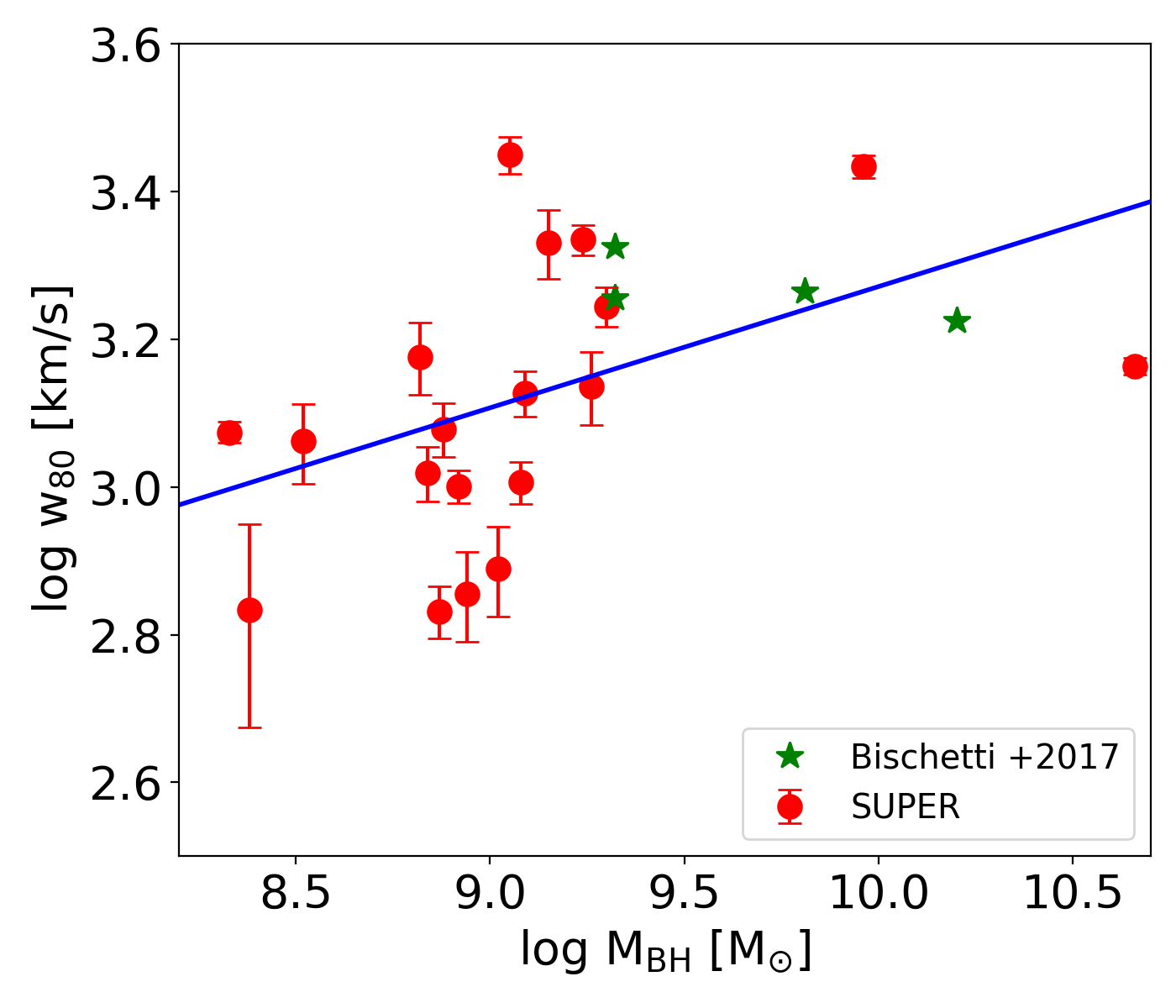}}
\subfloat{\includegraphics[scale=0.4]{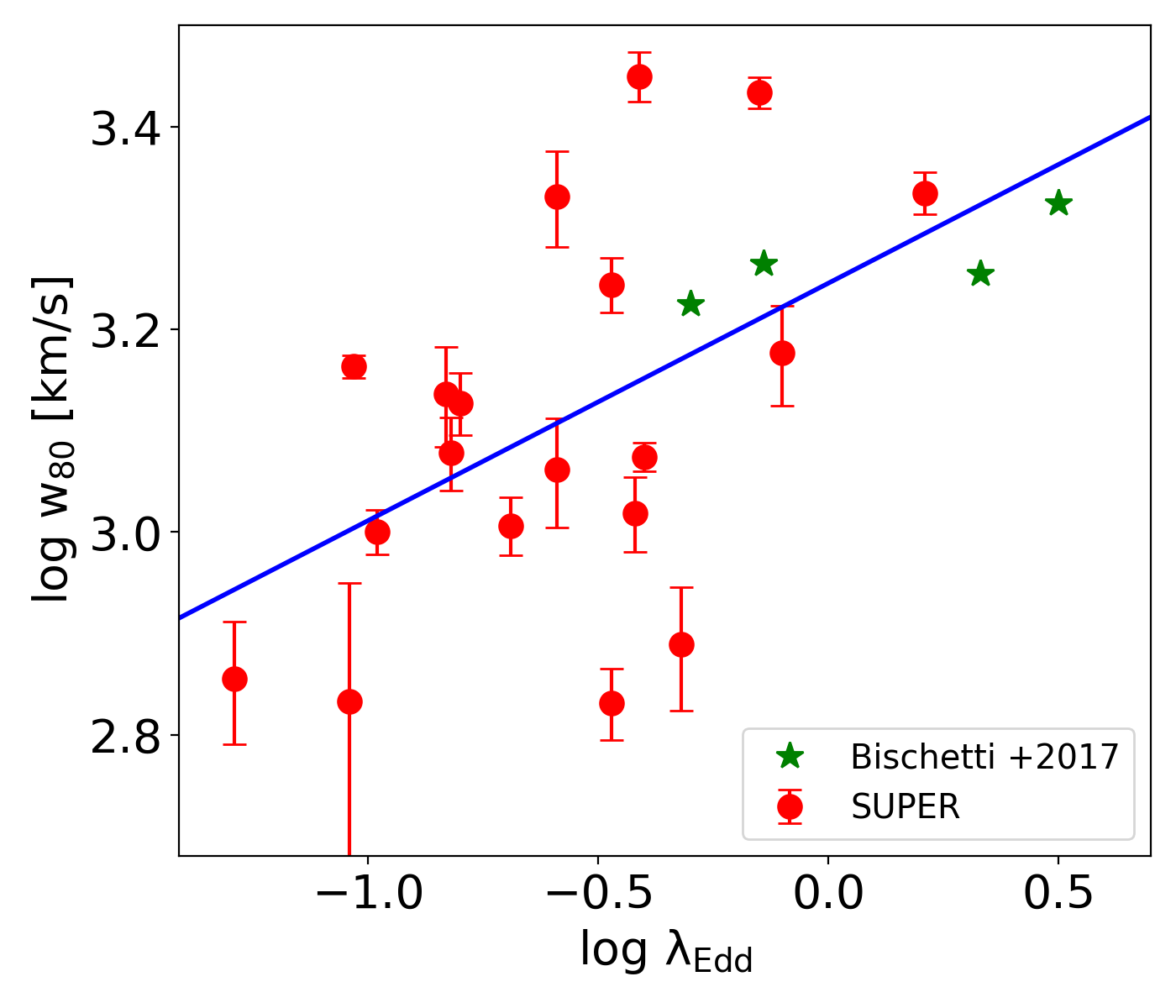}}
\caption{Top left panel: Non-parametric velocity width, $\mathrm{w_{80}}$ derived from the integrated spectrum vs. X-ray luminosity measured in the 2-10 keV band. The red data points represent the SUPER sample, the grey data points show the luminosity and redshift-matched sample from the KASHz survey and the green data points show the high luminosity sample from WISSH survey. The dashed red line shows the best fit relation for SUPER galaxies, the dashed grey line shows the best fit relation for matched KASHz sample and the blue line shows the overall relation for SUPER, KASHz and the WISSH sample \citep{bischetti17}. Top right panel: $\mathrm{w_{80}}$ vs. bolometric luminosity. Bottom left panel: $\mathrm{w_{80}}$ vs. black hole mass, $\mathrm{M_{BH}}$. Bottom right panel: $\mathrm{w_{80}}$ vs. Eddington ratio ($\mathrm{L_{bol}/L_{Edd}}$). In all the plots, the blue line shows the best fit to all the data points within the plot. The values plotted for the SUPER sample are obtained from Tables \ref{table:SUPER_properties} and \ref{table:SINFONI_properties1}.\label{fig:w80-scaling}}
\end{figure*}

\begin{figure}
\centering
\subfloat{\includegraphics[scale=0.4]{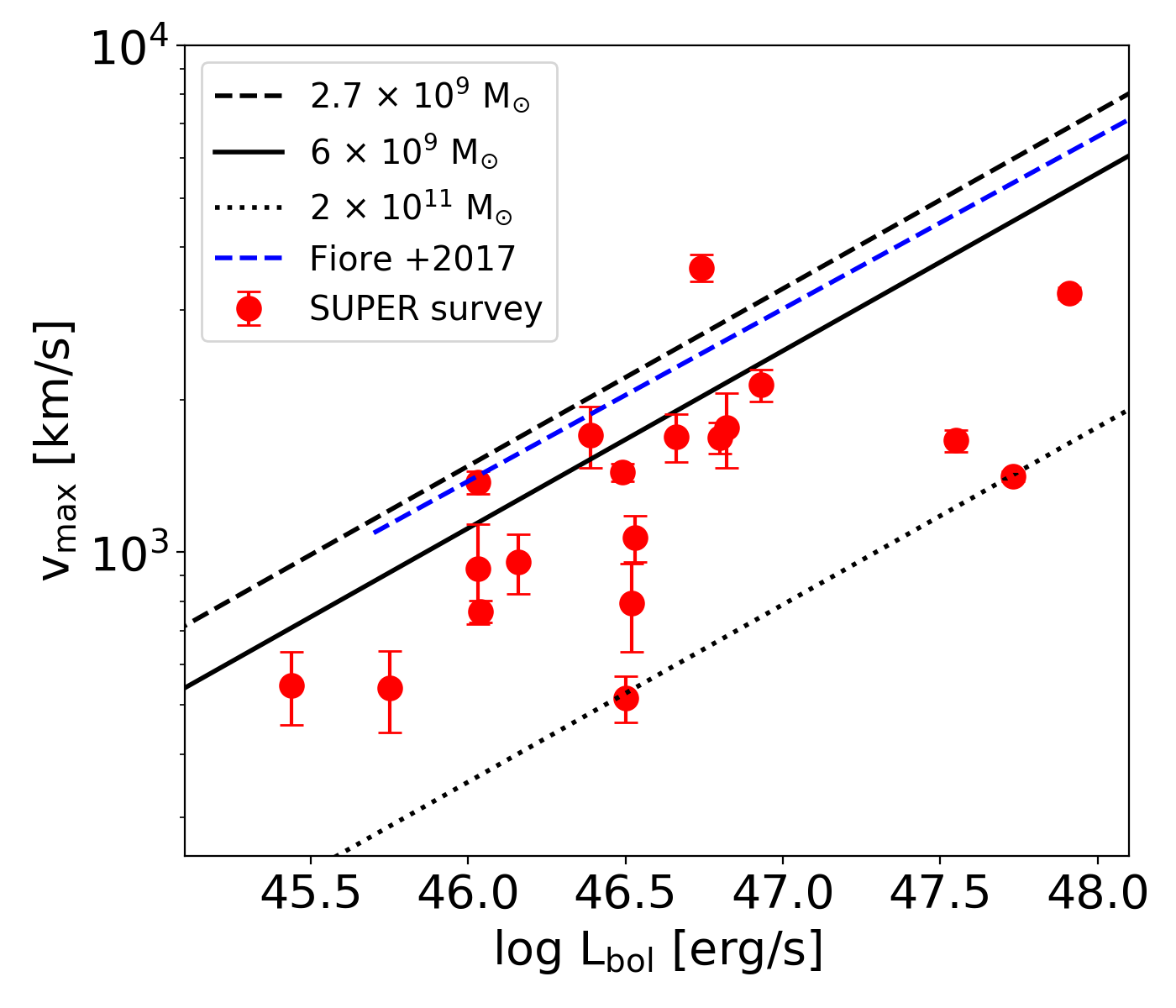}}
\caption{Maximum velocity, $\mathrm{v_{max}}$, defined in Sect. \ref{sect6}, vs. AGN bolometric luminosity. The red circles represent the SUPER targets, the dashed blue line shows the best-fit relation for literature data compiled in \citet{fiore17}. The black curves are the \citet{menci19} predictions for the shock velocity (approximated here by the measured v$_{\rm max}$) vs. AGN luminosity for galaxies. The three curves correspond to molecular gas masses of 2.7 $\times 10^{9}$ M$_{\odot}$, 6 $\times 10^{9}$ M$_{\odot}$ and 2 $\times 10^{11}$ M$_{\odot}$ respectively. More details are given in Sect. \ref{sect6}. \label{fig:vmax-scaling}}
\end{figure}

\subsection{Mass outflow rates} \label{sect5.5}

To understand the impact that AGN feedback may have on their host galaxies it is important to estimate the mass of the ouflowing gas and the kinetic energy associated with it. Although the ionized gas phase is believed to trace only a fraction of the total outflow mass \citep[e.g.][]{cicone18}, the energy associated with these outflows can provide valuable information on its driving mechanism \citep[e.g.][]{fiore17, brusa18, husemann19, shimizu19}.

An accurate estimation of the mass outflow rate for high redshift galaxies is a major challenge due to the limited spatial resolution achievable with current instrumentation and the uncertainties on the physical properties of the outflowing gas. As a consequence, there is the need to make a number of assumptions which leads to large uncertainties in the derived mass outflow rate. The key elements that need to be constrained to estimate the mass outflow rate are: the geometry and the volume of the outflow; the physical properties of the gas (e.g. its electron density, temperature and associated metallicity); the velocity of the outflowing gas and its radial evolution. For the scope of this paper, we focus on reporting the range of mass outflow rates derived from our measured quantities by exploring the possible assumptions for key physical quantities, described as follows. 

\noindent
{\bf Geometry:} Both theoretical models \citep[e.g.][]{ishibashi19} and results from imaging and IFU observations of low redshift galaxies \citep[e.g.][]{crenshaw10,muller-sanchez11, liu13,wylezalek16,venturi18} provide evidence for the presence of spherical and conical outflow morphology. Recent observations of low redshift AGN host galaxies have also suggested the possibility of expanding shell-like shocks driving an outflow across the host galaxy \citep[e.g.][]{husemann19}. A shell-like expanding outflow model can take into account variable physical quantities such as the outflow luminosity, velocity, density and temperature as a function of distance from the AGN. Total outflow rate throughout the galaxy is then the sum over all the shells. As current observations with SINFONI do not allow us to infer the outflow morphology, we will calculate mass outflow rates assuming both bi-conical and thin shell geometry. A spherical geometry would change the overall normalization factor in the mass outflow rate formula but not the general trends, when compared with the bi-conical outflow geometry. We will consider the volume up to a distance of 2 kpc from the AGN (see further discussion on "Radius and Velocity" later in the section). For the thin-shell model, we will assume a shell width of $\sim$500 pc which is the upper end of the limits inferred in low redshift AGN host galaxies \citep[e.g.][]{husemann19}.

\noindent
{\bf Luminosity:} For an emission line modeled using multiple Gaussian components, the broad Gaussian component is often associated with the outflowing gas. However, as stated in earlier sections, the detection of an additional broad component depends on the S/N of the data and the model function used. Therefore, we will continue to use the non-parametric approach and define the outflow luminosity as the luminosity calculated from the flux of the \oiii$\lambda$5007 emission line channels at |v|>300 km/s. This definition is consistent with our outflow's classification, i.e. $\mathrm{w_{80}}$ > 600 km/s, presented in Sec. 5.1. The outflow luminosity as defined above is reported in Table \ref{table:outflow_properties}. We note that the average ratio between the broad component luminosity (Table \ref{table:SUPER_properties}) and the outflow luminosity ranges from 0.56--4.07 with a mean ratio of  1.80$\pm$0.89. Therefore, on average the luminosity of the broad Gaussian component is higher than that estimated from the non-parametric procedure described above. We finally note that the luminosity calculated in this paper is not corrected for extinction and therefore represents a lower limit on the outflow luminosity.

\noindent
{\bf Radius and velocity:} Several spatially resolved observations of ionized gas at low redshift clearly show a radial dependence of outflow properties \citep[e.g.][]{kakkad18, venturi18, husemann19}. It is therefore necessary to take into account the variations in the velocity across the outflowing medium while calculating the mass outflow rates. Due to the relatively low spatial resolution of high redshift targets compared to the low redshift AGN, deriving spatial variations in the physical properties of the outflow is not always feasible. Observations at low redshift report outflow radius values up to few kiloparsecs \citep[e.g.][]{liu13, husemann16}. For the purpose of this paper, we will calculate the mass outflow rate up to a fixed radius of $\sim$2 kpc, which is also the maximum radius for unresolved targets (see PSF values in Table \ref{table:SUPER_observations}). The outflow velocity at 2 kpc will be derived from the $\mathrm{w_{80}}$ maps in case of resolved targets (Figs. \ref{fig:flux_velocity1} and \ref{fig:flux_velocity2}) and the integrated $\mathrm{w_{80}}$ value in case of unresolved targets. Other definitions for outflow velocity have also been used in the literature such as broad Gaussian widths and $\mathrm{v_{max}}$, which can introduce a systematic uncertainty up to a factor of $\sim$2 in the calculation of mass outflow rate \citep[e.g.][]{kakkad16, fiore17}. 

\noindent
{\bf Density:} The electron density represents the major source of uncertainty in the calculation of mass outflow rates \citep[e.g.][]{harrison18} since it is very challenging to measure it and in fact there is a wide range of values reported in the literature for the gas density in the NLR \citep{perna17,kakkad18, baron19, davies20}. Similarly to the outflow's velocity, variations in the electron density across the outflowing medium make the calculation of outflow rates even more challenging. Moreover, key diagnostic emission lines to estimate electron density such as \sii$\lambda$6716, 6731 are too faint or remain undetected for the SUPER targets. Therefore, instead of assuming a single value for the gas density, we will consider two values that cover the wide range observed in the literature, i.e. from 500 cm$^{-3}$ -- 10,000 cm$^{-3}$, which would return maximum and minimum possible outflow rates given the rest of the model assumptions described above.

Since H$\beta$ is relatively faint, we will calculate the mass of the outflowing ionized gas from the \oiii$\lambda$5007 line as follows \citep[see e.g.][]{carniani15, kakkad16}: 

\begin{equation}
\mathrm{
M_{\oiii}^{out} = 0.8\cdot 10^{8} M_{\odot} \left(
\frac{1}{10^{[O/H]-[O/H]_{\odot}}} \right) \left(
\frac{L_{\oiii}}{10^{44}~erg/s} \right)\left(
\frac{<n_{e}>}{500 cm^{-3}}\right)^{-1}
}
\label{eq:outflow_mass}
\end{equation}

\noindent
where all the oxygen is assumed to be ionized to O$^{2+}$, the electron temperature is 10,000 K and [O/H], the metallicity of the outflowing material, is assumed solar. For a uniformly filled bi-conical outflow, the mass outflow rate takes the form \citep{fiore17}:

\begin{equation}
\mathrm{
\dot{M}_{\oiii, cone}^{out} = 3\cdot v_{out} \frac{M_{out}}{R_{out}}
}
\label{eq:mass_outflow_rate_cone}
\end{equation}

\noindent
where $\mathrm{v_{out}}$ is the outflow velocity, $\mathrm{M_{out}}$ is obtained from Eq. \ref{eq:outflow_mass} and $\mathrm{R_{out}}$ is the distance at which the outflow rate is computed. We are making the assumption here that the bi-conical outflow region is uniformly filled with ionized gas with density <$\mathrm{n_e}$>. Eq. \ref{eq:mass_outflow_rate_cone} will give the instantaneous outflow rate of the ionized gas crossing the spherical sector at distance $\mathrm{R_{out}}$ from the AGN. On the other hand, if the outflow is assumed to propagate in thin-shells of thickness $\mathrm{\Delta R}$ (assumed 500 pc here) which shock against the ISM, the outflow rate is simply \citep{husemann19}:

\begin{equation}
\mathrm{
\dot{M}_{\oiii, thin-shell}^{out} = v_{out} \frac{M_{out}}{\Delta R}
}
\label{eq:mass_outflow_rate_shell}
\end{equation}

\noindent 
where $\mathrm{v_{out}}$ is the outflow velocity at the location of the shell (assumed to be at 2 kpc here). Eq. \ref{eq:mass_outflow_rate_shell} will provide the outflow rate within the shell of thickness $\mathrm{\Delta R}$ instead of an integrated value. 

We compute the mass outflow rates using both formulas described above and for two values of the electron density: 500 cm$^{-3}$ and 10$^{4}$ cm$^{-3}$ \citep[e.g.][]{baron19, davies20}. The values obtained are reported in Table \ref{table:outflow_properties}. Note that most of the SUPER targets have mass outflow rate values below 10 $\mathrm{M_{\odot}}$/yr for the density range 500 cm$^{-3}$ to 10$^{4}$ cm$^{-3}$. At the high luminosity end, however, the mass outflow rate values exceed 10 $\mathrm{M_{\odot}}$/yr even for a conservative electron density value of 10$^{4}$ cm$^{-3}$. In Fig. \ref{fig:outflow_rates} we show the range of mass outflow rates vs. the bolometric luminosities covered by our sample, from literature ionized gas outflow rates compiled in \citet{fiore17}, and from the low-redshift X-ray AGN studied in \citet{davies20}. The mass outflow rates values from \citet{fiore17} and \citet{davies20} have been updated to match the electron density range used in this study. In addition, these literature mass outflow rates from \citet{fiore17} and \citet{davies20} have also been re-scaled to match the methodology used in this paper. \citet{fiore17} used the luminosity of the broad Gaussian component as a proxy for the outflow luminosity, while, as described above, we considered only channels with |v|>300 km/s. We verified that for our objects the mean ratio between the luminosity calculated from the broad Gaussian and the luminosity from \oiii channels at |v|>300 km/s is $\approx$2 and re-scaled by this factor the values from \citet{fiore17}. A similar scaling was used for \citet{davies20} who use the \oiii luminosity to calculate the outflow mass. The $\mathrm{\dot{M_{out}}-L_{bol}}$ relation in \cite{fiore17} also used, when available, either $\ha$ or $\hb$ to calculate the total mass outflow rates for the ionized gas. Outflow rates for sources with only \oiii detection were scaled by a factor of three which was consistent with the mean ratio between the $\ha$ and \oiii luminosities in their sample. We have therefore applied the same factor to their relation to make it consistent with us using the \oiii luminosities. After taking these factors into account, the \citet{fiore17} relation roughly match the mass outflow rates observed for the Type-1 sample in this paper. Fig. \ref{fig:outflow_rates} clearly illustrates that the different assumptions on the outflow geometry (i.e. by-conical outflow or thin spherical shell) have, by far, a smaller impact on the derived value for the mass outflow rate than the assumed value of the gas density. The latter is responsible for more than an order of magnitude difference in the range on n$_{\rm e}$ explored. Finally, combining the SUPER Type-1 sample presented in this paper with literature data we confirm a correlation of the mass outflow rate with the AGN bolometric luminosity across five orders of magnitude. This relation could be potentially used to put constraints on the different AGN outflow models, but we caution that the real scatter and the slope of this relation is uncertain as long as we do not have better constraints on the gas density object by objects (see also \citet{davies20}).

Finally, as argued in \citet{baron19}, the ionized gas is most likely associated with a significant amount of neutral dusty gas that takes part in the same outflow. Along with the fact that outflows can exist in multiple gas phases \citep[e.g.][]{perna19,serafinelli19}, this suggests that the actual mass outflow rates can be substantially larger than the ones deduced here.

\section{Scaling Relations}	\label{sect6}

We now focus on the relation between the properties of the detected outflows (e.g. velocity and escape fraction) and the properties of the central super-massive black (e.g. bolometric luminosity and black hole mass). Theoretical models of AGN outflows predict fast winds to originate from the accretion disk which later impact on the ISM resulting in a forward shock that expand within the host galaxy \citep[e.g.][]{zubovas12,zubovas14,faucher12,wagner13}. This would naturally predict positive correlations between outflow properties (e.g.  velocity, mass outflow rate) and AGN properties \citep[e.g.][]{king_pounds15}. The slope of such relation will in addition depends on the density profile of the medium \citep[e.g.][]{faucher12}. While there are already a few studies exploring such scaling relations for literature samples of AGN on a wide redshift range \citep[e.g.][]{fiore17, fluetsch19},  we will perform a similar analysis on the SUPER sample for which we have a homogeneous and detailed characterization of both the outflows properties as well as those of the central black hole. 

\begin{table}
\centering
\caption{Correlation functions and statistical parameters for the scaling relations in Fig. \ref{fig:w80-scaling}. \label{table:correlation_test}}
\begin{tabular}{cccc}
\hline 
 Correlation & Slope & Pearson coefficient & $p$-value\\ \hline\hline
$\mathrm{w_{80}}$ vs. $\mathrm{L_{X}}$ $^{\dagger}$     & 0.21$\pm$0.01 & 0.59 & <1.0e-5\\
$\mathrm{w_{80}}$ vs. $\mathrm{L_{X}}$ $^{\ast}$     & 0.21$\pm$0.01 & 0.51 & 0.024\\
$\mathrm{w_{80}}$ vs. $\mathrm{L_{bol}}$ $^{\ast\ast}$  & 0.16$\pm$0.04 & 0.70 & 2.1e-4 \\
$\mathrm{w_{80}}$ vs. $\mathrm{M_{BH}}$ $^{\ast\ast}$ & 0.16$\pm$0.06 & 0.50 & 0.02\\
$\mathrm{w_{80}}$ vs. $\mathrm{\lambda_{edd}}$ $^{\ast\ast}$ & 0.23$\pm$0.07 & 0.57 & 4.0e-3\\
$\mathrm{f_{esc}}$ vs. $\mathrm{L_{bol}}$ $^{\ast\ast}$ & 23.5$\pm$6.7 & 0.65 & 0.003\\
\hline
\end{tabular}
\vspace{1ex}\newline
{\raggedright{\bf Notes:}\newline
$^{\dagger}$The values reported for the $\mathrm{w_{80}}$ vs. $\mathrm{L_X}$ relation are for the SUPER, KASHz and \citet{bischetti17} targets combined. $^{\ast}$The reported values for the $\mathrm{w_{80}}$ vs. $\mathrm{L_X}$ relation consider the SUPER sample. $^{\ast\ast}$The reported values for the $\mathrm{w_{80}}$ vs. $\mathrm{L_X}$ relation consider the SUPER and \citet{bischetti17} sample. \par}
\end{table}

We first investigate the scaling relations for $\mathrm{w_{80}}$ (Fig. \ref{fig:w80-scaling}) versus the physical properties of the central SMBH. In the top left panel of Fig. \ref{fig:w80-scaling}, we plot $\mathrm{w_{80}}$ against the AGN X-ray luminosity $\mathrm{L_{X}(2-10 keV)}$. A clear positive trend is present i.e. higher X-ray luminosity targets show larger \oiii width. A similar positive correlation is observed if we consider the AGN bolometric luminosity (top right panel in Fig. \ref{fig:w80-scaling}) instead of the X-ray luminosity. To cover a wider range in luminosity, we add the AGN sample observed in the KASHz survey matched in redshift (z$\sim$2--2.5) and the high luminosity targets from the WISSH survey \citep[e.g.][]{bischetti17, martocchia17, zappacosta20}. We derived the Pearson coefficient and the p-value for each of these relations and the results are reported in Table \ref{table:correlation_test}. For the SUPER data alone, the Pearson coefficient for $\mathrm{w_{80}}$ vs. $\mathrm{L_{X}}$ relation is 0.51 while for the redshift matched KASHz sample the coefficient is 0.37 with a null hypothesis probability (for a non-correlation) of $\sim$2.4\% and 2\% respectively. For the combined datasets of SUPER, KASHz and WISSH sample, the Pearson coefficient and the null hypothesis probability is 0.59 and <0.001\% respectively, suggesting a strong correlation between $\mathrm{w_{80}}$ and $\mathrm{L_{X}}$.

The statistical tests also confirm strong positive correlation for $\mathrm{w_{80}}$ vs. $\mathrm{L_{bol}}$ (Fig. \ref{fig:w80-scaling}, null-hypothesis probability <0.1\%), which may be interpreted as an increase in the turbulence of the gas within the AGN host galaxies with the overall radiation power of the central black hole. We note that there is a positive correlation also for $\mathrm{v_{10}}$ vs. $\mathrm{L_{X}}$ and $\mathrm{L_{bol}}$, although they are weaker, according to the Pearson test, compared to those presented above for $\mathrm{w_{80}}$.

Further exploring the correlation of the gas kinematics with the AGN properties, $\mathrm{w_{80}}$ has a relatively weak correlation with the black hole mass (bottom left panel in Fig. \ref{fig:w80-scaling}) with a null hypothesis of $\sim$2\%. Similar correlation is observed between $\mathrm{w_{80}}$ and the Eddington ratio ($\mathrm{\lambda_{Edd} = L_{bol}/L_{Edd}}$, where $\mathrm{L_{Edd} = 1.3 \times 10^{38} (M_{BH}/M_{\odot}}$ erg/s)) with a null-hypothesis probability of $\sim$0.4\%. These values suggest that the outflow velocity correlate weakly with the black hole mass when compared with the correlation with bolometric luminosity.

Apart from $\mathrm{w_{80}}$, we tested the presence of a  correlation in the $\mathrm{L_{bol}}$ vs. maximum velocity of the \oiii line (Fig. \ref{fig:vmax-scaling}), defined in the literature as $\mathrm{v_{max} = \Delta\lambda_{broad} + 2\sigma_{broad}}$ i.e. the sum of the difference between the centroid velocity of the broad and the narrow component and twice the velocity dispersion of the broad component \citep[e.g.][]{rupke13}. The Pearson coefficient for the relation between $\mathrm{v_{max}}$ and $\mathrm{L_{bol}}$ is 0.68 with a null hypothesis probability of 0.1\%, confirming also in this case a positive correlation between these two quantities. Note that despite the similar proportionality, the normalization of the curve in the SUPER sample and that in the \citet{fiore17} sample are different. We expect that the velocity of the outflowing gas depends on the properties of the medium it is moving through, which would result in some intrinsic scatter in any correlation between such velocity and the AGN bolometric luminosity. The difference in the normalization can also stem from differences in the gas mass of the host galaxies as explained below. 

To test the latter hypothesis we consider the \citet{menci19} model which provides a two-dimensional description of the expansion of an AGN-driven shock in a disk with an exponential gas density profile. Their predictions are expressed in terms of the total molecular gas mass and circular velocity of the host galaxy. In particular, from Eq. 14 in \citet{menci19}, we can express the velocity of the shock, $\mathrm{V_{S}}$, as a function of the AGN bolometric luminosity, $\mathrm{L_{bol}}$ and host galaxy gas mass, $\mathrm{M_{gas}}$, in the following way:

\begin{multline}
\mathrm{log\Bigg ( 
\mathrm{\frac{V_{S}}{km/s}}\Bigg )=2.65+0.35\,log\Bigg (  
\frac{L_{bol}}{10^{45}~erg s^{-1}}\Bigg )-\frac{1}{3} log \Bigg(
\frac{M_{gas}}{10^{10} ~M_{\odot}}\Bigg)}+ \\ \mathrm{log \Bigg (
\frac{V_c}{200~km/s}\Bigg )-0.65\,log \Bigg(
\frac{R_{S}}{1~kpc}\Bigg)}
\label{eq:vmax_lbol}
\end{multline}

\begin{figure}
\centering
\subfloat{\includegraphics[scale=0.4]{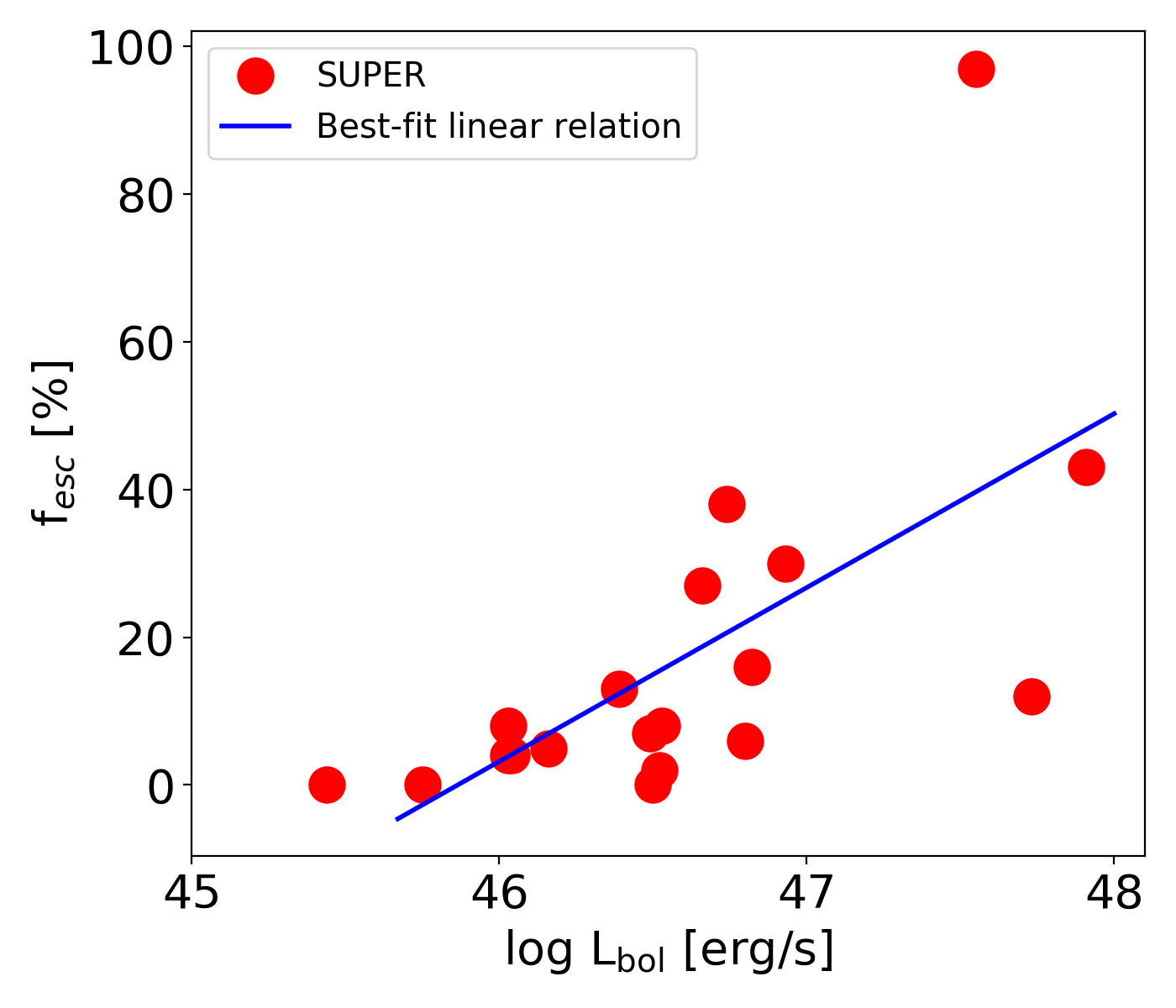}}
\caption{The plot shows the escape fraction as a function of the bolometric luminosity of the AGN for the Type-1 sample presented in this paper. The escape velocity is calculated for a distance of 2 kpc from Eq. \ref{eq:escape_velocity}. The blue curve shows the best-fit relation.\label{fig:fesc}}
\end{figure}

We use the above formula to predict the scaling of the outflow velocity versus the AGN bolometric luminosity for a range of host galaxies properties that encompass, to the best of our knowledge, those of our sample. We assume that the circular velocity ($\mathrm{V_{c}}$) of the galaxy is $\approx$300 km/s which is a reasonable assumption from the Tully-Fischer relation for galaxies with M$_{\ast} \approx 10^{9} - 10^{11} {\rm M}_{\odot}$ \citep[e.g.][]{cresci09}. We fix the radius of the shock (R$_{\rm S}$) to $\approx 1 $kpc which is where the bulk of the outflowing gas is located following our spectroastrometry analysis in Sect. \ref{sect5.4}. We use our CO(3-2) ALMA follow-up observations (Circosta et al. in prep.) to estimate the range of molecular gas masses for our AGN hosts: the mean gas mass for the sample is M$_{\rm gas}\approx 6\times10^{9}$ M$_\odot$, with 90$\%$ of the sample above M$_{\rm gas}\approx $ 2.7$\times 10^{9}$ M$_\odot$ and maximum measured gas mass is M$_{\rm gas}\approx 2\times10^{11} $M$_\odot$. As it can be seen in Fig. \ref{fig:vmax-scaling} the range of model tracks (shown by the black relations) is able to reproduce most of our observed values (shown as red circles), where we have assumed that $\mathrm{v_{max}}$ is a good approximation of V$_{\rm S}$. Therefore the \citet{menci19} description of an AGN-driven shock moving through a galaxy disk is able to reproduce our observations once the observed range of molecular gas masses for the host galaxies is used. The data points lying outside the model predictions from \citet{menci19} can be explained if galaxies with lower or higher molecular gas mass are taken into account. 

AGN-driven outflows are suggested as a physical mechanism able to effectively remove gas from their host galaxies, which is a very important process to reproduce the observed mass build-up of galaxies in the local universe and the chemical composition of the ISM \citep[e.g.][]{muratov15}. We can use the \oiii velocity profile to estimate the fraction of the ionized gas which has the ability to escape the gravitational potential of  AGN host galaxy. We first calculate the escape velocity of a cloud which is at a distance r from the AGN location \citep[see][]{rupke02}: 

\begin{equation}
\mathrm{V_{esc} = \sqrt{2}V_{c}\left[
1 + ln\left(
\frac{r_{max}}{r}
\right)
\right]^{1/2}},
\label{eq:escape_velocity}
\end{equation}

\noindent
where $\mathrm{r_{max}}$ is the extent of the dark matter halo (assumed $\sim$100 kpc here) and $\mathrm{V_{c}}$ is the circular velocity of the galaxy, assumed 300 km/s as Eq. \ref{eq:vmax_lbol} from the Tully-Fischer relation for the range of stellar mass in SUPER galaxies \citep[see][]{cresci09}. At 2 kpc, which is roughly the spatial resolution of the H-band SINFONI observations, the escape velocity calculated using Eq. \ref{eq:escape_velocity} is on average $\sim$950 km/s. We then estimate the escape fraction of ionized gas as the ratio between the flux of the \oiii$\lambda$5007 channels in the integrated spectra with |v| > $\mathrm{V_{esc}}$ and the total flux of the \oiii$\lambda$5007 line. The escape fraction of the ionized gas calculated for the Type-1 SUPER sample presented in this paper is reported in Table \ref{table:outflow_properties}. For most galaxies, $\lesssim$10\% of the outflowing ionized gas has the ability to escape the host galaxy, while the rest of the gas is expected to re-accrete back onto the host galaxy. The highest escape fractions of $\sim$43\% and $\sim$97\% are observed in J1333+1649 and J1441+0454 which are also the targets with high bolometric luminosity. Note that in J1441+0454, the exceptional high escape fraction is a result of the extremely blue-shifted \oiii profile which is also blended with the iron emission as explained in Sect. \ref{sect5.1}. Fig. \ref{fig:fesc} shows the escape fraction as a function of the bolometric luminosity which suggests a positive correlation between the two quantities, but with a large scatter. The Pearson correlation coefficient between the escape fraction and the bolometric luminosity is 0.64 with a null hypothesis probability of $\sim$0.2\%. Although the correlation is relatively weak compared to the $\mathrm{w_{80}}-L_{bol}$ relation in Fig. \ref{fig:w80-scaling}, the data might suggest that outflows hosted in high luminosity AGN have a higher fraction of escaping gas. 

We also note that SINFONI data only trace the ionized phase of the outflow, while a large fraction of the outflow might be present in the molecular gas phase \citep[e.g.][]{cicone18, fluetsch19}. Therefore, follow-up studies in other gas phases such as the warm and cold molecular phase and neutral gas phase are required to complete the picture of the effects of radiation pressure on the ISM of the galaxies.

\section{Summary and Conclusions}	\label{sect7}

We present near-infrared IFU spectroscopy for 21 X-ray selected Type-1 AGN from the SUPER survey (half of the survey size). We trace and characterize the velocity and extension of the ionized gas, and its outflow component, in the Narrow Line Region using the \oiii$\lambda$5007 transition. The main conclusions from the work presented in this paper are summarized below.

\begin{itemize}
\item[1.] Using a cut on the non-parametric velocity width, $\mathrm{w_{80}}>600$ km/s, in the integrated spectra, we find that all the Type-1 AGN in the SUPER survey show  the presence of ionized outflows. We consider the selected threshold in $\mathrm{w_{80}}$ a conservative choice, based on the very diverse distributions of low-z and high-z mass matched sample of start forming galaxies compared to the $\mathrm{w_{80}}$ distribution of AGN host galaxies (Sect. \ref{sect5.1}, Fig. \ref{fig:w80_distribution}). We also confirm a strong linear correlation between the \oiii luminosity and the X-ray luminosity, as previously observed in the literature (Sect. \ref{sect5.1}, Fig. \ref{fig:Loiii-Lx-relation}). 
\item[2.] Using  three different  methods (curve  of  growth  analysis, half-light radii and PSF-subtraction method) we find evidence of kiloparsec-scale extended ionized gas emission in seven out of the eleven targets for which \oiii is detected at a S/N$>5$ in the integrated spectrum (Sect. \ref{sect5.2}, Figs. \ref{fig:growth_curves} and \ref{fig:astrometry1}). Flux and velocity maps of these resolved targets reveal outflows extended to $\sim6 $ kiloparsec (quantified by $\mathrm{D_{600}}$ value, Table \ref{table:outflow_properties} and Figs. \ref{fig:flux_velocity1} \& \ref{fig:flux_velocity2}), with indications of redshifted outflows in three objects (cid$\_346$, J1333+1649 and S82X1905).
\item[3.] We use the spectroastrometry method to determine the distance of the bulk of the gas moving at a given velocity. We find that the high velocity outflowing gas (|v| > 600 km/s) is contained in the central $\sim$1 kpc for $\sim$80\% of the targets for which the spectroastrometry analysis was performed (Sect. \ref{sect5.4}). For two objects (cid\_346 and S82X1905) the bulk motion of high velocity gas is extended to $\sim$3 kpc.
\item[4.] We explore a range of plausible assumptions on the physical properties of the outflow (its geometry, velocity and radius) and of the outflowing gas (i.e. its electron density) and report the range of derived mass outflow rates for each target (Sect. \ref{sect5.5}). The mass outflow rates of the Type-1 sample are in the range $\sim$0.01--1000 $\mathrm{M_{\odot}}$/yr (Table \ref{table:outflow_properties}). After factoring in the systematic uncertainties in the outflow models, these outflow rates seem to correlate with the bolometric luminosity of the AGN.
\item[5.] The non-parametric velocity, $\mathrm{w_{80}}$, strongly correlates with the X-ray luminosity, $\mathrm{L_{X}(2-10 keV)}$, and the bolometric luminosity of the central super-massive black hole (Fig. \ref{fig:w80-scaling}). The correlation is relatively weak for $\mathrm{w_{80}}$ vs. $\mathrm{M_{BH}}$ (or $\mathrm{L_{Edd}}$). The maximum velocity vs. bolometric luminosity plot for the SUPER sample agree with model predictions for an AGN-driven shock driving an outflow through a galaxy disk (Fig. \ref{fig:vmax-scaling}).
\item[6.] For most galaxies, <10\% of the outflows in the ionized gas have the potenial to escape the gravitational potential of the host galaxy. The escape fraction also increases with the bolometric luminosity of the AGN (Fig. \ref{fig:fesc}).
\end{itemize}

While this paper focused on  presenting the ionized gas kinematics of the Type-1 sample, an upcoming publication will present the SINFONI and ALMA results for the overall (Type-1 and Type-2) AGN SUPER sample.
The different types of morphology of the ionized gas and the associated outflows presented in this paper clearly show the advantages of performing high spatial resolution observations. Even better spatial and spectral resolution and higher sensitivity is expected with future facilities such as ELT/HARMONI. Such observations will enable us to resolve the extended redshifted and blueshifted outflowing gas which might reveal sub-structures within the outflow, and trace the radial evolution of the outflowing gas, which is certainly not possible at z$\sim 2$ with today's instrumentation. Finally, we are currently limited to study ionized outflows from the ground up to z$\approx 4$, observing the \oiii line in the K-band. With the IFU capabilities of NIRSpec on-board of JWST, a study similar to the one presented in this paper could be performed with comparable resolution at z$>4$ and will allow to further constrain the importance of AGN feedback at earlier comic epochs.

\vspace{0.5cm}
\noindent
Acknowledgements: We thank the referee for the useful and constructive comments. We thank Michele Cirasuolo, Alice Concas and Dominika Wylezalek for providing the comparison data for the $\mathrm{w_{80}}$ distribution. MP is supported by the Programa Atracci\'on de Talento de la Comunidad de Madrid via grant 2018-T2/TIC-11715. Based on observations collected at the European organisation for Astronomical Research in the Southern Hemisphere under ESO program 196.A-0377.

\bibliographystyle{aa}
\bibliography{reference.bib}

\begin{thebibliography}{152}
\expandafter\ifx\csname natexlab\endcsname\relax\def\natexlab#1{#1}\fi

\bibitem[{{Aalto} {et~al.}(2019){Aalto}, {Muller}, {K{\"o}nig}, {Falstad},
  {Mangum}, {Sakamoto}, {Privon}, {Gallagher}, {Combes}, {Garc{\'\i}a-Burillo},
  {Mart{\'\i}n}, {Viti}, {van der Werf}, {Evans}, {Black}, {Varenius},
  {Beswick}, {Fuller}, {Henkel}, {Kohno}, {Alatalo}, \& {M{\"u}hle}}]{aalto19}
{Aalto}, S., {Muller}, S., {K{\"o}nig}, S., {et~al.} 2019, \aap, 627, A147

\bibitem[{{Aladro} {et~al.}(2018){Aladro}, {K{\"o}nig}, {Aalto},
  {Gonz{\'a}lez-Alfonso}, {Falstad}, {Mart{\'\i}n}, {Muller},
  {Garc{\'\i}a-Burillo}, {Henkel}, \& {van der Werf}}]{aladro18}
{Aladro}, R., {K{\"o}nig}, S., {Aalto}, S., {et~al.} 2018, \aap, 617, A20

\bibitem[{{Alexander} {et~al.}(2010){Alexander}, {Swinbank}, {Smail},
  {McDermid}, \& {Nesvadba}}]{alexander10}
{Alexander}, D.~M., {Swinbank}, A.~M., {Smail}, I., {McDermid}, R., \&
  {Nesvadba}, N.~P.~H. 2010, \mnras, 402, 2211

\bibitem[{{Bae} {et~al.}(2017){Bae}, {Woo}, {Karouzos}, {Gallo}, {Flohic},
  {Shen}, \& {Yoon}}]{bae17}
{Bae}, H.-J., {Woo}, J.-H., {Karouzos}, M., {et~al.} 2017, \apj, 837, 91

\bibitem[{{Baron} \& {Netzer}(2019)}]{baron19}
{Baron}, D. \& {Netzer}, H. 2019, \mnras, 486, 4290

\bibitem[{{Batiste} {et~al.}(2017){Batiste}, {Bentz}, {Raimundo},
  {Vestergaard}, \& {Onken}}]{batiste17}
{Batiste}, M., {Bentz}, M.~C., {Raimundo}, S.~I., {Vestergaard}, M., \&
  {Onken}, C.~A. 2017, \apjl, 838, L10

\bibitem[{{Begelman}(2003)}]{begelman03}
{Begelman}, M.~C. 2003, Science, 300, 1898

\bibitem[{{Bennert} {et~al.}(2002){Bennert}, {Falcke}, {Schulz}, {Wilson}, \&
  {Wills}}]{bennert02}
{Bennert}, N., {Falcke}, H., {Schulz}, H., {Wilson}, A.~S., \& {Wills}, B.~J.
  2002, \apjl, 574, L105

\bibitem[{{Bischetti} {et~al.}(2017){Bischetti}, {Piconcelli}, {Vietri},
  {Bongiorno}, {Fiore}, {Sani}, {Marconi}, {Duras}, {Zappacosta}, {Brusa},
  {Comastri}, {Cresci}, {Feruglio}, {Giallongo}, {La Franca}, {Mainieri},
  {Mannucci}, {Martocchia}, {Ricci}, {Schneider}, {Testa}, \&
  {Vignali}}]{bischetti17}
{Bischetti}, M., {Piconcelli}, E., {Vietri}, G., {et~al.} 2017, \aap, 598, A122

\bibitem[{{Boroson} \& {Green}(1992)}]{boroson92}
{Boroson}, T.~A. \& {Green}, R.~F. 1992, \apjs, 80, 109

\bibitem[{{Bouch{\'e}} {et~al.}(2013){Bouch{\'e}}, {Murphy}, {Kacprzak},
  {P{\'e}roux}, {Contini}, {Martin}, \& {Dessauges-Zavadsky}}]{bouche13}
{Bouch{\'e}}, N., {Murphy}, M.~T., {Kacprzak}, G.~G., {et~al.} 2013, Science,
  341, 50

\bibitem[{{Brandt} \& {Alexander}(2015)}]{brandt15}
{Brandt}, W.~N. \& {Alexander}, D.~M. 2015, \aapr, 23, 1

\bibitem[{{Brusa} {et~al.}(2015){Brusa}, {Bongiorno}, {Cresci}, {Perna},
  {Marconi}, {Mainieri}, {Maiolino}, {Salvato}, {Lusso}, {Santini}, {Comastri},
  {Fiore}, {Gilli}, {La Franca}, {Lanzuisi}, {Lutz}, {Merloni}, {Mignoli},
  {Onori}, {Piconcelli}, {Rosario}, {Vignali}, \& {Zamorani}}]{brusa15a}
{Brusa}, M., {Bongiorno}, A., {Cresci}, G., {et~al.} 2015, \mnras, 446, 2394

\bibitem[{{Brusa} {et~al.}(2018){Brusa}, {Cresci}, {Daddi}, {Paladino},
  {Perna}, {Bongiorno}, {Lusso}, {Sargent}, {Casasola}, {Feruglio},
  {Fraternali}, {Georgiev}, {Mainieri}, {Carniani}, {Comastri}, {Duras},
  {Fiore}, {Mannucci}, {Marconi}, {Piconcelli}, {Zamorani}, {Gilli}, {La
  Franca}, {Lanzuisi}, {Lutz}, {Santini}, {Scoville}, {Vignali}, {Vito},
  {Rabien}, {Busoni}, \& {Bonaglia}}]{brusa18}
{Brusa}, M., {Cresci}, G., {Daddi}, E., {et~al.} 2018, \aap, 612, A29

\bibitem[{{Brusa} {et~al.}(2016){Brusa}, {Perna}, {Cresci}, {Schramm},
  {Delvecchio}, {Lanzuisi}, {Mainieri}, {Mignoli}, {Zamorani}, {Berta},
  {Bongiorno}, {Comastri}, {Fiore}, {Kakkad}, {Marconi}, {Rosario}, {Contini},
  \& {Lamareille}}]{brusa16}
{Brusa}, M., {Perna}, M., {Cresci}, G., {et~al.} 2016, \aap, 588, A58

\bibitem[{{Bryant} {et~al.}(2015){Bryant}, {Owers}, {Robotham}, {Croom},
  {Driver}, {Drinkwater}, {Lorente}, {Cortese}, {Scott}, {Colless}, {Schaefer},
  {Taylor}, {Konstantopoulos}, {Allen}, {Baldry}, {Barnes}, {Bauer},
  {Bland-Hawthorn}, {Bloom}, {Brooks}, {Brough}, {Cecil}, {Couch}, {Croton},
  {Davies}, {Ellis}, {Fogarty}, {Foster}, {Glazebrook}, {Goodwin}, {Green},
  {Gunawardhana}, {Hampton}, {Ho}, {Hopkins}, {Kewley}, {Lawrence},
  {Leon-Saval}, {Leslie}, {McElroy}, {Lewis}, {Liske}, {L{\'o}pez-S{\'a}nchez},
  {Mahajan}, {Medling}, {Metcalfe}, {Meyer}, {Mould}, {Obreschkow}, {O'Toole},
  {Pracy}, {Richards}, {Shanks}, {Sharp}, {Sweet}, {Thomas}, {Tonini}, \&
  {Walcher}}]{bryant15}
{Bryant}, J.~J., {Owers}, M.~S., {Robotham}, A.~S.~G., {et~al.} 2015, \mnras,
  447, 2857

\bibitem[{{Bundy} {et~al.}(2015){Bundy}, {Bershady}, {Law}, {Yan}, {Drory},
  {MacDonald}, {Wake}, {Cherinka}, {S{\'a}nchez-Gallego}, {Weijmans}, {Thomas},
  {Tremonti}, {Masters}, {Coccato}, {Diamond-Stanic}, {Arag{\'o}n-Salamanca},
  {Avila-Reese}, {Badenes}, {Falc{\'o}n-Barroso}, {Belfiore}, {Bizyaev},
  {Blanc}, {Bland-Hawthorn}, {Blanton}, {Brownstein}, {Byler}, {Cappellari},
  {Conroy}, {Dutton}, {Emsellem}, {Etherington}, {Frinchaboy}, {Fu}, {Gunn},
  {Harding}, {Johnston}, {Kauffmann}, {Kinemuchi}, {Klaene}, {Knapen},
  {Leauthaud}, {Li}, {Lin}, {Maiolino}, {Malanushenko}, {Malanushenko}, {Mao},
  {Maraston}, {McDermid}, {Merrifield}, {Nichol}, {Oravetz}, {Pan}, {Parejko},
  {Sanchez}, {Schlegel}, {Simmons}, {Steele}, {Steinmetz}, {Thanjavur},
  {Thompson}, {Tinker}, {van den Bosch}, {Westfall}, {Wilkinson}, {Wright},
  {Xiao}, \& {Zhang}}]{bundy14}
{Bundy}, K., {Bershady}, M.~A., {Law}, D.~R., {et~al.} 2015, \apj, 798, 7

\bibitem[{{Caglar} {et~al.}(2020){Caglar}, {Burtscher}, {Brandl}, {Brinchmann},
  {Davies}, {Hicks}, {Koss}, {Lin}, {Maciejewski}, {M{\"u}ller-S{\'a}nchez},
  {Riffel}, {Riffel}, {Rosario}, {Schartmann}, {Schnorr-M{\"u}ller}, {Taro
  Shimizu}, {Storchi-Bergmann}, {Veilleux}, {de Xivry}, \&
  {Bennert}}]{caglar20}
{Caglar}, T., {Burtscher}, L., {Brandl}, B., {et~al.} 2020, \aap, 634, A114

\bibitem[{{Cano-D{\'{\i}}az} {et~al.}(2012){Cano-D{\'{\i}}az}, {Maiolino},
  {Marconi}, {Netzer}, {Shemmer}, \& {Cresci}}]{cano-diaz12}
{Cano-D{\'{\i}}az}, M., {Maiolino}, R., {Marconi}, A., {et~al.} 2012, \aap,
  537, L8

\bibitem[{{Carniani} {et~al.}(2016){Carniani}, {Marconi}, {Maiolino},
  {Balmaverde}, {Brusa}, {Cano-D{\'\i}az}, {Cicone}, {Comastri}, {Cresci}, \&
  {Fiore}}]{carniani16}
{Carniani}, S., {Marconi}, A., {Maiolino}, R., {et~al.} 2016, \aap, 591, A28

\bibitem[{{Carniani} {et~al.}(2015){Carniani}, {Marconi}, {Maiolino},
  {Balmaverde}, {Brusa}, {Cano-D{\'{\i}}az}, {Cicone}, {Comastri}, {Cresci},
  {Fiore}, {Feruglio}, {La Franca}, {Mainieri}, {Mannucci}, {Nagao}, {Netzer},
  {Piconcelli}, {Risaliti}, {Schneider}, \& {Shemmer}}]{carniani15}
{Carniani}, S., {Marconi}, A., {Maiolino}, R., {et~al.} 2015, \aap, 580, A102

\bibitem[{{Cazzoli} {et~al.}(2016){Cazzoli}, {Arribas}, {Maiolino}, \&
  {Colina}}]{cazzoli16}
{Cazzoli}, S., {Arribas}, S., {Maiolino}, R., \& {Colina}, L. 2016, \aap, 590,
  A125

\bibitem[{{Choi} {et~al.}(2014){Choi}, {Naab}, {Ostriker}, {Johansson}, \&
  {Moster}}]{choi14}
{Choi}, E., {Naab}, T., {Ostriker}, J.~P., {Johansson}, P.~H., \& {Moster},
  B.~P. 2014, \mnras, 442, 440

\bibitem[{{Cicone} {et~al.}(2018){Cicone}, {Brusa}, {Ramos Almeida}, {Cresci},
  {Husemann}, \& {Mainieri}}]{cicone18}
{Cicone}, C., {Brusa}, M., {Ramos Almeida}, C., {et~al.} 2018, Nature
  Astronomy, 2, 176

\bibitem[{{Cicone} {et~al.}(2020){Cicone}, {Maiolino}, {Aalto}, {Muller}, \&
  {Feruglio}}]{cicone20}
{Cicone}, C., {Maiolino}, R., {Aalto}, S., {Muller}, S., \& {Feruglio}, C.
  2020, \aap, 633, A163

\bibitem[{{Circosta} {et~al.}(2018){Circosta}, {Mainieri}, {Padovani},
  {Lanzuisi}, {Salvato}, {Harrison}, {Kakkad}, {Puglisi}, {Vietri}, {Zamorani},
  {Cicone}, {Husemann}, {Vignali}, {Balmaverde}, {Bischetti}, {Bongiorno},
  {Brusa}, {Carniani}, {Civano}, {Comastri}, {Cresci}, {Feruglio}, {Fiore},
  {Fotopoulou}, {Karim}, {Lamastra}, {Magnelli}, {Mannucci}, {Marconi},
  {Merloni}, {Netzer}, {Perna}, {Piconcelli}, {Rodighiero}, {Schinnerer},
  {Schramm}, {Schulze}, {Silverman}, \& {Zappacosta}}]{circosta18}
{Circosta}, C., {Mainieri}, V., {Padovani}, P., {et~al.} 2018, \aap, 620, A82

\bibitem[{{Civano} {et~al.}(2016){Civano}, {Marchesi}, {Comastri}, {Urry},
  {Elvis}, {Cappelluti}, {Puccetti}, {Brusa}, {Zamorani}, {Hasinger},
  {Aldcroft}, {Alexand er}, {Allevato}, {Brunner}, {Capak}, {Finoguenov},
  {Fiore}, {Fruscione}, {Gilli}, {Glotfelty}, {Griffiths}, {Hao}, {Harrison},
  {Jahnke}, {Kartaltepe}, {Karim}, {LaMassa}, {Lanzuisi}, {Miyaji}, {Ranalli},
  {Salvato}, {Sargent}, {Scoville}, {Schawinski}, {Schinnerer}, {Silverman},
  {Smolcic}, {Stern}, {Toft}, {Trakhtenbrot}, {Treister}, \&
  {Vignali}}]{civano16}
{Civano}, F., {Marchesi}, S., {Comastri}, A., {et~al.} 2016, \apj, 819, 62

\bibitem[{{Coatman} {et~al.}(2019){Coatman}, {Hewett}, {Banerji}, {Richards},
  {Hennawi}, \& {Prochaska}}]{coatman19}
{Coatman}, L., {Hewett}, P.~C., {Banerji}, M., {et~al.} 2019, \mnras, 486, 5335

\bibitem[{{Concas} {et~al.}(2019){Concas}, {Popesso}, {Brusa}, {Mainieri}, \&
  {Thomas}}]{concas19}
{Concas}, A., {Popesso}, P., {Brusa}, M., {Mainieri}, V., \& {Thomas}, D. 2019,
  \aap, 622, A188

\bibitem[{{Crenshaw} {et~al.}(2010){Crenshaw}, {Kraemer}, {Schmitt},
  {Jaff{\'e}}, {Deo}, {Collins}, \& {Fischer}}]{crenshaw10}
{Crenshaw}, D.~M., {Kraemer}, S.~B., {Schmitt}, H.~R., {et~al.} 2010, \aj, 139,
  871

\bibitem[{{Cresci} {et~al.}(2009){Cresci}, {Hicks}, {Genzel}, {Schreiber},
  {Davies}, {Bouch{\'e}}, {Buschkamp}, {Genel}, {Shapiro}, {Tacconi},
  {Sommer-Larsen}, {Burkert}, {Eisenhauer}, {Gerhard}, {Lutz}, {Naab},
  {Sternberg}, {Cimatti}, {Daddi}, {Erb}, {Kurk}, {Lilly}, {Renzini},
  {Shapley}, {Steidel}, \& {Caputi}}]{cresci09}
{Cresci}, G., {Hicks}, E.~K.~S., {Genzel}, R., {et~al.} 2009, \apj, 697, 115

\bibitem[{{Cresci} {et~al.}(2015){Cresci}, {Mainieri}, {Brusa}, {Marconi},
  {Perna}, {Mannucci}, {Piconcelli}, {Maiolino}, {Feruglio}, {Fiore},
  {Bongiorno}, {Lanzuisi}, {Merloni}, {Schramm}, {Silverman}, \&
  {Civano}}]{cresci15}
{Cresci}, G., {Mainieri}, V., {Brusa}, M., {et~al.} 2015, \apj, 799, 82

\bibitem[{{Curran}(2019)}]{curran19}
{Curran}, S.~J. 2019, \mnras, 484, 3911

\bibitem[{{Curti} {et~al.}(2020){Curti}, {Maiolino}, {Cirasuolo}, {Mannucci},
  {Williams}, {Auger}, {Mercurio}, {Hayden-Pawson}, {Cresci}, {Marconi},
  {Belfiore}, {Cappellari}, {Cicone}, {Cullen}, {Meneghetti}, {Ota}, {Peng},
  {Pettini}, {Swinbank}, \& {Troncoso}}]{curti20}
{Curti}, M., {Maiolino}, R., {Cirasuolo}, M., {et~al.} 2020, \mnras, 492, 821

\bibitem[{{Davies} {et~al.}(2020{\natexlab{a}}){Davies}, {Baron}, {Shimizu},
  {Netzer}, {Burtscher}, {de Zeeuw}, {Genzel}, {Hicks}, {Koss}, {Lin}, {Lutz},
  {Maciejewski}, {M{\"u}ller-S{\'a}nchez}, {Orban de Xivry}, {Ricci}, {Riffel},
  {Riffel}, {Rosario}, {Schartmann}, {Schnorr-M{\"u}ller}, {Shangguan},
  {Sternberg}, {Sturm}, {Storchi-Bergmann}, {Tacconi}, \&
  {Veilleux}}]{davies20}
{Davies}, R., {Baron}, D., {Shimizu}, T., {et~al.} 2020{\natexlab{a}}, arXiv
  e-prints, arXiv:2003.06153

\bibitem[{{Davies}(2007)}]{davies07}
{Davies}, R.~I. 2007, \mnras, 375, 1099

\bibitem[{{Davies} {et~al.}(2014){Davies}, {Maciejewski}, {Hicks}, {Emsellem},
  {Erwin}, {Burtscher}, {Dumas}, {Lin}, {Malkan}, {M{\"u}ller-S{\'a}nchez},
  {Orban de Xivry}, {Rosario}, {Schnorr-M{\"u}ller}, \& {Tran}}]{davies14}
{Davies}, R.~I., {Maciejewski}, W., {Hicks}, E.~K.~S., {et~al.} 2014, \apj,
  792, 101

\bibitem[{{Davies} {et~al.}(2019){Davies}, {F{\"o}rster Schreiber},
  {{\"U}bler}, {Genzel}, {Lutz}, {Renzini}, {Tacchella}, {Tacconi}, {Belli}, \&
  {Burkert}}]{davies19}
{Davies}, R.~L., {F{\"o}rster Schreiber}, N.~M., {{\"U}bler}, H., {et~al.}
  2019, \apj, 873, 122

\bibitem[{{Davies} {et~al.}(2020{\natexlab{b}}){Davies}, {Schreiber}, {Lutz},
  {Genzel}, {Belli}, {Shimizu}, {Contursi}, {Davies}, {Herrera-Camus}, {Lee},
  {Naab}, {Price}, {Renzini}, {Schruba}, {Sternberg}, {Tacconi}, {{\"U}bler},
  {Wisnioski}, \& {Wuyts}}]{daviesR20}
{Davies}, R.~L., {Schreiber}, N.~M.~F., {Lutz}, D., {et~al.}
  2020{\natexlab{b}}, \apj, 894, 28

\bibitem[{{Dempsey} \& {Zakamska}(2018)}]{dempsey18}
{Dempsey}, R. \& {Zakamska}, N.~L. 2018, \mnras, 477, 4615

\bibitem[{{den Brok} {et~al.}(2020){den Brok}, {Carollo}, {Erroz-Ferrer},
  {Fagioli}, {Brinchmann}, {Emsellem}, {Krajnovi{\'c}}, {Marino}, {Onodera},
  {Tacchella}, {Weilbacher}, \& {Woo}}]{den-brok20}
{den Brok}, M., {Carollo}, C.~M., {Erroz-Ferrer}, S., {et~al.} 2020, \mnras,
  491, 4089

\bibitem[{{Dimitrijevi{\'c}} {et~al.}(2007){Dimitrijevi{\'c}}, {Popovi{\'c}},
  {Kova{\v c}evi{\'c}}, {Da{\v c}i{\'c}}, \& {Ili{\'c}}}]{dimitrijevic07}
{Dimitrijevi{\'c}}, M.~S., {Popovi{\'c}}, L.~{\v C}., {Kova{\v c}evi{\'c}}, J.,
  {Da{\v c}i{\'c}}, M., \& {Ili{\'c}}, D. 2007, \mnras, 374, 1181

\bibitem[{{Eisenhauer} {et~al.}(2003){Eisenhauer}, {Abuter}, {Bickert},
  {Biancat-Marchet}, {Bonnet}, {Brynnel}, {Conzelmann}, {Delabre}, {Donaldson},
  {Farinato}, {Fedrigo}, {Genzel}, {Hubin}, {Iserlohe}, {Kasper},
  {Kissler-Patig}, {Monnet}, {Roehrle}, {Schreiber}, {Stroebele}, {Tecza},
  {Thatte}, \& {Weisz}}]{eisenhauer03}
{Eisenhauer}, F., {Abuter}, R., {Bickert}, K., {et~al.} 2003, in Society of
  Photo-Optical Instrumentation Engineers (SPIE) Conference Series, Vol. 4841,
  Instrument Design and Performance for Optical/Infrared Ground-based
  Telescopes, ed. M.~{Iye} \& A.~F.~M. {Moorwood}, 1548--1561

\bibitem[{{Emonts} {et~al.}(2017){Emonts}, {Colina}, {Piqueras-L{\'o}pez},
  {Garcia-Burillo}, {Pereira-Santaella}, {Arribas}, {Labiano}, \&
  {Alonso-Herrero}}]{emonts17}
{Emonts}, B.~H.~C., {Colina}, L., {Piqueras-L{\'o}pez}, J., {et~al.} 2017,
  \aap, 607, A116

\bibitem[{{Faucher-Gigu{\`e}re} \& {Quataert}(2012)}]{faucher12}
{Faucher-Gigu{\`e}re}, C.-A. \& {Quataert}, E. 2012, \mnras, 425, 605

\bibitem[{{Feruglio} {et~al.}(2017){Feruglio}, {Ferrara}, {Bischetti},
  {Downes}, {Neri}, {Ceccarelli}, {Cicone}, {Fiore}, {Gallerani}, \&
  {Maiolino}}]{feruglio17}
{Feruglio}, C., {Ferrara}, A., {Bischetti}, M., {et~al.} 2017, \aap, 608, A30

\bibitem[{{Fiore} {et~al.}(2017){Fiore}, {Feruglio}, {Shankar}, {Bischetti},
  {Bongiorno}, {Brusa}, {Carniani}, {Cicone}, {Duras}, \& {Lamastra}}]{fiore17}
{Fiore}, F., {Feruglio}, C., {Shankar}, F., {et~al.} 2017, \aap, 601, A143

\bibitem[{{Fluetsch} {et~al.}(2019){Fluetsch}, {Maiolino}, {Carniani},
  {Marconi}, {Cicone}, {Bourne}, {Costa}, {Fabian}, {Ishibashi}, \&
  {Venturi}}]{fluetsch19}
{Fluetsch}, A., {Maiolino}, R., {Carniani}, S., {et~al.} 2019, \mnras, 483,
  4586

\bibitem[{{F{\"o}rster Schreiber} {et~al.}(2018){F{\"o}rster Schreiber},
  {Renzini}, {Mancini}, {Genzel}, {Bouch{\'e}}, {Cresci}, {Hicks}, {Lilly},
  {Peng}, \& {Burkert}}]{forster-schreiber18}
{F{\"o}rster Schreiber}, N.~M., {Renzini}, A., {Mancini}, C., {et~al.} 2018,
  \apjs, 238, 21

\bibitem[{{F{\"o}rster Schreiber} {et~al.}(2019){F{\"o}rster Schreiber},
  {{\"U}bler}, {Davies}, {Genzel}, {Wisnioski}, {Belli}, {Shimizu}, {Lutz},
  {Fossati}, \& {Herrera-Camus}}]{forster-schreiber19}
{F{\"o}rster Schreiber}, N.~M., {{\"U}bler}, H., {Davies}, R.~L., {et~al.}
  2019, \apj, 875, 21

\bibitem[{{Fruchter} \& {Hook}(2002)}]{fruchter02}
{Fruchter}, A.~S. \& {Hook}, R.~N. 2002, \pasp, 114, 144

\bibitem[{{Gallagher} {et~al.}(2019){Gallagher}, {Maiolino}, {Belfiore},
  {Drory}, {Riffel}, \& {Riffel}}]{gallagher19}
{Gallagher}, R., {Maiolino}, R., {Belfiore}, F., {et~al.} 2019, \mnras, 485,
  3409

\bibitem[{{Garc{\'\i}a-Burillo} {et~al.}(2014){Garc{\'\i}a-Burillo}, {Combes},
  {Usero}, {Aalto}, {Krips}, {Viti}, {Alonso-Herrero}, {Hunt}, {Schinnerer}, \&
  {Baker}}]{garcia-burillo14}
{Garc{\'\i}a-Burillo}, S., {Combes}, F., {Usero}, A., {et~al.} 2014, \aap, 567,
  A125

\bibitem[{{Gebhardt} {et~al.}(2000){Gebhardt}, {Bender}, {Bower}, {Dressler},
  {Faber}, {Filippenko}, {Green}, {Grillmair}, {Ho}, \&
  {Kormendy}}]{gebhardt00}
{Gebhardt}, K., {Bender}, R., {Bower}, G., {et~al.} 2000, \apj, 539, L13

\bibitem[{{Georgakakis} \& {Nandra}(2011)}]{georgakakis11}
{Georgakakis}, A. \& {Nandra}, K. 2011, \mnras, 414, 992

\bibitem[{{Hainline} {et~al.}(2014){Hainline}, {Hickox}, {Greene}, {Myers},
  {Zakamska}, {Liu}, \& {Liu}}]{hainline14}
{Hainline}, K.~N., {Hickox}, R.~C., {Greene}, J.~E., {et~al.} 2014, \apj, 787,
  65

\bibitem[{{Harrison} {et~al.}(2016){Harrison}, {Alexander}, {Mullaney},
  {Stott}, {Swinbank}, {Arumugam}, {Bauer}, {Bower}, {Bunker}, \&
  {Sharples}}]{harrison16}
{Harrison}, C.~M., {Alexander}, D.~M., {Mullaney}, J.~R., {et~al.} 2016,
  \mnras, 456, 1195

\bibitem[{{Harrison} {et~al.}(2014){Harrison}, {Alexander}, {Mullaney}, \&
  {Swinbank}}]{harrison14}
{Harrison}, C.~M., {Alexander}, D.~M., {Mullaney}, J.~R., \& {Swinbank}, A.~M.
  2014, \mnras, 441, 3306

\bibitem[{{Harrison} {et~al.}(2012){Harrison}, {Alexander}, {Swinbank},
  {Smail}, {Alaghband-Zadeh}, {Bauer}, {Chapman}, {Del Moro}, {Hickox},
  {Ivison}, {Men{\'e}ndez-Delmestre}, {Mullaney}, \& {Nesvadba}}]{harrison12}
{Harrison}, C.~M., {Alexander}, D.~M., {Swinbank}, A.~M., {et~al.} 2012,
  \mnras, 426, 1073

\bibitem[{{Harrison} {et~al.}(2018){Harrison}, {Costa}, {Tadhunter},
  {Fl{\"u}tsch}, {Kakkad}, {Perna}, \& {Vietri}}]{harrison18}
{Harrison}, C.~M., {Costa}, T., {Tadhunter}, C.~N., {et~al.} 2018, Nature
  Astronomy, 2, 198

\bibitem[{{Hill} \& {Zakamska}(2014)}]{hill14}
{Hill}, M.~J. \& {Zakamska}, N.~L. 2014, \mnras, 439, 2701

\bibitem[{{Hopkins} {et~al.}(2016){Hopkins}, {Torrey}, {Faucher-Gigu{\`e}re},
  {Quataert}, \& {Murray}}]{hopkins16}
{Hopkins}, P.~F., {Torrey}, P., {Faucher-Gigu{\`e}re}, C.-A., {Quataert}, E.,
  \& {Murray}, N. 2016, \mnras, 458, 816

\bibitem[{{Husemann} {et~al.}(2017){Husemann}, {Davis}, {Jahnke},
  {Dannerbauer}, {Urrutia}, \& {Hodge}}]{husemann17}
{Husemann}, B., {Davis}, T.~A., {Jahnke}, K., {et~al.} 2017, \mnras, 470, 1570

\bibitem[{{Husemann} {et~al.}(2014){Husemann}, {Jahnke}, {S{\'a}nchez},
  {Wisotzki}, {Nugroho}, {Kupko}, \& {Schramm}}]{husemann14}
{Husemann}, B., {Jahnke}, K., {S{\'a}nchez}, S.~F., {et~al.} 2014, \mnras, 443,
  755

\bibitem[{{Husemann} {et~al.}(2016){Husemann}, {Scharw{\"a}chter}, {Bennert},
  {Mainieri}, {Woo}, \& {Kakkad}}]{husemann16}
{Husemann}, B., {Scharw{\"a}chter}, J., {Bennert}, V.~N., {et~al.} 2016, \aap,
  594, A44

\bibitem[{{Husemann} {et~al.}(2019){Husemann}, {Scharw{\"a}chter}, {Davis},
  {P{\'e}rez-Torres}, {Smirnova-Pinchukova}, {Tremblay}, {Krumpe}, {Combes},
  {Baum}, {Busch}, {Connor}, {Croom}, {Gaspari}, {Kraft}, {O'Dea}, {Powell},
  {Singha}, \& {Urrutia}}]{husemann19}
{Husemann}, B., {Scharw{\"a}chter}, J., {Davis}, T.~A., {et~al.} 2019, \aap,
  627, A53

\bibitem[{{Husemann} {et~al.}(2013){Husemann}, {Wisotzki}, {S{\'a}nchez}, \&
  {Jahnke}}]{husemann13}
{Husemann}, B., {Wisotzki}, L., {S{\'a}nchez}, S.~F., \& {Jahnke}, K. 2013,
  \aap, 549, A43

\bibitem[{{Ishibashi} \& {Fabian}(2016)}]{ishibashi16}
{Ishibashi}, W. \& {Fabian}, A.~C. 2016, \mnras, 457, 2864

\bibitem[{{Ishibashi} {et~al.}(2019){Ishibashi}, {Fabian}, \&
  {Reynolds}}]{ishibashi19}
{Ishibashi}, W., {Fabian}, A.~C., \& {Reynolds}, C.~S. 2019, \mnras, 486, 2210

\bibitem[{{Jahnke} {et~al.}(2004){Jahnke}, {Wisotzki}, {S{\'a}nchez},
  {Christensen}, {Becker}, {Kelz}, \& {Roth}}]{jahnke04}
{Jahnke}, K., {Wisotzki}, L., {S{\'a}nchez}, S.~F., {et~al.} 2004,
  Astronomische Nachrichten, 325, 128

\bibitem[{{Kakkad} {et~al.}(2018){Kakkad}, {Groves}, {Dopita}, {Thomas},
  {Davies}, {Mainieri}, {Kharb}, {Scharw{\"a}chter}, {Hampton}, \&
  {Ho}}]{kakkad18}
{Kakkad}, D., {Groves}, B., {Dopita}, M., {et~al.} 2018, \aap, 618, A6

\bibitem[{{Kakkad} {et~al.}(2016){Kakkad}, {Mainieri}, {Padovani}, {Cresci},
  {Husemann}, {Carniani}, {Brusa}, {Lamastra}, {Lanzuisi}, {Piconcelli}, \&
  {Schramm}}]{kakkad16}
{Kakkad}, D., {Mainieri}, V., {Padovani}, P., {et~al.} 2016, \aap, 592, A148

\bibitem[{{Karouzos} {et~al.}(2016){Karouzos}, {Woo}, \& {Bae}}]{karouzos16}
{Karouzos}, M., {Woo}, J.-H., \& {Bae}, H.-J. 2016, \apj, 819, 148

\bibitem[{{King}(2003)}]{king03}
{King}, A. 2003, \apjl, 596, L27

\bibitem[{{King} \& {Pounds}(2015)}]{king_pounds15}
{King}, A. \& {Pounds}, K. 2015, \araa, 53, 115

\bibitem[{{Krug} {et~al.}(2010){Krug}, {Rupke}, \& {Veilleux}}]{krug10}
{Krug}, H.~B., {Rupke}, D. S.~N., \& {Veilleux}, S. 2010, \apj, 708, 1145

\bibitem[{{LaMassa} {et~al.}(2016){LaMassa}, {Urry}, {Cappelluti},
  {B{\"o}hringer}, {Comastri}, {Glikman}, {Richards}, {Ananna}, {Brusa},
  {Cardamone}, {Chon}, {Civano}, {Farrah}, {Gilfanov}, {Green}, {Komossa},
  {Lira}, {Makler}, {Marchesi}, {Pecoraro}, {Ranalli}, {Salvato}, {Schawinski},
  {Stern}, {Treister}, \& {Viero}}]{lamassa16}
{LaMassa}, S.~M., {Urry}, C.~M., {Cappelluti}, N., {et~al.} 2016, \apj, 817,
  172

\bibitem[{{L{\"a}sker} {et~al.}(2016){L{\"a}sker}, {Greene}, {Seth}, {van de
  Ven}, {Braatz}, {Henkel}, \& {Lo}}]{laesker16}
{L{\"a}sker}, R., {Greene}, J.~E., {Seth}, A., {et~al.} 2016, \apj, 825, 3

\bibitem[{{Leighly}(1999)}]{leighly99}
{Leighly}, K.~M. 1999, \apjs, 125, 317

\bibitem[{{Leung} {et~al.}(2019){Leung}, {Coil}, {Aird}, {Azadi}, {Kriek},
  {Mobasher}, {Reddy}, {Shapley}, {Siana}, {Fetherolf}, {Fornasini}, {Freeman},
  {Price}, {Sanders}, {Shivaei}, \& {Zick}}]{leung19}
{Leung}, G. C.~K., {Coil}, A.~L., {Aird}, J., {et~al.} 2019, \apj, 886, 11

\bibitem[{{Liu} {et~al.}(2013){Liu}, {Zakamska}, {Greene}, {Nesvadba}, \&
  {Liu}}]{liu13}
{Liu}, G., {Zakamska}, N.~L., {Greene}, J.~E., {Nesvadba}, N.~P.~H., \& {Liu},
  X. 2013, \mnras, 436, 2576

\bibitem[{{Liu} {et~al.}(2016){Liu}, {Merloni}, {Georgakakis}, {Menzel},
  {Buchner}, {Nandra}, {Salvato}, {Shen}, {Brusa}, \& {Streblyanska}}]{liu16}
{Liu}, Z., {Merloni}, A., {Georgakakis}, A., {et~al.} 2016, \mnras, 459, 1602

\bibitem[{{Luo} {et~al.}(2017){Luo}, {Brandt}, {Xue}, {Lehmer}, {Alexander},
  {Bauer}, {Vito}, {Yang}, {Basu-Zych}, {Comastri}, {Gilli}, {Gu},
  {Hornschemeier}, {Koekemoer}, {Liu}, {Mainieri}, {Paolillo}, {Ranalli},
  {Rosati}, {Schneider}, {Shemmer}, {Smail}, {Sun}, {Tozzi}, {Vignali}, \&
  {Wang}}]{luo17}
{Luo}, B., {Brandt}, W.~N., {Xue}, Y.~Q., {et~al.} 2017, \apjs, 228, 2

\bibitem[{{Madau} \& {Dickinson}(2014)}]{madau14}
{Madau}, P. \& {Dickinson}, M. 2014, \araa, 52, 415

\bibitem[{{Magorrian} {et~al.}(1998){Magorrian}, {Tremaine}, {Richstone},
  {Bender}, {Bower}, {Dressler}, {Faber}, {Gebhardt}, {Green}, {Grillmair},
  {Kormendy}, \& {Lauer}}]{magorrian98}
{Magorrian}, J., {Tremaine}, S., {Richstone}, D., {et~al.} 1998, \aj, 115, 2285

\bibitem[{{Maiolino} {et~al.}(2017){Maiolino}, {Russell}, {Fabian}, {Carniani},
  {Gallagher}, {Cazzoli}, {Arribas}, {Belfiore}, {Bellocchi}, {Colina},
  {Cresci}, {Ishibashi}, {Marconi}, {Mannucci}, {Oliva}, \&
  {Sturm}}]{maiolino17}
{Maiolino}, R., {Russell}, H.~R., {Fabian}, A.~C., {et~al.} 2017, \nat, 544,
  202

\bibitem[{{Martocchia} {et~al.}(2017){Martocchia}, {Piconcelli}, {Zappacosta},
  {Duras}, {Vietri}, {Vignali}, {Bianchi}, {Bischetti}, {Bongiorno}, {Brusa},
  {Lanzuisi}, {Marconi}, {Mathur}, {Miniutti}, {Nicastro}, {Bruni}, \&
  {Fiore}}]{martocchia17}
{Martocchia}, S., {Piconcelli}, E., {Zappacosta}, L., {et~al.} 2017, \aap, 608,
  A51

\bibitem[{{May} {et~al.}(2018){May}, {Rodr{\'\i}guez-Ardila}, {Prieto},
  {Fern{\'a}ndez-Ontiveros}, {Diaz}, \& {Mazzalay}}]{may18}
{May}, D., {Rodr{\'\i}guez-Ardila}, A., {Prieto}, M.~A., {et~al.} 2018, \mnras,
  481, L105

\bibitem[{{McElroy} {et~al.}(2015){McElroy}, {Croom}, {Pracy}, {Sharp}, {Ho},
  \& {Medling}}]{mcelroy15}
{McElroy}, R., {Croom}, S.~M., {Pracy}, M., {et~al.} 2015, \mnras, 446, 2186

\bibitem[{{Menci} {et~al.}(2019){Menci}, {Fiore}, {Feruglio}, {Lamastra},
  {Shankar}, {Piconcelli}, {Giallongo}, \& {Grazian}}]{menci19}
{Menci}, N., {Fiore}, F., {Feruglio}, C., {et~al.} 2019, \apj, 877, 74

\bibitem[{{Menci} {et~al.}(2008){Menci}, {Fiore}, {Puccetti}, \&
  {Cavaliere}}]{menci08}
{Menci}, N., {Fiore}, F., {Puccetti}, S., \& {Cavaliere}, A. 2008, \apj, 686,
  219

\bibitem[{{Menzel} {et~al.}(2016){Menzel}, {Merloni}, {Georgakakis}, {Salvato},
  {Aubourg}, {Brandt}, {Brusa}, {Buchner}, {Dwelly}, {Nandra}, {P{\^a}ris},
  {Petitjean}, \& {Schwope}}]{menzel16}
{Menzel}, M.~L., {Merloni}, A., {Georgakakis}, A., {et~al.} 2016, \mnras, 457,
  110

\bibitem[{{Michiyama} {et~al.}(2018){Michiyama}, {Iono}, {Sliwa}, {Bolatto},
  {Nakanishi}, {Ueda}, {Saito}, {Ando}, {Yamashita}, \& {Yun}}]{michiyama18}
{Michiyama}, T., {Iono}, D., {Sliwa}, K., {et~al.} 2018, \apj, 868, 95

\bibitem[{{M{\"u}ller-S{\'a}nchez} {et~al.}(2011){M{\"u}ller-S{\'a}nchez},
  {Prieto}, {Hicks}, {Vives-Arias}, {Davies}, {Malkan}, {Tacconi}, \&
  {Genzel}}]{muller-sanchez11}
{M{\"u}ller-S{\'a}nchez}, F., {Prieto}, M.~A., {Hicks}, E.~K.~S., {et~al.}
  2011, \apj, 739, 69

\bibitem[{{Muratov} {et~al.}(2015){Muratov}, {Kere{\v{s}}},
  {Faucher-Gigu{\`e}re}, {Hopkins}, {Quataert}, \& {Murray}}]{muratov15}
{Muratov}, A.~L., {Kere{\v{s}}}, D., {Faucher-Gigu{\`e}re}, C.-A., {et~al.}
  2015, \mnras, 454, 2691

\bibitem[{{Nesvadba} {et~al.}(2017){Nesvadba}, {De Breuck}, {Lehnert}, {Best},
  \& {Collet}}]{nesvadba17}
{Nesvadba}, N.~P.~H., {De Breuck}, C., {Lehnert}, M.~D., {Best}, P.~N., \&
  {Collet}, C. 2017, \aap, 599, A123

\bibitem[{{Nesvadba} {et~al.}(2006){Nesvadba}, {Lehnert}, {Eisenhauer},
  {Gilbert}, {Tecza}, \& {Abuter}}]{nesvadba06}
{Nesvadba}, N.~P.~H., {Lehnert}, M.~D., {Eisenhauer}, F., {et~al.} 2006, \apj,
  650, 693

\bibitem[{{Nims} {et~al.}(2015){Nims}, {Quataert}, \&
  {Faucher-Gigu{\`e}re}}]{nims15}
{Nims}, J., {Quataert}, E., \& {Faucher-Gigu{\`e}re}, C.-A. 2015, \mnras, 447,
  3612

\bibitem[{{Padovani} {et~al.}(2017){Padovani}, {Alexander}, {Assef}, {De
  Marco}, {Giommi}, {Hickox}, {Richards}, {Smol{\v{c}}i{\'c}},
  {Hatziminaoglou}, {Mainieri}, \& {Salvato}}]{padovani17}
{Padovani}, P., {Alexander}, D.~M., {Assef}, R.~J., {et~al.} 2017, \aapr, 25, 2

\bibitem[{{Perna} {et~al.}(2015){Perna}, {Brusa}, {Cresci}, {Comastri},
  {Lanzuisi}, {Lusso}, {Marconi}, {Salvato}, {Zamorani}, {Bongiorno},
  {Mainieri}, {Maiolino}, \& {Mignoli}}]{perna15}
{Perna}, M., {Brusa}, M., {Cresci}, G., {et~al.} 2015, \aap, 574, A82

\bibitem[{{Perna} {et~al.}(2019){Perna}, {Cresci}, {Brusa}, {Lanzuisi},
  {Concas}, {Mainieri}, {Mannucci}, \& {Marconi}}]{perna19}
{Perna}, M., {Cresci}, G., {Brusa}, M., {et~al.} 2019, \aap, 623, A171

\bibitem[{{Perna} {et~al.}(2017){Perna}, {Lanzuisi}, {Brusa}, {Cresci}, \&
  {Mignoli}}]{perna17}
{Perna}, M., {Lanzuisi}, G., {Brusa}, M., {Cresci}, G., \& {Mignoli}, M. 2017,
  \aap, 606, A96

\bibitem[{{Perrotta} {et~al.}(2019){Perrotta}, {Hamann}, {Zakamska}, {Alexand
  roff}, {Rupke}, \& {Wylezalek}}]{perrotta19}
{Perrotta}, S., {Hamann}, F., {Zakamska}, N.~L., {et~al.} 2019, \mnras, 488,
  4126

\bibitem[{{Petric} {et~al.}(2018){Petric}, {Armus}, {Flagey}, {Guillard},
  {Howell}, {Inami}, {Charmand aris}, {Evans}, {Stierwalt}, {Diaz-Santos},
  {Lu}, {Spoon}, {Mazzarella}, {Appleton}, {Chan}, {Chu}, {Hand}, {Privon},
  {Sand ers}, {Surace}, {Xu}, \& {Zhao}}]{petric18}
{Petric}, A.~O., {Armus}, L., {Flagey}, N., {et~al.} 2018, \aj, 156, 295

\bibitem[{{Radovich} {et~al.}(2019){Radovich}, {Poggianti}, {Jaff{\'e}},
  {Moretti}, {Bettoni}, {Gullieuszik}, {Vulcani}, \& {Fritz}}]{radovich19}
{Radovich}, M., {Poggianti}, B., {Jaff{\'e}}, Y.~L., {et~al.} 2019, \mnras,
  486, 486

\bibitem[{{Rakshit} \& {Woo}(2018)}]{rakshit18}
{Rakshit}, S. \& {Woo}, J.-H. 2018, \apj, 865, 5

\bibitem[{{Revalski} {et~al.}(2018){Revalski}, {Dashtamirova}, {Crenshaw},
  {Kraemer}, {Fischer}, {Schmitt}, {Gnilka}, {Schmidt}, {Elvis}, {Fabbiano},
  {Storchi-Bergmann}, {Maksym}, \& {Gandhi}}]{revalski18}
{Revalski}, M., {Dashtamirova}, D., {Crenshaw}, D.~M., {et~al.} 2018, \apj,
  867, 88

\bibitem[{{Riffel} {et~al.}(2015){Riffel}, {Storchi-Bergmann}, \&
  {Riffel}}]{riffel15}
{Riffel}, R.~A., {Storchi-Bergmann}, T., \& {Riffel}, R. 2015, \mnras, 451,
  3587

\bibitem[{{Riffel} {et~al.}(2013){Riffel}, {Storchi-Bergmann}, \&
  {Winge}}]{riffel13}
{Riffel}, R.~A., {Storchi-Bergmann}, T., \& {Winge}, C. 2013, \mnras, 430, 2249

\bibitem[{{Riffel} {et~al.}(2020){Riffel}, {Zakamska}, \& {Riffel}}]{riffel20}
{Riffel}, R.~A., {Zakamska}, N.~L., \& {Riffel}, R. 2020, \mnras, 491, 1518

\bibitem[{{Roberts-Borsani}(2020)}]{roberts-borsani20}
{Roberts-Borsani}, G.~W. 2020, \mnras

\bibitem[{{Rodr{\'\i}guez-Ardila} {et~al.}(2000){Rodr{\'\i}guez-Ardila},
  {Binette}, {Pastoriza}, \& {Donzelli}}]{rodriguez-ardila00}
{Rodr{\'\i}guez-Ardila}, A., {Binette}, L., {Pastoriza}, M.~G., \& {Donzelli},
  C.~J. 2000, \apj, 538, 581

\bibitem[{{Rupke} {et~al.}(2002){Rupke}, {Veilleux}, \& {Sanders}}]{rupke02}
{Rupke}, D.~S., {Veilleux}, S., \& {Sanders}, D.~B. 2002, \apj, 570, 588

\bibitem[{{Rupke} {et~al.}(2017){Rupke}, {G{\"u}ltekin}, \&
  {Veilleux}}]{rupke17}
{Rupke}, D. S.~N., {G{\"u}ltekin}, K., \& {Veilleux}, S. 2017, \apj, 850, 40

\bibitem[{{Rupke} \& {Veilleux}(2011)}]{rupke11}
{Rupke}, D. S.~N. \& {Veilleux}, S. 2011, \apj, 729, L27

\bibitem[{{Rupke} \& {Veilleux}(2013)}]{rupke13}
{Rupke}, D.~S.~N. \& {Veilleux}, S. 2013, \apj, 768, 75

\bibitem[{{S{\'a}nchez} {et~al.}(2012){S{\'a}nchez}, {Kennicutt}, {Gil de Paz},
  {van de Ven}, {V{\'\i}lchez}, {Wisotzki}, {Walcher}, {Mast}, {Aguerri},
  {Albiol-P{\'e}rez}, {Alonso-Herrero}, {Alves}, {Bakos}, {Bart{\'a}kov{\'a}},
  {Bland-Hawthorn}, {Boselli}, {Bomans}, {Castillo-Morales}, {Cortijo-Ferrero},
  {de Lorenzo-C{\'a}ceres}, {Del Olmo}, {Dettmar}, {D{\'\i}az}, {Ellis},
  {Falc{\'o}n-Barroso}, {Flores}, {Gallazzi}, {Garc{\'\i}a-Lorenzo},
  {Gonz{\'a}lez Delgado}, {Gruel}, {Haines}, {Hao}, {Husemann},
  {Igl{\'e}sias-P{\'a}ramo}, {Jahnke}, {Johnson}, {Jungwiert}, {Kalinova},
  {Kehrig}, {Kupko}, {L{\'o}pez-S{\'a}nchez}, {Lyubenova}, {Marino},
  {M{\'a}rmol-Queralt{\'o}}, {M{\'a}rquez}, {Masegosa}, {Meidt},
  {Mendez-Abreu}, {Monreal-Ibero}, {Montijo}, {Mour{\~a}o}, {Palacios-Navarro},
  {Papaderos}, {Pasquali}, {Peletier}, {P{\'e}rez}, {P{\'e}rez}, {Quirrenbach},
  {Rela{\~n}o}, {Rosales-Ortega}, {Roth}, {Ruiz-Lara},
  {S{\'a}nchez-Bl{\'a}zquez}, {Sengupta}, {Singh}, {Stanishev}, {Trager},
  {Vazdekis}, {Viironen}, {Wild}, {Zibetti}, \& {Ziegler}}]{sanchez12}
{S{\'a}nchez}, S.~F., {Kennicutt}, R.~C., {Gil de Paz}, A., {et~al.} 2012,
  \aap, 538, A8

\bibitem[{{Scholtz} {et~al.}(2020){Scholtz}, {Harrison}, {Rosario}, {Alexand
  er}, {Chen}, {Kakkad}, {Mainieri}, {Tiley}, {Turner}, {Cirasuolo},
  {Sharples}, \& {Stach}}]{scholtz20}
{Scholtz}, J., {Harrison}, C.~M., {Rosario}, D.~J., {et~al.} 2020, \mnras, 492,
  3194

\bibitem[{{Schutte} {et~al.}(2019){Schutte}, {Reines}, \& {Greene}}]{schutte19}
{Schutte}, Z., {Reines}, A.~E., \& {Greene}, J.~E. 2019, \apj, 887, 245

\bibitem[{{Serafinelli} {et~al.}(2019){Serafinelli}, {Tombesi}, {Vagnetti},
  {Piconcelli}, {Gaspari}, \& {Saturni}}]{serafinelli19}
{Serafinelli}, R., {Tombesi}, F., {Vagnetti}, F., {et~al.} 2019, \aap, 627,
  A121

\bibitem[{{Shankar} {et~al.}(2009){Shankar}, {Weinberg}, \&
  {Miralda-Escud{\'e}}}]{shankar09}
{Shankar}, F., {Weinberg}, D.~H., \& {Miralda-Escud{\'e}}, J. 2009, \apj, 690,
  20

\bibitem[{{Shimizu} {et~al.}(2019){Shimizu}, {Davies}, {Lutz}, {Burtscher},
  {Lin}, {Baron}, {Davies}, {Genzel}, {Hicks}, {Koss}, {Maciejewski},
  {M{\"u}ller-S{\'a}nchez}, {Orban de Xivry}, {Price}, {Ricci}, {Riffel},
  {Riffel}, {Rosario}, {Schartmann}, {Schnorr-M{\"u}ller}, {Sternberg},
  {Sturm}, {Storchi-Bergmann}, {Tacconi}, \& {Veilleux}}]{shimizu19}
{Shimizu}, T.~T., {Davies}, R.~I., {Lutz}, D., {et~al.} 2019, \mnras, 490, 5860

\bibitem[{{Skrutskie} {et~al.}(2006){Skrutskie}, {Cutri}, {Stiening},
  {Weinberg}, {Schneider}, {Carpenter}, {Beichman}, {Capps}, {Chester},
  {Elias}, {Huchra}, {Liebert}, {Lonsdale}, {Monet}, {Price}, {Seitzer},
  {Jarrett}, {Kirkpatrick}, {Gizis}, {Howard}, {Evans}, {Fowler}, {Fullmer},
  {Hurt}, {Light}, {Kopan}, {Marsh}, {McCallon}, {Tam}, {Van Dyk}, \&
  {Wheelock}}]{skrutskie06}
{Skrutskie}, M.~F., {Cutri}, R.~M., {Stiening}, R., {et~al.} 2006, \aj, 131,
  1163

\bibitem[{{Steidel} {et~al.}(2010){Steidel}, {Erb}, {Shapley}, {Pettini},
  {Reddy}, {Bogosavljevi{\'c}}, {Rudie}, \& {Rakic}}]{steidel10}
{Steidel}, C.~C., {Erb}, D.~K., {Shapley}, A.~E., {et~al.} 2010, \apj, 717, 289

\bibitem[{{Storey} \& {Zeippen}(2000)}]{storey00}
{Storey}, P.~J. \& {Zeippen}, C.~J. 2000, \mnras, 312, 813

\bibitem[{{Stott} {et~al.}(2016){Stott}, {Swinbank}, {Johnson}, {Tiley},
  {Magdis}, {Bower}, {Bunker}, {Bureau}, {Harrison}, {Jarvis}, {Sharples},
  {Smail}, {Sobral}, {Best}, \& {Cirasuolo}}]{stott16}
{Stott}, J.~P., {Swinbank}, A.~M., {Johnson}, H.~L., {et~al.} 2016, \mnras,
  457, 1888

\bibitem[{{Swinbank} {et~al.}(2019){Swinbank}, {Harrison}, {Tiley}, {Johnson},
  {Smail}, {Stott}, {Best}, {Bower}, {Bureau}, \& {Bunker}}]{swinbank19}
{Swinbank}, A.~M., {Harrison}, C.~M., {Tiley}, A.~L., {et~al.} 2019, \mnras,
  487, 381

\bibitem[{{Tacconi} {et~al.}(2020){Tacconi}, {Genzel}, \&
  {Sternberg}}]{tacconi20}
{Tacconi}, L.~J., {Genzel}, R., \& {Sternberg}, A. 2020, arXiv e-prints,
  arXiv:2003.06245

\bibitem[{{Tadhunter} {et~al.}(2014){Tadhunter}, {Morganti}, {Rose}, {Oonk}, \&
  {Oosterloo}}]{tadhunter14}
{Tadhunter}, C., {Morganti}, R., {Rose}, M., {Oonk}, J.~B.~R., \& {Oosterloo},
  T. 2014, \nat, 511, 440

\bibitem[{{Thomas} {et~al.}(2017){Thomas}, {Dopita}, {Shastri}, {Davies},
  {Hampton}, {Kewley}, {Banfield}, {Groves}, {James}, {Jin}, {Juneau}, {Kharb},
  {Sairam}, {Scharw{\"a}chter}, {Shalima}, {Sundar}, {Sutherland}, \&
  {Zaw}}]{thomas17}
{Thomas}, A.~D., {Dopita}, M.~A., {Shastri}, P., {et~al.} 2017, \apjs, 232, 11

\bibitem[{{Tsuzuki} {et~al.}(2006){Tsuzuki}, {Kawara}, {Yoshii}, {Oyabu},
  {Tanab{\'e}}, \& {Matsuoka}}]{tsuzuki06}
{Tsuzuki}, Y., {Kawara}, K., {Yoshii}, Y., {et~al.} 2006, \apj, 650, 57

\bibitem[{{Vayner} {et~al.}(2017){Vayner}, {Wright}, {Murray}, {Armus},
  {Larkin}, \& {Mieda}}]{vayner17}
{Vayner}, A., {Wright}, S.~A., {Murray}, N., {et~al.} 2017, \apj, 851, 126

\bibitem[{{Veilleux} {et~al.}(2020){Veilleux}, {Maiolino}, {Bolatto}, \&
  {Aalto}}]{veilleux20}
{Veilleux}, S., {Maiolino}, R., {Bolatto}, A.~D., \& {Aalto}, S. 2020, \aapr,
  28, 2

\bibitem[{{Veilleux} {et~al.}(2009){Veilleux}, {Rupke}, \&
  {Swaters}}]{veilleux09}
{Veilleux}, S., {Rupke}, D. S.~N., \& {Swaters}, R. 2009, \apjl, 700, L149

\bibitem[{{Venturi} {et~al.}(2018){Venturi}, {Nardini}, {Marconi}, {Carniani},
  {Mingozzi}, {Cresci}, {Mannucci}, {Risaliti}, {Maiolino}, {Balmaverde},
  {Bongiorno}, {Brusa}, {Capetti}, {Cicone}, {Ciroi}, {Feruglio}, {Fiore},
  {Gallazzi}, {La Franca}, {Mainieri}, {Matsuoka}, {Nagao}, {Perna},
  {Piconcelli}, {Sani}, {Tozzi}, \& {Zibetti}}]{venturi18}
{Venturi}, G., {Nardini}, E., {Marconi}, A., {et~al.} 2018, \aap, 619, A74

\bibitem[{{V{\'e}ron-Cetty} {et~al.}(2004){V{\'e}ron-Cetty}, {Joly}, \&
  {V{\'e}ron}}]{veron-cetty04}
{V{\'e}ron-Cetty}, M.~P., {Joly}, M., \& {V{\'e}ron}, P. 2004, \aap, 417, 515

\bibitem[{{Vietri}(2020)}]{vietri20}
{Vietri}, G. 2020, et. al. in prep

\bibitem[{{Vietri} {et~al.}(2018){Vietri}, {Piconcelli}, {Bischetti}, {Duras},
  {Martocchia}, {Bongiorno}, {Marconi}, {Zappacosta}, {Bisogni}, {Bruni},
  {Brusa}, {Comastri}, {Cresci}, {Feruglio}, {Giallongo}, {La Franca},
  {Mainieri}, {Mannucci}, {Ricci}, {Sani}, {Testa}, {Tombesi}, {Vignali}, \&
  {Fiore}}]{vietri18}
{Vietri}, G., {Piconcelli}, E., {Bischetti}, M., {et~al.} 2018, \aap, 617, A81

\bibitem[{{Villar-Mart{\'\i}n} {et~al.}(2016){Villar-Mart{\'\i}n}, {Arribas},
  {Emonts}, {Humphrey}, {Tadhunter}, {Bessiere}, {Cabrera Lavers}, \& {Ramos
  Almeida}}]{villar-martin16}
{Villar-Mart{\'\i}n}, M., {Arribas}, S., {Emonts}, B., {et~al.} 2016, \mnras,
  460, 130

\bibitem[{{Wagner} {et~al.}(2013){Wagner}, {Umemura}, \& {Bicknell}}]{wagner13}
{Wagner}, A.~Y., {Umemura}, M., \& {Bicknell}, G.~V. 2013, \apjl, 763, L18

\bibitem[{{Whitaker} {et~al.}(2014){Whitaker}, {Rigby}, {Brammer}, {Gladders},
  {Sharon}, {Teng}, \& {Wuyts}}]{whitaker14}
{Whitaker}, K.~E., {Rigby}, J.~R., {Brammer}, G.~B., {et~al.} 2014, \apj, 790,
  143

\bibitem[{{Wilkins} {et~al.}(2019){Wilkins}, {Lovell}, \&
  {Stanway}}]{wilkins19}
{Wilkins}, S.~M., {Lovell}, C.~C., \& {Stanway}, E.~R. 2019, \mnras, 490, 5359

\bibitem[{{Wisnioski} {et~al.}(2019){Wisnioski}, {F{\"o}rster Schreiber},
  {Fossati}, {Mendel}, {Wilman}, {Genzel}, {Bender}, {Wuyts}, {Davies},
  {{\"U}bler}, {Bandara}, {Beifiori}, {Belli}, {Brammer}, {Chan}, {Davies},
  {Fabricius}, {Galametz}, {Lang}, {Lutz}, {Nelson}, {Momcheva}, {Price},
  {Rosario}, {Saglia}, {Seitz}, {Shimizu}, {Tacconi}, {Tadaki}, {van Dokkum},
  \& {Wuyts}}]{wisnioski19}
{Wisnioski}, E., {F{\"o}rster Schreiber}, N.~M., {Fossati}, M., {et~al.} 2019,
  \apj, 886, 124

\bibitem[{{Wylezalek} {et~al.}(2020){Wylezalek}, {Flores}, {Zakamska},
  {Greene}, \& {Riffel}}]{wylezalek20}
{Wylezalek}, D., {Flores}, A.~M., {Zakamska}, N.~L., {Greene}, J.~E., \&
  {Riffel}, R.~A. 2020, \mnras, 492, 4680

\bibitem[{{Wylezalek} {et~al.}(2016){Wylezalek}, {Zakamska}, {Liu}, \&
  {Obied}}]{wylezalek16}
{Wylezalek}, D., {Zakamska}, N.~L., {Liu}, G., \& {Obied}, G. 2016, \mnras,
  457, 745

\bibitem[{{Zakamska} \& {Greene}(2014)}]{zakamska14}
{Zakamska}, N.~L. \& {Greene}, J.~E. 2014, \mnras, 442, 784

\bibitem[{{Zakamska} {et~al.}(2016){Zakamska}, {Hamann}, {P{\^a}ris}, {Brandt},
  {Greene}, {Strauss}, {Villforth}, {Wylezalek}, {Alexand roff}, \&
  {Ross}}]{zakamska16}
{Zakamska}, N.~L., {Hamann}, F., {P{\^a}ris}, I., {et~al.} 2016, \mnras, 459,
  3144

\bibitem[{{Zappacosta} {et~al.}(2020){Zappacosta}, {Piconcelli}, {Giustini},
  {Vietri}, {Duras}, {Miniutti}, {Bischetti}, {Bongiorno}, {Brusa},
  {Chiaberge}, {Comastri}, {Feruglio}, {Luminari}, {Marconi}, {Ricci},
  {Vignali}, \& {Fiore}}]{zappacosta20}
{Zappacosta}, L., {Piconcelli}, E., {Giustini}, M., {et~al.} 2020, \aap, 635,
  L5

\bibitem[{{Zschaechner} {et~al.}(2018){Zschaechner}, {Bolatto}, {Walter},
  {Leroy}, {Herrera}, {Krieger}, {Kruijssen}, {Meier}, {Mills}, \&
  {Ott}}]{zschaechner18}
{Zschaechner}, L.~K., {Bolatto}, A.~D., {Walter}, F., {et~al.} 2018, \apj, 867,
  111

\bibitem[{{Zubovas}(2018)}]{zubovas18}
{Zubovas}, K. 2018, \mnras, 479, 3189

\bibitem[{{Zubovas} \& {King}(2012)}]{zubovas12}
{Zubovas}, K. \& {King}, A. 2012, \apjl, 745, L34

\bibitem[{{Zubovas} \& {Nayakshin}(2014)}]{zubovas14}
{Zubovas}, K. \& {Nayakshin}, S. 2014, \mnras, 440, 2625

\end{thebibliography}
\begin{appendix}
\section{Integrated H-band Spectra and \oiii channel maps of Type-1 SUPER sample} \label{sect:app}
\begin{figure*}
\centering
\subfloat{\includegraphics[scale=0.45]{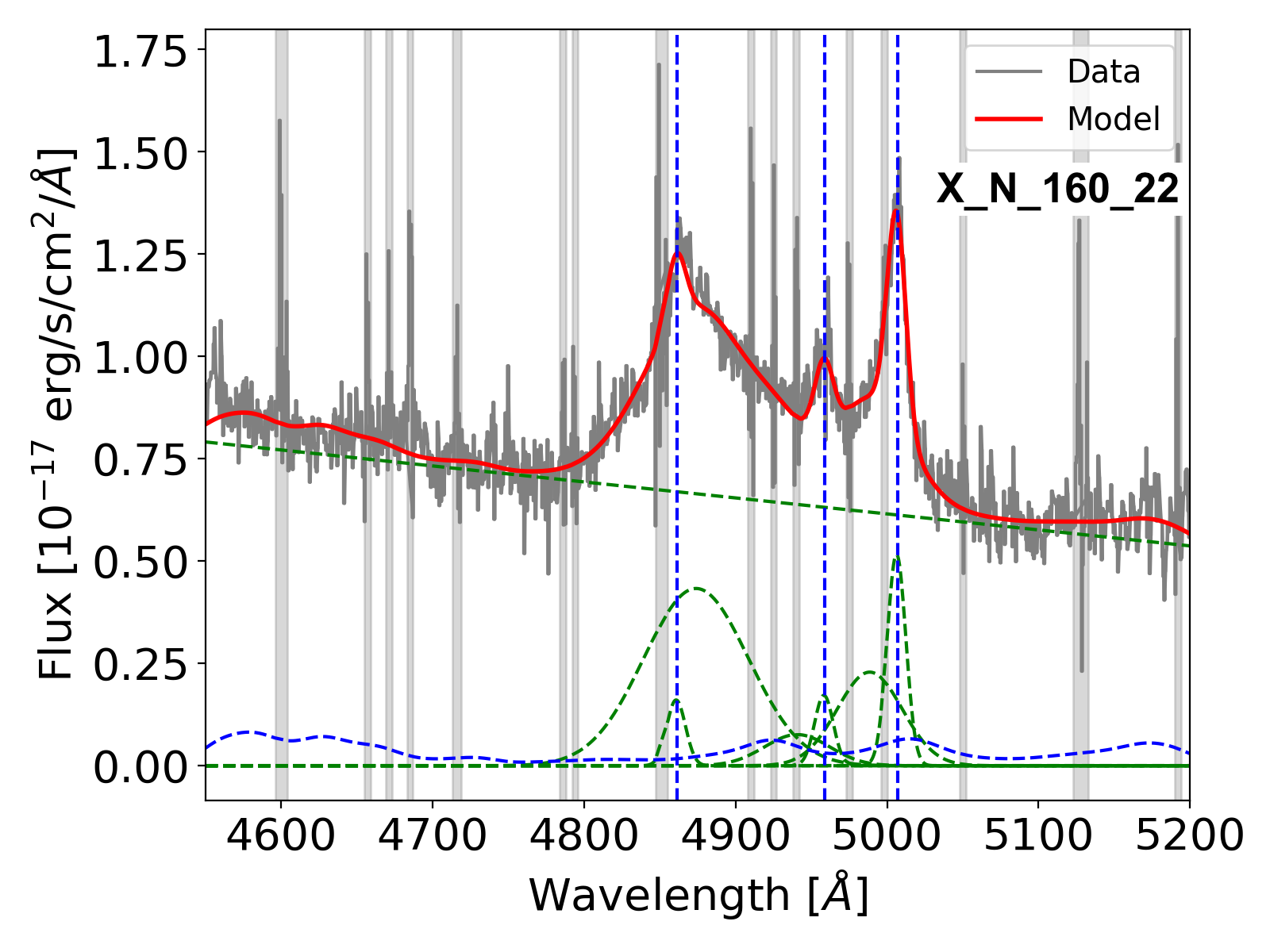}}
\subfloat{\includegraphics[scale=0.45]{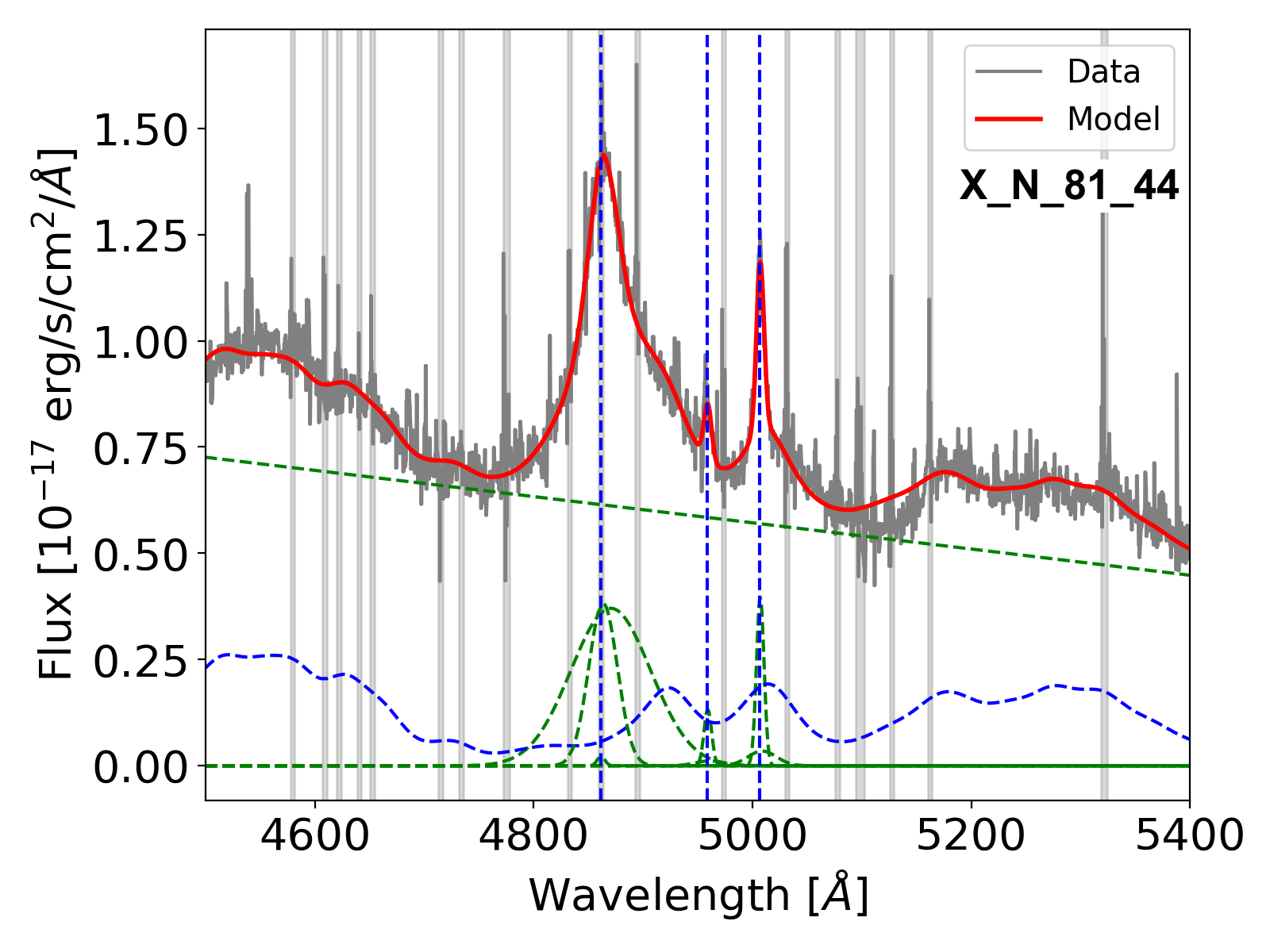}}\\
\subfloat{\includegraphics[scale=0.45]{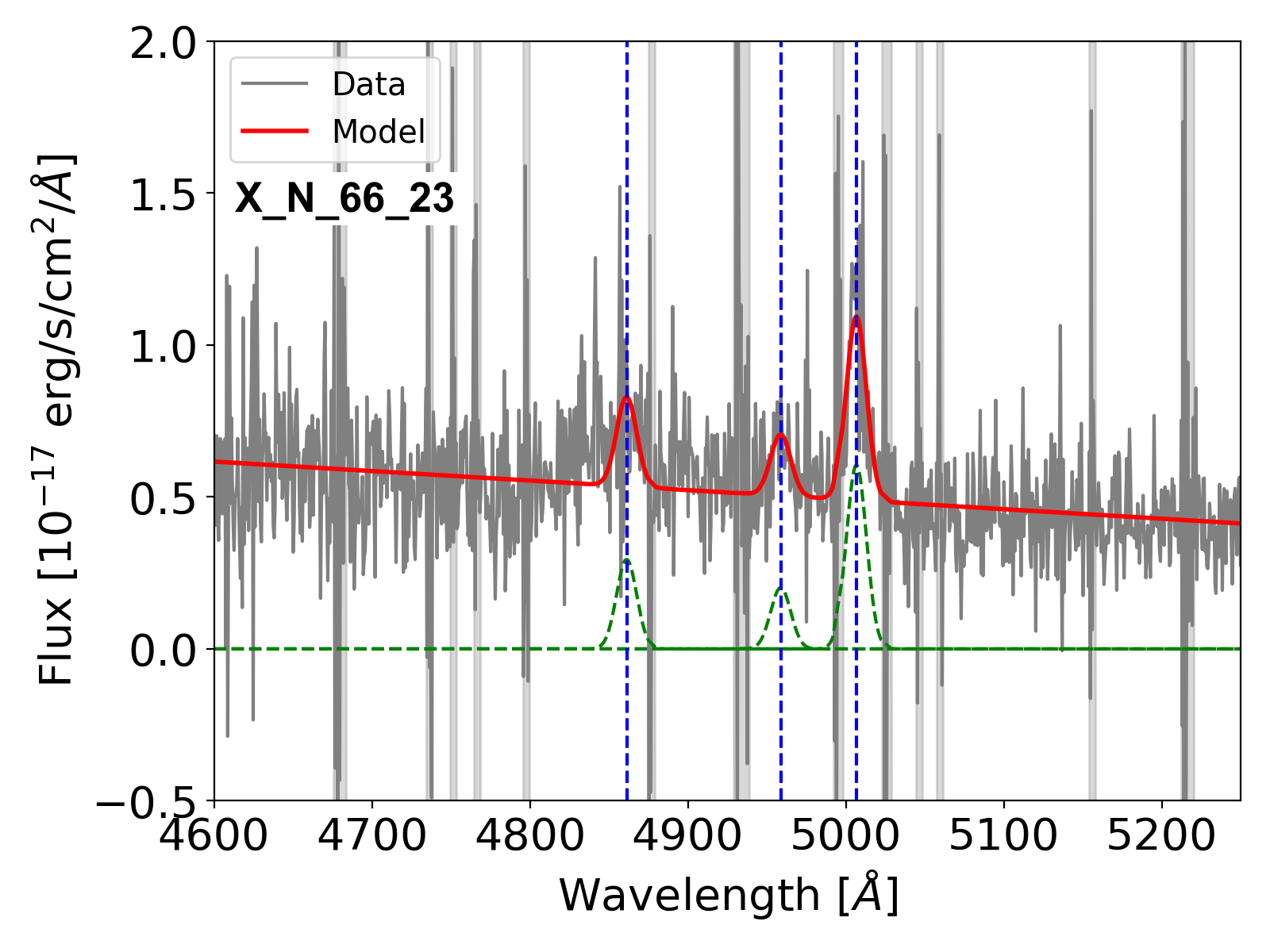}}
\subfloat{\includegraphics[scale=0.45]{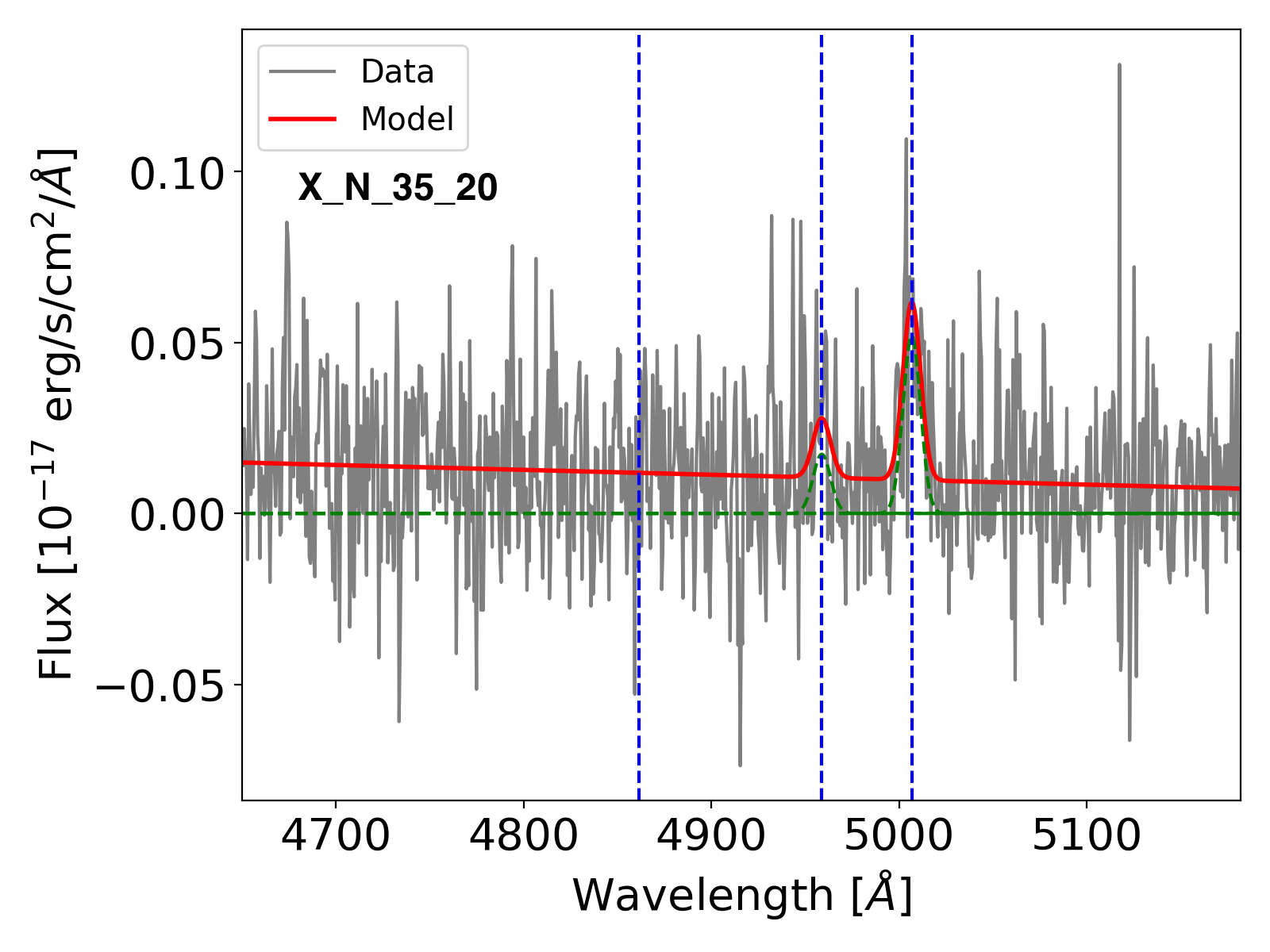}}\\
\subfloat{\includegraphics[scale=0.45]{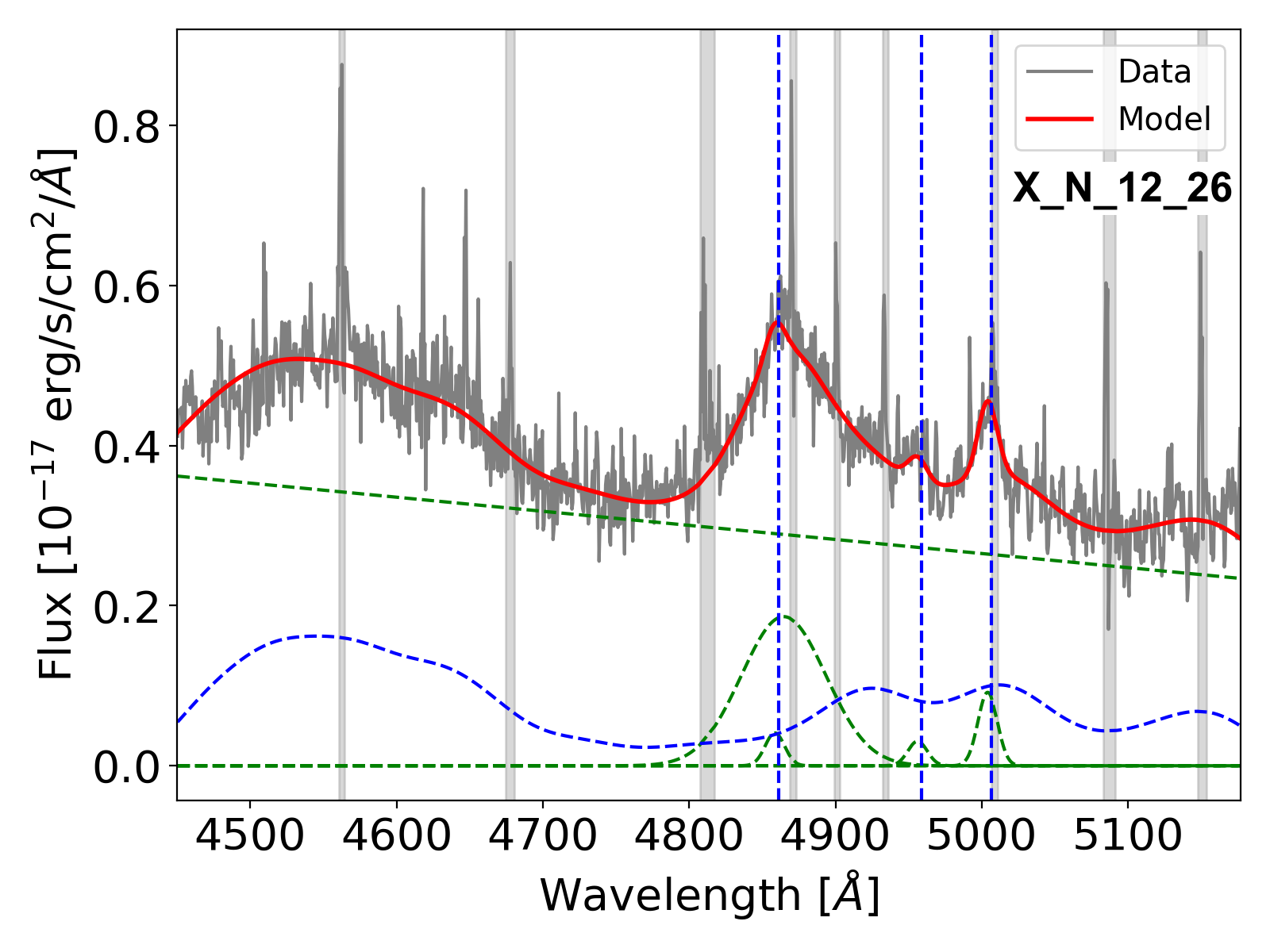}}
\subfloat{\includegraphics[scale=0.45]{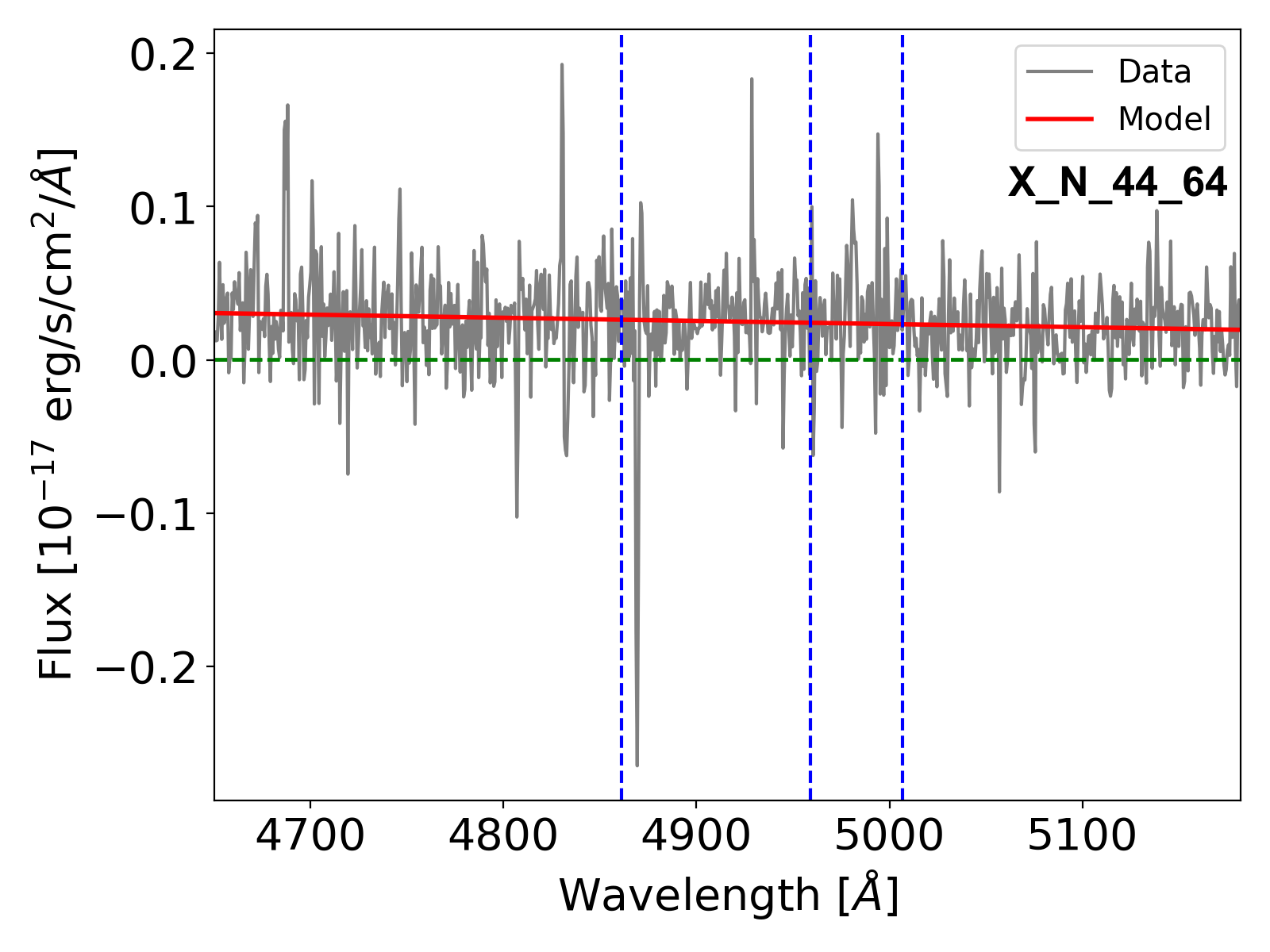}}\\
\caption{Integrated H-band spectrum of the SUPER targets X\_N\_160\_22, X\_N\_81\_44, X\_N\_66\_23, X\_N\_35\_20, X\_N\_12\_26 and X\_N\_44\_64. The grey curve shows the observed spectrum, the red curve shows the reproduced overall emission line model,  the blue dashed curve shows the iron emission and the dashed green curves show the continuum emission and the individual Gaussian components (Narrow, Broad and BLR) used to reproduce the profiles of various emission lines. The blue vertical lines indicate the location of $\hb$, \oiii$\lambda$4959 and \oiii$\lambda$5007. The vertical grey regions mark the channels with strong skylines which were masked during the fitting procedure. The X-axis shows the rest frame wavelength after correcting for the redshift of the target and the Y-axis shows the observed flux. \label{fig:intspec_alltargets1}}
\end{figure*}

\begin{figure*}
\centering
\subfloat{\includegraphics[scale=0.45]{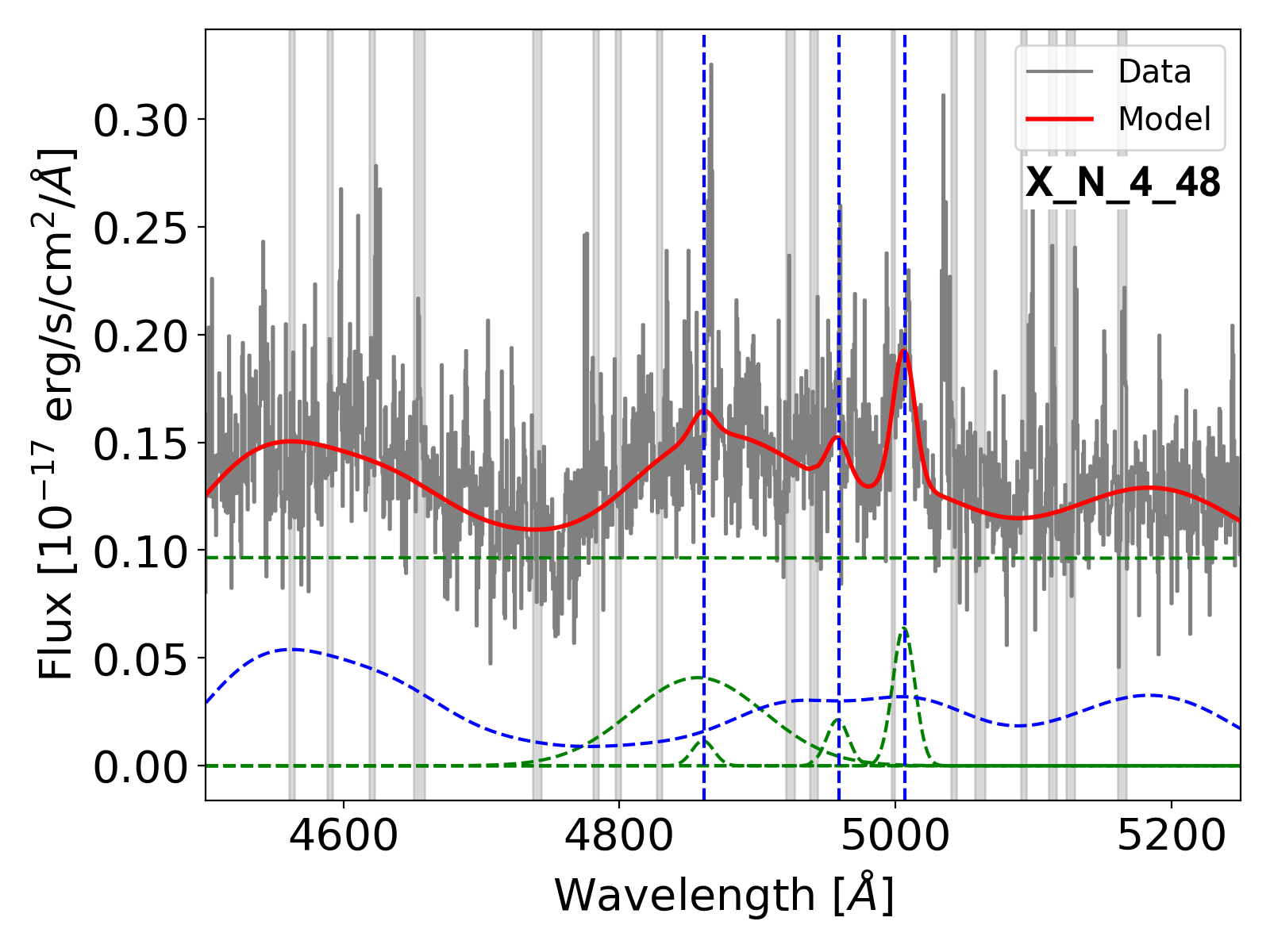}}
\subfloat{\includegraphics[scale=0.45]{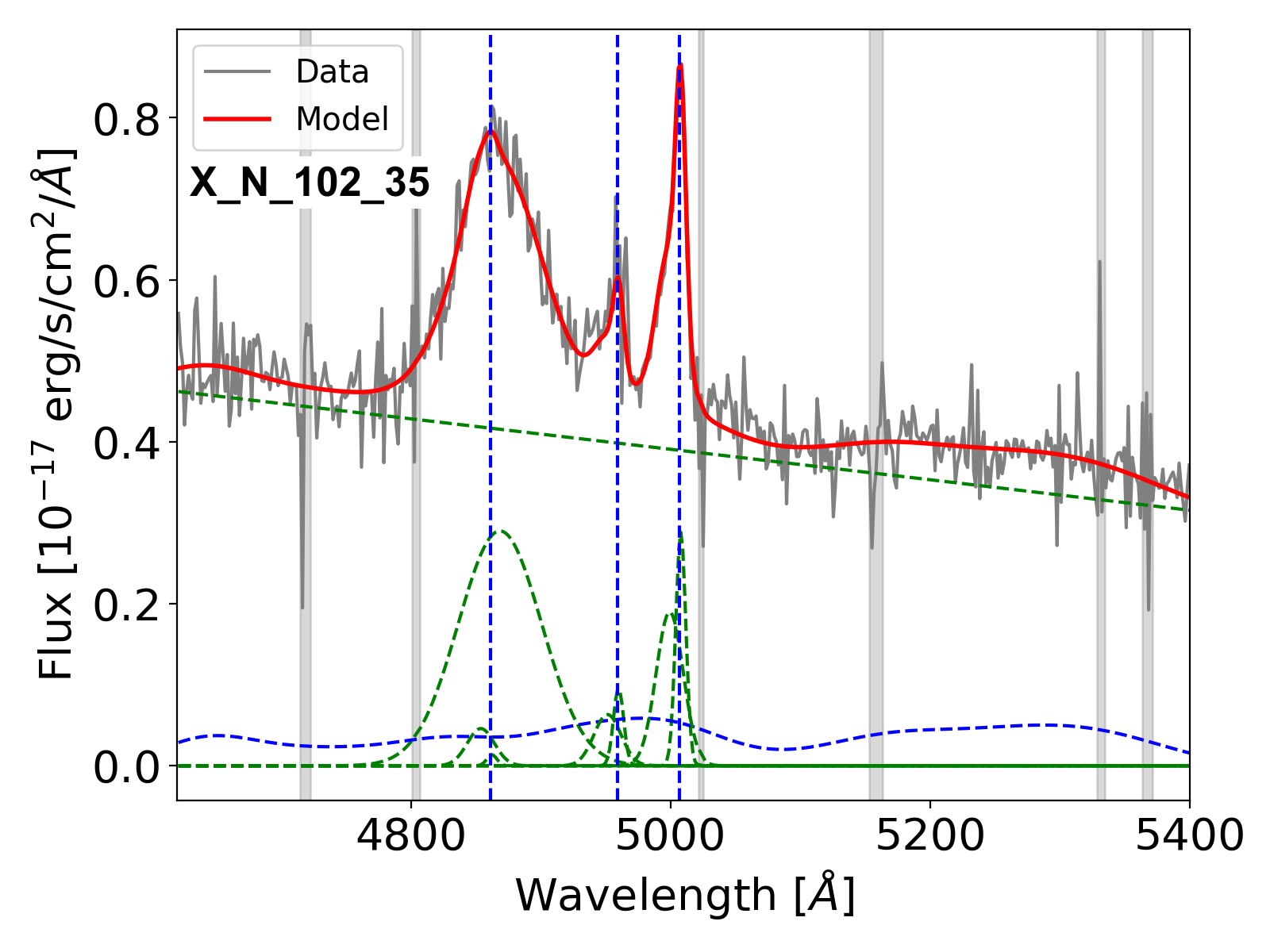}}\\
\subfloat{\includegraphics[scale=0.45]{intspec_X_N_115_23_H_updated.png}}
\subfloat{\includegraphics[scale=0.45]{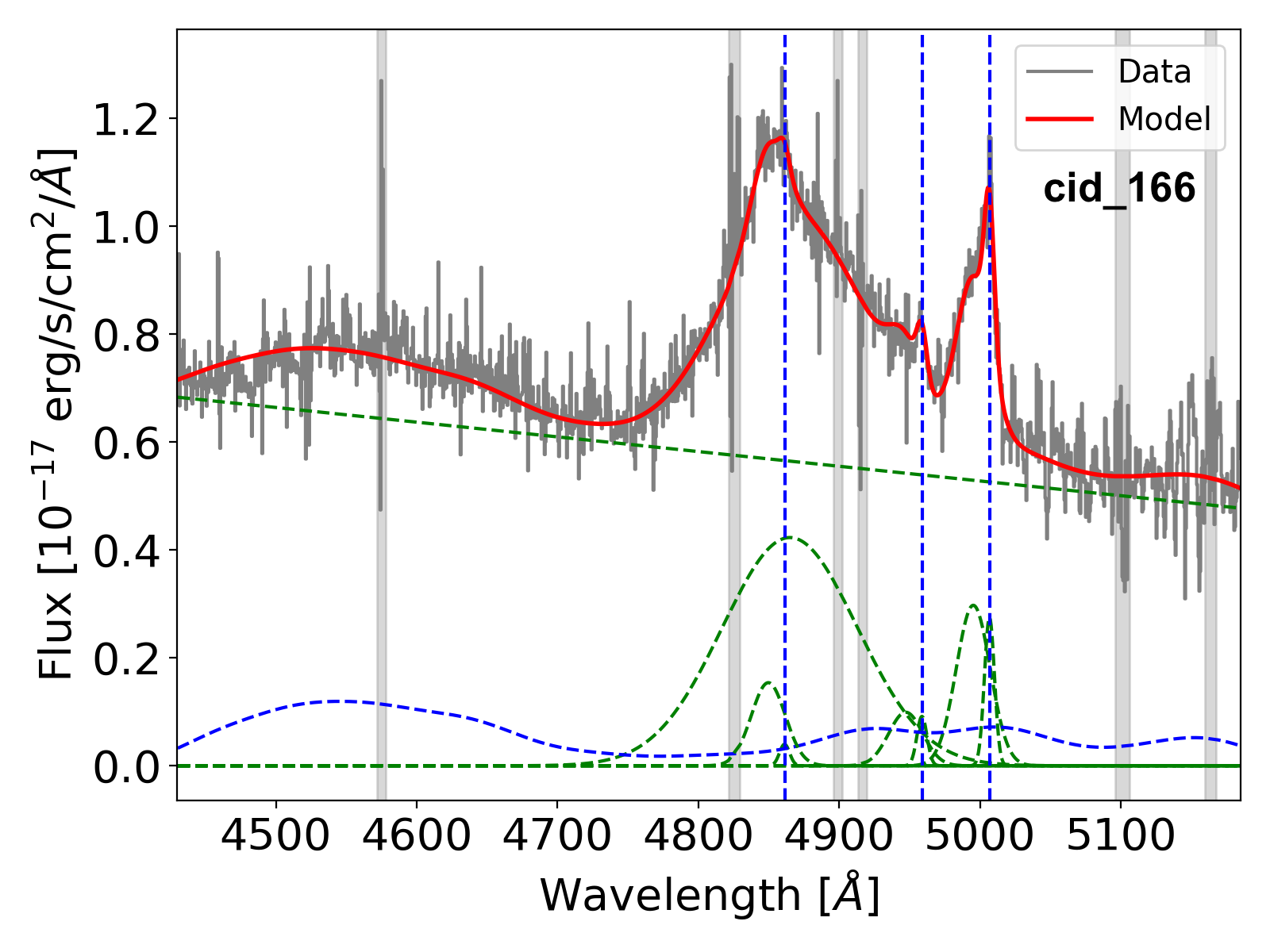}}\\
\subfloat{\includegraphics[scale=0.45]{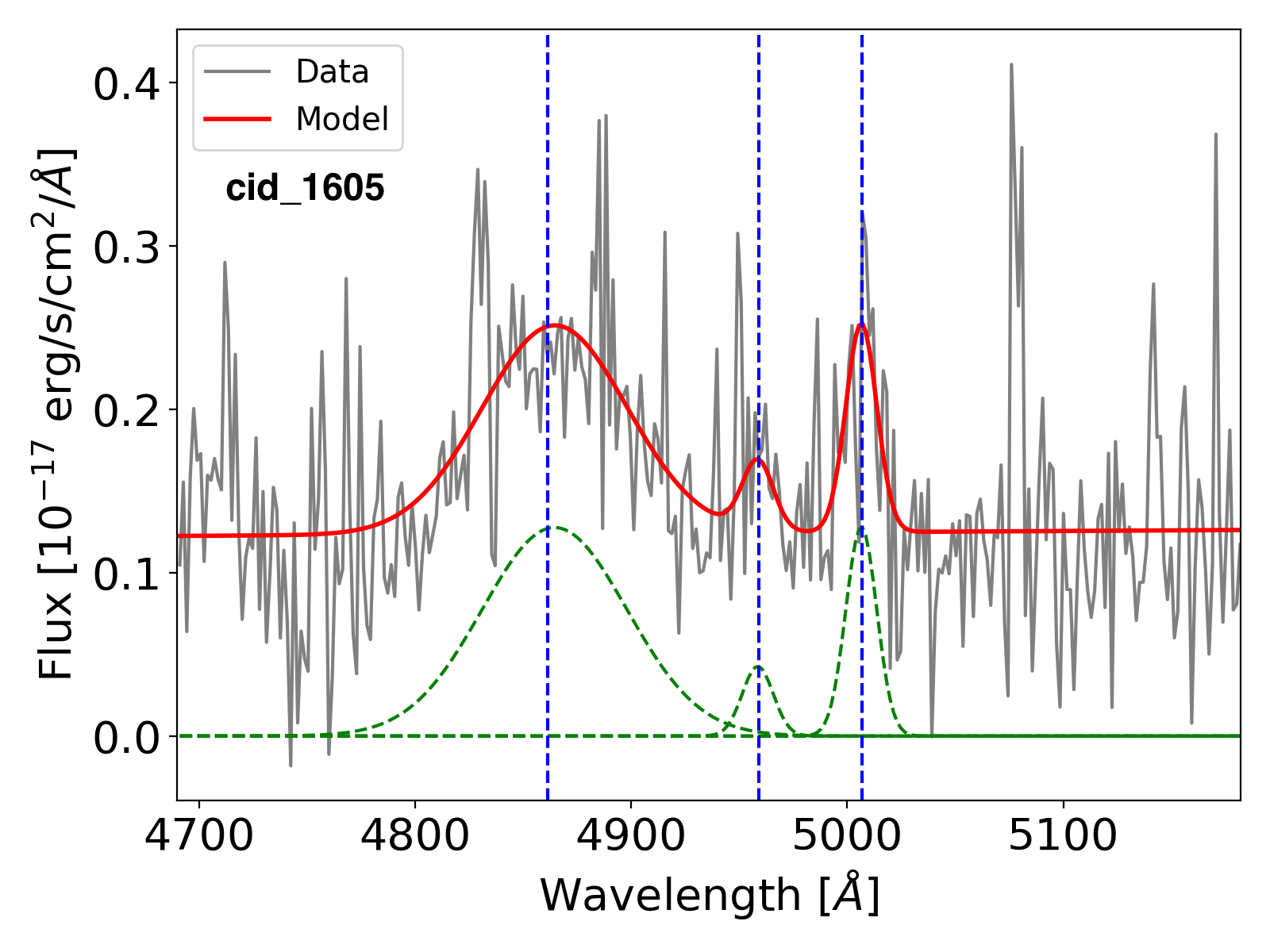}}
\subfloat{\includegraphics[scale=0.45]{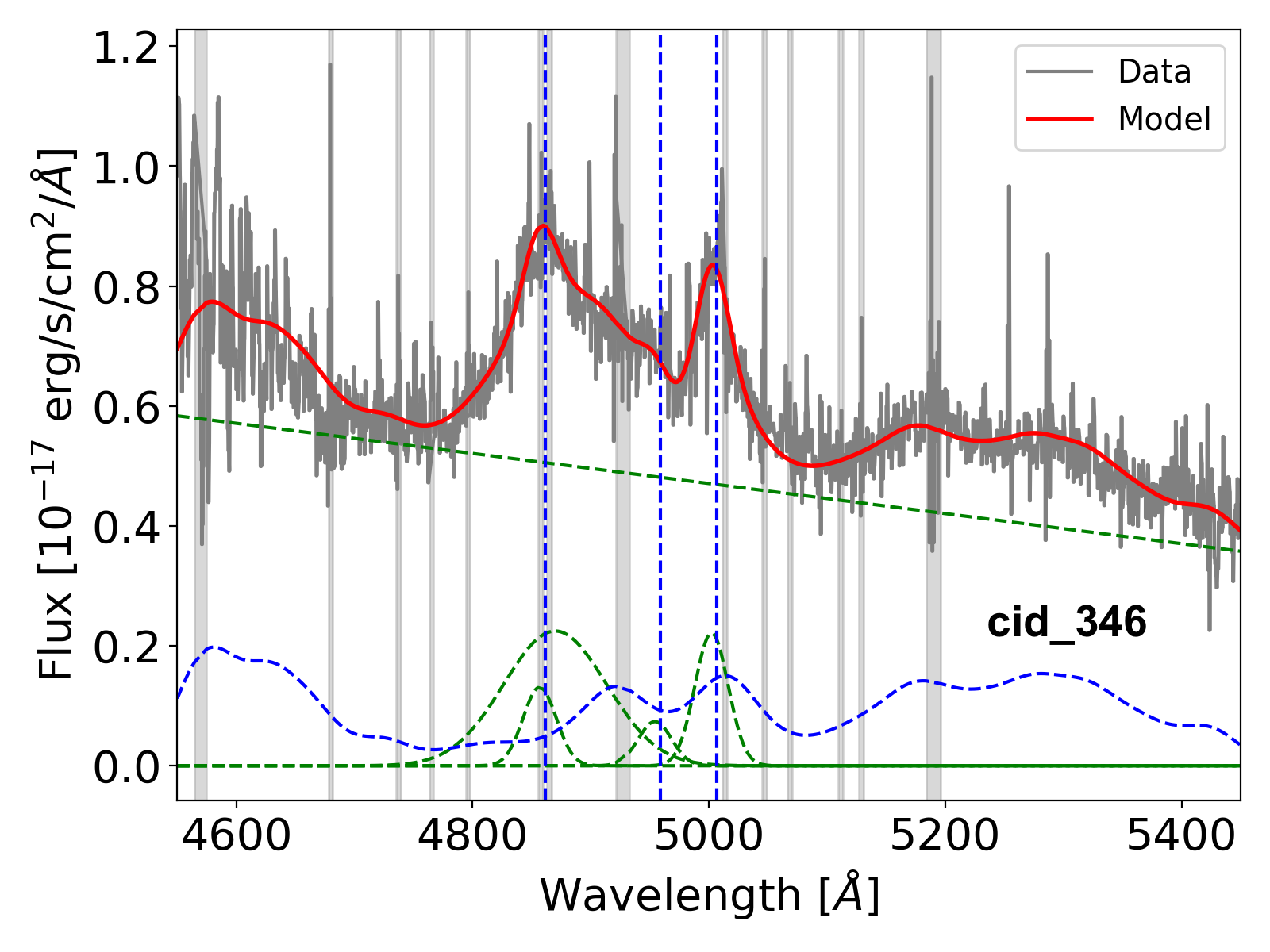}}\\
\caption{Same as Fig. \ref{fig:intspec_alltargets1} for X\_N\_4\_48, X\_N\_102\_35, X\_N\_115\_23, cid\_166, cid\_1605 and cid\_346. \label{fig:intspec_alltargets2}}
\end{figure*}

\begin{figure*}
\centering
\subfloat{\includegraphics[scale=0.45]{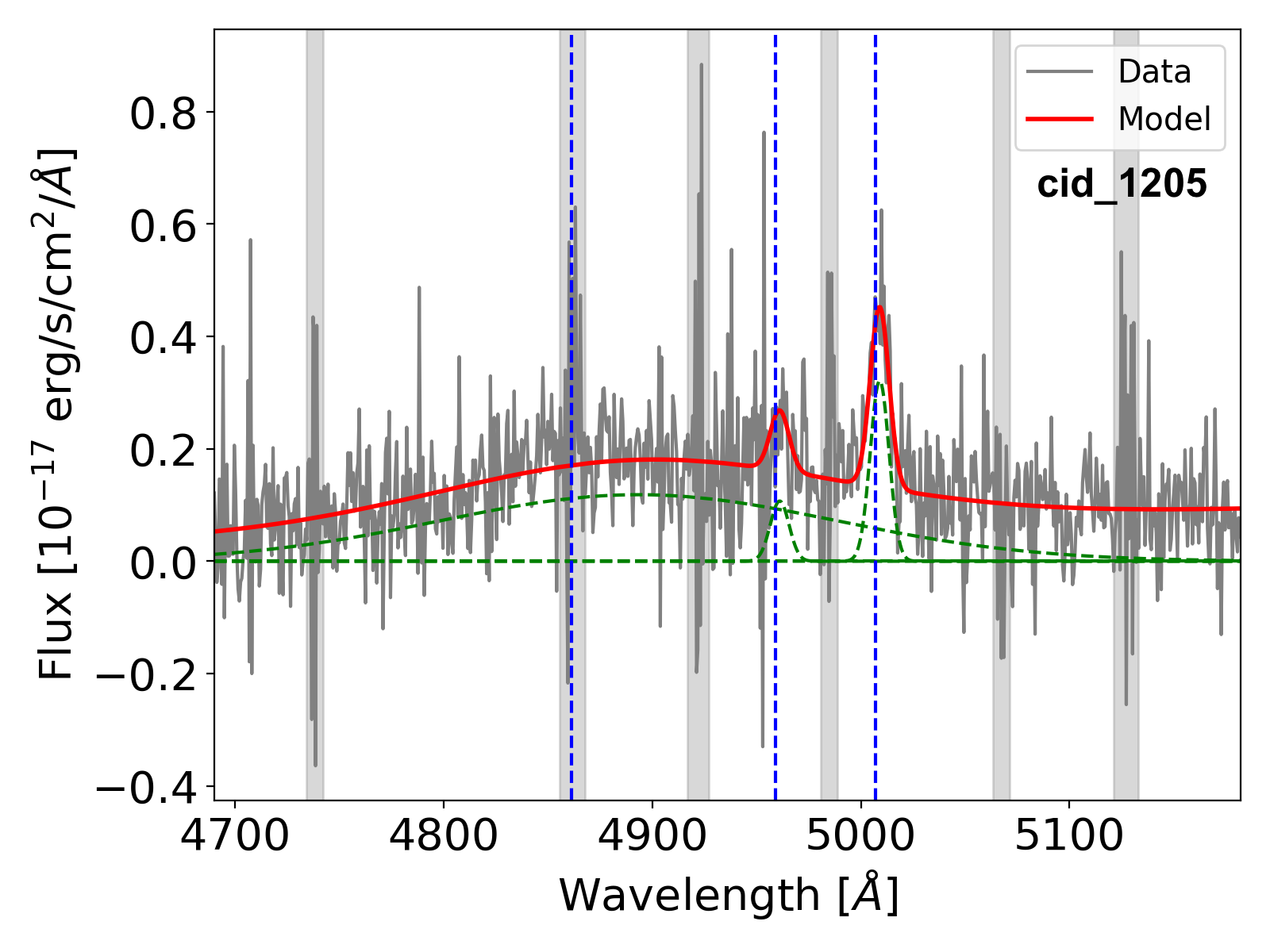}}
\subfloat{\includegraphics[scale=0.45]{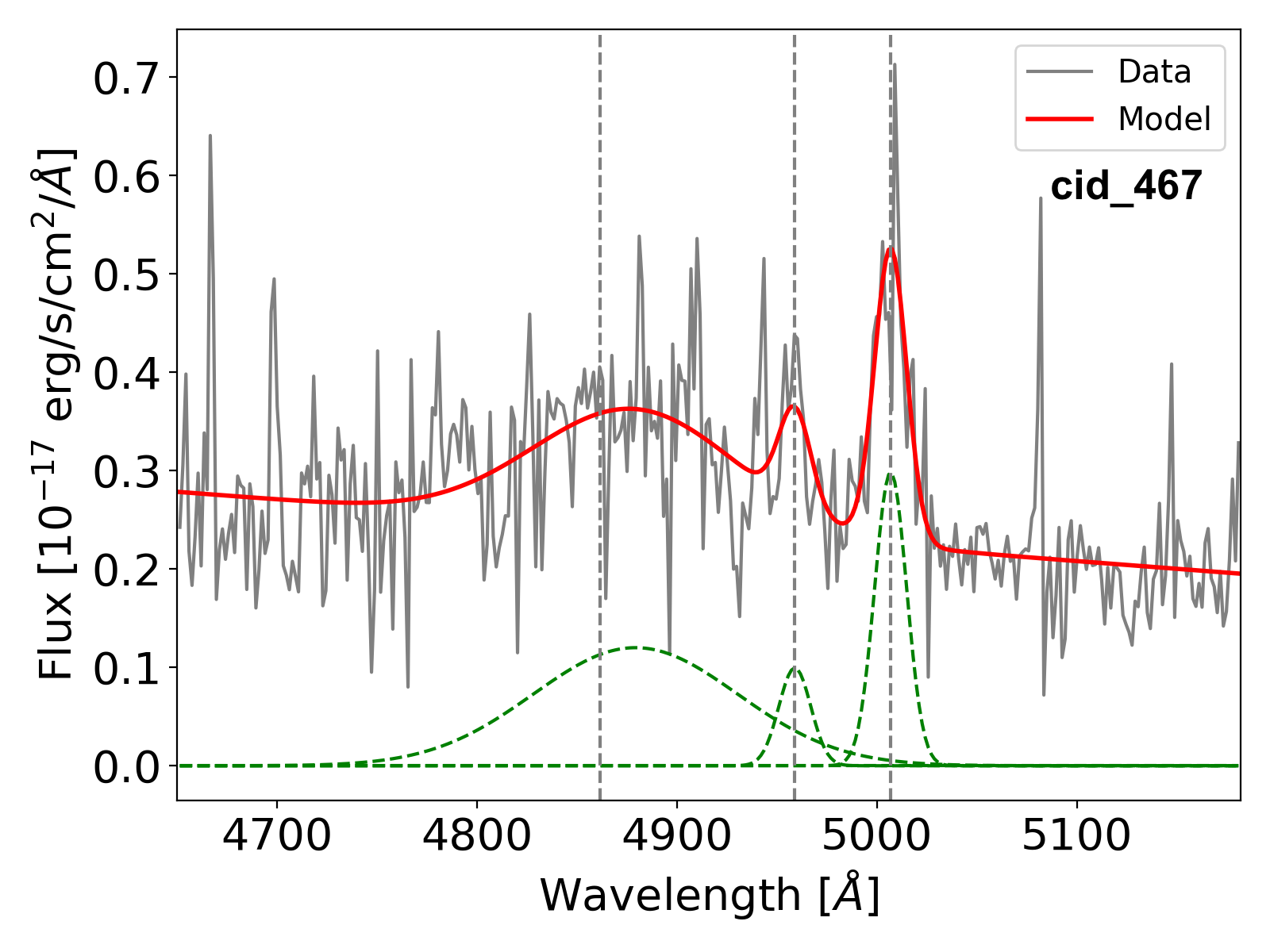}}\\
\subfloat{\includegraphics[scale=0.45]{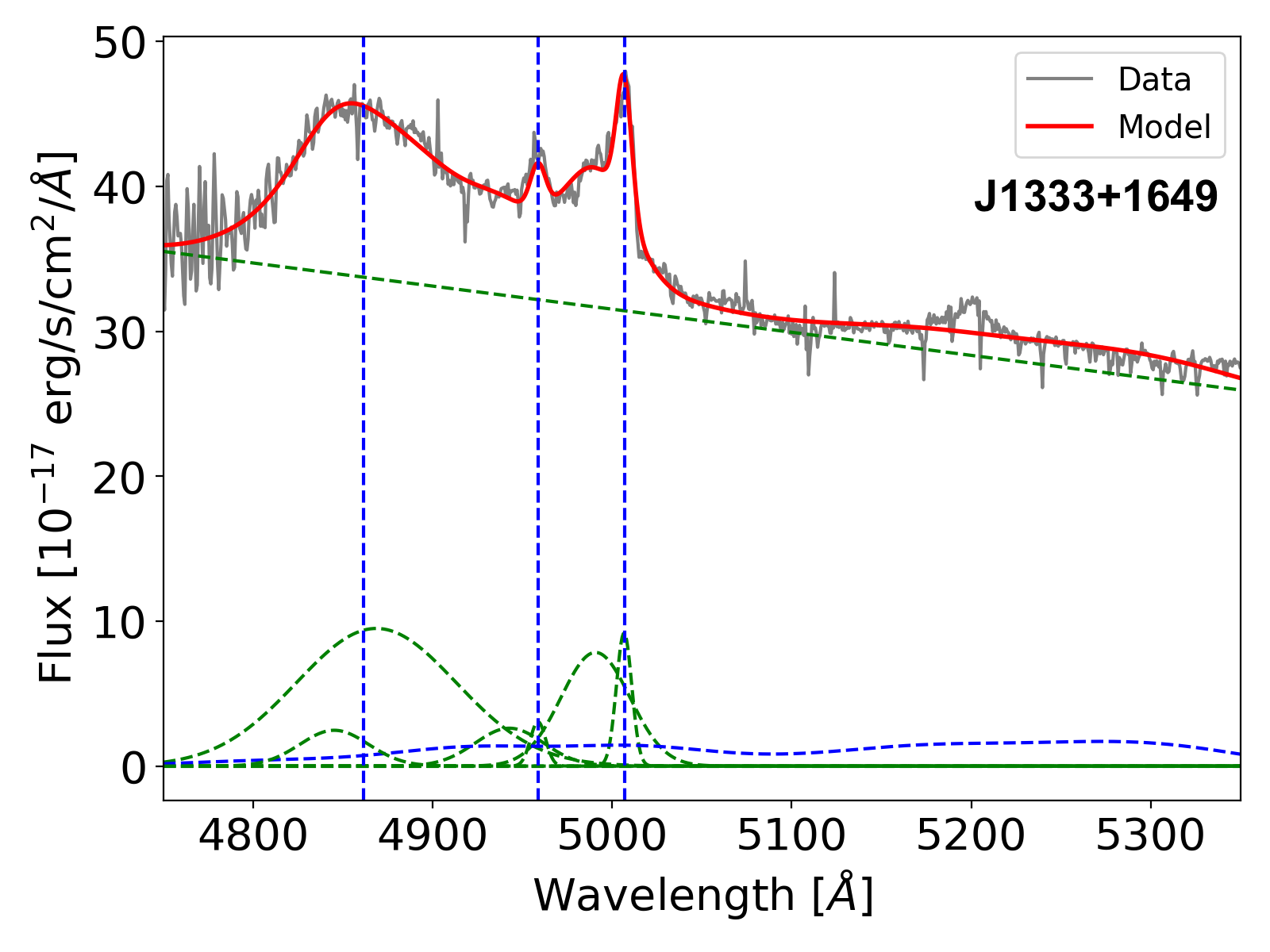}}
\subfloat{\includegraphics[scale=0.45]{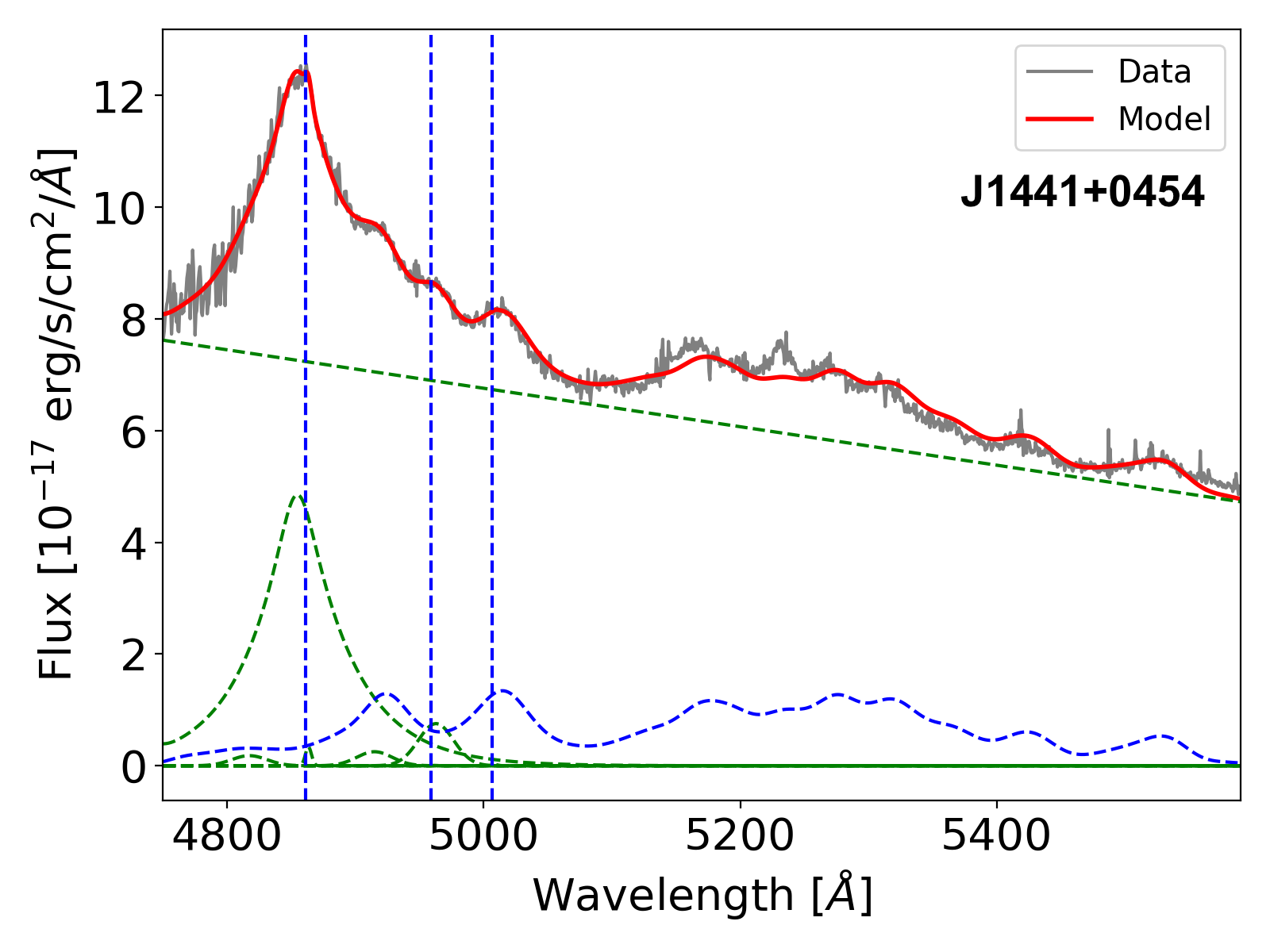}}\\
\subfloat{\includegraphics[scale=0.45]{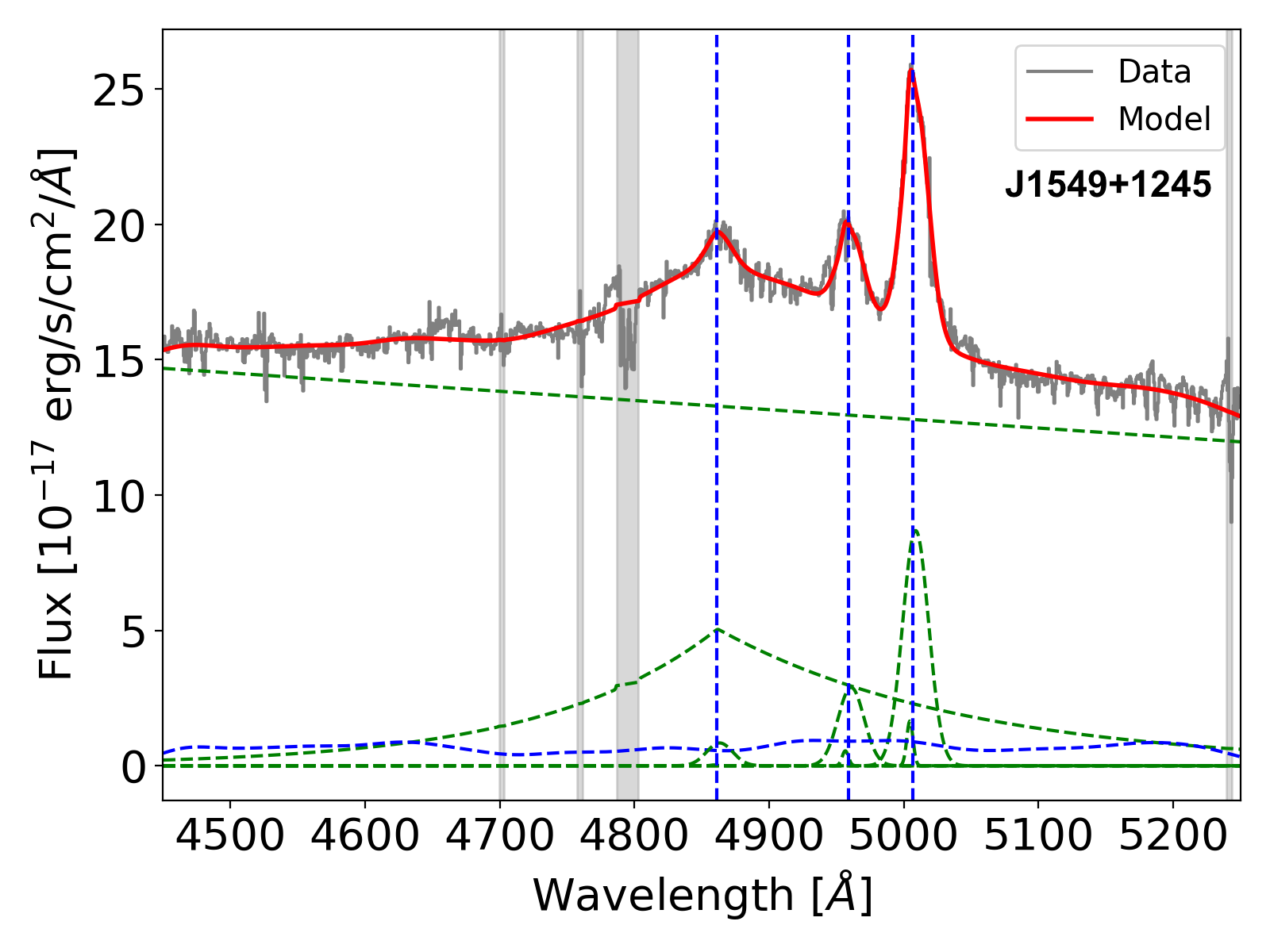}}
\subfloat{\includegraphics[scale=0.45]{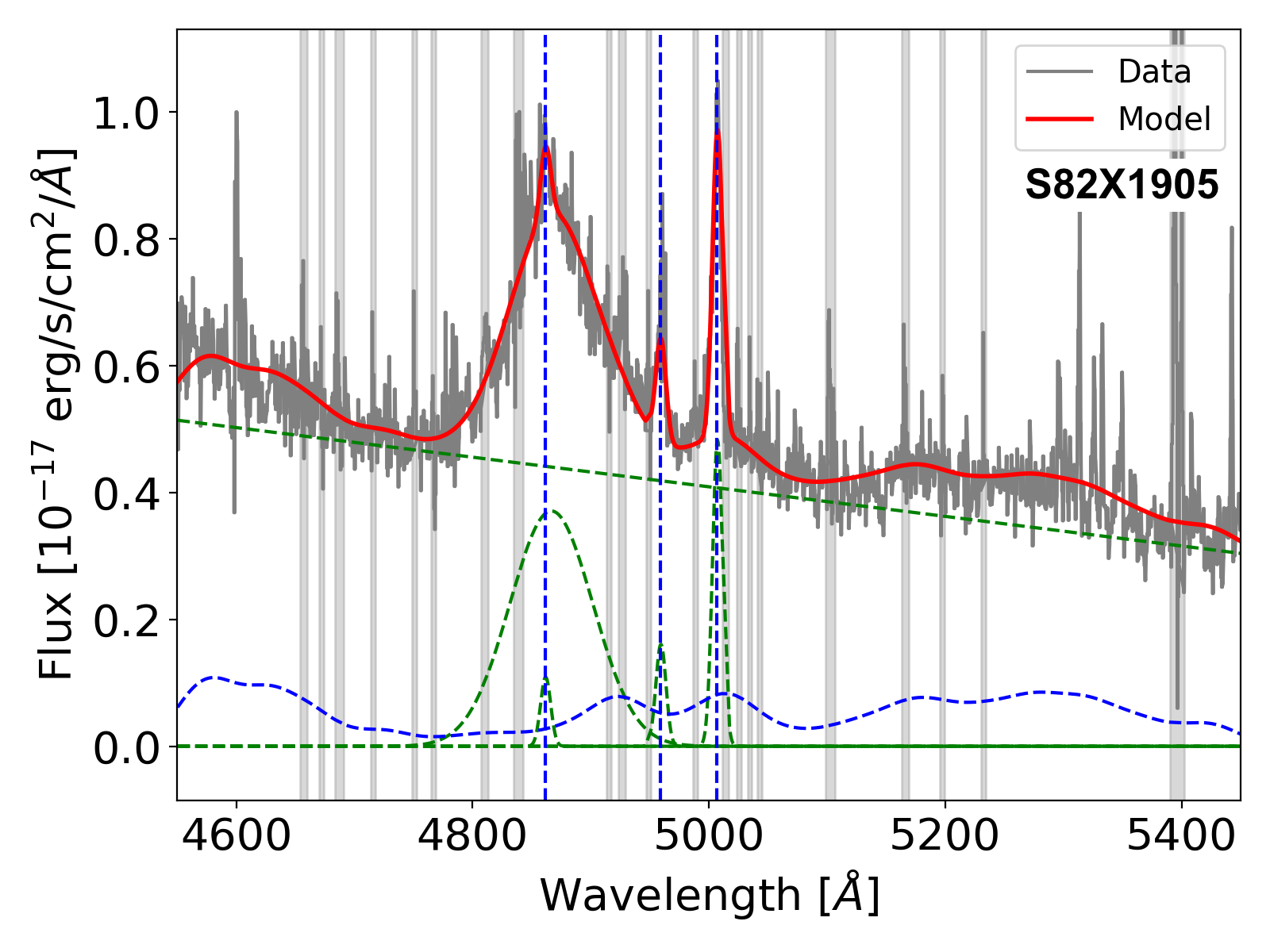}}\\
\caption{Same as Fig. \ref{fig:intspec_alltargets1} for cid\_1205, cid\_467, J1333+1649, J1441+0454, J1549+1245 and S82X1905. \label{fig:intspec_alltargets3}}
\end{figure*}

\begin{figure*}
\centering
\subfloat{\includegraphics[scale=0.45]{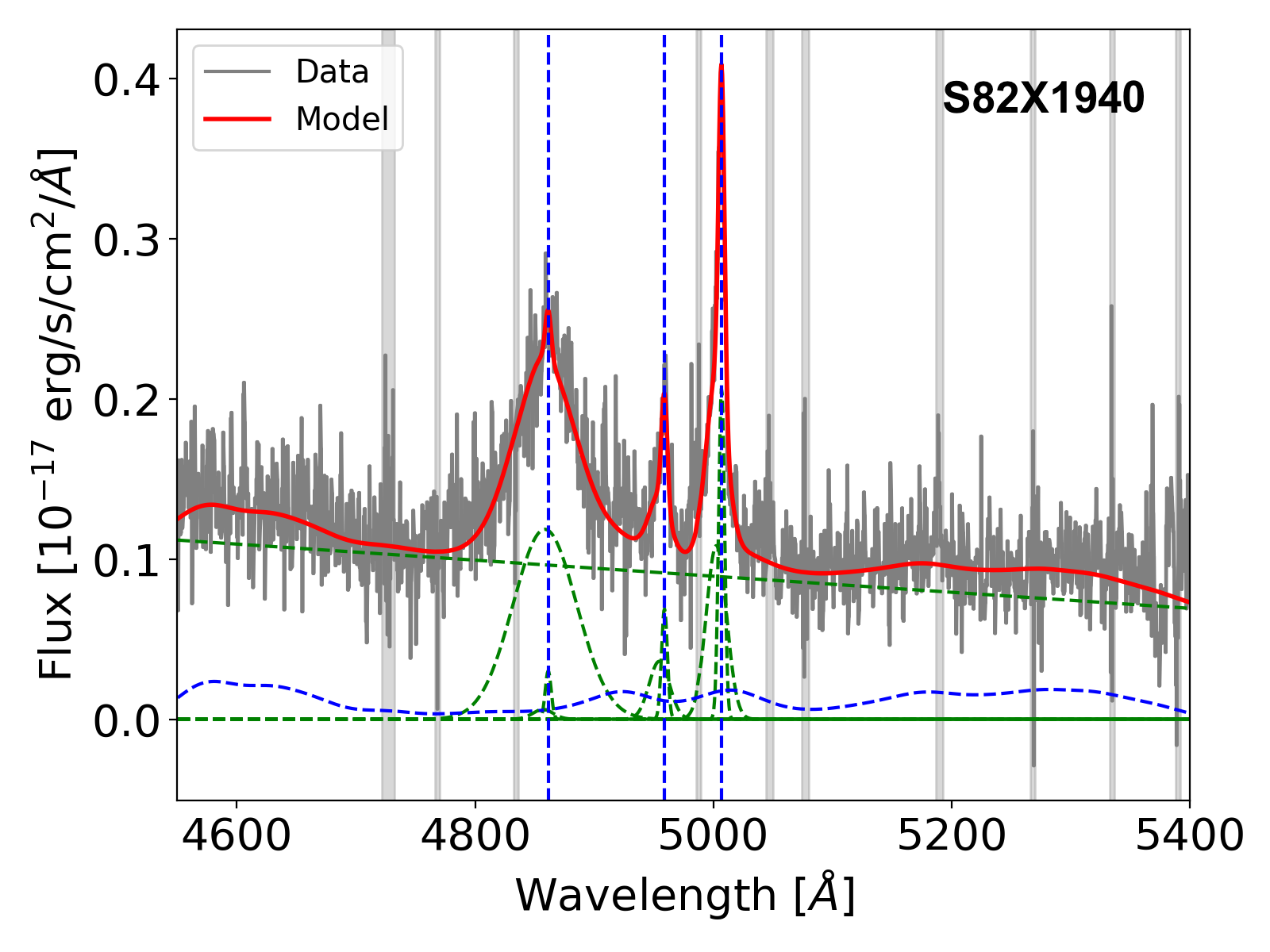}}\\
\subfloat{\includegraphics[scale=0.45]{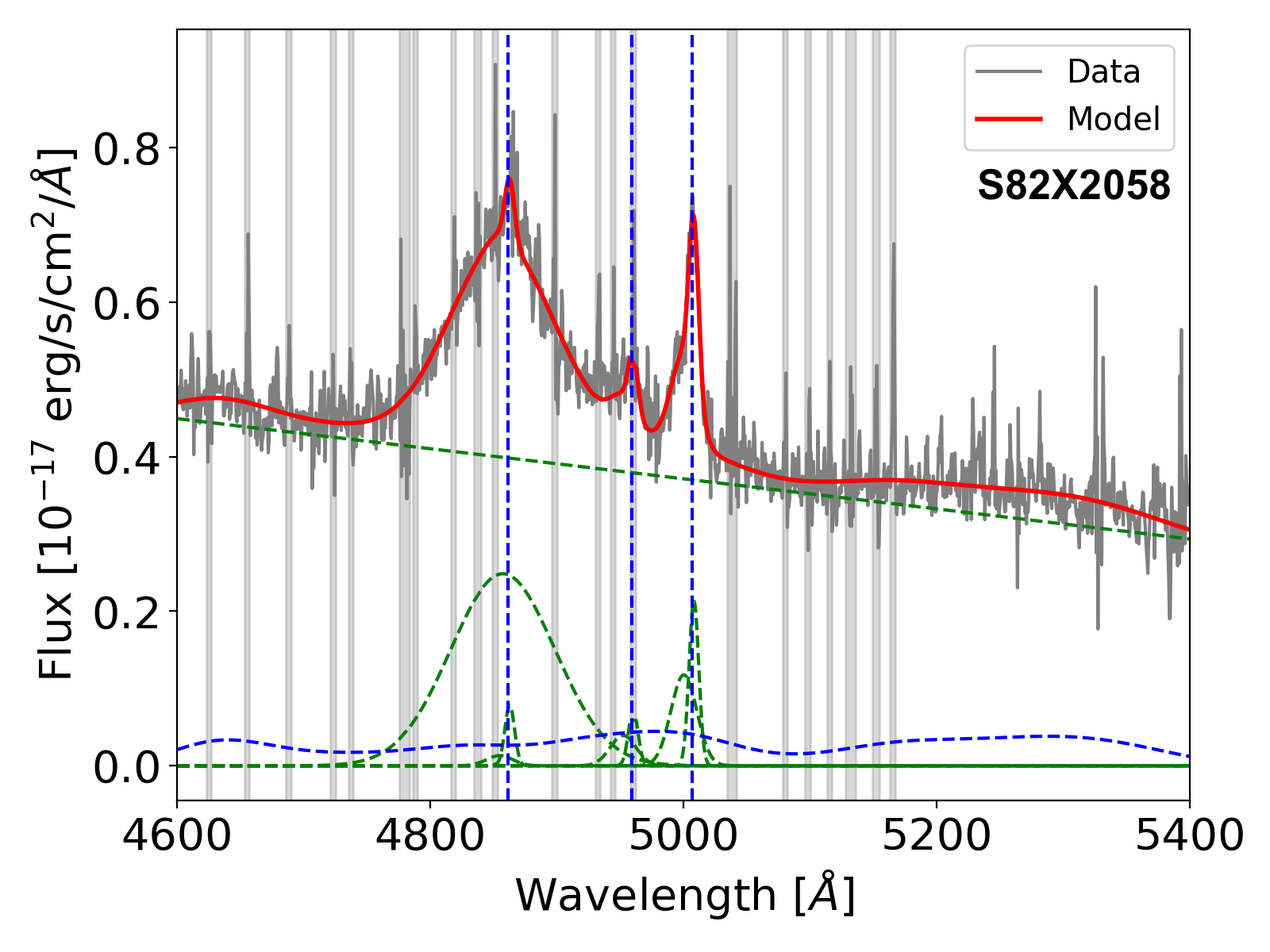}}
\caption{Same as Fig. \ref{fig:intspec_alltargets1} for S82X1940 and S82X2058. \label{fig:intspec_alltargets4}}
\end{figure*}

\begin{figure*}
\centering
\subfloat{\includegraphics[scale=0.7]{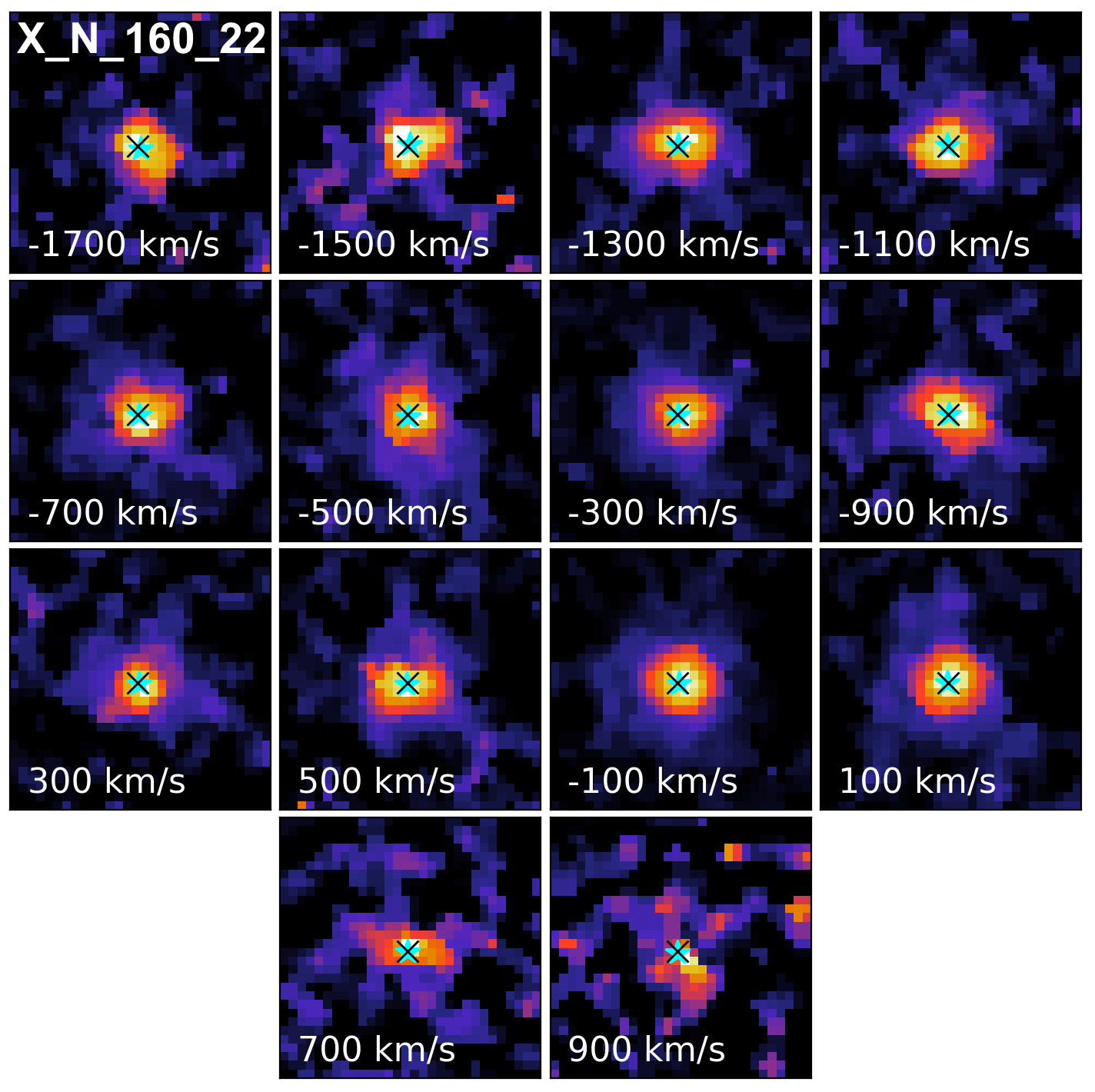}}
\caption{1.5$\arcsec$ x 1.5$\arcsec$ \oiii$\lambda$5007 channel maps of the SUPER target X\_N\_160\_22 at different velocity slices, after subtracting the $\hb$, \oiii$\lambda$4959 and iron models from the raw cube. Each velocity slice is 200 km/s wide and the displayed value is the center velocity of the respective channel. Black cross marks the location of the H-band continuum peak, used as a proxy for the AGN position. North is up and East is to left. \label{fig:all_spectroastrometry_images1}}
\end{figure*}

\begin{figure*}
\centering
\subfloat{\includegraphics[scale=0.5]{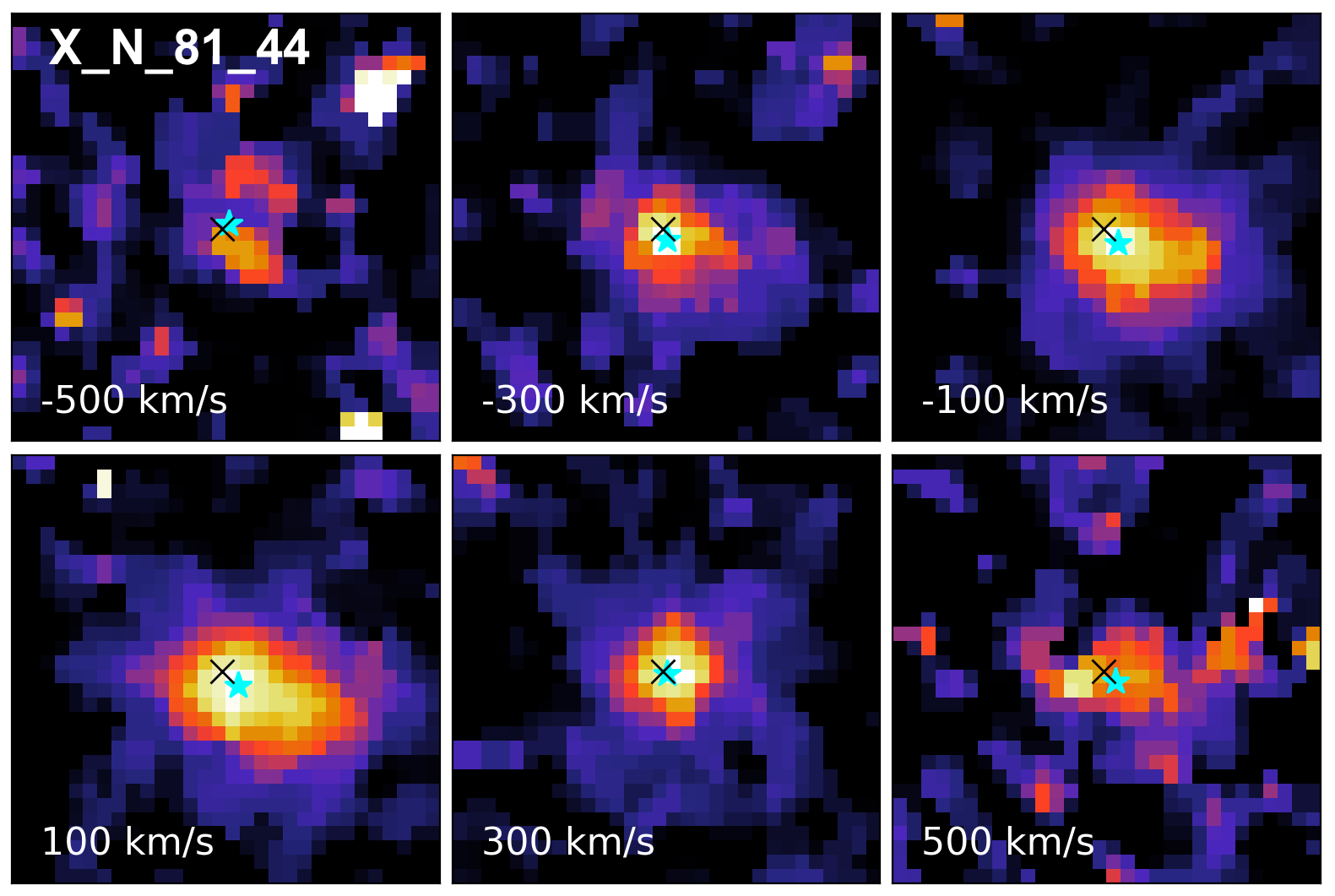}}\\
\subfloat{\includegraphics[scale=0.7]{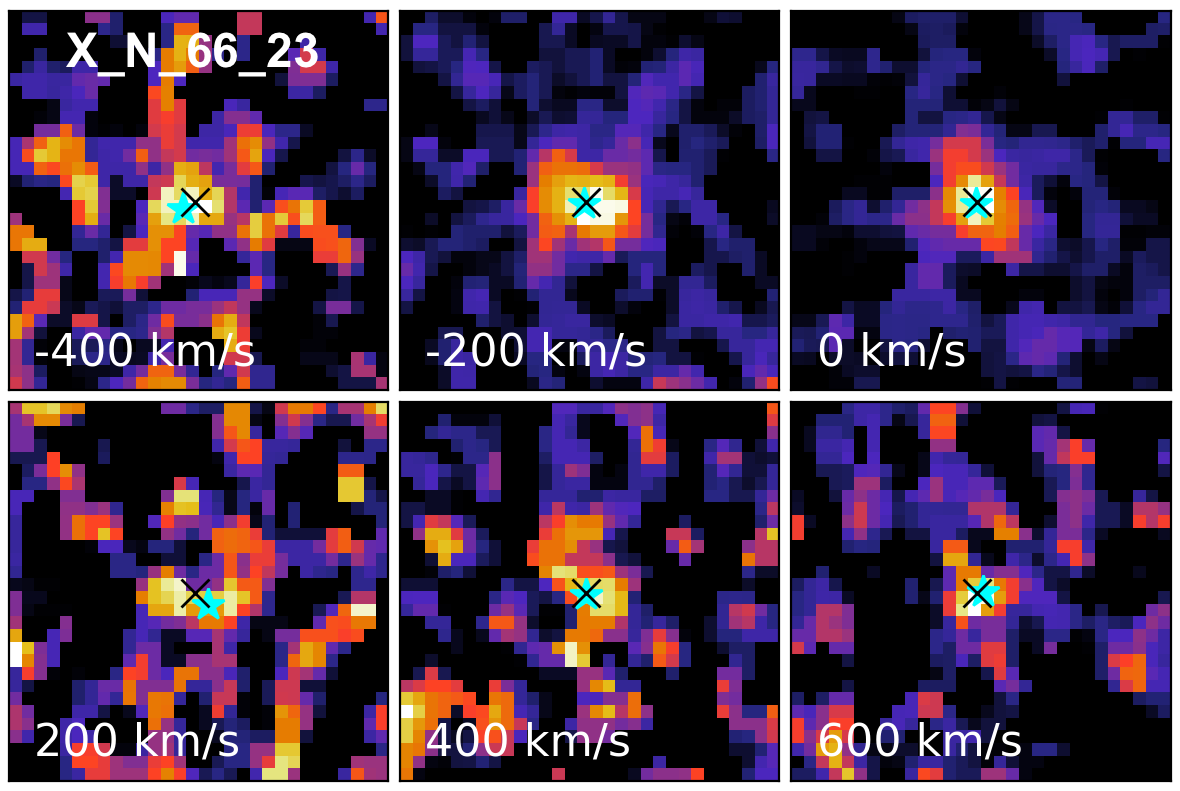}}\\
\subfloat{\includegraphics[scale=0.7]{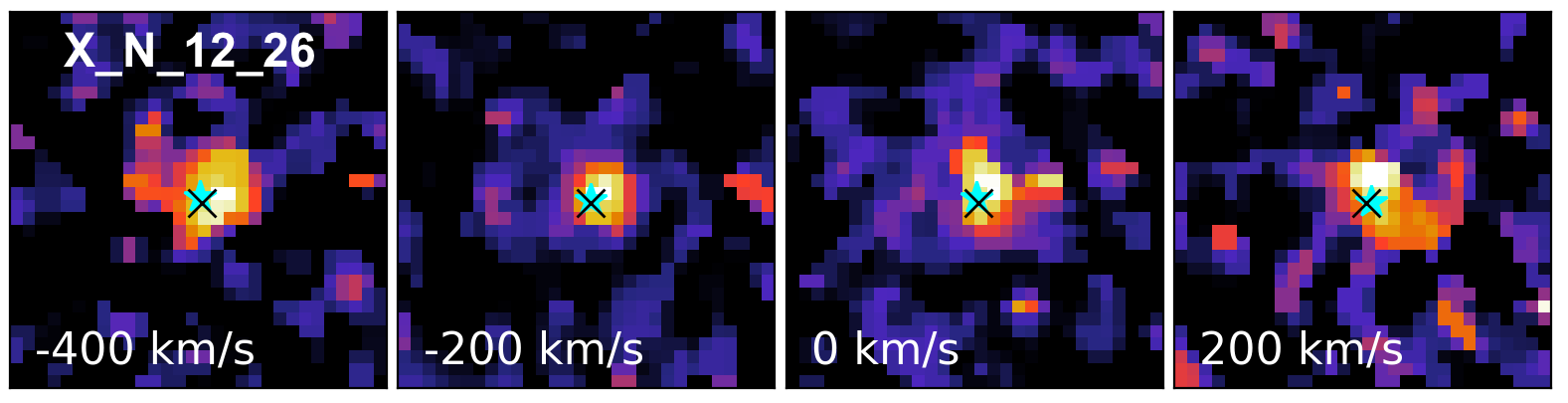}}\\
\caption{Same as Fig. \ref{fig:all_spectroastrometry_images1} for X\_N\_81\_44, X\_N\_66\_23 and X\_N\_12\_26.  \label{fig:all_spectroastrometry_images2}}
\end{figure*}

\begin{figure*}
\centering
\subfloat{\includegraphics[scale=0.7]{spectroastrometry_images_X_N_115_23_updated.png}}\\
\subfloat{\includegraphics[scale=0.7]{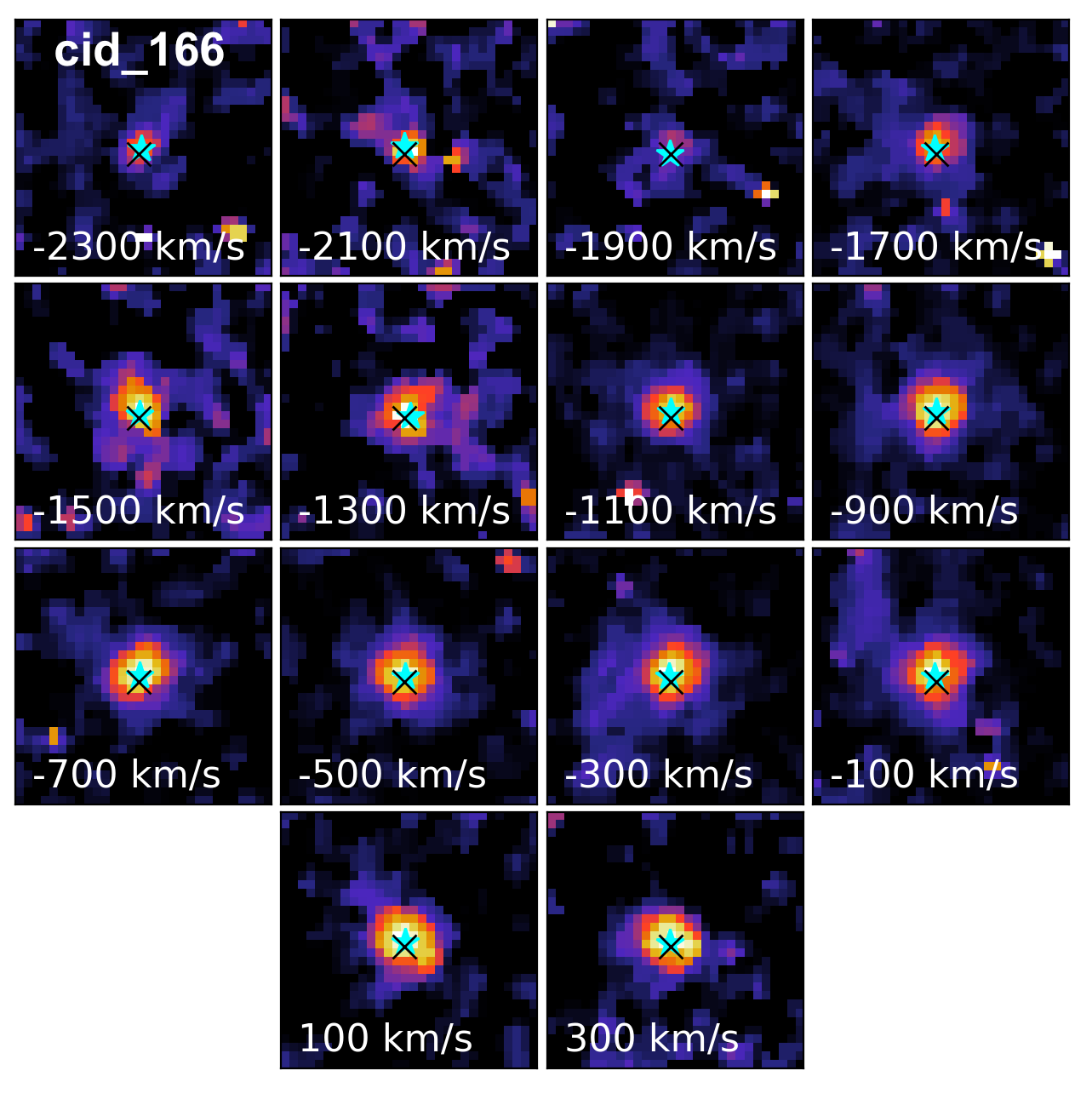}}\\
\caption{Same as Fig. \ref{fig:all_spectroastrometry_images1} for X\_N\_115\_23 and cid\_166.  \label{fig:all_spectroastrometry_images3}}
\end{figure*}

\begin{figure*}
\centering
\subfloat{\includegraphics[scale=0.7]{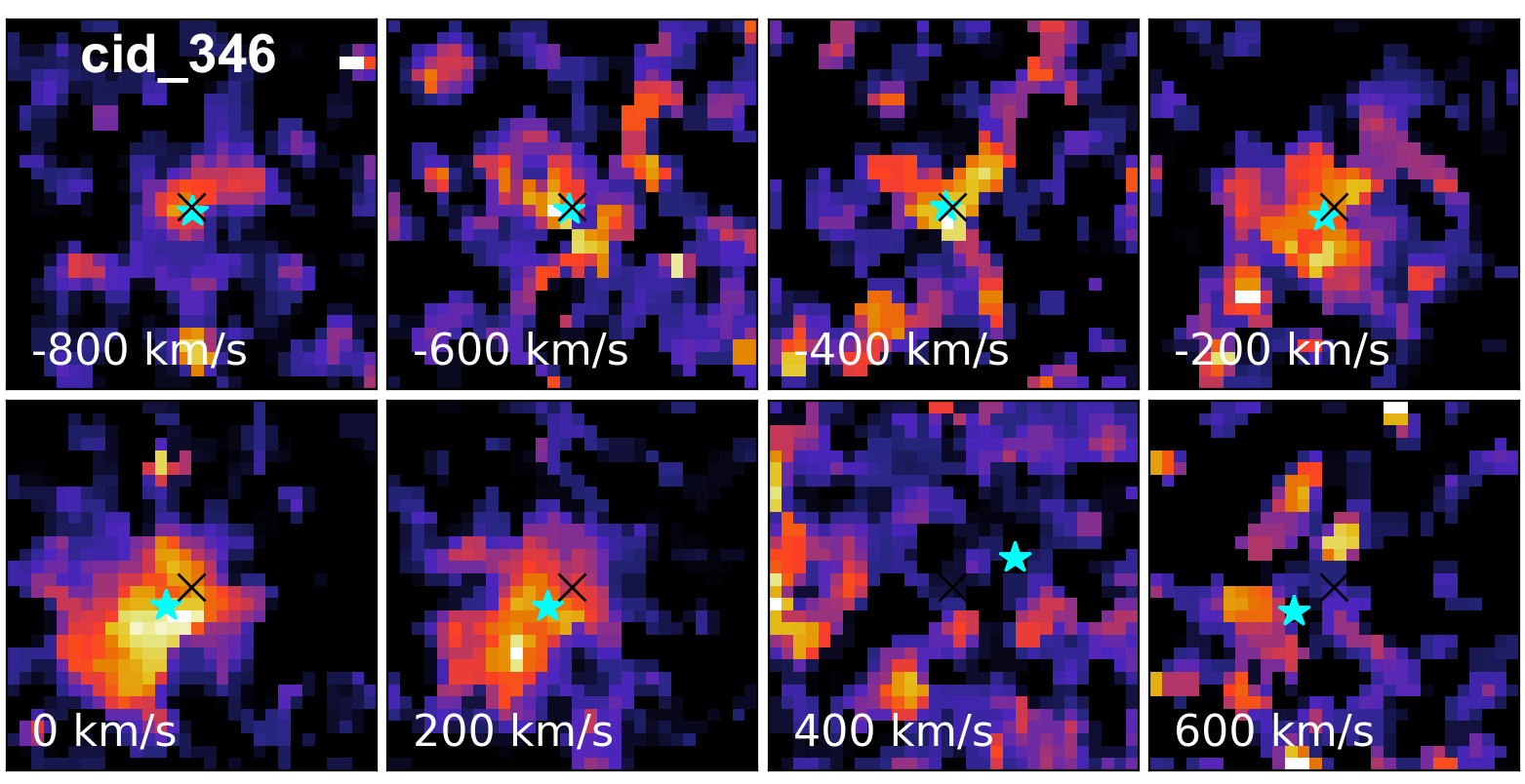}}\\
\subfloat{\includegraphics[scale=0.75]{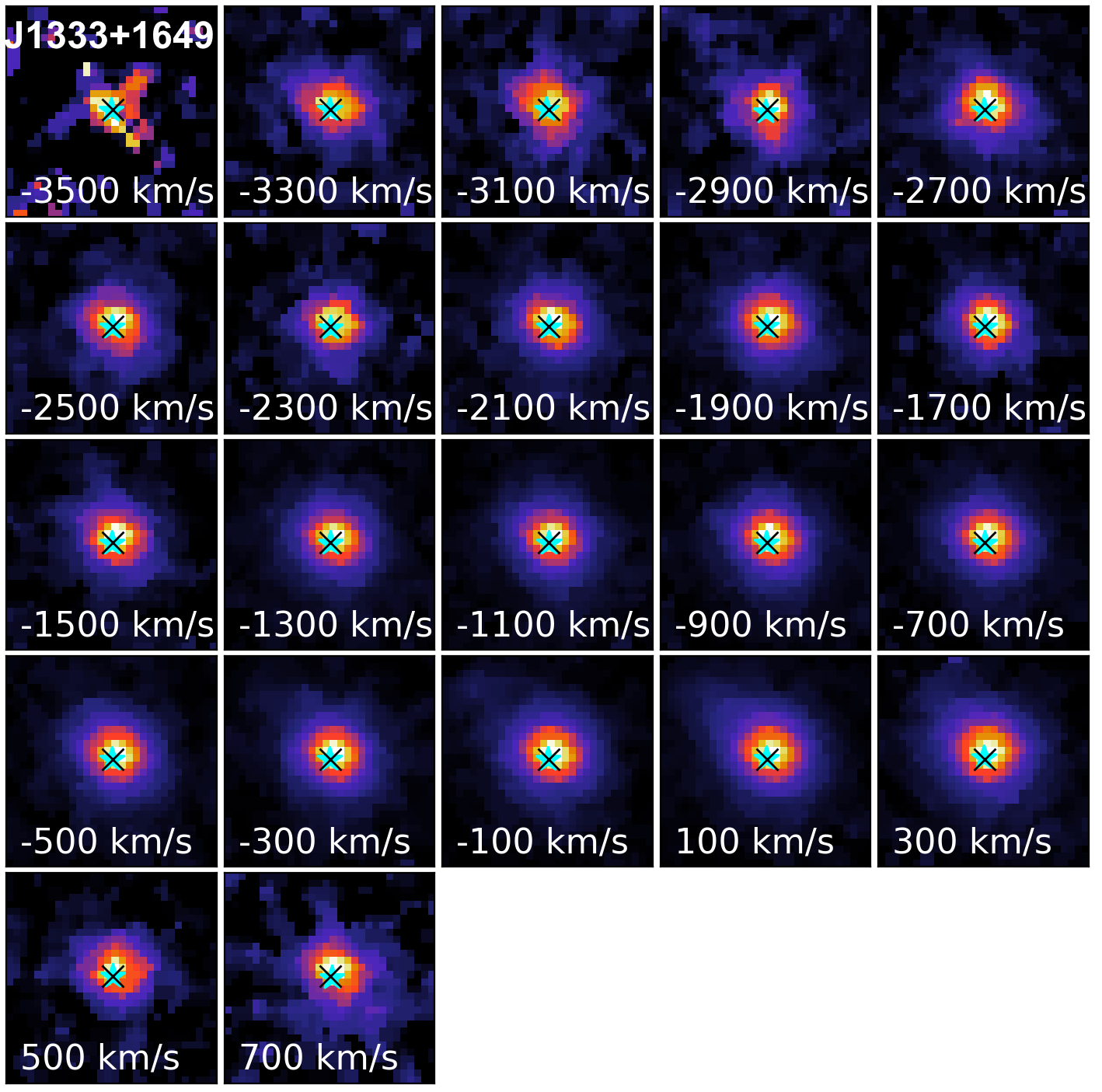}}\\
\caption{Same as Fig. \ref{fig:all_spectroastrometry_images1} for cid\_346 and J1333+1649.  \label{fig:all_spectroastrometry_images4}}
\end{figure*}

\begin{figure*}
\centering
\subfloat{\includegraphics[scale=0.6]{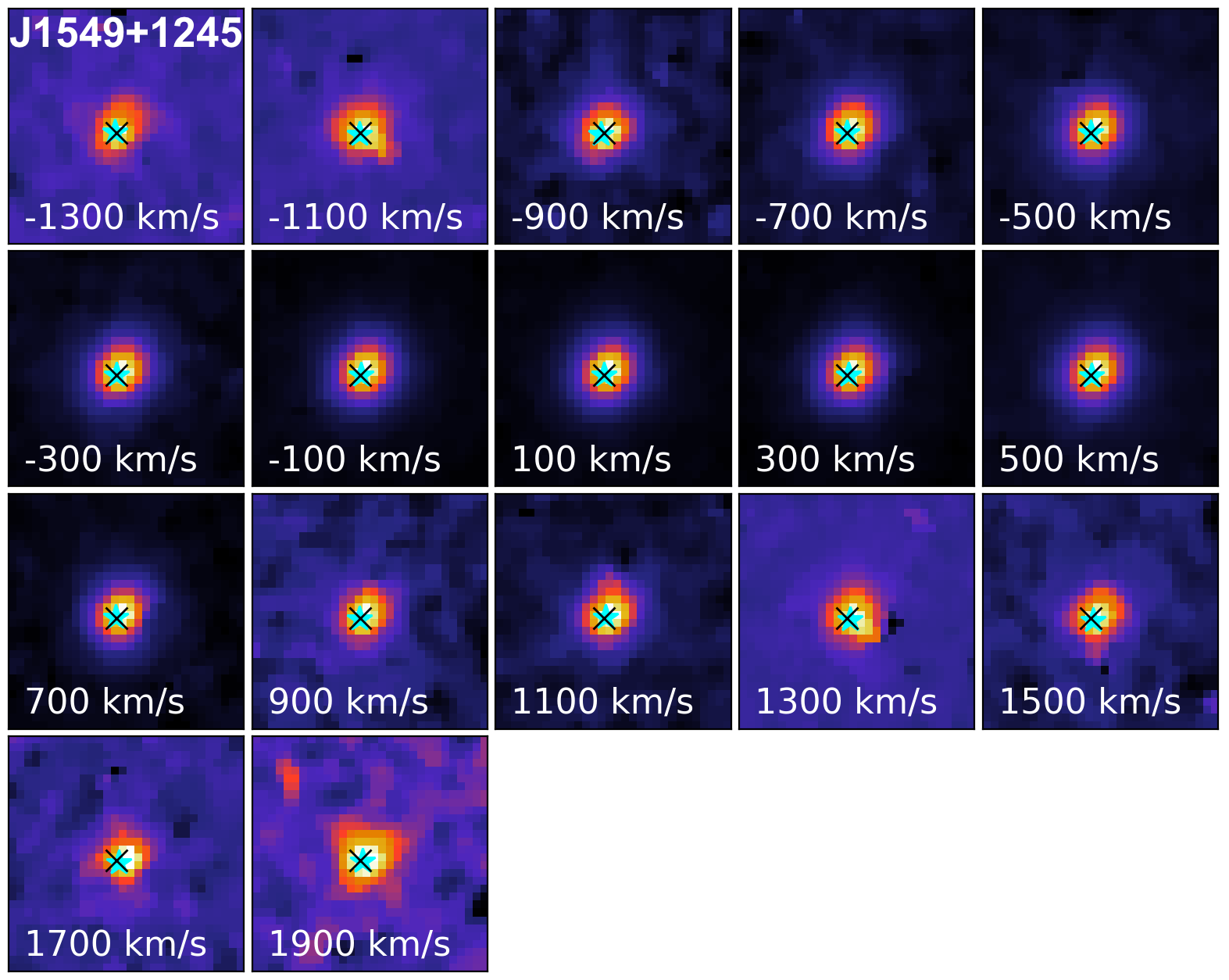}}\\
\subfloat{\includegraphics[scale=0.7]{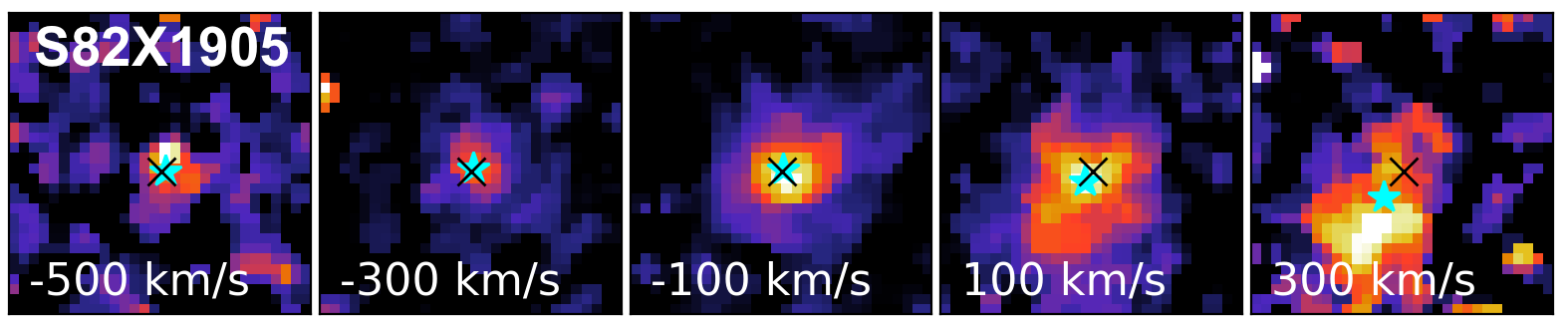}}\\
\subfloat{\includegraphics[scale=0.5]{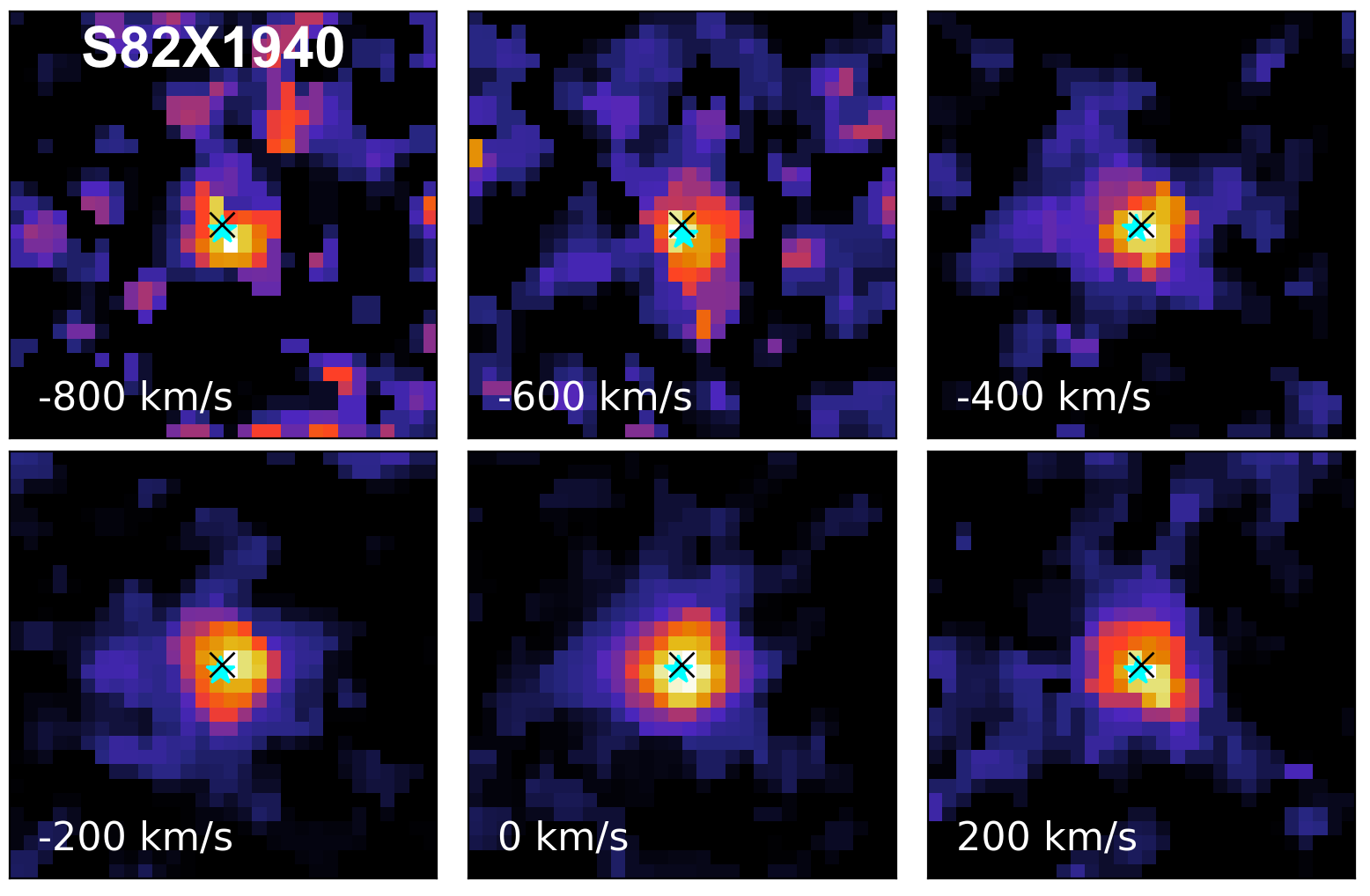}}\\
\caption{Same as Fig. \ref{fig:all_spectroastrometry_images1} for J1549+1245, S82X1905 and S82X1940.  \label{fig:all_spectroastrometry_images5}}
\end{figure*}

\end{appendix}
\end{document}